\documentclass[aps,prc,twoside,twocolumn,nofootinbib,10pt,showpacs,floatfix]{revtex4-1}
\usepackage{amsmath,amssymb}
\usepackage{graphicx,bm}
\usepackage{slashed}
\usepackage{epstopdf}
\usepackage{ulem} 
\usepackage[usenames]{color}
\usepackage{float}
\usepackage{hyperref}
\usepackage{subfigure}
\usepackage{rotating}
\usepackage{color}
\usepackage{multirow}
\usepackage{dcolumn}
\usepackage{overpic}
\usepackage{booktabs}
\usepackage{makecell}
\usepackage{diagbox}

\usepackage{booktabs}
\usepackage{colortbl}
\usepackage{xcolor}
\definecolor{tabcolor1}{rgb}{.105,.410,.113}
\definecolor{tabcolor2}{rgb}{.425,.130,.303}
\definecolor{tabcolor3}{rgb}{1.00,.0,.0}

\usepackage{array}
\newcolumntype{|}{!{\vline}}

\renewcommand\sout{\bgroup \color{red} \ULdepth=-.5ex \ULset}

\newsavebox{\tablebox}
\begin{document}
\title{Spectroscopy behavior of fully heavy tetraquarks}
\author{Hong-Tao An$^{1,2}$}\email{anht14@lzu.edu.cn}
\author{Si-Qiang Luo$^{1,2}$}\email{luosq15@lzu.edu.cn}
\author{Zhan-Wei Liu$^{1,2,3}$}\email{liuzhanwei@lzu.edu.cn}
\author{Xiang Liu$^{1,2,3}$}\email{xiangliu@lzu.edu.cn}
\affiliation{
$^1$School of Physical Science and Technology, Lanzhou University, Lanzhou 730000, China\\
$^2$Research Center for Hadron and CSR Physics, Lanzhou University and Institute of Modern Physics of CAS, Lanzhou 730000, China\\
$^3$Lanzhou Center for Theoretical Physics, Lanzhou University, Lanzhou, Gansu 730000, China}

\date{\today}
\begin{abstract}
Stimulated by the observation of the $X(6900)$ from LHCb in 2020 and the recent results from CMS and ATLAS in the di-$J/\psi$ invariant mass spectrum,
in this work we systematically study all possible configurations for the ground fully heavy tetraquark states in the constituent quark model.
By our calculation, we present their spectroscopy behaviors such as binding energy, lowest meson-meson thresholds, specific wave function, magnetic moment, transition magnetic moment, radiative decay width, rearrangement strong width ratio, internal mass contributions, relative lengths between (anti)quarks, and the spatial distribution of four valence (anti)quarks.
We cannot find a stable S-wave state for the fully heavy tetraquark system.
We hope that our results will be valuable to further experimental exploration of fully heavy tetraquark states.
\end{abstract}
\maketitle

\section{Introduction}\label{sec1}
At the birth of the quark model \cite{Gell-Mann:1964ewy,Zweig:1964ruk,Zweig:1964jf}, exotic states beyond conventional hadrons were proposed.
The search for exotic hadronic states is full of challenges and opportunities.
Since the $X(3872)$ was first reported by the Belle Collaboration in 2003 \cite{Choi:2003ue,Acosta:2003zx,Abazov:2004kp},
a series of charmonium-like or bottomonium-like exotic states \cite{Swanson:2006st,Zhu:2007wz,Voloshin:2007dx,Drenska:2010kg,Esposito:2014rxa,Chen:2016qju,Chen:2016heh,
Hosaka:2016pey,Richard:2016eis,Lebed:2016hpi,Esposito:2016noz,Ali:2017jda,Brambilla:2019esw,Liu:2019zoy,Chen:2022asf} and $P_{c}$ states \cite{Aaij:2015tga,Aaij:2016phn,Aaij:2019vzc} have been observed experimentally, which stimulates
extensive discussions of their properties by introducing the assignments of conventional hadron, compact multiquark states, molecular state, hybrid, glueball, and kinematic effects \cite{Swanson:2006st,Zhu:2007wz,Voloshin:2007dx,Drenska:2010kg,Esposito:2014rxa,Chen:2016qju,Chen:2016heh,
Hosaka:2016pey,Richard:2016eis,Lebed:2016hpi,Esposito:2016noz,Ali:2017jda,Brambilla:2019esw,Liu:2019zoy,Chen:2022asf}.

In 2003, the BaBar Collaboration observed a narrow heavy-light state $D^{*}_{s0}(2317)$ in the $D_s^+\pi^0$ invariant mass spectrum \cite{Aubert:2003fg}.
However, since the mass of the observed $D_{s0}^*(2317)$ is about 100 MeV below the quark model predictions in Ref.\cite{Godfrey:1985xj}, it is difficult to understand the $D_{s0}^*(2317)$ in conventional quark model directly, which is the ``low-mass puzzle" of $D_{s0}^*(2317)$. In order to solve the low-mass puzzle,
the tetraquark explanation with the $Qq\bar{q}\bar{q}$ configuration was proposed in Refs. \cite{Barnes:2003dj,Cheng:2003kg,Szczepaniak:2003vy}.
Later, the CLEO Collaboration \cite{Besson:2003cp} confirmed the $D^{*}_{s0}(2317)$ and announced another narrow resonance $D_{s1}(2460)$ in the $D_s^{*+}\pi^0$ final states. Low-mass puzzle also happens to the $D_{s1}(2460)$ \cite{Godfrey:1985xj}. The discussion of the $D_{s1}(2460)$ as tetraquark state can be found in Refs. \cite{Dmitrasinovic:2004cu,Chen:2004dy,Liu:2004kd,Faessler:2007us,Ebert:2010af,Xiao:2016hoa,Kong:2021ohg,Fu:2021wde}.
In particular, the LHCb Collaboration reported the discovery of two new exotic structures $X_0(2900)$ and $X_1(2900)$ \cite{Aaij:2020hon,LHCb:2020pxc}, which inspired the study of exotic charmed tetraquarks
\cite{Hu:2020mxp,Karliner:2020vsi,Liu:2020orv,Wang:2020prk,Xue:2020vtq,Burns:2020epm,Huang:2020ptc,Wang:2020xyc,Agaev:2021knl,Ozdem:2022ydv}.
In addition, the theorists also began to study the doubly charmed tetraquark states early
\cite{DelFabbro:2004ta,Ebert:2007rn,Hyodo:2012pm,Ikeda:2013vwa,Esposito:2013fma}.
In 2017, the LHCb Collaboration observed a doubly charmed baryon $\Xi_{cc}^{++}(3620)$ in the $\Lambda_{c}^{+}K^{-}\pi^{+}\pi^{+}$ decay mode \cite{Aaij:2017ueg}. Using the $\Xi_{cc}^{++}(3620)$ as the scaling point,
the theorists further explored the possible stable doubly charmed tetraquark states with the $QQ\bar{q}\bar{q}$ configuration \cite{Luo:2017eub,Karliner:2017qjm,Eichten:2017ffp,Ali:2018xfq,Park:2018wjk,Guo:2021yws,
Lu:2020rog,Noh:2021lqs,Park:2013fda,Deng:2018kly}.
Surprisingly, as a candidate of doubly charmed tetraquark, the $T^{+}_{cc}$ was detected by LHCb in the
$D^{0}D^{0}\pi^{+}$ invariant mass spectrum, which has a minimal quark configuration of $cc\bar{u}\bar{d}$ \cite{LHCb:2021vvq}.
In addition to these singly and doubly charmed tetraquarks, there should be a triply charmed tetraquark. To our knowledge, the triply charmed tetraquark states with the $QQ\bar{Q}\bar{q}$ configuration have also been studied by various approaches \cite{Chen:2016ont,Xing:2019wil,Jiang:2017tdc,Lu:2021kut,Weng:2021ngd}.

Briefly reviewing the status of heavy flavor tetraquark states, we must mention the fully heavy  tetraquark with the $QQ\bar{Q}\bar{Q}$ configuration, which has attracted the attention of both theorists and experimentalists.
Chao et al. suggested that the peculiar resonance-like structures of $R(e^{+}e^{-}\rightarrow$ hadrons$)$ for $\sqrt{s}=6-7$ GeV may be due to the production of the predicted P-wave $(cc)$-$(\bar{c}\bar{c})$ states in the energy range of $6.4-6.8$ GeV, which could dominantly decay into charmed mesons \cite{Chao:1980dv}.
The calculation of the fully heavy  tetraquark was then carried out using the potential model \cite{Ader:1981db, Badalian:1985es} and the MIT bag model with the Born-Oppenheimer approximation \cite{Heller:1985cb}.
This system has also been studied in a non-relativistic potential model, where no $QQ\bar{Q}\bar{Q}$ bound state can be found \cite{Zouzou:1986qh}. However, Lloyd {\it et al.} adopted a parameterized non-relativistic Hamiltonian to study such system \cite{Lloyd:2003yc}, where they found several closely-lying bound states with a large oscillator basis.
Later, Karliner {\it et al.} estimated the masses of the fully heavy tetraquark states by a simple quark model, and obtained $\rm M(X_{cc\bar{c}\bar{c}})=6192\pm25$ MeV and $\rm M(X_{bb\bar{b}\bar{b}})=18826\pm25$ MeV for the fully charmed and fully bottom tetraquarks with the $J^{PC}=0^{++}$ quantum number, respectively \cite{Karliner:2016zzc}. Anwar {\it et al.} have calculated the ground-state energy of the $bb\bar{b}\bar{b}$ bound state in a nonrelativistic effective field theory with one-gluon-exchange (OGE) color Coulomb interaction,
and the ground state  $bb\bar{b}\bar{b}$ tetraquark mass is predicted to be ($18.72\pm0.02$) GeV \cite{Anwar:2017toa}.
In Ref. \cite{Bai:2016int}, Bai {\it et al.} presented a calculation of the $bb\bar{b}\bar{b}$ tetraquark ground-state energy using a diffusion Monte Carlo method to solve the non-relativistic many-body system. Debastiani {\it et al.} extended the updated Cornell model to study the fully charmed tetraquark in a diquark-antidiquark configuration \cite{Debastiani:2017xcr}.
Chen {\it et al.} used a moment QCD sum rule method to give the existence of the exotic states $cc\bar{c}\bar{c}$ and $bb\bar{b}\bar{b}$ in the compact diquark-antidiquark configuration, where they suggested to search for them in the $J/\psi J/\psi$ and $\eta_{c}(1S)\eta_{1S}$ channels \cite{Chen:2016jxd}.

With the accumulation of experimental data,
many collaborations have tried to search for it.
The CMS Collaboration reported the first observation of the $\Upsilon(1S)$ pair production in $pp$ collisions, where there is evidence for a structure around 18.4 GeV with a global significance of 3.6 $\sigma$ exists in the four-lepton channel, which is probably a fully-bottom tetraquark state \cite{CMS:2016liw}.
However, this structure was not confirmed by the later CMS analysis \cite{CMS:2020qwa}.
Subsequently, the LHCb Collaboration studied the $\Upsilon(1S)_{\mu^{+}\mu^{-}}$ invariant mass distribution to search for a possible $bb\bar{b}\bar{b}$ exotic meson, but they did not see any significant excess in the range $17.5-20.0$ GeV \cite{LHCb:2018uwm}.
By 2020, the LHCb Collaboration declared a narrow resonance $X(6900)$ in the di-$J/\psi$ mass spectrum with a significance of more than $5\sigma$ \cite{LHCb:2020bwg}.
In addition, a broad structure ranging from 6.2 to 6.8 GeV and an underlying peak near 7.3 GeV were also reported at the same time \cite{LHCb:2020bwg}.
Recently, the ATLAS and CMS collaborations published their measurements on the di-$J/\psi$ invariant mass spectrum. Here,
they not only confirmed the existence of the $X(6900)$, but also found some new peaks \cite{Xu:2022rnl,Zhang:2022toq,CMS:2023owd}.
There have been extensive discussions about the observed $X(6900)$ from different approaches and with different assignments \cite{Zhao:2020zjh,Zhao:2020nwy,Lu:2020cns,liu:2020eha,Chen:2020xwe,Bedolla:2019zwg,Liu:2021rtn,Faustov:2020qfm,Karliner:2020dta,Ke:2021iyh,Wan:2020fsk,Gong:2020bmg,Zhu:2020snb,Wang:2020wrp,Wang:2022jmb,Santowsky:2021bhy,Zhuang:2021pci,Lotstedt:2023crt}.

The problem of the stability of the fully heavy tetraquark state has been debated for a long time.
Debastiani {\it et al.} found that the lowest S-wave $cc\bar{c}\bar{c}$ tetraquarks may be below the dicharmonium thresholds in their updated Cornell model \cite{Debastiani:2017msn}.
The $1^{+}$ $bb\bar{b}\bar{c}$ state is thought to be a narrow state in the extended chromomagnetic model \cite{Weng:2020jao}.
However, many other studies have suggested that the ground state of fully heavy tetraquarks is above the di-meson threshold.
Wang {\it et al.} also calculated the fully-heavy tetraquark state in two nonrelativistic quark models with different OGE Coulomb, linear confinement and hyperfine potentials \cite{Wang:2019rdo}. 
Based on the numerical calculations, they suggested that the ground states should be located about $300-450$ MeV above the lowest scattering states, indicating that there is no bound tetraquark state.
The lattice nonrelativistic QCD method was applied to study the lowest energy eigenstate of the $bb\bar{b}\bar{b}$ system, and no state was found below the lowest bottomonium-pair threshold \cite{Hughes:2017xie}.
Furthermore, Richard {\it et al.} also claimed that the fully heavy configuration $QQ\bar{Q}\bar{Q}$ is not stable if one adopts a standard quark model and treats the four-body problem appropriately \cite{Richard:2018yrm}.
Jin {\it et al.} studied full-charm and full-bottom tetraquarks using the quark delocalization color screening
model and the chiral quark model, respectively, and the results within the quantum numbers $J^{P}=0^{+}, 1^{+}$, and $2^{+}$ show that the bound state exists in both models \cite{Jin:2020jfc}.
Frankly speaking, theorists have not come to an agreement on the stability of the fully heavy tetraquark state.

Facing the present status of fully heavy tetraquark, in this work
we adopt the variational method to systematically study the fully heavy tetraquark states, where
the mass spectrum of the fully heavy tetraquark is given in the framework of nonrelativistic quark model associated with a potential containing Coulomb, linear, and hyperfine terms. The constructed total wave functions involved in these discussed systems satisfy the requirement of the Pauli principle. We should emphasize that
we can also reproduce the masses of these conventional hadrons with the same parameters, which is a test of our adopted framework.
With this preparation, we calculate the binding energies, the lowest meson-meson thresholds, and the rearrangement strong width ratio, and study the stability of the fully heavy tetraquark states against the decay into two meson states.
Furthermore, we discuss whether the discussed tetraquarks have a compact configuration based on the eigenvalue of the hyperfine potential matrix.
According to specific wave functions, we could obtain the magnetic moments, transition magnetic moments, and radiative decay widths, which may reflect their electromagnetic properties and internal structures.
We also give the the size of the tetraquarks, the relative distances between (anti)quarks, and the spatial distribution of the four valence (anti)quarks for each state.
Through the present systematic work, we can test whether compact bound fully heavy tetraquarks exist within the given Hamiltonian.

This paper is organized as follows. After the introduction,
we present the Hamiltonian of the constituent quark model and list the corresponding parameters in Section \ref{sec2}.
Then we give the spatial function with a simple Gaussian form and construct the flavor, color, and spin wave functions of the fully heavy tetraquark states (see Section \ref{sec3}).
In Section \ref{sec4},
we show the numerical results obtained by the variational method and further calculate their magnetic moment, transition magnetic moment, radiative decay width, rearrangement strong width ratio, the internal mass contributions and relative lengths between (anti) quarks.
In addition, a comparison of our results with those of other theoretical groups is made in Section \ref{sec5}.
Finally, the paper ends with a short summary in Section \ref{sec6}.

\section{Hamiltonian}\label{sec2}

We choose a nonrelativistic Hamiltonian for the fully heavy tetraquark system, which is written as,
\begin{eqnarray}\label{Eq1}
H=\sum_{i=1}^{4}\left(m_{i}+\frac{\textbf{p}^{2}_{i}}{2m_{i}}\right)-\frac{3}{4}\sum_{i<j}^{4}\frac{\lambda^{c}_{i}}{2}.\frac{\lambda^{c}_{j}}{2}(V^{Con}_{ij}+V^{SS}_{ij}).
\end{eqnarray}
Here, $m_{i}$ is the (anti)quark mass, $\lambda^{c}_{i}$ is the $SU(3)$ color operator for the $i$-th quark, and for the antiquark, $\lambda^{c}_{i}$ is replaced by $-\lambda^{c*}_{i}$.
The internal quark potentials $V^{Con}_{ij}$ and $V_{ij}^{SS}$ have the following forms:
\begin{eqnarray}\label{Eq2}
V^{Con}_{ij}&=&-\frac{\kappa}{r_{ij}}+\frac{r_{ij}}{a^{2}_{0}}-D,\nonumber\\
V^{SS}_{ij}&=&\frac{\kappa'}{m_{i}m_{j}}\frac{1}{r_{0ij}r_{ij}}e^{-r^{2}_{ij}/r^{2}_{0ij}}{\vec\sigma}_{i}\cdot{\vec\sigma}_{j},
\end{eqnarray}
where $r_{ij}=|\textbf{r}_{i}-\textbf{r}_{j}|$ is the distance between the $i$-th (anti)quark and the $j$-th (anti)quark, and
the $\sigma_{i}$ is the $SU(2)$ spin operator for the $i$-th quark.
As for the $r_{0ij}$ and $\kappa'$, we have

\begin{eqnarray}\label{Eq3}
r_{0ij}&=&1/\left(\alpha+\beta\frac{m_{i}m_{j}}{m_{i}+m_{j}}\right),\nonumber\\
\kappa'&=&\kappa_{0}\left(1+\gamma\frac{m_{i}m_{j}}{m_{i}+m_{j}}\right).
\end{eqnarray}

The corresponding parameters appearing in Eqs. (\ref{Eq2}-\ref{Eq3}) are shown in Table \ref{para}.
Here, $\kappa$ and $\kappa'$ are the couplings of the Coulomb and hyperfine potentials, respectively,
and they are proportional to the running coupling constant $\alpha_{s}(r)$ of QCD.
The Coulomb and hyperfine interactions can be deduced from the one-gluon-exchange model.
$1/a^{2}_{0}$ represents the strength of the linear potential.
$r_{0ij}$ is the Gaussian-smearing parameter.
Furthermore, we introduce $\kappa_{0}$ and $\gamma$ in $\kappa'$ to better describe the interaction between different quark pairs \cite{An:2022fvs}.

\begin{table}[htp]
\caption{Parameters of the Hamiltonian.}\label{para}
\begin{lrbox}{\tablebox}
\renewcommand\arraystretch{1.5}
\renewcommand\tabcolsep{2.8pt}
\begin{tabular}{c|c|c|c}
\midrule[1.5pt]
\toprule[0.5pt]
Parameter&$\kappa$&$a_{0}$&$D$\\
\toprule[0.5pt]
Value&120.0 MeV fm&0.0318119 $\rm (MeV^{-1}fm)^{1/2}$& 983 MeV\\
\toprule[1.0pt]
Parameter&$\alpha$&$\beta$&$m_{c}$\\
\toprule[0.5pt]
Value&1.0499 $\rm fm^{-1}$& 0.0008314 $\rm (MeV fm)^{-1}$&1918 MeV\\
\toprule[1.0pt]
Parameter&$\kappa_{0}$&$\gamma$&$m_{b}$\\
\toprule[0.5pt]
Value&194.144 MeV& 0.00088 $\rm MeV^{-1}$& 5343 MeV\\
\toprule[0.5pt]
\toprule[1.0pt]
\end{tabular}
\end{lrbox}\scalebox{0.91}{\usebox{\tablebox}}
\end{table}

\section{WAVE FUNCTIONS}\label{sec3}

Here, we focus on the ground fully heavy tetraquark states.
We present the flavor, spatial, and color-spin parts of the total wave function for fully-heavy tetraquark system.
In order to consider the constraint by the Pauli principle, we use a diquark-antidiquark picture to analyze this tetraquark system.

\subsection{Flavor Part}

First we discuss the flavor part.
Here, we list all the possible flavor combinations for the fully-heavy tetraquark system in Table \ref{flavor}.

\begin{table}[t]
\centering \caption{All possible flavor combinations for the fully-heavy tetraquark system.
}\label{flavor}
\begin{lrbox}{\tablebox}
\renewcommand\arraystretch{1.4}
\begin{tabular}{c|ccccc}
\bottomrule[1.5pt]
\bottomrule[0.5pt]
System&\multicolumn{5}{c}{Flavor combinations}\\
\bottomrule[0.5pt]
\multirow{2}*{$QQ\bar{Q}\bar{Q}$}&$cc\bar{c}\bar{c}$&\quad&$bb\bar{b}\bar{b}$&\quad&$cb\bar{c}\bar{b}$\\
&$cc\bar{b}\bar{b}$ ($bb\bar{c}\bar{c}$)&\quad\quad&$cc\bar{c}\bar{b}$  ($bc\bar{c}\bar{c}$)&\quad\quad&$bb\bar{b}\bar{c}$  ($cb\bar{b}\bar{b}$)\\
\bottomrule[0.5pt]
\bottomrule[1.5pt]
\end{tabular}
\end{lrbox}\scalebox{1.12}{\usebox{\tablebox}}
\end{table}

In Table \ref{flavor}, the three flavor combinations in the first row are purely neutral particles and the C parity is a ``good" quantum number.
For the other six states in the second row, each state has a charge conjugation anti-partner,
and their masses, internal mass contributions, relative distances between (anti)quarks are absolutely identical, so we only need to discuss one of the pair.

Furthermore, the $cc\bar{c}\bar{c}$, $bb\bar{b}\bar{b}$, and $cc\bar{b}\bar{b}$ states have the two pairs of (anti)quarks which are identical, but
only the first two quarks in the $cc\bar{c}\bar{b}$ and $bb\bar{b}\bar{c}$ states are identical.

\subsection{Spatial Part}

In this part, we construct the wave function for the spatial part in a simple Gaussian form.
We denote the fully heavy tetraquark state as the $Q(1)Q(2)\bar{Q}(3)\bar{Q}(4)$ configuration, and choose the Jacobian coordinate system as follows:
\begin{eqnarray}\label{Eq4}
\textbf{x}_{1}&=&\sqrt{1/2}(\textbf{r}_{1}-\textbf{r}_{2}),\nonumber\\
\textbf{x}_{2}&=&\sqrt{1/2}(\textbf{r}_{3}-\textbf{r}_{4}),\nonumber\\
\textbf{x}_{3}&=&\left[\left(\frac{m_{1}\textbf{r}_{1}+m_{2}\textbf{r}_{2}}{m_{1}+m_{2}}\right)-\left(\frac{m_{3}\textbf{r}_{3}+m_{4}\textbf{r}_{4}}{m_{3}+m_{4}}\right)\right].
\end{eqnarray}

Here, we set the Jacobi coordinates with the following conditions:
\begin{eqnarray}\label{Eq5}
m_{1}=m_{2}=m_{3}=m_{4}=m_{c}, \quad \rm{for} &\quad& cc\bar{c}\bar{c}, \nonumber\\
m_{1}=m_{2}=m_{3}=m_{4}=m_{b}, \quad \rm{for} &\quad& bb\bar{b}\bar{b}, \nonumber\\
m_{1}=m_{2}=m_{c}, m_{3}=m_{4}=m_{b}, \quad \rm{for} &\quad& cc\bar{b}\bar{b}, \nonumber\\
m_{1}=m_{2}=m_{3}=m_{c}, m_{4}=m_{b}, \quad \rm{for} &\quad& cc\bar{c}\bar{b}, \nonumber\\
m_{1}=m_{2}=m_{3}=m_{b}, m_{4}=m_{c}, \quad \rm{for} &\quad& bb\bar{b}\bar{c}, \nonumber\\
m_{1}=m_{c}, m_{2}=m_{b}, m_{3}=m_{c}, m_{4}=m_{b}, \quad \rm{for} &\quad& cb\bar{c}\bar{b}. \nonumber
\end{eqnarray}

Based on this, we construct the spatial wave functions of the $QQ\bar{Q}\bar{Q}$ states in a single Gaussian form.
The spatial wave function can satisfy the required symmetry property:
\begin{eqnarray}\label{Eq6}
R^{s}=\rm exp[-C_{11}\textbf{x}^{2}_{1}-C_{22}\textbf{x}^{2}_{2}-C_{33}\textbf{x}^{2}_{3}],
\end{eqnarray}
where $C_{11}$, $C_{22}$, and $C_{33}$ are the variational parameters.

It is also useful to introduce the center of mass frame so that the kinetic term in the Hamiltonian of
Eq. (\ref{Eq1}) can be reduced appropriately for our calculations.
The kinetic term, denoted by $T_{c}$, is as follows
\begin{eqnarray}\label{Eq10}
T_{c}=\sum^{4}_{i=1}\frac{\textbf{p}^{2}_{i}}{2m_{i}}- \frac{\textbf{p}^{2}_{rC}}{2M}= \frac{\textbf{p}^{2}_{x_{1}}}{2m'_{1}}+\frac{\textbf{p}^{2}_{x_{2}}}{2m'_{2}}+\frac{\textbf{p}^{2}_{x_{3}}}{2m'_{3}},
\end{eqnarray}
where different states have different reduced masses $m'_{i}$, which are listed in Table \ref{mass}.

\begin{table}[htp]
\caption{The reduced mass $m'_{i}$ in different states.}\label{mass}
\begin{lrbox}{\tablebox}
\renewcommand\arraystretch{1.9}
\renewcommand\tabcolsep{2.5pt}
\begin{tabular}{c|ccc|c|cccc}
\toprule[1.0pt]
\toprule[0.5pt]
States&$m'_{1}$&$m'_{2}$&$m'_{3}$&States&$m'_{1}$&$m'_{2}$&$m'_{3}$\\
\toprule[0.5pt]
$cc\bar{c}\bar{c}$&$m_{c}$&$m_{c}$&$m_{c}$&$cc\bar{c}\bar{b}$&$m_{c}$&$\frac{2m_{c}m_{b}}{m_{c}+m_{b}}$&$\frac{(m_{c}+m_{b})m_{c}}{2(3m_{c}+m_{b})}$\\
$bb\bar{b}\bar{b}$&$m_{b}$&$m_{b}$&$m_{b}$&$bb\bar{b}\bar{c}$&$m_{b}$&$\frac{2m_{c}m_{b}}{m_{c}+m_{b}}$&$\frac{(m_{c}+m_{b})m_{b}}{2(3m_{b}+m_{c})}$\\
$cc\bar{b}\bar{b}$&$m_{c}$&$m_{b}$&$\frac{2m_{c}m_{b}}{m_{c}+m_{b}}$&$cb\bar{c}\bar{b}$&$\frac{2m_{c}m_{b}}{m_{c}+m_{b}}$&$\frac{2m_{c}m_{b}}{m_{c}+m_{b}}$&$\frac{m_{c}+m_{b}}{2}$\\
\toprule[0.5pt]
\toprule[1.0pt]
\end{tabular}
\end{lrbox}\scalebox{0.96}{\usebox{\tablebox}}
\end{table}

\subsection{Color-spin Part}
In the color space, the color wave functions can be analyzed using the SU(3) group theory, where the direct product of the diquark and antidiquark components reads
\begin{eqnarray}\label{Eq11}
(3_{c}\otimes3_{c})\otimes(\bar{3}_{c}\otimes\bar{3}_{c})=(6_{c}\oplus\bar{3}_{c})\otimes(\bar{6}_{c}\oplus3_{c}).
\end{eqnarray}

Based on this, we get two types of color-singlet states:
\begin{eqnarray}\label{Eq12}
\phi_{1}=|(Q_{1}Q_{2})^{\bar{3}}(\bar{Q}_{3}\bar{Q}_{4})^{3}\rangle, \phi_{2}=|(Q_{1}Q_{2})^{6}(\bar{Q}_{3}\bar{Q}_{4})^{\bar{6}}\rangle.
\end{eqnarray}

In the spin space, the allowed wave functions are in the diquark-antidiquark picture:
\begin{eqnarray}\label{Eq13}
\chi_{1}&=&|(Q_{1}Q_{2})_{1}(\bar{Q}_{3}\bar{Q}_{4})_{1}\rangle_{2}, \chi_{2}=|(Q_{1}Q_{2})_{1}(\bar{Q}_{3}\bar{Q}_{4})_{1}\rangle_{1}, \nonumber\\
\chi_{3}&=&|(Q_{1}Q_{2})_{1}(\bar{Q}_{3}\bar{Q}_{4})_{0}\rangle_{1}, \chi_{4}=|(Q_{1}Q_{2})_{0}(\bar{Q}_{3}\bar{Q}_{4})_{1}\rangle_{1}, \nonumber\\
\chi_{5}&=&|(Q_{1}Q_{2})_{1}(\bar{Q}_{3}\bar{Q}_{4})_{1}\rangle_{0}, \chi_{6}=|(Q_{1}Q_{2})_{0}(\bar{Q}_{3}\bar{Q}_{4})_{0}\rangle_{0}.\nonumber\\
\end{eqnarray}
In the notation $|(Q_{1}Q_{2})_{\rm{spin1}}(\bar{Q}_{3}\bar{Q}_{4})_{\rm{spin2}}\rangle_{\rm{spin3}}$, the spin1, spin2, and spin3 represent the spin of the diquark, the spin of the antidiquark, and the total spin of the tetraquark state, respectively.

Since the flavor part and spatial parts are chosen to be fully symmetric for the (anti)diquark, the color-spin part of the total wave function should be fully antisymmetric.
Combining the flavor part, we show all possible color-spin part satisfying the Pauli principle with $J^{PC}$ in Table \ref{cs}.

In addition, it is convenient to consider the strong decay properties, we use the meson-meson configuration to represent color-singlet and spin wave functions, again.
The color wave functions in the meson-meson configuration can be derived from the following direct product:
\begin{eqnarray}\label{Eq101}
(3_{c}\otimes\bar{3}_{c})\otimes(3_{c}\otimes\bar{3}_{c})=(1_{c}\oplus8_{c})\otimes(1_{c}\oplus8_{c}).
\end{eqnarray}

Based on Eq. (\ref{Eq101}), they can be expressed as
\begin{eqnarray}\label{Eq102}
\psi_{1}=|(Q_{1}\bar{Q}_{3})^{1}(Q_{2}\bar{Q}_{4})^{1}\rangle,
\psi_{2}=|(Q_{1}\bar{Q}_{3})^{8}(Q_{2}\bar{Q}_{4})^{8}\rangle.
\end{eqnarray}

Similarly, the spin wave functions in the meson-meson configuration read as
\begin{eqnarray}\label{Eq103}
\zeta_{1}&=&|(Q_{1}\bar{Q}_{3})_{1}(Q_{2}\bar{Q}_{4})_{1}\rangle_{2},
\zeta_{2}=|(Q_{1}\bar{Q}_{3})_{0}(Q_{2}\bar{Q}_{4})_{1}\rangle_{1}, \nonumber\\
\zeta_{3}&=&|(Q_{1}\bar{Q}_{3})_{1}(Q_{2}\bar{Q}_{4})_{0}\rangle_{1},
\zeta_{4}=|(Q_{1}\bar{Q}_{3})_{1}(Q_{2}\bar{Q}_{4})_{1}\rangle_{1}, \nonumber\\
\zeta_{5}&=&|(Q_{1}\bar{Q}_{3})_{1}(Q_{2}\bar{Q}_{4})_{1}\rangle_{0},
\zeta_{6}=|(Q_{1}\bar{Q}_{3})_{0}(Q_{2}\bar{Q}_{4})_{0}\rangle_{0}.\nonumber\\
\end{eqnarray}

\begin{table}[h!]
\caption{The allowed color-spin parts for each flavor configuration.}\label{cs}
\begin{lrbox}{\tablebox}
\renewcommand\arraystretch{1.5}
\renewcommand\tabcolsep{3.5pt}
\begin{tabular}{c|c|cccc}
\toprule[1.0pt]
\toprule[0.5pt]
Type&$J^{P(C)}$&\multicolumn{4}{c}{Color-spin Part}\\
\toprule[0.5pt]
\multirow{3}*{\makecell[c]{$cc\bar{c}\bar{c}$\\ $bb\bar{b}\bar{b}$\\$cc\bar{b}\bar{b}$}}&$2^{+(+)}$&$\phi_{1}\chi_{1}$\\
&$1^{+(-)}$&$\phi_{1}\chi_{2}$\\
&$0^{+(+)}$&$\phi_{1}\chi_{5}$&$\phi_{2}\chi_{6}$\\
\toprule[0.5pt]
\multirow{3}*{\makecell[c]{$cc\bar{c}\bar{b}$\\\\$bb\bar{b}\bar{c}$}}&$2^{+}$&$\phi_{1}\chi_{1}$\\
&$1^{+}$&$\phi_{1}\chi_{2}$&$\phi_{1}\chi_{3}$&$\phi_{2}\chi_{4}$&\\
&$0^{+}$&$\phi_{1}\chi_{5}$&$\phi_{2}\chi_{6}$\\
\toprule[0.5pt]
\multirow{5}*{\makecell[c]{$cb\bar{c}\bar{b}$}}&$2^{++}$&$\phi_{1}\chi_{1}$&$\phi_{2}\chi_{1}$\\
&\multirow{2}*{$1^{+-}$}&$\phi_{1}\chi_{2}$&$\phi_{2}\chi_{2}$&\multicolumn{2}{c}{$\frac{1}{\sqrt{2}}(\phi_{1}\chi_{3}+\phi_{1}\chi_{4})$}\\
&&\multicolumn{2}{c}{$\frac{1}{\sqrt{2}}(\phi_{2}\chi_{3}+\phi_{2}\chi_{4})$}\\
&$1^{++}$&\multicolumn{2}{c}{$\frac{1}{\sqrt{2}}(\phi_{1}\chi_{3}-\phi_{1}\chi_{4})$}&\multicolumn{2}{c}{$\frac{1}{\sqrt{2}}(\phi_{2}\chi_{3}-\phi_{2}\chi_{4})$}\\
&$0^{++}$&$\phi_{1}\chi_{5}$&$\phi_{2}\chi_{5}$&$\phi_{1}\chi_{6}$&$\phi_{2}\chi_{6}$\\
\toprule[0.5pt]
\toprule[1.0pt]
\end{tabular}
\end{lrbox}\scalebox{1.05}{\usebox{\tablebox}}
\end{table}

\section{NUMERICAL ANALYSIS}\label{sec4}



\subsubsection{Mass spectrum, internal contribution, and  spatial size}

In this subsection, we check the consistency between the experimental masses and the obtained masses of traditional hadrons using the variational method based
on the Hamiltonian of Eq. (\ref{Eq1}) and the parameters in Table \ref{para}.
We show the results in Table \ref{meson} and note that our values are relatively reliable since the deviations for most states are less than 20 MeV.

In addition, we have systematically constructed the total wave function satisfied by the Pauli principle in the previous section.
The corresponding total wave function could be expanded as follows:
\begin{eqnarray}\label{Eq451}
|\Psi_{\alpha}\rangle=\sum_{ij}C^{\alpha}_{ij}|F\rangle|R^{s}\rangle|[\phi_{i}\chi_{j}]\rangle.
\end{eqnarray}

To study the mass of the fully heavy tetraquarks with the variational method, we calculate the Schr\"{o}dinger equation
$H|\Psi_{\alpha}\rangle=E_{\alpha}|\Psi_{\alpha}\rangle$, diagonalize the corresponding matrix, and then determine the ground state masses for the fully heavy tetraquarks.
According to the corresponding variational parameters, we also give the internal mass contributions, including the quark mass part, the kinetic energy part, the confinement potential part, and the hyperfine potential part.
For comparison, we also show the lowest meson-meson thresholds for the tetraquarks with different quantum numbers and their internal contributions.
This is how we define the binding energy:
\begin{eqnarray}\label{Eq14}
B_{T}=M_{tetraquark}-M_{meson1}-M_{meson2},
\end{eqnarray}
where $M_{tetraquark}$, $M_{meson1}$, and $M_{meson2}$ are the masses of the tetraquark and the two mesons at the lowest threshold allowed in the rearrangement decay of the tetraquark, respectively.
To facilitate the discussion in the next subsection, we also define the $V^{C}$, which is the sum of the Coulomb potential and the linear potential.

Here, it is also useful to investigate the spatial size of the tetraquarks, which is strongly related to the magnitude of the various kinetic energies and the potential energies between the quarks.
It is also important to understand the relative lengths between the quarks in the tetraquarks and their lowest thresholds, and the relative distance between the heavier quarks is generally shorter than that between the lighter quarks \cite{Park:2018wjk}.
This tendency is also maintained in each tetraquark state according to the corresponding tables.


\subsubsection{Magnetic moments, transition magnetic moments, and radiative decay widths}

The magnetic moment of hadrons is a physical quantity that reflects their internal structures \cite{Zhou:2022gra}.
The total magnetic moment $\vec{\mu}_{total}$ of a compound system contains the spin magnetic moment $\vec{\mu}_{spin}$ and the orbital magnetic moment $\vec{\mu}_{orbital}$ from all of its constituent quarks.
For ground hadron states, their contribution of the orbital magnetic moment $\vec{\mu}_{orbital}$ is zero, and so we
only concentrate on the spin magnetic moment $\vec{\mu}_{spin}$.
The explicit expression for the spin magnetic moment $\vec{\mu}_{spin}$ is written as
\begin{eqnarray}\label{magn}
\vec{\mu}_{spin}=\sum_{i}\mu_{i}\vec{\sigma}_{i}=\sum_{i}\frac{Q^{eff}_{i}}{2M^{eff}_{i}}\vec{\sigma}_{i},
\end{eqnarray}
where $Q^{eff}_{i}$ and $M^{eff}_{i}$ are the effective charge and effective mass of the $i$-th constituent quark, respectively.
The $\vec{\sigma}_{i}$ denotes the Pauli's spin matrix of the $i$-th constituent quark.
According to Ref. \cite{Kumar:2005ei}, the effective charge of the quark is affected by other quarks in the inner hadron.
We now assume that the
effective charge is linearly dependent on the charge of the shielding quarks.
So the effective charge $Q^{eff}_{i}$ is defined as
\begin{eqnarray}
Q^{eff}_{i}=Q_{i}+\sum_{i\neq j}\alpha_{ij}Q_{j},
\end{eqnarray}
where $Q_{i}$ is the bare charge of the $i$-th constituent quark, the $\alpha_{ij}$ is a corrected parameter that  reflects how much the charge of other quarks affects the charge of the
$i$-th quark.
To simplify the calculation, we also set $\alpha_{ij}$ always equal to 0.033 according to the Ref. \cite{Kumar:2005ei}. The effective quark masses $M^{eff}_{i}$ contain the contributions from both the bare quark mass terms and the interaction terms in the chromomagnetic model, and their values are taken from Ref. \cite{Wu:2016vtq}.

To obtain the magnetic moment of the discussed hadron,
we calculate the $z$-component of the magnetic moment operator $\hat{\mu}^{z}$ sandwiched by the corresponding total wave function $\Psi_{\alpha}$ (Eq.[\ref{Eq13}]).
Now, only the spin part of the total wave function is involved.
The total spin wave functions of the discussed hadrons are written as
\begin{eqnarray}
|\chi_{total}\rangle=C_{1}\chi_{1}+C_{2}\chi_{2}+...
\end{eqnarray}
Based on this, we can  quantitatively obtain the magnetic moment of the discussed hadron
\begin{eqnarray}\label{Eq105}
\mu&=&\langle\Psi_{\alpha}|\hat{\mu}^{z}|\Psi_{\alpha}\rangle
=\langle\chi_{\alpha}|\hat{\mu}^{z}|\chi_{\alpha}\rangle \nonumber\\
&=&C^{2}_{1}\mu(\chi_{1})+C^{2}_{2}\mu(\chi_{2})+...+2C_{1}C_{2}\mu^{tr}(\chi_{1},\chi_{2})+...\nonumber
\end{eqnarray}
where $\mu^{tr}$ is the cross-term representing the  transition moment, and
$C_{1}$, $C_{2}$ are the eigenvectors of the given mixing state \cite{Zhang:2021yul}.
Similarly, the transition magnetic moments between the hadrons can be obtained as
$\mu_{H' \rightarrow H\gamma}=\langle\Psi_{H_{f}}|\hat{\mu}^{z}|\Psi_{H_{i}}\rangle$.

According to Eq. (\ref{Eq105}), the numerical values for the magnetic moments of the traditional hadrons have been listed in Table \ref{meson}.
Here, $\mu_{N}=e/2m_{N}$ is the
nuclear magnetic moment with $m_{N}=938$ MeV as the nuclear mass, which is the unit of the magnetic moment.
For comparison, we also show the experimental values and other theoretical results from Refs. \cite{Zhou:2022gra,Kumar:2005ei,Zhang:2021yul,Wang:2016dzu,Wang:2022tib,Wang:2022ugk}.
Because of the $\mu_{Q}=-\mu_{\bar{Q}}$,
the magnetic moment of all of the $J^{P}=0^{+}$ ground mesons and tetraquarks and the ground states with certain $C$-parity is 0.

The decay property is another important aspect to investigate the nature of the exotic hadron. According to the transition magnetic moments in the above subsection, we can further obtain the radiative decay widths around fully heavy tetraquarks
\cite{Li:1997gd,Brodsky:1968ea,Zhao:2002id,Li:1994cy,Deng:2016ktl,Deng:2016stx,Xiao:2017udy,Lu:2017meb,Wang:2017kfr,Yao:2018jmc}.

\begin{equation}\label{eq:Gammaradiative}
\Gamma=\frac{|{\bf k}|^2}{\pi}\frac{2}{2J_i+1}\frac{M_f}{M_i}\sum\limits_{M_{J_f},M_{J_i}}|{\cal M}_{M_{J_f},M_{J_i}}|^2,
\end{equation}
where the $J_i$ and $J_f$ are the total angular momentums of the initial and final hadrons, respectively. The $M_i$ and $M_f$ in Eq.~(\ref{eq:Gammaradiative}) represent initial and final hadron masses, respectively.

\subsubsection{Relative decay widths of tetraquarks}

In addition to radiative decay, we also consider the rearrangement strong decay properties for fully heavy tetraquarks.
Based on Eqs. (\ref{Eq101}-\ref{Eq103}),  the color wave function also falls into two categories:
the color-singlet $\psi_{1}=|(Q_{1}\bar{Q}_{3})^{1}(Q_{2}\bar{Q}_{4})^{1}\rangle$ which can
easily decay into two S-wave mesons, and the
color-octet $\psi_{2}=|(Q_{1}\bar{Q}_{3})^{8}(Q_{2}\bar{Q}_{4})^{8}\rangle$ which can only fall apart by the gluon exchange.
Thus we transform the total wave functions $\Psi_{\alpha}$ into the new configuration,
\begin{eqnarray}\label{Eq106}
|\Psi_{\alpha}\rangle=\sum_{ij}C'^{\alpha}_{ij}|F\rangle|R^{s}\rangle|[\psi_{i}\zeta_{j}]\rangle.
\end{eqnarray}

Among the decay behaviors of the tetraquarks, one decay mode is that the quarks simply fall apart into the final decay channels without quark pair creations or annihilations, which is donated as `` Okubo-Zweig-Iizuka (OZI)-superallowd" decays. In this part, we will only focus this type of decay channels.
For two body decay by $L$-wave, the partial decay width reads as \cite{Weng:2021ngd,Weng:2020jao,An:2020vku,Guo:2021mja,Weng:2021hje}:
\begin{eqnarray}\label{width}
\Gamma_{i}=\gamma_{i}\alpha\frac{k^{2L+1}}{m^{2L}}\,|c_{i}|^{2},
\end{eqnarray}
where $\alpha$ is an effective coupling constant, $c_{i}$ is the overlap corresponding exactly to $C'^{\alpha}_{ij}$ of Eq. (\ref{Eq106}),
$m$ is the mass of the initial state, $k$ is the momentum of the final state in the rest frame of the initial state.
For the decays of the S-wave tetraquarks, $(k/m)^{-2}$ is of order $\mathcal{O}(10^{-2})$ or even smaller, so all higher-wave decays are suppressed.
So we only need to consider the S-wave decays.
As for $\gamma_{i}$, it is determined by the
spatial wave functions of the initial and final states, which
are different for each decay process.
In the quark model in the heavy quark limit, the spatial wave functions of the ground S-wave pseudoscalar and the vector meson are the same.
The relations of $\gamma_{i}$ for fully heavy tetraquarks are given in Table \ref{ship}.
Based on this, the branching fraction is proportional to the square of the
coefficient of the corresponding component in the eigenvectors, and the strong decay phase space, i.e., $k\cdot|c_{i}|^{2}$, for each decay mode.
From the value of $k\cdot|c_{i}|^{2}$, one can roughly estimate the ratios of the relative decay widths between different decay processes of different initial tetraquarks.

\begin{table}[htp]
\caption{The approximate relation about $\gamma_{i}$ for the $QQ\bar{Q}\bar{Q}$
system.}\label{ship}
\begin{lrbox}{\tablebox}
\renewcommand\arraystretch{1.6}
\renewcommand\tabcolsep{9.pt}
\begin{tabular}{c|l}
\toprule[1.0pt]
\toprule[0.5pt]
States&\multicolumn{1}{c}{$\gamma_{i}$}\\
\toprule[0.5pt]
$cc\bar{c}\bar{c}$&$\gamma_{J/\psi J/\psi}=\gamma_{\eta_{c} J/\psi}=\gamma_{\eta_{c}\eta_{c}}$\\
$bb\bar{b}\bar{b}$&$\gamma_{\Upsilon\Upsilon}=\gamma_{\eta_{b} \Upsilon}=\gamma_{\eta_{b}\eta_{b}}$\\
\multirow{1}*{\makecell[c]{$cc\bar{b}\bar{b}$}}&$\gamma_{B^{*}_{c}B^{*}_{c}}=\gamma_{B_{c}B^{*}_{c}}
=\gamma_{B_{c}B_{c}}$\\
\toprule[0.5pt]
$cc\bar{c}\bar{b}$&$\gamma_{J/\psi B^{*}_{c}}=\gamma_{J/\psi B_{c}}=
\gamma_{\eta_{c} B^{*}_{c}}=\gamma_{\eta_{c} B_{c}}$\\
$bb\bar{b}\bar{c}$&$\gamma_{\Upsilon \bar{B}^{*}_{c}}=\gamma_{\Upsilon \bar{B}_{c}}=
\gamma_{\eta_{b} \bar{B}^{*}_{c}}=\gamma_{\eta_{b} \bar{B}_{c}}$\\
\toprule[0.5pt]
\multirow{2}*{\makecell[c]{$cb\bar{c}\bar{b}$}}&$\gamma_{J/\psi \Upsilon}=\gamma_{J/\psi \eta_{b}}=\gamma_{\eta_{c}\Upsilon}=\gamma_{\eta_{c}\eta_{b}}$\\
&$\gamma_{B^{*}_{c}\bar{B}^{*}_{c}}=\gamma_{B^{*}_{c}\bar{B}_{c}}
=\gamma_{B_{c}\bar{B}^{*}_{c}}=\gamma_{B_{c}\bar{B}_{c}}$\\
\toprule[0.5pt]
\toprule[1.0pt]
\end{tabular}
\end{lrbox}\scalebox{1.2}{\usebox{\tablebox}}
\end{table}

\begin{table*}[htp]
\caption{
Masses and magnetic moments of some ground hadrons obtained from the theoretical calculations.
$M_{result}$, $\mu_{results}$, $\mu_{bag}$, $\mu_{the(1)}$, and $\mu_{the(2)}$ are theoretical masses and magnetic moments for Eq. (\ref{Eq1}), Eq. (\ref{magn}), and Refs. \cite{Zhang:2021yul,Zhou:2022gra,Kumar:2005ei}, respectively.
$M_{exp}$ and $\mu_{exp}$ are the observed values of masses and magnetic moments.
The masses and errors are in units of MeV.
The magnetic moment is in units of the nuclear magnetic moment $\mu_{N}$.
The variational parameter is in units of $\rm fm^{-2}$.
}\label{meson}
\begin{center}
\begin{lrbox}{\tablebox}
\renewcommand\arraystretch{1.6}
\renewcommand\tabcolsep{2.5pt}
\begin{tabular}{l|c|c|c|c|c|c|c|c|c|c|c|c|c|c|c|c|c}
\midrule[1.5pt]
\toprule[0.5pt]
Hadron&$\Sigma^{+}$&$\Sigma^{0}$&$\Sigma^{-}$&$\Xi^{0}$&$\Xi^{-}$&$\Sigma_{c}^{++}$&$\Sigma_{c}^{+}$&$\Sigma_{c}^{0}$&$\Sigma_{c}^{*++}$&$\Sigma_{c}^{*+}$&$\Sigma_{c}^{*0}$&$\Sigma_{b}^{+}$&$\Sigma_{b}^{0}$&$\Sigma_{b}^{-}$&$\Sigma_{b}^{*+}$&$\Sigma_{b}^{*0}$&$\Sigma_{b}^{*-}$\\
\toprule[0.5pt]
$M_{result}$ &\multicolumn{3}{c|}{1187.7}&\multicolumn{2}{c|}{1295.4}&\multicolumn{3}{c|}{2445.2}&\multicolumn{3}{c|}{2518.3}&\multicolumn{3}{c|}{5832.1}&\multicolumn{3}{c}{5860.8}\\
\multirow{2}*{\makecell[c]{Parameters}} &\multicolumn{3}{c|}{$2.1$}&\multicolumn{2}{c|}{3.3}&\multicolumn{3}{c|}{2.1}&\multicolumn{3}{c|}{2.0}&\multicolumn{3}{c|}{2.1}&\multicolumn{3}{c}{2.0}\\
&\multicolumn{3}{c|}{$3.1$}&\multicolumn{2}{c|}{2.9}&\multicolumn{3}{c|}{3.7}&\multicolumn{3}{c|}{3.4}&\multicolumn{3}{c|}{4.0}&\multicolumn{3}{c}{3.4}\\
$M_{exp}$ &\multicolumn{3}{c|}{1189.4}&\multicolumn{2}{c|}{1314.9}&\multicolumn{3}{c|}{2454.0}&\multicolumn{3}{c|}{2851.4}&\multicolumn{3}{c|}{5811.3}&\multicolumn{3}{c}{5832.1}\\
Error&\multicolumn{3}{c|}{-1.7}&\multicolumn{2}{c|}{-19.5}&\multicolumn{3}{c|}{-8.8}&\multicolumn{3}{c|}{-0.1}&\multicolumn{3}{c|}{20.8}&\multicolumn{3}{c}{28.7}\\
$\mu_{result}$&2.53&0.75&-1.04&-1.31&-0.52&2.34&0.55&-1.24&4.09&1.38&-1.32&2.37&0.58&-1.21&3.48&0.78&-1.92\\
$\mu_{bag}$ \cite{Zhang:2021yul}&2.72&0.86&-1.01&-1.58&-0.64&2.13&0.41&-1.31&4.07&1.39&-1.29&2.23&0.58&-1.07&3.29&0.76&-1.77\\
$\mu_{the(1)}$\cite{Zhou:2022gra}&2.74&0.84&-1.06&-1.47&-0.52&2.36&0.50&-1.37&4.09&1.30&-1.49&&&&&\\
\Xcline{13-18}{0.6pt}
$\mu_{the(2)}$ \cite{Kumar:2005ei}&2.46&0.47&-1.10&-1.61&-0.65&3.57&1.96&0.04&&&\\
\Xcline{7-12}{0.5pt}\Xcline{14-18}{1.3pt}
$\mu_{exp}$&2.46&&-1.16&-1.25&-0.65&\multicolumn{7}{c!{\color{black}\vrule width 1.5pt}}{}&\multicolumn{1}{c|}{6.14}&2.70&\multicolumn{2}{c|}{}&-2.02\\
\Xcline{1-13}{1.3pt}\Xcline{14-18}{0.5pt}
Hadron&$D^{*0}$&$D^{*+}$&$D^{*+}_{s}$&$B^{*0}$&$B^{*+}$&$B^{*0}_{s}$&$J/\psi$&$\eta_{c}$&$\Upsilon$&$\eta_{b}$&$B_{c}$&$B^{*}_{c}$&$\Delta^{++}$&$\Delta^{+}$&$\Delta^{0}$&$\Delta^{0}$&$\Omega^{-}$\\
\Xcline{1-18}{0.5pt}
$M_{result}$ &\multicolumn{2}{c|}{1996.9}&2093.3&\multicolumn{2}{c|}{5363.6}&5434.7&3092.2&2998.5&9468.9&9389.0&6287.9&6350.5&\multicolumn{4}{c|}{1245.6}&1675.8\\
Parameters&\multicolumn{2}{c|}{3.8}&6.2&\multicolumn{2}{c|}{4.2}&7.5&12.5&15.0&49.7&57.4&22.9&20.2&\multicolumn{4}{c|}{$1.8$}&3.3\\
$M_{exp}$&\multicolumn{2}{c|}{2010.3}&2112.2&\multicolumn{2}{c|}{5324.7}&5415.4&3096.9&2983.9&9460.3&9399.0&6274.9&(6332)&\multicolumn{4}{c|}{1232.0}&1672.5\\
Error&\multicolumn{2}{c|}{-13.4}&-18.9&\multicolumn{2}{c|}{38.8}&19.3&-4.7&14.6&8.6&10.0&13.0&(17.5)&\multicolumn{4}{c|}{13.6}&3.1\\
$\mu_{result}$&-1.37&1.24&1.00&-0.78&1.83&0.51&0&-&0&-&-&0.44&5.57&-2.78&0&-2.78&-1.86\\
$\mu_{bag}$ \cite{Zhang:2021yul}&-0.98&1.21&1.08&-0.53&1.21&1.01&0&-&0&-&-&0.52&5.70&2.85&0&-2.85&-2.20\\
$\mu_{the(1)}$ \cite{Zhou:2022gra}&-1.49&1.30&1.07&&&&0&-&&&&&5.58&2.79&0&-2.79&-1.88\\
\toprule[0.5pt]
\toprule[1.0pt]
\end{tabular}
\end{lrbox}\scalebox{0.94}{\usebox{\tablebox}}
\end{center}
\end{table*}



In the following subsections, we concretely discuss all possible configurations for fully heavy tetraquarks.
\\

\subsection{$cc\bar{c}\bar{c}$ and $bb\bar{b}\bar{b}$ states}

First we investigate the $cc\bar{c}\bar{c}$ and $bb\bar{b}\bar{b}$ systems.
There are two $J^{PC}=0^{++}$ states, one $J^{PC}=1^{+-}$ state, and one $J^{PC}=2^{++}$ state according to Table \ref{cs}.
We show the masses of the ground states, the variational parameters, the internal mass contributions, the relative
lengths between the quarks, their lowest meson-meson thresholds, the specific wave function,  the magnetic moments, the transition magnetic moments, the radiative decay widths, and the rearrangement strong width ratios in Tables \ref{cccc}-\ref{bbbb}, respectively.

Here, we take the $J^{PC}=0^{++}$ $bb\bar{b}\bar{b}$ ground state as an example, and others have similar discussions according to Tables \ref{cccc}-\ref{bbbb}.
We now analyze the numerical results obtained from the variational method.
For the $J^{PC}=0^{++}$ $bb\bar{b}\bar{b}$ ground state, its mass is 19240.0 MeV and the corresponding binding energy $B_{T}$ is +461.9 MeV.
Its variational parameters are given as $C_{11}=7.7 ~{\rm fm}^{-2}$, $C_{22}=7.7 ~{\rm fm}^{-2}$, and $C_{33}=11.4 ~{\rm fm}^{-2}$, giving roughly the inverse ratios of the size for the diquark, the antidiquark, and between the center of the diquark and the antidiquark, respectively.
We naturally find that the $C_{11}$ is equal to $C_{22}$, so the distance of $(b-b)$ would be equal to that of $(\bar{b}-\bar{b})$, and the reason is that the $bb\bar{b}\bar{b}$ system is a neutral system.

The total wave function in the diquark-antidiquark configuration is given by
\begin{eqnarray}\label{eq:wave1}
|\Psi_{tot}\rangle=-0.936|F\rangle|R^{s}\rangle|[\phi_{2}\chi_{6}]\rangle+0.352|F\rangle|R^{s}\rangle|[\phi_{1}\chi_{5}]\rangle.\nonumber\\
\end{eqnarray}
The meson-meson configuration is connected to the diquark-antidiquark configuration by a linear transformation.
We then obtain the total wave function in the meson-meson configuration:
\begin{eqnarray}\label{Eq215}
|\Psi_{tot}\rangle&=&0.558|F\rangle|R^{s}\rangle|[\psi_{1}\zeta_{5}]\rangle
+0.560|F\rangle|R^{s}\rangle|[\psi_{1}\zeta_{6}]\rangle \nonumber\\
&+&0.021|F\rangle|R^{s}\rangle|[\psi_{2}\zeta_{5}]\rangle+
0.612|F\rangle|R^{s}\rangle|[\psi_{2}\zeta_{6}]\rangle.\nonumber\\
\end{eqnarray}

According to Eq. (\ref{Eq215}), we are sure that the overlaps $c_{i}$ of $\eta_{b}\eta_{b}$ and $\Upsilon\Upsilon$ are 0.560 and 0.558, respectively.
Then, based on Eq. (\ref{width}), the rearrangement strong width ratios are
\begin{eqnarray}
\frac{\Gamma_{T_{b^{2}\bar{b}^{2}}(19240.0,0^{++})\rightarrow \Upsilon\Upsilon}}{\Gamma_{T_{b^{2}\bar{b}^{2}}(19240.0,0^{++})\rightarrow\eta_{b}\eta_{b}}}=1:1.2,
\end{eqnarray}
i.e., both the $\Upsilon\Upsilon$ and $\eta_{b}\eta_{b}$ are dominant decay channels for the $T_{b^{2}\bar{b}^{2}}(19240.0,\Upsilon\Upsilon)$ state.

As for the magnetic moments of the $cc\bar{c}\bar{c}$ and $bb\bar{b}\bar{b}$ ground states,
their values are all 0, because the same quark and antiquark have exactly opposite magnetic moments, which cancel each other out.
We also discuss the transition magnetic moment of the $T_{b^{2}\bar{b}^{2}}(19303.9,1^{+-})\rightarrow T_{b^{2}\bar{b}^{2}}(19240.0,0^{++})\gamma$ process.
We construct their flavor $\otimes$ spin wave functions as
\begin{eqnarray}
|\Psi\rangle^{S=1;S_{s}=0}_{T_{b^{2}\bar{b}^{2}}(1^{+-})}&=&|R^{s}\rangle|\psi\rangle|bb\bar{b}\bar{b}\rangle|\frac{1}{\sqrt{2}}(\uparrow\uparrow\downarrow\downarrow-\downarrow\downarrow\uparrow\uparrow)\rangle\nonumber\\
|\Psi\rangle^{S=0;S_{s}=0}_{T_{b^{2}\bar{b}^{2}}(0^{++})}&=&|R^{s}\rangle|\psi\rangle|bb\bar{b}\bar{b}\rangle|0.352\frac{1}{\sqrt{3}}(\uparrow\uparrow\downarrow\downarrow+\downarrow\downarrow\uparrow\uparrow)+...\rangle.\nonumber\\
\end{eqnarray}
And then, the transition magnetic momentum of the $T_{b^{2}\bar{b}^{2}}(19303.9,1^{+-})\rightarrow T_{b^{2}\bar{b}^{2}}(19240.0,0^{++})\gamma$ process can be given by the $z$-component of the magnetic moment operator $\hat{\mu^{z}}$  sandwiched by the
flavor-spin wave functions of the $T_{b^{2}\bar{b}^{2}}(19303.9,1^{+-})$ and $T_{b^{2}\bar{b}^{2}}(19240.0,0^{++})$.
So, the corresponding transition magnetic momentum is
\begin{eqnarray}\label{Eq217}
&&\mu_{T_{b^{2}\bar{b}^{2}}(1^{+-})\rightarrow T_{b^{2}\bar{b}^{2}}(0^{++})}=\langle\Psi^{1^{+-}}_{tot}|\hat{\mu^{z}}|\Psi^{0^{++}}_{tot}\rangle\nonumber\\
&&=0.352\times\frac{1}{\sqrt{6}}(4\mu_{b}-4\mu_{\bar{b}})=-0.072~\mu_{N}.
\end{eqnarray}
As for the transition magnetic moment of the
$T_{b^{2}\bar{b}^{2}}(19327.9,2^{++})\rightarrow T_{b^{2}\bar{b}^{2}}(19240.0,0^{++})\gamma$ process, its value is 0 due to the C parity conservation restriction.

Furthermore, according to
Eq. (\ref{eq:Gammaradiative}) and Eq. (\ref{Eq217}), we also obtain the corresponding radiative decay widths
\begin{eqnarray}\label{Eq218}
\Gamma_{T_{b^{2}\bar{b}^{2}}(19303.9,1^{+-})\rightarrow T_{c^{2}\bar{b}^{2}}(19240.0,0^{++})\gamma}&=&2.8\, \rm{keV},\\
\Gamma_{T_{b^{2}\bar{b}^{2}}(19327.9,2^{++})\rightarrow T_{c^{2}\bar{b}^{2}}(19240.0,0^{++})\gamma}&=&0\, \rm{keV}.
\end{eqnarray}

\subsubsection{Relative distances and symmetry}

Here, we concentrate on the the relative distances between the (anti)quarks in tetraquarks.
Looking at the relative distances in Table \ref{bbbb}, we find that the relative distances of (1,2) and (3,4) pairs are the same, and other relative distances are the same in all the $cc\bar{c}\bar{c}$ and $bb\bar{b}\bar{b}$ states
This is due to the permutation symmetry for the ground state wave function in each tetraquark \cite{Noh:2021lqs}.
For the $c_{1}c_{2}\bar{c}_{3}\bar{c}_{4}$ and $b_{1}b_{2}\bar{b}_{3}\bar{b}_{4}$ states, they need to satisfy the Pauli principle for identical particles are as follows:
\begin{eqnarray}\label{Eq161}
A_{12}|\Psi_{tot}\rangle=A_{34}|\Psi_{tot}\rangle&=&-|\Psi_{tot}\rangle,
\end{eqnarray}
where the operator $A_{ij}$ means exchanging the coordinate of $Q_{i}$ ($\bar{Q}_i$) and $Q_{j}$ ($\bar{Q}_j$).

Meanwhile, they are pure neutral particles with definite C-parity, so the permutation symmetries for total wave functions are as follows:
\begin{eqnarray}\label{Eq162}
A_{12-34}|\Psi_{tot}\rangle&=&\pm|\Psi_{tot}\rangle,
\end{eqnarray}
where $A_{12-34}$ means that the coordinates of the diquark and the antidiquark are exchanged.

Based on this, the relationship of the relative distances for all the $c_{1}c_{2}\bar{c}_{3}\bar{c}_{4}$ and $b_{1}b_{2}\bar{b}_{3}\bar{b}_{4}$ states can be obtained as follows:
\begin{small}
\begin{eqnarray}\label{Eq17}
&&\langle\Psi_{tot}|\textbf{r}_{1}-\textbf{r}_{3}|\Psi_{tot}\rangle\nonumber\\
&&=\langle\Psi_{tot}|A^{-1}_{12}A_{12}|\textbf{r}_{1}-\textbf{r}_{3}|A^{-1}_{12}A_{12}|\Psi_{tot}\rangle=\langle\Psi_{tot}|\textbf{r}_{2}-\textbf{r}_{3}|\Psi_{tot}\rangle\nonumber\\
&&=\langle\Psi_{tot}|A^{-1}_{34}A_{34}|\textbf{r}_{2}-\textbf{r}_{3}|A^{-1}_{34}A_{34}|\Psi_{tot}\rangle=\langle\Psi_{tot}|\textbf{r}_{2}-\textbf{r}_{4}|\Psi_{tot}\rangle\nonumber\\
&&=\langle\Psi_{tot}|A^{-1}_{12}A_{12}|\textbf{r}_{2}-\textbf{r}_{4}|A^{-1}_{12}A_{12}|\Psi_{tot}\rangle=\langle\Psi_{tot}|\textbf{r}_{1}-\textbf{r}_{4}|\Psi_{tot}\rangle,\nonumber\\
\end{eqnarray}
\end{small}
and
\begin{eqnarray}\label{Eq18}
&&\langle\Psi_{tot}|\textbf{r}_{1}-\textbf{r}_{2}|\Psi_{tot}\rangle\nonumber\\
&&=\langle\Psi_{tot}|A^{-1}_{12-34}A_{12-34}|\textbf{r}_{1}-\textbf{r}_{2}||A^{-1}_{12-34}A_{12-34}|\Psi_{tot}\rangle\nonumber\\
&&=\langle\Psi_{tot}|\textbf{r}_{3}-\textbf{r}_{4}|\Psi_{tot}\rangle.
\end{eqnarray}
Obviously, our theoretical derivations are in perfect agreement with the calculated results in Table \ref{bbbb}.

We can also prove three Jacobi coordinates, $\textbf{R}_{1,2}=\textbf{r}_{1}-\textbf{r}_{2}$, $\textbf{R}_{3,4}=\textbf{r}_{3}-\textbf{r}_{4}$, and
$\textbf{R}'=1/2(\textbf{r}_{1}+\textbf{r}_{2}-\textbf{r}_{3}-\textbf{r}_{4})$, are orthogonal to each other for all the $cc\bar{c}\bar{c}$ and $bb\bar{b}\bar{b}$ states:
\begin{eqnarray}\label{Eq19}
&&\langle\Psi_{tot}|(\textbf{R}_{1,2}\cdot\textbf{R}_{3,4})|\Psi_{tot}\rangle\nonumber\\
&&=\langle\Psi_{tot}|(34)^{-1}(34)|(\textbf{R}_{1,2}\cdot\textbf{R}_{3,4})|(34)^{-1}(34)|\Psi_{tot}\rangle\nonumber\\
&&=-\langle\Psi_{tot}|(\textbf{R}_{1,2}\cdot\textbf{R}_{3,4})|\Psi_{tot}\rangle=0,
\end{eqnarray}
\begin{eqnarray}\label{Eq20}
&&\langle\Psi_{tot}|(\textbf{R}_{1,2}\cdot\textbf{R}')|\Psi_{tot}\rangle\nonumber\\
&&=\langle\Psi_{tot}|(12)^{-1}(12)|(\textbf{R}_{1,2}\cdot\textbf{R}')|(12)^{-1}(12)|\Psi_{tot}\rangle\nonumber\\
&&=-\langle\Psi_{tot}|(\textbf{R}_{1,2}\cdot\textbf{R}')|\Psi_{tot}\rangle=0,
\end{eqnarray}
and
\begin{eqnarray}\label{Eq21}
&&\langle\Psi_{tot}|(\textbf{R}_{3,4}\cdot\textbf{R}')|\Psi_{tot}\rangle\nonumber\\
&&=\langle\Psi_{tot}|(34)^{-1}(34)|(\textbf{R}_{1,2}\cdot\textbf{R}')|(34)^{-1}(34)|\Psi_{tot}\rangle\nonumber\\
&&=-\langle\Psi_{tot}|(\textbf{R}_{3,4}\cdot\textbf{R}')|\Psi_{tot}\rangle=0.
\end{eqnarray}

\begin{figure}[t]
\begin{tabular}{c}
\includegraphics[width=8.6cm]{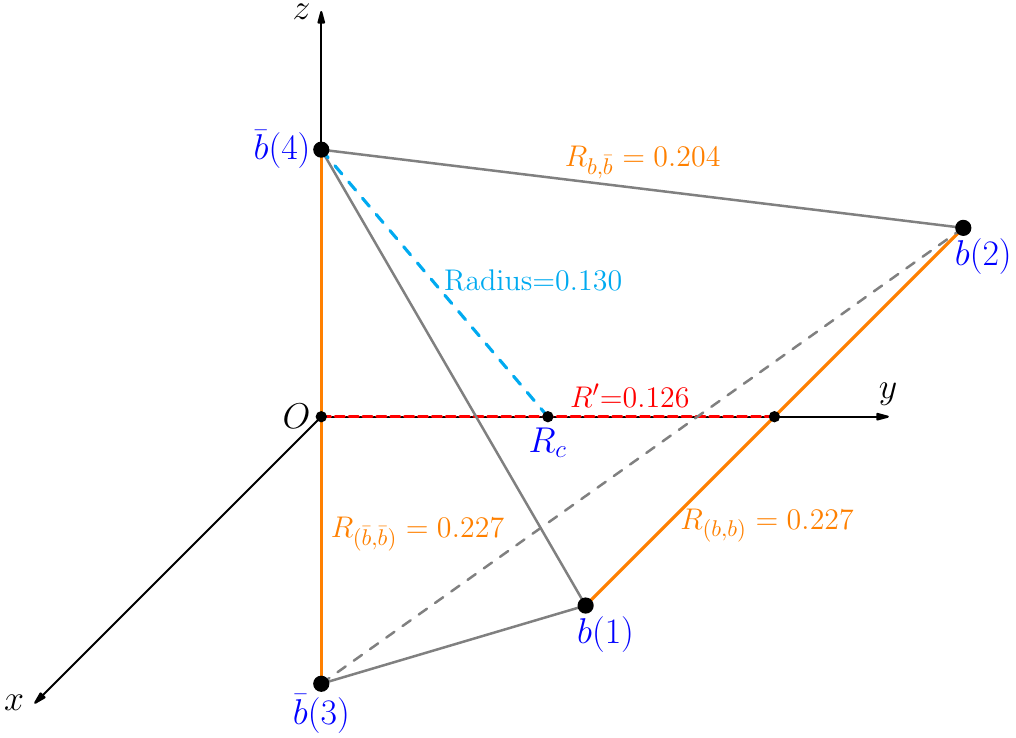}\\
\end{tabular}
\caption{
Relative positions for four valence quarks and $R_{c}$ in the $J^{PC}=0^{++}$ $bb\bar{b}\bar{b}$ ground state.
Meanwhile, we label the relative distances of $R_{b,b}$, $R_{b,\bar{b}}$, $R_{\bar{b},\bar{b}}$, $R'$, and the radius (units: fm).
}
\label{1}
\end{figure}

According to the relative distances in Table \ref{bbbb} and the relationship of Eqs. (\ref{Eq161}-\ref{Eq21}), we can well describe the relative positions of the four valence quarks for all the $cc\bar{c}\bar{c}$ and $bb\bar{b}\bar{b}$ states.
Meanwhile, using the relative distances between (anti)quarks and the orthogonal relation, one can also determine the relative distance of $(12)-(34)$, which is consistent with our results in Table \ref{bbbb}.
We can also give the relative position of $R_{c}$ and the spherical radius of the tetraquarks.
Here, we define $R_{c}$ to be the geometric center of the four quarks (the center of the sphere).
Based on these results, we show the spatial distribution of the four valence quarks for the $J^{PC}=0^{++}$ $bb\bar{b}\bar{b}$ ground state in Fig. \ref{1}.

In quark model, a compact tetraquark state has no color-singlet substruture, while a hadronic molecule is a loosely bound state which contains several color-singlet hadrons.
According to Table \ref{bbbb}, we easily find the relative distances of (1,2), (1,3), (1,4), (2,3), (2,4), and (3,4) quark pairs are all 0.227 or 0.204 fm.
Meanwhile, the radius of the state is only 0.130 fm.
Thus, in this state, all the distances between the quark pairs are roughly the same order of magnitude apart.
If it is a molecular configuration, the distances between two quarks and two antiquarks should be much greater than the distances in the compact multiquark scheme.
And the radius of molecular configuration can reach several femtometers.
So, our calculations are consistent with the compact tetraquark expectations.

\subsubsection{The internal contribution}

Let us now turn our discussion to the internal mass contribution for the $J^{PC}=0^{++}$ $bb\bar{b}\bar{b}$ ground state.

First, for the kinetic energy, this $bb\bar{b}\bar{b}$ state has 814.0 MeV, which can be understood as the sum
of three internal kinetic energies: kinetic energies of two pairs of the $b-\bar{b}$, and the $(b\bar{b})-(b\bar{b})$ pair.
Accordingly, the sum of the internal kinetic energies of the $\eta_{b}\eta_{b}$ state only comes from the two pairs of the $b-\bar{b}$.
Therefore, this $bb\bar{b}\bar{b}$ state has an additional kinetic energy needed to bring the $\eta_{b}\eta_{b}$ into a compact configuration.
The actual kinetic energies of two pairs of the $b-\bar{b}$ in the the $J^{PC}=0^{++}$ $bb\bar{b}\bar{b}$ ground state are smaller than those
in the $\eta_{b}\eta_{b}$ state.
This is because, as can be seen in Table \ref{bbbb}, the distance of $b-\bar{b}$ is larger  in the tetraquark state  than in the meson: the distance of $b-\bar{b}$ is 0.204 fm in this $bb\bar{b}\bar{b}$ state while it is 0.148 fm in $\eta_{b}$.
Meanwhile, we find that even if we consider the additional kinetic energy between the $(b\bar{b})-(b\bar{b})$ pair, the total kinetic energies in this $bb\bar{b}\bar{b}$ state are still smaller than that in the $\eta_{b}\eta_{b}$ state.
However, this does not lead the ground $J^{PC}=0^{++}$ $bb\bar{b}\bar{b}$ state to a stable state due to the confinement potential part.

As for the confinement potential part, the contributions from $V^{C}$ for the $J^{PC}=0^{++}$ $bb\bar{b}\bar{b}$ ground state in Table \ref{bbbb} are all attractive. Thus, this state has a large positive binding energy.
However, it is still above the meson-meson threshold because the $V^{C}(b\bar{b})$ in $\eta_{b}$ is very attractive.
As for the other internal contributions, the quark contents of this state are the same as the corresponding rearrangement decay threshold.
Moreover, the mass contribution from the hyperfine potential term is negligible compared to the contributions from other terms.

\subsubsection{Comparison with two models of chromomagnetic interaction}

Now, we compare the numerical values for the $cc\bar{c}\bar{c}$ and $bb\bar{b}\bar{b}$ systems between the constituent quark model and two CMI models \cite{Weng:2020jao,Wu:2016vtq} in Tables \ref{cccc}-\ref{bbbb}.
The comparisons of the values for $cc\bar{c}\bar{c}$ and $bb\bar{b}\bar{b}$ states, which are in the last three columns of Tables \ref{cccc}-\ref{bbbb}, can be summarized in the following important conclusions.

First, we find that there is no stable state below the lowest heavy quarkonium pair thresholds in any of the three different models.
In all three models, we consider two possible color configurations, the color-sextet $|(QQ)^{3_{c}}(\bar{Q}\bar{Q})^{\bar{3}_{c}}\rangle$ and the color-triplet $|(QQ)^{6_{c}}(\bar{Q}\bar{Q})^{\bar{6}_{c}}\rangle$.
According to the extended chromomagnetic model \cite{Weng:2020jao}, the ground state is always dominated
by the color-sextet configuration.
This view is consistent with the specific wave function of the ground state in Eq.(\ref{eq:wave1}) given by the constituent quark model.

In contrast, the obtained masses from the
constituent quark model are systematically larger than
those from the extended CMI model \cite{Weng:2020jao} according to Tables \ref{cccc}-\ref{bbbb}.
Meanwhile, the obtained masses from the
the CMI model \cite{Wu:2016vtq} are obviously larger than
those of the constituent quark model.
Their mass differences are mainly due to the effective quark masses as given in the last three columns of Tables \ref{cccc}-\ref{bbbb}.
The effective quark masses are the sum of the quark mass, the relevant kinetic term, and
all the relevant interaction terms in the constituent quark model,
which indeed seems to approximately reproduce the effective quark mass from two CMI model \cite{Weng:2020jao,Wu:2016vtq}.
We compare the subtotal values of $c$ and $\bar{c}$ quark part in Table \ref{cccc2}.
The $c$ effective quark mass in constituent quark model is 3225 MeV, which is  about 100 MeV larger than that of the extended CMI model in the $J^{PC}=0^{++}$ $cc\bar{c}\bar{c}$ state.
Correspondingly, we also find that the $c$ effective quark mass in the CMI model \cite{Wu:2016vtq} is 3450 MeV and about 200 MeV larger than that of constituent quark model.
The effective quark masses in extended CMI model depend on the parameters of the traditional hadron.
However, the effective quark masses should be different
depending on whether they are inside a meson, a baryon, or a tetraquark.
The effective quark masses trend to be large when they are inside configurations with larger constituents
in the extended CMI model, as can be seen from the comparisons of the last three columns in Tables \ref{cccc}-\ref{bbbb}.
Moreover, we notice the similar situation also occurs in the $ud\bar{c}\bar{c}$ state in Table X of Ref. \cite{Park:2018wjk}.

\begin{table*}
\caption{ The masses, binding energies,  variational parameters, the internal contribution, total wave functions, magnetic moments, transition magnetic moments, radiative decay widths, rearrangement strong width ratios, and the relative lengths between the quarks for the $J^{PC}=0^{++}$, $1^{+-}$ $cc\bar{c}\bar{c}$ states and their lowest meson-meson thresholds.
Here, $(i,j)$ denotes the contribution of the $i$-th and $j$-th quarks.
The number is given as i=1 and 2 for the quarks, 3 and 4 for the antiquark.
The masses and corresponding contributions are given in units of MeV, and the relative lengths (variational parameters) are in units of fm ($\rm fm^{-2}$).
Meanwhile, we give a comparison with the other two CMI models \cite{Weng:2020jao, Wu:2016vtq}, to further secure the effective quark mass.
}\label{cccc}
\begin{lrbox}{\tablebox}
\renewcommand\arraystretch{1.42}
\renewcommand\tabcolsep{2.3pt}
\begin{tabular}{c|c|ccc !{\color{black}\vrule width 1.5pt} ccc !{\color{black}\vrule width 1.5pt}c|cc|c|cc}
\midrule[1.5pt]
\toprule[0.5pt]
\multicolumn{1}{c}{$cc\bar{c}\bar{c}$}&\multicolumn{4}{r!{\color{black}\vrule width 1.5pt}}{The contribution from each term}&\multicolumn{3}{c !{\color{black}\vrule width 1.5pt}}{Relative Lengths (fm)}&\multirow{2}*{Overall}&\multicolumn{2}{c|}{Present Work}&\multicolumn{2}{c}{CMI Model}\\
\Xcline{1-8}{0.5pt}\Xcline{10-13}{0.5pt}
\multicolumn{1}{r}{$J^{PC}=0^{++}$}&&Value&\multicolumn{1}{r}{$\eta_{c}\eta_{c}$}&
\multicolumn{1}{r!{\color{black}\vrule width 1.5pt}}{Difference}&$(i,j)$&\multicolumn{1}{r}{Value}
&\multicolumn{1}{c!{\color{black}\vrule width 1.5pt}}{$\eta_{c}\eta_{c}$}&&Contribution&Value&Ref. \cite{Weng:2020jao}&Ref. \cite{Wu:2016vtq}\\ \Xcline{1-13}{0.5pt}
\multicolumn{2}{c}{Mass/$B_{T}$}&\multicolumn{1}{r}{6384.4}&5997.0&387.4&(1,2)&0.406&&
\multirow{8}*{\makecell[c]{$c$-quark:\\ \\$m^{eff}_{c}$}}&2$m_{c}$&3836.0&\\
\Xcline{1-5}{0.5pt}
\multirow{2}*{\makecell[c]{Variational\\ Parameters\\ (fm$^{-2}$)}}&\multirow{2}*{\makecell[c]{$C_{11}$\\$C_{22}$ \\ $C_{33}$}}&\multicolumn{1}{r}{\multirow{2}*{\makecell[c]{7.7\\7.7\\11.4}}}&\multicolumn{1}{r}{\multirow{2}*{\makecell[c]{15.0\\15.9\\-}}}&&(1.3)&0.371&\multicolumn{1}{c!{\color{black}\vrule width 1.5pt}}{0.290($\eta_{c}$)}&
&$\frac{\textbf{p}^{2}_{x_{1}}}{2m'_{1}}$&233.9&\\
&&&&&(2,3)&0.371&&&
$\frac{m_{\bar{c}}}{m_{c}+m_{\bar{c}}}\frac{\textbf{p}^{2}_{x_{3}}}{2m'_{3}}$&123.1&\\
\Xcline{1-5}{0.5pt}
\multicolumn{2}{c}{Quark Mass}&\multicolumn{1}{r}{7672.0}&\multicolumn{1}{r}{7672.0}&0.0&(1,4)&0.371&&&$V^{C}(12)$&\multirow{1}*{-6.8}&$-\frac{1}{4}$$m_{cc}$\\
\multicolumn{2}{c}{\multirow{1}{*}{Confinement Potential}}&\multirow{1}{*}{-2083.8}&\multirow{1}{*}{-2440.4}&\multirow{1}{*}{356.6}&(2,4)&0.371&
\multicolumn{1}{c !{\color{black}\vrule width 1.5pt}}{0.290($\eta_{c}$)}&&\multirow{1}*{\makecell[l]{$\frac{1}{2}[V^{C}(13)+V^{C}(14)$}}&
\multirow{2}*{\makecell[l]{-52.1}}&-792.9\\
\Xcline{1-5}{0.5pt}
\multicolumn{2}{c}{\multirow{1}*{Kinetic Energy}}&\multicolumn{1}{r}{\multirow{1}*{814.0}}&\multicolumn{1}{r}{915.1}&\multirow{1}*{-101.1}
&(3,4)&0.406&&&$+V^{C}(23)+V^{C}(24)]$&&$\frac54m_{c\bar{c}}$&$2m_{c}$\\
\Xcline{6-8}{0.5pt}
\multicolumn{2}{c}{\multirow{1}*{CS Interaction}}&\multicolumn{1}{r}{\multirow{1}*{22.7}}&\multirow{1}*{-150.0}&\multirow{1}*{172.7}&\multicolumn{2}{r}{(1,2)-(3,4):}&0.235 fm
&&$-D$&-983.0&3835.6&3449.6\\
\Xcline{10-13}{0.5pt}\Xcline{1-5}{0.5pt}
\multicolumn{1}{c|}{\multirow{4}{*}{$V^{C}$}}&\multicolumn{1}{c}{\multirow{1}{*}{(1,2)}}&\multicolumn{1}{r}{-6.8}&&
&\multicolumn{2}{r}{Radius:}&0.235 fm&&\multirow{1}*{Subtotal}&\multirow{1}*{3151.1}&
\multirow{1}*{3042.7}&\multirow{1}*{3449.6}\\
\Xcline{9-13}{0.5pt}\Xcline{6-8}{1.6pt}
&\multicolumn{1}{c}{\multirow{1}{*}{(2,3)}}&\multicolumn{1}{r}{-26.1}&&\multicolumn{1}{c}{}&&&&
\multirow{8}*{\makecell[c]{$\bar{c}$-quark:\\ \\$m^{eff}_{\bar{c}}$}}&$2m_{c}$&3836.0&\\
&\multicolumn{1}{c}{\multirow{1}{*}{(1,4)}}&\multicolumn{1}{r}{-26.1}&&\multicolumn{1}{r}{}
&(1,3)&-26.1&$-237.2(\eta_{c})$&&$\frac{\textbf{p}^{2}_{x_{2}}}{2m'_{2}}$&233.9&\\
\Xcline{2-5}{0.5pt}
\multicolumn{1}{c|}{\multirow{1}{*}{}}&\multicolumn{1}{c}{\multirow{1}{*}{Subtotal}}&\multicolumn{1}{r}{-117.8}&\multicolumn{1}{r}{-474.4}&\multicolumn{1}{c|}{356.6}
&\multicolumn{1}{c}{(3,4)}&-6.8&&&
$\frac{m_{\bar{c}}}{m_{c}+m_{\bar{c}}}\frac{\textbf{p}^{2}_{x_{3}}}{2m'_{3}}$&123.1&\\
\Xcline{1-5}{0.5pt}
\multicolumn{2}{c}{\multirow{1}*{Total Contribution}}&\multicolumn{1}{r}{\multirow{1}*{718.9}}&\multicolumn{1}{r}{291.0}&\multicolumn{1}{c|}{\multirow{1}*{427.9}}&(2,4)&-26.1&$-237.2(\eta_{c})$&&$V^{C}(34)$&\multirow{1}*{-6.8}&$-\frac{1}{4}$$m_{cc}$\\
\Xcline{1-8}{1.6pt}
\multicolumn{7}{l}{Total Wave function:}&&
&\multirow{1}*{\makecell[l]{$\frac{1}{2}[V^{C}(13)+V^{C}(14)$}}&
\multirow{2}*{\makecell[l]{-52.1}}&-792.4\\
\multicolumn{8}{r!{\color{black}\vrule width 1.5pt}}{$\Psi_{tot}=0.535|F\rangle|R^{s}\rangle|[\phi_{1}\chi_{5}]\rangle-0.845|F\rangle|R^{s}\rangle|[\phi_{2}\chi_{6}]\rangle=0.612|F\rangle|R^{s}\rangle|[\psi_{1}\zeta_{5}]\rangle$}&&$+V^{C}(23)+V^{C}(24)]$&&$\frac54m_{c\bar{c}}$&$2m_{c}$\\
\multicolumn{8}{r!{\color{black}\vrule width 1.5pt}}{$+0.443|F\rangle|R^{s}\rangle|[\psi_{1}\zeta_{6}]\rangle=0.612|F\rangle|R^{s}\rangle|[\psi_{1}\zeta_{5}]\rangle+0.443|F\rangle|R^{s}\rangle|[\psi_{1}\zeta_{6}]\rangle$}&
&$-D$&-983.0&3835.6&3449.6\\
\Xcline{1-8}{0.5pt}\Xcline{10-13}{0.5pt}
\multicolumn{8}{l!{\color{black}\vrule width 1.5pt}}{The rearrangement strong width ratios:}
&&
\multirow{1}*{Subtotal}&\multirow{1}*{3151.1}&3042.7&\multirow{1}*{3449.6}\\
\Xcline{9-13}{0.5pt}
\multicolumn{8}{l!{\color{black}\vrule width 1.5pt}}{$\Gamma_{T_{c^{2}\bar{c}^{2}}(6384.3,0^{++})\rightarrow J/\psi J/\psi}:\Gamma_{T_{c^{2}\bar{c}^{2}}(6384.3,0^{++})\rightarrow\eta_{c}\eta_{c}}=1:2.8$}
&\multirow{5}*{\makecell[c]{CS \\Interaction}}&\multirow{2}*{$\frac{3}{4}V^{SS}(12)$}&\multirow{2}*{11.4}&$4v_{cc}$&$4C_{cc}$\\
\Xcline{1-8}{0.5pt}
\multicolumn{8}{l!{\color{black}\vrule width 1.5pt}}{The radiative decay widths: $\Gamma_{T_{c^{2}\bar{c}^{2}}(6482.7,2^{++})\rightarrow T_{c^{2}\bar{c}^{2}}(6384.3,0^{++})\Upsilon}=0$ keV}&&&&14.2&21.2\\
\multicolumn{8}{r!{\color{black}\vrule width 1.5pt}}{$\Gamma_{T_{c^{2}\bar{c}^{2}}(6451.5,1^{+-})\rightarrow T_{c^{2}\bar{c}^{2}}(6384.3,0^{++})\Upsilon}=238.1$ keV \quad }&&\multirow{2}*{$\frac{3}{4}V^{SS}(12)$}&\multirow{2}*{11.4}&$4v_{\bar{c}\bar{c}}$&$4C_{\bar{c}\bar{c}}$\\
\Xcline{1-8}{0.5pt}
\multicolumn{3}{l}{The magnetic moments:}& \multicolumn{5}{r !{\color{black}\vrule width 1.5pt}}{$\mu_{T_{c^{2}\bar{c}^{2}}(6384.3,0^{++})}=\langle\Psi^{0^{++}}_{tot}|\hat{\mu^{z}}|\Psi^{0^{++}}_{tot}\rangle=0$ \quad}&&&&14.2&21.2\\
\Xcline{10-13}{0.5pt}\Xcline{1-8}{0.5pt}
\multicolumn{8}{l !{\color{black}\vrule width 1.5pt}}{The transition magnetic moments:}
&&\multirow{1}*{Subtotal}&\multirow{1}*{22.7}&\multirow{1}*{28.4}&42.4\\
\Xcline{9-13}{0.5pt}
\multicolumn{8}{l !{\color{black}\vrule width 1.5pt}}{$\mu_{T_{c^{2}\bar{c}^{2}}(6451.5,1^{+-})\rightarrow T_{c^{2}\bar{c}^{2}}(6384.3,0^{++})\gamma}=\langle\Psi^{1^{+-}}_{tot}|\hat{\mu^{z}}|\Psi^{0^{++}}_{tot}\rangle=0.671$ $\mu_{N}$ }&\multicolumn{2}{l|}{\multirow{1}*{Matrix nondiagonal element}}&\multirow{1}*{-40.5}&-60.9&159.2\\
\Xcline{9-13}{0.5pt}
\multicolumn{8}{l !{\color{black}\vrule width 1.5pt}}{$\mu_{T_{c^{2}\bar{c}^{2}}(6482.7,2^{++})\rightarrow T_{c^{2}\bar{c}^{2}}(6384.3,0^{++})\gamma}=\langle\Psi^{2^{++}}_{tot}|\hat{\mu^{z}}|\Psi^{0^{++}}_{tot}\rangle=0$ }&
\multicolumn{2}{l|}{\multirow{1}*{Total contribution}}&\multirow{1}*{6384.4}&\multirow{1}*{6044.9}&\multirow{1}*{7016.0}\\
\toprule[1.5pt]
\multicolumn{1}{r}{$J^{PC}=1^{+-}$}&&Value&\multicolumn{1}{r}{$J/\psi\eta_{c}$}&
\multicolumn{1}{r!{\color{black}\vrule width 1.5pt}}{Difference}&\multicolumn{3}{c!{\color{black}\vrule width 1.5pt}}{Relative Lengths (fm)}&&Contribution&Value&Ref. \cite{Weng:2020jao}&Ref. \cite{Wu:2016vtq}\\ \Xcline{1-13}{0.5pt}
\multicolumn{2}{c}{Mass/$B_{T}$}&\multicolumn{1}{r}{6451.5}&6090.7&360.8&(i,j)&
\multicolumn{1}{r}{Value}&\multicolumn{1}{c !{\color{black}\vrule width 1.5pt}}{$J/\psi\eta_{c}$}&
\multirow{7}*{\makecell[c]{$c$-quark:\\ \\$m^{eff}_{c}$}}&2$m_{c}$&3836.0&\\
\Xcline{1-8}{0.5pt}
\multirow{2}*{\makecell[c]{Variational\\ Parameters\\ (fm$^{-2}$)}}&\multirow{2}*{\makecell[c]{$C_{11}$\\$C_{22}$ \\ $C_{33}$}}&\multicolumn{1}{r}{\multirow{2}*{\makecell[c]{9.1\\9.1\\7.3}}}&
\multicolumn{1}{r}{\multirow{2}*{\makecell[c]{15.0\\12.5\\-}}}&&(1.2)&0.373&&
&$\frac{\textbf{p}^{2}_{x_{1}}}{2m'_{1}}$&277.2&\\
&&&&&(1,3)&0.395&\multicolumn{1}{c !{\color{black}\vrule width 1.5pt}}{0.290($\eta_{c}$)}&&
$\frac{m_{\bar{c}}}{m_{c}+m_{\bar{c}}}\frac{\textbf{p}^{2}_{x_{3}}}{2m'_{3}}$&111.2&$\frac{1}{2}$$m_{cc}$\\
\Xcline{1-5}{0.5pt}
\multicolumn{2}{c}{Quark Mass}&\multicolumn{1}{r}{7672.0}&7672.0&0.0&(2,3)&0.395&&
&\multirow{2}*{\makecell[c]{$V^{C}(12)$\\$\frac{1}{2}[V^{C}(13)+V^{C}(14)$\\$+V^{C}(23)+V^{C}(24)]$}}
&\multirow{2}*{\makecell[c]{-19.4\\\\3.0}}&1585.8\\
\multicolumn{2}{c}{\multirow{1}{*}{Confinement Potential}}&\multirow{1}{*}{-1998.8}&\multirow{1}{*}{-2367.4}&\multirow{1}{*}{368.6}&(1,4)&0.395&&&&&$\frac12m_{c\bar{c}}$&$2m_{c}$\\
\Xcline{1-5}{0.5pt}
\multicolumn{2}{c}{\multirow{1}*{Kinetic Energy}}&\multicolumn{1}{r}{\multirow{1}*{767.2}}&\multirow{1}*{839.0}&\multicolumn{1}{c!{\color{black}\vrule width 1.5pt}}{-71.8}&(2,4)&0.395&
\multicolumn{1}{c!{\color{black}\vrule width 1.5pt}}{0.318($J/\psi$)}&&$-D$&-983.0&1534.3&3449.6\\
\Xcline{10-13}{0.5pt}
\multicolumn{2}{c}{\multirow{1}*{CS Interaction}}&\multicolumn{1}{r}{\multirow{1}*{1.5}}&\multirow{1}*{-53.9}&
\multirow{1}*{54.4}&(3,4)&0.373&&&\multirow{1}*{Subtotal}&\multirow{1}*{3225.0}&
\multirow{1}*{3120.0}&\multirow{1}*{3449.6}\\
\Xcline{6-8}{0.5pt}\Xcline{9-13}{0.5pt}\Xcline{1-5}{0.5pt}
\multicolumn{1}{c|}{\multirow{4}{*}{$V^{C}$}}&\multicolumn{1}{c}{\multirow{1}{*}{(1,2)}}&\multicolumn{1}{r}{-19.4}&&
&\multicolumn{2}{r}{(1,2)-(3,4):}&0.294 fm&\multirow{8}*{\makecell[c]{$\bar{c}$-quark:\\ \\$m^{eff}_{\bar{c}}$}}&$2m_{c}$&3836.0&\\
&\multicolumn{1}{c}{\multirow{1}{*}{(2,3)}}&\multicolumn{1}{r}{1.5}&&\multicolumn{1}{c!{\color{black}\vrule width 1.5pt}}{}
&\multicolumn{2}{r}{Radius:}&0.235 fm&&$\frac{\textbf{p}^{2}_{x_{2}}}{2m'_{2}}$&277.2&\\
\Xcline{6-8}{1.6pt}
&\multicolumn{1}{c}{\multirow{1}{*}{(1,4)}}&\multicolumn{1}{r}{1.5}&&\multicolumn{1}{r}{}
&(1,3)&1.5&$-237.2(\eta_{c})$&&
$\frac{m_{\bar{c}}}{m_{c}+m_{\bar{c}}}\frac{\textbf{p}^{2}_{x_{3}}}{2m'_{3}}$&111.2&$\frac12 m_{cc}$\\
\Xcline{2-5}{0.5pt}
\multicolumn{1}{c|}{\multirow{1}{*}{}}&\multicolumn{1}{c}{\multirow{1}{*}{Subtotal}}&\multicolumn{1}{r}{-32.8}&\multicolumn{1}{r}{-401.4}&\multicolumn{1}{c|}{368.6}
&\multicolumn{1}{c}{(3,4)}&-19.4&&
&\multirow{2}*{\makecell[c]{$V^{C}(12)$\\$\frac{1}{2}[V^{C}(13)+V^{C}(14)$\\$+V^{C}(23)+V^{C}(24)]$}}
&\multirow{2}*{\makecell[c]{-19.4\\\\3.0}}&1585.8\\
\Xcline{1-5}{0.5pt}
\multicolumn{2}{c}{Total Contribution}&\multicolumn{1}{r}{\multirow{1}*{718.9}}&\multicolumn{1}{r}{291.0}&\multicolumn{1}{c|}{\multirow{1}*{427.9}}&(2,4)&-1.5&$-164.2(J/\psi)$&
&&&$\frac{1}{2}$$m_{c\bar{c}}$&$2m_{c}$\\
\Xcline{1-8}{1.6pt}
\multicolumn{2}{l}{Total Wave function:}&
\multicolumn{6}{r!{\color{black}\vrule width 1.5pt}}{
$\Psi_{tot}=|F\rangle|R^{s}\rangle|[\phi_{1}\chi_{2}]\rangle=-0.408|F\rangle|R^{s}\rangle|[\psi_{1}\zeta_{2}]\rangle$
}
&$-D$&-983.0&1534.3&3449.6\\
\Xcline{10-13}{0.5pt}
\multicolumn{8}{r!{\color{black}\vrule width 1.5pt}}{$-0.408|F\rangle|R^{s}\rangle|[\psi_{1}\zeta_{3}]\rangle+0.577|F\rangle|R^{s}\rangle|[\psi_{2}\zeta_{5}]\rangle+0.577|F\rangle|R^{s}\rangle|[\psi_{2}\zeta_{6}]\rangle$}&
&\multirow{1}*{Subtotal}&\multirow{1}*{3225.0}&
\multirow{1}*{3120.0}&\multirow{1}*{3449.6}\\
\Xcline{9-13}{0.5pt}\Xcline{1-8}{0.5pt}
\multicolumn{8}{l!{\color{black}\vrule width 1.5pt}}{The rearrangement strong decay channel: \quad $J/\psi\eta_{c}$}&\multirow{7}*{\makecell[c]{CS \\Interaction}}&\multirow{2}*{\makecell[c]{$\frac{1}{2}V^{SS}(12)$
}}&\multirow{2}*{8.7}&$\frac83v_{cc}$&$\frac83C_{cc}$\\
\Xcline{1-8}{0.5pt}
\multicolumn{8}{l!{\color{black}\vrule width 1.5pt}}{
The radiative decay widths:}&&&&9.5&14.1\\
\multicolumn{8}{l !{\color{black}\vrule width 1.5pt}}{$\Gamma_{T_{c^{2}\bar{c}^{2}}(6482.7,2^{++})\rightarrow T_{c^{2}\bar{c}^{2}}(6451.5,1^{+-})\Upsilon}=70.4$ keV}&&\multirow{2}*{\makecell[c]{$\frac{1}{2}V^{SS}(34)$
}}&\multirow{2}*{8.7}&$\frac83v_{\bar{c}\bar{c}}$&$\frac83C_{\bar{c}\bar{c}}$\\
\multicolumn{8}{l !{\color{black}\vrule width 1.5pt}}{$\Gamma_{T_{c^{2}\bar{c}^{2}}(6451.5,1^{+-})\rightarrow T_{c^{2}\bar{c}^{2}}(6384.3,0^{++})\Upsilon}=238.1$ keV}&&\multirow{2}*{}&\multirow{2}*{}
&9.5&14.1\\
\Xcline{1-8}{0.3pt}
\multicolumn{8}{l !{\color{black}\vrule width 1.5pt}}{The magnetic moments: \quad $\mu_{T_{c^{2}\bar{c}^{2}}(6451.5,1^{+-})}=\langle\Psi^{1^{+-}}_{tot}|\hat{\mu^{z}}|\Psi^{1^{+-}}_{tot}\rangle=0$}&&\multirow{2}*{\makecell[c]{
$-\frac{1}{4}(V^{SS}(13)+V^{SS}(14)$\\ $+V^{SS}(23)+V^{SS}(24)$}}&\multirow{2}*{-15.8}&$-\frac{16}{3}v_{c\bar{c}}$&$-\frac{16}{3}C_{c\bar{c}}$\\
\Xcline{1-8}{0.3pt}
\multicolumn{8}{l!{\color{black}\vrule width 1.5pt}}{The transition magnetic moments:}
&&&\multirow{1}*{}&\multirow{1}*{-28.4}&-28.2\\
\Xcline{10-13}{0.5pt}
\multicolumn{8}{l!{\color{black}\vrule width 1.5pt}}{$\mu_{T_{c^{2}\bar{c}^{2}}(6482.7,2^{++})\rightarrow T_{c^{2}\bar{c}^{2}}(6451.4,1^{+-})\gamma}=\langle\Psi^{2^{++}}_{tot}|\hat{\mu^{z}}|\Psi^{1^{+-}}_{tot}\rangle=0.750 \mu_{N}$}&&\multirow{1}*{Subtotal}&\multirow{1}*{1.5}&\multirow{1}*{-9.5}&0.0\\
\Xcline{9-13}{0.5pt}
\multicolumn{8}{l!{\color{black}\vrule width 1.5pt}}{$\mu_{T_{c^{2}\bar{c}^{2}}(6451.5,1^{+-})\rightarrow T_{c^{2}\bar{c}^{2}}(6384.3,0^{++})\gamma}=\langle\Psi^{1^{+-}}_{tot}|\hat{\mu^{z}}|\Psi^{0^{++}}_{tot}\rangle=0.335\mu_{N}$}&
\multicolumn{2}{l|}{\multirow{1}*{Total contribution}}&\multirow{1}*{6451.5}&\multirow{1}*{6231.0}&\multirow{1}*{6899.0}\\
\toprule[0.5pt]
\toprule[1.5pt]
\end{tabular}
\end{lrbox}\scalebox{0.867}{\usebox{\tablebox}}
\end{table*}

\begin{table*}
\caption{
The masses, binding energies,  variational parameters, the internal contribution, total wave functions, magnetic moments, transition magnetic moments, radiative decay widths, rearrangement strong width ratios, and the relative lengths between quarks for the $J^{PC}=2^{++}$ $cc\bar{c}\bar{c}$ and $bb\bar{b}\bar{b}$ states and their lowest meson-meson thresholds. The notation is the  same as that in Table \ref{cccc}.
}\label{cccc2}
\begin{lrbox}{\tablebox}
\renewcommand\arraystretch{1.45}
\renewcommand\tabcolsep{2.3pt}
\begin{tabular}{c|c|ccc  !{\color{black}\vrule width 1.5pt} ccc  !{\color{black}\vrule width 1.5pt} c|cc|c|cc}
\midrule[1.5pt]
\toprule[0.5pt]
\multicolumn{1}{c}{$cc\bar{c}\bar{c}$}&\multicolumn{4}{r  !{\color{black}\vrule width 1.5pt}}{The contribution from each term}&\multicolumn{3}{c  !{\color{black}\vrule width 1.5pt}}{Relative Lengths (fm)}&\multirow{2}*{Overall}&\multicolumn{2}{c}{Present Work}&\multicolumn{2}{c}{CMI Model}\\
\Xcline{1-8}{0.5pt}\Xcline{10-13}{0.5pt}
\multicolumn{1}{r}{$J^{PC}=2^{++}$}&&Value&\multicolumn{1}{r}{$J/\psi J/\psi$}&
\multicolumn{1}{r !{\color{black}\vrule width 1.5pt}}{Difference}&$(i,j)$&\multicolumn{1}{r}{Vaule}
&\multicolumn{1}{c  !{\color{black}\vrule width 1.5pt}}{$J/\psi J/\psi$}&&Contribution&Value&Ref. \cite{Weng:2020jao}&Ref. \cite{Wu:2016vtq}\\ \Xcline{1-13}{0.5pt}
\multicolumn{2}{c}{Mass/$B_{T}$}&\multicolumn{1}{r}{6482.7}&6184.5&298.2&(1,2)&0.377&&
\multirow{7}*{\makecell[c]{$c$-quark:\\ \\$m^{eff}_{c}$}}&2$m_{c}$&3836.0&\\
\Xcline{1-5}{0.5pt}
\multirow{2}*{\makecell[c]{Variational\\ Parameters\\ (fm$^{-2}$)}}&\multirow{2}*{\makecell[c]{$C_{11}$\\$C_{22}$ \\ $C_{33}$}}&\multicolumn{1}{r}{\multirow{2}*{\makecell[c]{8.9\\8.9\\6.9}}}&
\multicolumn{1}{r}{\multirow{2}*{\makecell[c]{12.5\\12.5\\-}}}&&(1.3)&0.403&
\multicolumn{1}{c  !{\color{black}\vrule width 1.5pt}}{0.318($J/\psi$)}&
&$\frac{\textbf{p}^{2}_{x_{1}}}{2m'_{1}}$&270.4&\\
&&&&&(2,3)&0.403&&&
$\frac{m_{\bar{c}}}{m_{c}+m_{\bar{c}}}\frac{\textbf{p}^{2}_{x_{3}}}{2m'_{3}}$&105.6&$\frac{1}{2}$$m_{cc}$\\
\Xcline{1-5}{0.5pt}
\multicolumn{2}{c}{Quark Mass}&\multicolumn{1}{r}{7672.0}&7672.0&0.0&(1,4)&0.403&&
&\multirow{2}*{\makecell[c]{$V^{C}(12)$\\$\frac{1}{2}[V^{C}(13)+V^{C}(14)$\\$+V^{C}(23)+V^{C}(24)]$}}
&\multirow{2}*{\makecell[c]{-14.6\\\\10.8}}&1585.8\\
\multicolumn{2}{c}{\multirow{1}{*}{Confinement Potential}}&\multirow{1}{*}{-1973.6}&\multirow{1}{*}{-2294.4}&\multirow{1}{*}{320.8}&(2,4)&0.403&
\multicolumn{1}{c !{\color{black}\vrule width 1.5pt}}{0.318($J/\psi$)}&&&&$\frac12m_{c\bar{c}}$&$2m_{c}$\\
\Xcline{1-5}{0.5pt}
\multicolumn{2}{c}{\multirow{1}*{Kinetic Energy}}&\multicolumn{1}{r}{\multirow{1}*{752.0}}&\multirow{1}*{769.9}&\multirow{1}*{-10.9}
&(3,4)&0.377&&&$-D$&-983.0&1534.3&3449.6\\
\Xcline{6-8}{0.5pt}\Xcline{10-13}{0.5pt}
\multicolumn{2}{c}{\multirow{1}*{CS Interaction}}&\multicolumn{1}{r}{\multirow{1}*{32.3}}&\multirow{1}*{43.9}&\multirow{1}*{-11.6}
&\multicolumn{2}{r}{(1,2)-(3,4):}&0.302 fm&&\multirow{1}*{Subtotal}&\multirow{1}*{3225.2}&
\multirow{1}*{3120.0}&\multirow{1}*{3449.6}\\
\Xcline{1-5}{0.5pt}\Xcline{9-13}{0.5pt}
\multicolumn{1}{c|}{\multirow{4}{*}{$V^{C}$}}&\multicolumn{1}{c}{\multirow{1}{*}{(1,2)}}&\multicolumn{1}{r}{-14.6}&&
&\multicolumn{2}{r}{Radius:}&0.241 fm&\multirow{8}*{\makecell[c]{$\bar{c}$-quark:\\ \\$m^{eff}_{\bar{c}}$}}&$2m_{c}$&3836.0&\\
\Xcline{6-8}{1.6pt}
&\multicolumn{1}{c}{\multirow{1}{*}{(2,3)}}&\multicolumn{1}{r}{5.4}&&\multicolumn{1}{c}{}&&&&
&$\frac{\textbf{p}^{2}_{x_{2}}}{2m'_{2}}$&270.4&\\
&\multicolumn{1}{c}{\multirow{1}{*}{(1,4)}}&\multicolumn{1}{r}{5.5}&&\multicolumn{1}{r}{}
&(1,3)&5.4&$-164.2(J/\psi)$&&
$\frac{m_{\bar{c}}}{m_{c}+m_{\bar{c}}}\frac{\textbf{p}^{2}_{x_{3}}}{2m'_{3}}$&105.6&$\frac12 m_{cc}$\\
\Xcline{2-5}{0.5pt}
\multicolumn{1}{c|}{\multirow{1}{*}{}}&\multicolumn{1}{c}{\multirow{1}{*}{Subtotal}}&\multicolumn{1}{r}{-7.6}&\multicolumn{1}{r}{-328.4}&\multicolumn{1}{c|}{320.8}
&\multicolumn{1}{c}{(3,4)}&-14.6&
&&\multirow{2}*{\makecell[c]{$V^{C}(12)$\\$\frac{1}{2}[V^{C}(13)+V^{C}(14)$\\$+V^{C}(23)+V^{C}(24)]$}}
&\multirow{2}*{\makecell[c]{-14.6\\\\10.8}}&1585.8\\
\Xcline{1-5}{0.5pt}
\multicolumn{2}{c}{\multirow{1}*{Total Contribution}}&\multicolumn{1}{r}{\multirow{1}*{776.7}}&\multicolumn{1}{r}{478.5}&\multicolumn{1}{c|}{\multirow{1}*{298.2}}&(2,4)&5.4&$-164.2(J/\psi)$&&&
&$\frac{1}{2}$$m_{c\bar{c}}$&$2m_{c}$\\
\Xcline{1-8}{1.6pt}
\multicolumn{7}{l}{Total Wave function:}&
&&$-D$&-983.0&1534.3&3449.6\\
\Xcline{10-13}{0.5pt}
\multicolumn{8}{l!{\color{black}\vrule width 1.5pt} }{$\Psi_{tot}=|F\rangle|R^{s}\rangle|[\phi_{1}\chi_{1}]\rangle=0.577|F\rangle|R^{s}\rangle|[\psi_{1}\zeta_{1}]\rangle-0.816|F\rangle|R^{s}\rangle|[\psi_{2}\zeta_{1}]\rangle$}
&\multirow{1}*{Subtotal}&\multirow{1}*{3225.2}&
\multirow{1}*{3120.0}&\multirow{1}*{3449.6}\\
\Xcline{9-13}{0.5pt}\Xcline{1-8}{0.3pt}
\multicolumn{8}{l!{\color{black}\vrule width 1.5pt}}{The rearrangement strong decay channel:  $J/\psi J/\psi$}&\multirow{7}*{\makecell[c]{CS \\Interaction}}&\multirow{2}*{\makecell[c]{$\frac{1}{2}V^{SS}(12)$
}}&\multirow{2}*{8.5}&$\frac83v_{cc}$&$\frac83C_{cc}$\\
\Xcline{1-8}{0.3pt}
\multicolumn{8}{l!{\color{black}\vrule width 1.5pt}}{The radiative decay widths:
}&&&&9.5&14.1\\
\multicolumn{8}{l!{\color{black}\vrule width 1.5pt}}{$\Gamma_{T_{c^{2}\bar{c}^{2}}(6482.7,2^{++})\rightarrow T_{c^{2}\bar{c}^{2}}(6384.3,0^{++})\Upsilon}=0$ keV }&&\multirow{2}*{\makecell[c]{$\frac{1}{2}V^{SS}(34)$
}}&\multirow{2}*{8.5}&$\frac83v_{\bar{c}\bar{c}}$&$\frac83C_{\bar{c}\bar{c}}$\\
\multicolumn{8}{l!{\color{black}\vrule width 1.5pt}}{$\Gamma_{T_{c^{2}\bar{c}^{2}}(6482.7,2^{++})\rightarrow T_{c^{2}\bar{c}^{2}}(6451.5,1^{+-})\Upsilon}=70.4$ keV}&&\multirow{2}*{}&\multirow{2}*{}
&9.5&14.1\\
\Xcline{1-8}{0.3pt}
\multicolumn{8}{l!{\color{black}\vrule width 1.5pt}}{The magnetic moments: \quad $\mu_{T_{c^{2}\bar{c}^{2}}(6451.5,1^{+-})}=\langle\Psi^{1^{+-}}_{tot}|\hat{\mu^{z}}|\Psi^{1^{+-}}_{tot}\rangle=0$.}&&\multirow{2}*{\makecell[c]{
$\frac{1}{4}(V^{SS}(13)+V^{SS}(14)$\\ $+V^{SS}(23)+V^{SS}(24)$}}&\multirow{2}*{15.3}&$\frac{16}{3}v_{c\bar{c}}$&$\frac{16}{3}C_{c\bar{c}}$\\
\Xcline{1-8}{0.3pt}
\multicolumn{8}{l!{\color{black}\vrule width 1.5pt}}{The transition magnetic moments:}
&&&\multirow{1}*{}&\multirow{1}*{28.4}&28.2\\
\Xcline{10-13}{0.5pt}
\multicolumn{8}{l!{\color{black}\vrule width 1.5pt}}{$\mu_{T_{c^{2}\bar{c}^{2}}(6482.7,2^{++})\rightarrow T_{c^{2}\bar{c}^{2}}(6384.3,0^{++})\gamma}=\langle\Psi^{2^{++}}_{tot}|\hat{\mu^{z}}|\Psi^{0^{++}}_{tot}\rangle=0$
}&&\multirow{1}*{Subtotal}&\multirow{1}*{32.3}&\multirow{1}*{47.4}&56.5\\
\Xcline{9-13}{0.5pt}
\multicolumn{8}{l!{\color{black}\vrule width 1.5pt}}{$\mu_{T_{c^{2}\bar{c}^{2}}(6482.7,2^{++})\rightarrow T_{c^{2}\bar{c}^{2}}(6451.4,1^{+-})\gamma}=\langle\Psi^{2^{++}}_{tot}|\hat{\mu^{z}}|\Psi^{1^{+-}}_{tot}\rangle=\mu_{c}-\mu_{\bar{c}}=0.750 \mu_{N}$}&
\multicolumn{2}{l|}{\multirow{1}*{Total contribution}}&\multirow{1}*{6482.7}&\multirow{1}*{6287.3}&\multirow{1}*{6956.0}\\
\toprule[1.5pt]
\multicolumn{2}{l|}{$bb\bar{b}\bar{b}$ \quad $J^{PC}=2^{++}$}&Value&\multicolumn{1}{c}{$\Upsilon\Upsilon$}&
\multicolumn{1}{r!{\color{black}\vrule width 1.5pt}}{Difference}&\multicolumn{3}{c!{\color{black}\vrule width 1.5pt}}{Relative Lengths (fm)}&Contribution&Value&Ref. \cite{Weng:2020jao}&Ref. \cite{Wu:2016vtq}\\ \Xcline{1-13}{0.5pt}
\multicolumn{2}{c}{Mass/$B_{T}$}&\multicolumn{1}{r}{19327.9}&18938.8&390.1&$(i,j)$&Value&\multicolumn{1}{c!{\color{black}\vrule width 1.5pt}}{$\Upsilon\Upsilon$}&
\multirow{7}*{\makecell[c]{$b$-quark:\\ \\$m^{eff}_{b}$}}&2$m_{b}$&10686.0&\\
\Xcline{1-8}{0.5pt}
\multirow{2}*{\makecell[c]{Variational\\ Parameters\\ (fm$^{-2}$)}}&\multirow{2}*{\makecell[c]{$C_{11}$\\$C_{22}$ \\ $C_{33}$}}&\multicolumn{1}{r}{\multirow{2}*{\makecell[c]{30.0\\30.0\\23.0}}}&
\multicolumn{1}{r}{\multirow{2}*{\makecell[c]{49.4\\49.4\\-}}}&&(1.2)&0.205&
&
&$\frac{\textbf{p}^{2}_{x_{1}}}{2m'_{1}}$&328.0&\\
&&&&&(1,3)&0.220
&\multicolumn{1}{c!{\color{black}\vrule width 1.5pt}}{0.160($\Upsilon$)}&&
$\frac{m_{\bar{b}}}{m_{b}+m_{\bar{b}}}\frac{\textbf{p}^{2}_{x_{3}}}{2m'_{3}}$&126.0&$\frac{1}{2}$$m_{bb}$\\
\Xcline{1-5}{0.5pt}
\multicolumn{2}{c}{Quark Mass}&\multicolumn{1}{r}{21372.0}&21372.0&0.0&(2,3)&0.220&&
&\multirow{2}*{\makecell[c]{$V^{C}(12)$\\$\frac{1}{2}[V^{C}(13)+V^{C}(14)$\\$+V^{C}(23)+V^{C}(24)]$}}
&\multirow{2}*{\makecell[c]{-269.1\\\\-236.6}}&4764.8\\
\multicolumn{2}{c}{\multirow{1}{*}{Confinement Potential}}&\multirow{1}{*}{-2977.3}&\multirow{1}{*}{-3559.5}&\multirow{1}{*}{582.2}&(1,4)&0.220&&&&&$\frac12m_{b\bar{b}}$&$2m_{b}$\\
\Xcline{1-5}{0.5pt}
\multicolumn{2}{c}{\multirow{1}*{Kinetic Energy}}&\multicolumn{1}{r}{\multirow{1}*{908.1}}&\multirow{1}*{1087.3}&\multirow{1}*{-179.2}
&(2,4)&0.220&
\multicolumn{1}{c!{\color{black}\vrule width 1.5pt}}{0.160($\Upsilon$)}&&$-D$&-983.0&4722.5&10105.8\\
\Xcline{10-13}{0.5pt}
\multicolumn{2}{c}{\multirow{1}*{CS Interaction}}&\multicolumn{1}{r}{\multirow{1}*{25.1}}&\multirow{1}*{38.0}&\multirow{1}*{-12.9}
&(3,4)&0.205&&&\multirow{1}*{Subtotal}&\multirow{1}*{9651.3}&
\multirow{1}*{9487.3}&\multirow{1}*{10105.8}\\
\Xcline{6-8}{0.5pt}\Xcline{9-13}{0.5pt}\Xcline{1-5}{0.5pt}
\multicolumn{1}{c|}{\multirow{4}{*}{$V^{C}$}}&\multicolumn{1}{c}{\multirow{1}{*}{(1,2)}}&\multicolumn{1}{r}{-269.1}&&
&\multicolumn{2}{r}{(1,2)-(3,4):}&0.165 fm&\multirow{8}*{\makecell[c]{$\bar{b}$-quark:\\ \\$m^{eff}_{\bar{b}}$}}&$2m_{b}$&10686.0&\\
&\multicolumn{1}{c}{\multirow{1}{*}{(2,3)}}&\multicolumn{1}{r}{-118.3}&&\multicolumn{1}{c!{\color{black}\vrule width 1.5pt}}{}&\multicolumn{2}{r}{Radius:}&0.132 fm&&$\frac{\textbf{p}^{2}_{x_{2}}}{2m'_{2}}$&328.0&\\\Xcline{6-8}{1.6pt}
&\multicolumn{1}{c}{\multirow{1}{*}{(1,4)}}&\multicolumn{1}{r}{-118.3}&&\multicolumn{1}{r}{}&(1,3)&-118.3&$-796.7(\Upsilon)$&
&
$\frac{m_{\bar{b}}}{m_{b}+m_{\bar{b}}}\frac{\textbf{p}^{2}_{x_{3}}}{2m'_{3}}$&126.0&$\frac12 m_{bb}$\\
\Xcline{2-5}{0.5pt}
\multicolumn{1}{c|}{\multirow{1}{*}{}}&\multicolumn{1}{c}{\multirow{1}{*}{Subtotal}}&\multicolumn{1}{r}{-1011.3}&\multicolumn{1}{r}{-1593.5}&\multicolumn{1}{c|}{582.2}
&\multicolumn{1}{c}{(3,4)}&-269.1&&
&\multirow{2}*{\makecell[c]{$V^{C}(12)$\\$\frac{1}{2}[V^{C}(13)+V^{C}(14)$\\$+V^{C}(23)+V^{C}(24)]$}}
&\multirow{2}*{\makecell[c]{-269.1\\\\-236.6}}&4764.8\\
\Xcline{1-5}{0.5pt}
\multicolumn{2}{c}{Total Contribution}&\multicolumn{1}{r}{\multirow{1}*{-78.1}}&\multicolumn{1}{r}{-468.2}&\multicolumn{1}{c|}{\multirow{1}*{390.1}}&(2,4)&-118.3&$-796.7(\Upsilon)$&
&&&$\frac{1}{2}$$m_{b\bar{b}}$&$2m_{b}$\\
\Xcline{1-8}{1.6pt}
\multicolumn{8}{l!{\color{black}\vrule width 1.5pt}}{Total Wave function:}&
&$-D$&-983.0&4722.5&10105.8\\
\Xcline{10-13}{0.5pt}
\multicolumn{8}{l!{\color{black}\vrule width 1.5pt}}{$\Psi_{tot}=|F\rangle|R^{s}\rangle|[\phi_{1}\chi_{2}]\rangle=0.577|F\rangle|R^{s}\rangle|[\psi_{1}\zeta_{1}]\rangle-0.816|F\rangle|R^{s}\rangle|[\psi_{2}\zeta_{1}]\rangle$}&
&\multirow{1}*{Subtotal}&\multirow{1}*{9651.3}&
\multirow{1}*{9487.3}&\multirow{1}*{10105.8}\\
\Xcline{1-8}{0.3pt}
\Xcline{9-13}{0.5pt}
\multicolumn{8}{l!{\color{black}\vrule width 1.5pt}}{The rearrangement strong decay channel:  $B^{*}_{c}B^{*}_{c}$}&\multirow{7}*{\makecell[c]{CS \\Interaction}}&\multirow{2}*{\makecell[c]{$\frac{1}{2}V^{SS}(12)$
}}&\multirow{2}*{6.6}&$\frac83v_{bb}$&$\frac83C_{bb}$\\
\Xcline{1-8}{0.3pt}
\multicolumn{8}{l!{\color{black}\vrule width 1.5pt}}{The radiative decay widths:}&&&&5.1&7.7\\
\multicolumn{8}{l!{\color{black}\vrule width 1.5pt}}{$\Gamma_{T_{b^{2}\bar{b}^{2}}(19327.9,2^{++})\rightarrow T_{b^{2}\bar{b}^{2}}(19240.0,0^{++})\gamma}=0$ keV}&&\multirow{2}*{\makecell[c]{$\frac{1}{2}V^{SS}(34)$
}}&\multirow{2}*{6.6}&$\frac83v_{\bar{b}\bar{b}}$&$\frac83C_{\bar{b}\bar{b}}$\\
\multicolumn{8}{l!{\color{black}\vrule width 1.5pt}}{$\Gamma_{T_{b^{2}\bar{b}^{2}}(19327.9,2^{++})\rightarrow T_{b^{2}\bar{b}^{2}}(19303.9,1^{+-})\gamma}=1.0$ keV}&&\multirow{2}*{}&\multirow{2}*{}
&5.1&7.7\\
\Xcline{1-8}{0.3pt}
\multicolumn{8}{l!{\color{black}\vrule width 1.5pt}}{The magnetic moments: \quad $\mu_{T_{b^{2}\bar{b}^{2}}(19303.9,1^{+-})}=\langle\Psi^{1^{+-}}_{tot}|\hat{\mu^{z}}|\Psi^{1^{+-}}_{tot}\rangle=0$}&&\multirow{2}*{\makecell[c]{
$\frac{1}{4}(V^{SS}(13)+V^{SS}(14)$\\
$+V^{SS}(23)+V^{SS}(24)$}}&\multirow{2}*{11.9}&$\frac{16}{3}v_{b\bar{b}}$&$\frac{16}{3}C_{b\bar{b}}$\\
\Xcline{1-8}{0.3pt}
\multicolumn{8}{l!{\color{black}\vrule width 1.5pt}}{The transition magnetic moments:}
&&&\multirow{1}*{}&\multirow{1}*{15.3}&15.5\\
\Xcline{10-13}{0.5pt}
\multicolumn{8}{l!{\color{black}\vrule width 1.5pt}}{$\mu_{T_{b^{2}\bar{b}^{2}}(19327.9,2^{++})\rightarrow T_{b^{2}\bar{b}^{2}}(19240.0,0^{++})\gamma}=\langle\Psi^{2^{++}}_{tot}|\hat{\mu^{z}}|\Psi^{0^{++}}_{tot}\rangle=0$}&&\multirow{1}*{Subtotal}&\multirow{1}*{25.1}&\multirow{1}*{25.5}&30.9\\
\Xcline{9-13}{0.5pt}
\multicolumn{8}{l!{\color{black}\vrule width 1.5pt}}{$\mu_{T_{b^{2}\bar{b}^{2}}(19327.9,2^{++})\rightarrow T_{b^{2}\bar{b}^{2}}(19303.9,1^{+-})\gamma}=\langle\Psi^{2^{++}}_{tot}|\hat{\mu^{z}}|\Psi^{1^{+-}}_{tot}\rangle=\mu_{b}-\mu_{\bar{b}}=-0.125 \mu_{N}$}&
\multicolumn{2}{l|}{\multirow{1}*{Total contribution}}&\multirow{1}*{19327.9}&\multirow{1}*{19000.1}&\multirow{1}*{20243.0}\\
\toprule[0.5pt]
\toprule[1.5pt]
\end{tabular}
\end{lrbox}\scalebox{0.87}{\usebox{\tablebox}}
\end{table*}

\begin{table*}
\caption{ The masses, binding energies,  variational parameters, the internal contribution, total wave functions, magnetic moments, transition magnetic moments, radiative decay widths, rearrangement strong width ratios, and the relative lengths between quarks for the $J^{PC}=0^{++}$, $1^{+-}$ $bb\bar{b}\bar{b}$ states and their lowest meson-meson thresholds. The notation is the same as that in Table \ref{cccc}.
}\label{bbbb}
\begin{lrbox}{\tablebox}
\renewcommand\arraystretch{1.43}
\renewcommand\tabcolsep{2.3pt}
\begin{tabular}{c|c|ccc !{\color{black}\vrule width 1.5pt} ccc !{\color{black}\vrule width 1.5pt} c|cc|c|cc}
\midrule[1.5pt]
\toprule[0.5pt]
\multicolumn{1}{c}{$bb\bar{b}\bar{b}$}&\multicolumn{4}{r!{\color{black}\vrule width 1.5pt}}{The contribution from each term}&\multicolumn{3}{c!{\color{black}\vrule width 1.5pt}}{Relative Lengths (fm)}&\multirow{2}*{Overall}&\multicolumn{2}{c}{Present Work}&\multicolumn{2}{c}{CMI Model}\\
\Xcline{1-8}{0.5pt}\Xcline{10-13}{0.5pt}
\multicolumn{1}{r}{$J^{PC}=0^{++}$}&&Value&\multicolumn{1}{r}{$\eta_{b}\eta_{b}$}&
\multicolumn{1}{r!{\color{black}\vrule width 1.5pt}}{Difference}&$(i,j)$&\multicolumn{1}{r}{Vaule}
&\multicolumn{1}{c!{\color{black}\vrule width 1.5pt}}{$\eta_{b}\eta_{b}$}&&Contribution&Value&Ref. \cite{Weng:2020jao}&Ref. \cite{Wu:2016vtq}\\ \Xcline{1-13}{0.5pt}
\multicolumn{2}{c}{Mass/$B_{T}$}&\multicolumn{1}{r}{19240.0}&18778.1&461.9&(1,2)&0.227
&&\multirow{8}*{\makecell[c]{$b$-quark:\\ \\$m^{eff}_{b}$}}&2$m_{b}$&10686.0&\\
\Xcline{1-5}{0.5pt}
\multirow{2}*{\makecell[c]{Variational\\ Parameters\\ (fm$^{-2}$)}}&\multirow{2}*{\makecell[c]{$C_{11}$\\$C_{22}$ \\ $C_{33}$}}&\multicolumn{1}{r}{\multirow{2}*{\makecell[c]{24.6\\24.6\\39.5}}}
&\multicolumn{1}{r}{\multirow{2}*{\makecell[c]{57.4\\57.4\\-}}}&&(1.3)&0.204
&\multicolumn{1}{c!{\color{black}\vrule width 1.5pt}}{0.148($\eta_{b}$)}&
&$\frac{\textbf{p}^{2}_{x_{1}}}{2m'_{1}}$&269.2&\\
&&&&&(2,3)&0.204&&&
$\frac{m_{\bar{c}}}{m_{b}+m_{\bar{b}}}\frac{\textbf{p}^{2}_{x_{3}}}{2m'_{3}}$&216.0&\\
\Xcline{1-5}{0.5pt}\multicolumn{2}{c}{Quark Mass}&\multicolumn{1}{r}{21372.0}&21372.0&0.0&(1,4)&0.204&&
&$V^{C}(12)$&\multirow{1}*{111.8}&$-\frac{1}{4}$$m_{bb}$\\
\multicolumn{2}{c}{\multirow{1}{*}{Confinement Potential}}&\multirow{1}{*}{-3101.0}&\multirow{1}{*}{-3724.2}&\multirow{1}{*}{623.2}&(2,4)&0.204&
\multicolumn{1}{c!{\color{black}\vrule width 1.5pt}}{0.148($\eta_{b}$)}&&\multirow{1}*{\makecell[l]{$\frac{1}{2}[V^{C}(13)+V^{C}(14)$}}&
\multirow{2}*{\makecell[l]{-697.4}}&-2382.4\\
\Xcline{1-5}{0.5pt}
\multicolumn{2}{c}{\multirow{1}*{Kinetic Energy}}&\multicolumn{1}{r}{\multirow{1}*{970.4}}&\multirow{1}*{1255.9}&\multirow{1}*{-285.5}
&(3,4)&0.227&&
&$+V^{C}(23)+V^{C}(24)]$&&$\frac54m_{b\bar{b}}$&$2m_{b}$\\
\Xcline{6-8}{0.5pt}
\multicolumn{2}{c}{\multirow{1}*{CS Interaction}}&\multicolumn{1}{r}{\multirow{1}*{17.0}}&\multirow{1}*{-125.5}&\multirow{1}*{142.5}&
\multicolumn{2}{r}{(1,2)-(3,4):}&0.126 fm&&$-D$&-983.0&11806.2&10105.8\\
\Xcline{10-13}{0.5pt}\Xcline{1-5}{0.5pt}
\multicolumn{1}{c|}{\multirow{4}{*}{$V^{C}$}}&\multicolumn{1}{c}{\multirow{1}{*}{(1,2)}}&\multicolumn{1}{r}{-111.8}&&
&\multicolumn{2}{r}{Radius:}&0.130 fm&&\multirow{1}*{Subtotal}&\multirow{1}*{9602.6}&
\multirow{1}*{9423.8}&\multirow{1}*{10105.8}\\
\Xcline{9-13}{0.5pt}\Xcline{6-8}{1.6pt}
&\multicolumn{1}{c}{\multirow{1}{*}{(2,3)}}&\multicolumn{1}{r}{-339.7}&&\multicolumn{1}{c}{}&&&
&\multirow{8}*{\makecell[c]{$\bar{b}$-quark:\\ \\$m^{eff}_{\bar{b}}$}}&$2m_{b}$&10686.0&\\
&\multicolumn{1}{c}{\multirow{1}{*}{(1,4)}}&\multicolumn{1}{r}{-339.7}&&\multicolumn{1}{r}{}
&(1,3)&-339.7&$-879.1(\eta_{b})$&
&$\frac{\textbf{p}^{2}_{x_{2}}}{2m'_{2}}$&216.0&\\
\Xcline{2-5}{0.5pt}
\multicolumn{1}{c|}{\multirow{1}{*}{}}&\multicolumn{1}{c}{\multirow{1}{*}{Subtotal}}&\multicolumn{1}{r}{-1135.0}&\multicolumn{1}{r}{-1758.2}&\multicolumn{1}{c|}{623.2}
&\multicolumn{1}{c}{(3,4)}&-111.8&&
&
$\frac{m_{\bar{c}}}{m_{b}+m_{\bar{b}}}\frac{\textbf{p}^{2}_{x_{3}}}{2m'_{3}}$&216.0&\\
\Xcline{1-5}{0.5pt}
\multicolumn{2}{c}{\multirow{1}*{Total Contribution}}&\multicolumn{1}{r}{\multirow{1}*{-147.6}}&\multicolumn{1}{r}{-627.9}&\multicolumn{1}{c|}{\multirow{1}*{480.3}}&(2,4)&-339.7&$-879.1(\eta_{b})$&
&$V^{C}(34)$&\multirow{1}*{111.8}&$-\frac{1}{4}$$m_{bb}$\\
\Xcline{1-8}{1.6pt}
\multicolumn{8}{l!{\color{black}\vrule width 1.5pt}}{Total Wave function:}
&&\multirow{1}*{\makecell[l]{$\frac{1}{2}[V^{C}(13)+V^{C}(14)$}}&
\multirow{2}*{\makecell[l]{-697.4}}&-2382.4\\
\multicolumn{8}{r!{\color{black}\vrule width 1.5pt}}{$\Psi_{tot}=0.352|F\rangle|R^{s}\rangle|[\phi_{1}\chi_{5}]\rangle-0.936|F\rangle|R^{s}\rangle|[\phi_{2}\chi_{6}]\rangle=0.558|F\rangle|R^{s}\rangle|[\psi_{1}\zeta_{5}]\rangle$}
&&$+V^{C}(23)+V^{C}(24)]$&&$\frac54m_{b\bar{b}}$&$2m_{b}$\\
\multicolumn{8}{r!{\color{black}\vrule width 1.5pt}}{$+0.560|F\rangle|R^{s}\rangle|[\psi_{1}\zeta_{6}]\rangle+0.021|F\rangle|R^{s}\rangle|[\psi_{2}\zeta_{5}]\rangle+0.612|F\rangle|R^{s}\rangle|[\psi_{2}\zeta_{6}]\rangle$}&&$-D$&-983.0&11806.2&10105.8\\
\Xcline{1-8}{0.3pt}\Xcline{10-13}{0.5pt}
\multicolumn{8}{l!{\color{black}\vrule width 1.5pt}}{The rearrangement strong width ratios:}&&
\multirow{1}*{Subtotal}&\multirow{1}*{9602.6}&9423.8&\multirow{1}*{10105.8}\\
\Xcline{9-13}{0.5pt}
\multicolumn{8}{l!{\color{black}\vrule width 1.5pt}}{$\Gamma_{T_{c^{2}\bar{c}^{2}}(19240.0,0^{++})\rightarrow \Upsilon\Upsilon}:\Gamma_{T_{c^{2}\bar{c}^{2}}(1924.0,0^{++})\rightarrow\eta_{b}\eta_{b}}=1:1.2$
}&\multirow{5}*{\makecell[c]{CS \\Interaction}}&\multirow{2}*{$\frac{3}{4}V^{SS}(12)$}&\multirow{2}*{8.5}&$4v_{bb}$&$4C_{bb}$\\
\Xcline{1-8}{0.3pt}
\multicolumn{8}{l!{\color{black}\vrule width 1.5pt}}{The radiative decay widths:}&&&&7.7&11.6\\
\multicolumn{8}{l!{\color{black}\vrule width 1.5pt}}{$\Gamma_{T_{b^{2}\bar{b}^{2}}(19327.9,2^{++})\rightarrow T_{b^{2}\bar{b}^{2}}(19240.0,0^{++})\gamma}=0\, \rm{keV}$}&&\multirow{2}*{$\frac{3}{4}V^{SS}(34)$}&\multirow{2}*{8.5}&$4v_{\bar{b}\bar{b}}$&$4C_{\bar{b}\bar{b}}$\\
\multicolumn{8}{l!{\color{black}\vrule width 1.5pt}}{$\Gamma_{T_{b^{2}\bar{b}^{2}}(19303.9,1^{+-})\rightarrow T_{b^{2}\bar{b}^{2}}(19240.0,0^{++})\gamma}=2.8\, \rm{keV}$}&&&&7.7&11.6\\
\Xcline{10-13}{0.5pt}\Xcline{1-8}{0.3pt}
\multicolumn{8}{l!{\color{black}\vrule width 1.5pt}}{The magnetic moments: \quad $\mu_{T_{b^{2}\bar{b}^{2}}(19240.0,0^{++})}=\langle\Psi^{0^{++}}_{tot}|\hat{\mu^{z}}|\Psi^{0^{++}}_{tot}\rangle=0$}
&&\multirow{1}*{Subtotal}&\multirow{1}*{17.0}&\multirow{1}*{15.4}&23.2\\
\Xcline{9-13}{0.5pt}
\Xcline{1-8}{0.3pt}
\multicolumn{8}{l!{\color{black}\vrule width 1.5pt}}{The transition magnetic moments:}&\multicolumn{2}{l|}{\multirow{1}*{Matrix nondiagonal element}}&\multirow{1}*{17.8}&-27.0&-40.2\\
\Xcline{9-13}{0.5pt}
\multicolumn{8}{l!{\color{black}\vrule width 1.5pt}}{$\mu_{T_{b^{2}\bar{b}^{2}}(19327.9,2^{++})\rightarrow T_{b^{2}\bar{b}^{2}}(19240.0,0^{++})\gamma}=\langle\Psi^{2^{++}}_{tot}|\hat{\mu^{z}}|\Psi^{0^{++}}_{tot}\rangle=0$}&
\multicolumn{2}{l|}{\multirow{1}*{Total contribution}}&\multirow{1}*{19240.0}&\multirow{1}*{18836.0}&\multirow{1}*{20275.0}\\
\Xcline{9-13}{1.6pt}
\multicolumn{13}{l}{$\mu_{T_{b^{2}\bar{b}^{2}}(19393.9,1^{+-})\rightarrow T_{b^{2}\bar{b}^{2}}(19240.0,0^{++})\gamma}=\langle\Psi^{1^{+-}}_{tot}|\hat{\mu^{z}}|\Psi^{0^{++}}_{tot}\rangle=0.352\times\frac{4}{\sqrt{6}}(\mu_{b}-\mu_{\bar{b}})=-0.072\mu_{N}$}\\
\toprule[1.5pt]
\multicolumn{1}{r}{$J^{PC}=1^{+-}$}&&Value&\multicolumn{1}{r}{$\Upsilon\eta_{b}$}&
\multicolumn{1}{r!{\color{black}\vrule width 1.5pt}}{Difference}&\multicolumn{3}{c!{\color{black}\vrule width 1.5pt}}{Relative Lengths (fm)}&&Contribution&Value&Ref. \cite{Weng:2020jao}&Ref. \cite{Wu:2016vtq}\\ \Xcline{1-13}{0.5pt}
\multicolumn{2}{c}{Mass/$B_{T}$}&\multicolumn{1}{r}{19303.9}&18857.9&446.0&$(i,j)$&\multicolumn{1}{r}{Value}&\multicolumn{1}{c!{\color{black}\vrule width 1.5pt}}{$\Upsilon\eta_{b}$}&
\multirow{7}*{\makecell[c]{$b$-quark:\\ \\$m^{eff}_{b}$}}&2$m_{b}$&10686.0&\\
\Xcline{1-8}{0.5pt}
\multirow{2}*{\makecell[c]{Variational\\ Parameters\\ (fm$^{-2}$)}}&\multirow{2}*{\makecell[c]{$C_{11}$\\$C_{22}$ \\ $C_{33}$}}&\multicolumn{1}{r}{\multirow{2}*{\makecell[c]{30.7\\30.7\\24.0}}}&
\multicolumn{1}{r}{\multirow{2}*{\makecell[c]{57.4\\49.4\\-}}}&&(1.2)&0.203&
&
&$\frac{\textbf{p}^{2}_{x_{1}}}{2m'_{1}}$&335.6&\\
&&&&&(1.3)&0.217&
\multicolumn{1}{c!{\color{black}\vrule width 1.5pt}}{0.148($\eta_{b}$)}&&
$\frac{m_{\bar{b}}}{m_{b}+m_{\bar{b}}}\frac{\textbf{p}^{2}_{x_{3}}}{2m'_{3}}$&131.4&$\frac{1}{2}$$m_{bb}$\\
\Xcline{1-5}{0.5pt}
\multicolumn{2}{c}{Quark Mass}&\multicolumn{1}{r}{21372.0}&21372.0&0.0&(2,3)&0.217&&
&\multirow{2}*{\makecell[c]{$V^{C}(12)$\\$\frac{1}{2}[V^{C}(13)+V^{C}(14)$\\$+V^{C}(23)+V^{C}(24)]$}}
&\multirow{2}*{\makecell[c]{-274.5\\\\-244.2}}&4764.8\\
\multicolumn{2}{c}{\multirow{1}{*}{Confinement Potential}}&\multirow{1}{*}{-3003.4}&\multirow{1}{*}{-3641.9}&\multirow{1}{*}{638.4}&(1,4)&0.217&
&&&&$\frac12m_{b\bar{b}}$&$2m_{b}$\\
\Xcline{1-5}{0.5pt}
\multicolumn{2}{c}{\multirow{1}*{Kinetic Energy}}&\multicolumn{1}{r}{\multirow{1}*{934.0}}&\multirow{1}*{1171.6}&\multirow{1}*{-237.6}
&(2,4)&0.217&
\multicolumn{1}{c!{\color{black}\vrule width 1.5pt}}{0.160($\Upsilon$)}&&$-D$&-983.0&4722.5&10105.8\\
\Xcline{10-13}{0.5pt}
\multicolumn{2}{c}{\multirow{1}*{CS Interaction}}&\multicolumn{1}{r}{\multirow{1}*{1.3}}&\multirow{1}*{-43.8}&\multirow{1}*{45.1}
&(3,4)&0.203&&&\multirow{1}*{Subtotal}&\multirow{1}*{9653.3}&
\multirow{1}*{9487.3}&\multirow{1}*{10105.8}\\
\Xcline{1-8}{0.5pt}\Xcline{9-13}{0.5pt}
\multicolumn{1}{c|}{\multirow{4}{*}{$V^{C}$}}&\multicolumn{1}{c}{\multirow{1}{*}{(1,2)}}&\multicolumn{1}{r}{-274.5}&&
&\multicolumn{2}{r}{(1,2)-(3,4):}&0.162 fm
&\multirow{8}*{\makecell[c]{$\bar{b}$-quark:\\ \\$m^{eff}_{\bar{b}}$}}&$2m_{b}$&10686.0&\\
&\multicolumn{1}{c}{\multirow{1}{*}{(2,3)}}&\multicolumn{1}{r}{-122.1}&&\multicolumn{1}{c!{\color{black}\vrule width 1.5pt}}{}&\multicolumn{2}{r}{Radius:}&0.130 fm&
&$\frac{\textbf{p}^{2}_{x_{2}}}{2m'_{2}}$&335.6&\\
\Xcline{6-8}{1.6pt}
&\multicolumn{1}{c}{\multirow{1}{*}{(1,4)}}&\multicolumn{1}{r}{-122.1}&&\multicolumn{1}{r}{}&(1,3)&-122.1&$-879.1(\eta_{b})$
&&
$\frac{m_{\bar{b}}}{m_{b}+m_{\bar{b}}}\frac{\textbf{p}^{2}_{x_{3}}}{2m'_{3}}$&131.4&$\frac12 m_{bb}$\\
\Xcline{2-5}{0.5pt}
\multicolumn{1}{c|}{\multirow{1}{*}{}}&\multicolumn{1}{c}{\multirow{1}{*}{Subtotal}}&\multicolumn{1}{r}{-1037.4}&\multicolumn{1}{r}{-1675.9}&\multicolumn{1}{c|}{638.5}
&\multicolumn{1}{c}{(3,4)}&-274.5&&
&\multirow{2}*{\makecell[c]{$V^{C}(12)$\\$\frac{1}{2}[V^{C}(13)+V^{C}(14)$\\$+V^{C}(23)+V^{C}(24)]$}}
&\multirow{2}*{\makecell[c]{-274.5\\\\-244.2}}&4764.8\\
\Xcline{1-5}{0.5pt}
\multicolumn{2}{c}{Total Contribution}&\multicolumn{1}{r}{\multirow{1}*{-102.1}}&\multicolumn{1}{r}{-548.1}&\multicolumn{1}{c|}{\multirow{1}*{446.0}}&(2,4)&-122.1&$-796.7(\Upsilon)$&&
&&$\frac{1}{2}$$m_{b\bar{b}}$&$2m_{b}$\\
\Xcline{1-8}{1.6pt}
\multicolumn{2}{l}{Total Wave function:}&\multicolumn{6}{r!{\color{black}\vrule width 1.5pt}}{$\Psi_{tot}=|F\rangle|R^{s}\rangle|[\phi_{1}\chi_{2}]\rangle=-0.408|F\rangle|R^{s}\rangle|[\psi_{1}\zeta_{2}]\rangle$}&&$-D$&-983.0&4722.5&10105.8\\
\Xcline{10-13}{0.5pt}
\multicolumn{8}{r!{\color{black}\vrule width 1.5pt}}{$-0.408|F\rangle|R^{s}\rangle|[\psi_{1}\zeta_{3}]\rangle+0.577|F\rangle|R^{s}\rangle|[\psi_{2}\zeta_{5}]\rangle+0.577|F\rangle|R^{s}\rangle|[\psi_{2}\zeta_{6}]\rangle$}&
&\multirow{1}*{Subtotal}&\multirow{1}*{9653.3}&
\multirow{1}*{9487.3}&\multirow{1}*{10105.8}\\
\Xcline{9-13}{0.5pt}\Xcline{1-8}{0.3pt}
\multicolumn{8}{l!{\color{black}\vrule width 1.5pt}}{The rearrangement strong decay channel: $J/\psi\eta_{c}$}&\multirow{7}*{\makecell[c]{CS \\Interaction}}&\multirow{2}*{\makecell[c]{$\frac{1}{2}V^{SS}(12)$
}}&\multirow{2}*{6.7}&$\frac83v_{bb}$&$\frac83C_{bb}$\\
\Xcline{1-8}{0.3pt}
\multicolumn{8}{l!{\color{black}\vrule width 1.5pt}}{The radiative decay widths:}&&&&5.1&7.7\\
\multicolumn{8}{l!{\color{black}\vrule width 1.5pt}}{$\Gamma_{T_{b^{2}\bar{b}^{2}}(19327.9,2^{++})\rightarrow T_{b^{2}\bar{b}^{2}}(19303.9,1^{+-})\gamma}=1.0\, \rm{keV}$}&&\multirow{2}*{\makecell[c]{$\frac{1}{2}V^{SS}(34)$
}}&\multirow{2}*{6.7}&$\frac83v_{\bar{b}\bar{b}}$&$\frac83C_{\bar{b}\bar{b}}$\\
\multicolumn{8}{l!{\color{black}\vrule width 1.5pt}}{$\Gamma_{T_{b^{2}\bar{b}^{2}}(19303.9,1^{+-})\rightarrow T_{b^{2}\bar{b}^{2}}(19240.0,0^{++})\gamma}=2.8\, \rm{keV}$}&&\multirow{2}*{}&\multirow{2}*{}
&5.1&7.7\\
\Xcline{1-8}{0.3pt}
\multicolumn{8}{l!{\color{black}\vrule width 1.5pt}}{The magnetic moments: \quad $\mu_{T_{b^{2}\bar{b}^{2}}(19303.9,1^{+-})}=\langle\Psi^{1^{+-}}_{tot}|\hat{\mu^{z}}|\Psi^{1^{+-}}_{tot}\rangle=0$}&&\multirow{2}*{\makecell[c]{
$-\frac{1}{4}(V^{SS}(13)+V^{SS}(14)$\\
$+V^{SS}(23)+V^{SS}(24)$}}&\multirow{2}*{-12.2}&$-\frac{16}{3}v_{b\bar{b}}$&$-\frac{16}{3}C_{b\bar{b}}$\\
\Xcline{1-8}{0.3pt}
\multicolumn{8}{l!{\color{black}\vrule width 1.5pt}}{The transition magnetic moments:}
&&&\multirow{1}*{}&\multirow{1}*{-15.3}&-15.5\\
\Xcline{10-13}{0.5pt}
\multicolumn{8}{l!{\color{black}\vrule width 1.5pt}}{$\mu_{T_{b^{2}\bar{b}^{2}}(19327.9,2^{++})\rightarrow T_{b^{2}\bar{b}^{2}}(19303.9,1^{+-})\gamma}=\langle\Psi^{2^{++}}_{tot}|\hat{\mu^{z}}|\Psi^{1^{+-}}_{tot}\rangle=-0.125 \mu_{N}$}&&\multirow{1}*{Subtotal}&\multirow{1}*{1.3}&\multirow{1}*{-5.1}&0.0\\
\Xcline{9-13}{0.5pt}
\multicolumn{8}{l!{\color{black}\vrule width 1.5pt}}{$\mu_{T_{b^{2}\bar{b}^{2}}(19393.9,1^{+-})\rightarrow T_{b^{2}\bar{b}^{2}}(19240.0,0^{++})\gamma}=\langle\Psi^{1^{+-}}_{tot}|\hat{\mu^{z}}|\Psi^{0^{++}}_{tot}\rangle=-0.072\mu_{N}$}&
\multicolumn{2}{l|}{\multirow{1}*{Total contribution}}&\multirow{1}*{19303.9}&\multirow{1}*{18696.4}&\multirow{1}*{20211.6}\\
\toprule[0.5pt]
\toprule[1.5pt]
\end{tabular}
\end{lrbox}\scalebox{0.869}{\usebox{\tablebox}}
\end{table*}

\subsection{$cc\bar{b}\bar{b}$ state}
Here, we will concentrate on the $cc\bar{b}\bar{b}$ system.
Similar to the $cc\bar{c}\bar{c}$ and $bb\bar{b}\bar{b}$ systems, the $cc\bar{b}\bar{b}$ system is also satisfied with fully antisymmetric for diquarks and antiquarks.
There are two $J^{P}=0^{+}$ states, one $J^{P}=1^{+}$ state, and one $J^{P}=2^{+}$ state in the $cc\bar{b}\bar{b}$ system.
We show the masses of the ground states, the variational parameters, the internal mass contributions, the relative
lengths between the quarks, their lowest meson-meson thresholds, the specific wave function,  the magnetic moments, the transition magnetic moments, the radiative decay widths, and the rearrangement strong width ratios in Tables \ref{ccbb} and \ref{ccbb2}.


First, we take the $J^{P}=0^{+}$ $cc\bar{b}\bar{b}$ ground state as an example to discuss its properties with the variational method.
A similar situation occurs in the other two quantum numbers according to Tables \ref{ccbb}, and \ref{ccbb2}.
The mass of the lowest $J^{P}=0^{+}$ $cc\bar{b}\bar{b}$ state is 12920.0 MeV, and the corresponding binding energy $B_{T}$ is +344.2 MeV according to Table \ref{ccbb}.
Thus, the state is obviously higher than the corresponding rearrangement meson-meson thresholds.
The wave function is given by
\begin{eqnarray}\label{Eq422}
|\Psi_{tot}\rangle=-0.966|F\rangle|R^{s}\rangle|[\phi_{1}\chi_{5}]\rangle+0.259|F\rangle|R^{s}\rangle|[\phi_{2}\chi_{6}]\rangle. \nonumber\\
\end{eqnarray}
Here, we see that the mass contribution of ground state mainly comes from the $|(Q_{1}Q_{2})^{\bar{3}}_{1}(\bar{Q}_{3}\bar{Q}_{4})^{3}_{1}\rangle_{0}$ component,
and the $|(Q_{1}Q_{2})^{6}_{0}(\bar{Q}_{3}\bar{Q}_{4})^{\bar{6}}_{0}\rangle_{0}$ component is negligible.
Its variational parameters are given as $C_{11}=23.9 {\rm fm}^{-2}$, $C_{22}=10.5 {\rm fm}^{-2}$, and $C_{33}=12.3 {\rm fm}^{-2}$.

The meson-meson configuration is connected to the diquark-antidiquark configuration by a linear transformation.
Then, we obtain the total wave function in the meson-meson configuration:
\begin{eqnarray}\label{Eq423}
|\Psi_{tot}\rangle&=&-0.589|F\rangle|R^{s}\rangle|[\psi_{1}\zeta_{5}]\rangle
+0.095|F\rangle|R^{s}\rangle|[\psi_{1}\zeta_{6}]\rangle \nonumber\\
&&+0.608|F\rangle|R^{s}\rangle|[\psi_{2}\zeta_{5}]\rangle+
0.524|F\rangle|R^{s}\rangle|[\psi_{2}\zeta_{6}]\rangle.
\nonumber\\
\end{eqnarray}

According to Eq. (\ref{Eq423}), we are sure that the overlaps $c_{i}$ of $B_{c}B_{c}$ and $B^{*}_{c}B^{*}_{c}$ are 0.095 and 0.589, respectively.
Then, based on Eq. (\ref{width}), the rearrangement strong width ratios is
\begin{small}
\begin{eqnarray}
\frac{\Gamma_{T_{c^{2}\bar{b}^{2}}(12920.0,0^{+})\rightarrow B^{*}_{c}B^{*}_{c}}}{\Gamma_{T_{c^{2}\bar{b}^{2}}(12920.0,0^{+})\rightarrow B_{c}B_{c}}}=1:50,
\end{eqnarray}
\end{small}
i.e., $B_{c}B_{c}$ is the dominant rearrangement decay channel for the  $T_{c^{2}\bar{b}^{2}}(12920.0,0^{+})$ state.

As for the magnetic moment of the $J^{P}=0^{+}$ $cc\bar{b}\bar{b}$ ground state,
its value is 0, while the magnetic moment of all $J^{P}=0^{+}$ tetraquark states is 0.
As for the $J^{P}=1^{+}$ $cc\bar{b}\bar{b}$ state,
we construct its flavor $\otimes$ spin wave functions as
\begin{normalsize}
\begin{eqnarray}\label{Eq324}
\hspace{-2cm}
&&|\Psi\rangle^{S=1;S_{s}=1}_{T_{c^{2}\bar{b}^{2}}(12939.9,1^{+})}\nonumber\\
&=&|R^{s}\rangle|\psi\rangle|cc\bar{b}\bar{b}\rangle|\frac{1}{2}(\uparrow\uparrow\uparrow\downarrow+\uparrow\uparrow\downarrow\uparrow-\downarrow\uparrow\uparrow\downarrow
-\downarrow\uparrow\downarrow\uparrow\rangle.
\end{eqnarray}
\end{normalsize}
So the corresponding transition magnetic momentum is
\begin{eqnarray}\label{Eq325}
\mu_{T_{c^{2}\bar{b}^{2}}(12939.9,1^{+})}=\langle\Psi^{1^{+}}_{tot}|\hat{\mu^{z}}|\Psi^{1^{+}}_{tot}\rangle=\mu_{c}+\mu_{\bar{b}}=0.490\mu_{N}.\nonumber\\
\end{eqnarray}

We also discuss the transition magnetic moment of the $T_{c^{2}\bar{b}^{2}}(12939.9,1^{+})\rightarrow T_{c^{2}\bar{b}^{2}}(12920.0,0^{+})\gamma$ process.
We still construct their flavor $\otimes$ spin wave functions as
\begin{eqnarray}
|\Psi\rangle^{S=1;S_{s}=0}_{T_{c^{2}\bar{b}^{2}}(1^{+})}&=&|R^{s}\rangle|\psi\rangle|cc\bar{b}\bar{b}\rangle|\frac{1}{\sqrt{2}}(\uparrow\uparrow\downarrow\downarrow-\downarrow\downarrow\uparrow\uparrow)\rangle,\nonumber\\
|\Psi\rangle^{S=0;S_{s}=0}_{T_{c^{2}\bar{b}^{2}}(0^{+})}&=&|R^{s}\rangle|\psi\rangle|cc\bar{b}\bar{b}\rangle|0.966\frac{1}{\sqrt{3}}(\uparrow\uparrow\downarrow\downarrow+\downarrow\downarrow\uparrow\uparrow)+...\rangle.\nonumber\\
\end{eqnarray}

And then, the transition magnetic momentum of the
$T_{c^{2}\bar{b}^{2}}(12939.9,1^{+})\rightarrow T_{c^{2}\bar{b}^{2}}(12920.0,0^{+})\gamma$
process can be described by the $z$-component of the magnetic moment operator $\hat{\mu^{z}}$  sandwiched by the
flavor-spin wave functions of the $T_{c^{2}\bar{b}^{2}}(12939.9,1^{+})$ and $T_{c^{2}\bar{b}^{2}}(12920.0,0^{+})$.
So the corresponding transition magnetic momentum is
\begin{eqnarray}\label{Eq326}
&&\mu_{T_{c^{2}\bar{b}^{2}}(12939.9,1^{+})\rightarrow T_{c^{2}\bar{b}^{2}}(12920.0,0^{+})\gamma}=\langle\Psi^{1^{+}}_{tot}|\hat{\mu^{z}}|\Psi^{0^{+}}_{tot}\rangle\nonumber\\
&&=0.966\times\frac{1}{\sqrt{6}}(4\mu_{c}-4\mu_{\bar{b}})=0.534\mu_{N}.
\end{eqnarray}

Further, according to
Eq. (\ref{eq:Gammaradiative}) and Eq. (\ref{Eq326}), we also obtain the radiative decay widths
\begin{eqnarray}\label{Eq327}
\Gamma_{T_{c^{2}\bar{b}^{2}}(12960.9,2^{+})\rightarrow T_{c^{2}\bar{b}^{2}}(12939.9,1^{+})\gamma}=3.6\, \rm{keV}.
\end{eqnarray}

Finally, we turn to the internal contribution for the  $cc\bar{b}\bar{b}$ ground state.
For the kinetic energy part, the $J^{P}=0^{+}$ $cc\bar{b}\bar{b}$ state receives 835.9 MeV, which is smaller than that of the meson-meson threshold $B_{c}B_{c}$.
The potential part of this state is much smaller than that of the lowest meson-meson threshold.
Furthermore, we find that all the $V^{C}$ for this state are attractive. However, compared to the $V^{C}$ of $B_{c}B_{c}$, these attractive values seem to trivial.
This is because the length between $c-\bar{b}$ in tetraquarks is longer than that in $B_{c}$ according to Table \ref{ccbb} and \ref{ccbb2}.
In summary, we tend to think that these $cc\bar{b}\bar{b}$ states are unstable compact states.
\\

\subsection{$cc\bar{c}\bar{b}$ and $bb\bar{b}\bar{c}$ states}

Here, we discuss the $cc\bar{c}\bar{b}$ and $bb\bar{b}\bar{c}$ systems.
For these two systems, they only need to satisfy the antisymmetry for the diquark.
Thus, compared to the above three systems, the $cc\bar{c}\bar{b}$ and $bb\bar{b}\bar{c}$ systems have more allowed states.
There are two $J^{P}=0^{+}$ states, three $J^{P}=1^{+}$ states, one $J^{P}=2^{+}$ state in the $cc\bar{c}\bar{b}$ and $bb\bar{b}\bar{c}$ systems.
We calculate the masses of the ground states, the corresponding variational parameters, the various internal contributions, the relative
lengths between the quarks, and their lowest meson-meson thresholds, their lowest meson-meson thresholds, specific wave functions,  magnetic moments, transition magnetic moments, radiative decay widths, and rearrangement strong width ratios in Tables \ref{cccb1}, \ref{bbbc1}, and \ref{cccb2}, respectively.


We now analyse the numerical results of the $J^{P}=1^{+}$ ground $bb\bar{b}\bar{c}$ state obtained from the variational method according to Table \ref{bbbc1}.
Other states would have similar discussions from Tables \ref{cccb1}-\ref{cccb2}.
The mass of the lowest $J^{P}=1^{+}$ $bb\bar{b}\bar{c}$ state is 16043.2 MeV, and the corresponding binding energy $B_{T}$ is +303.7 MeV.
Thus, the state is obviously above the lowest rearrangement meson-meson decay channel $B_{c}^{*}\eta_{b}$, and it is an unstable tetraquark state.
Its variational parameters are given as $C_{11}=12.4 {\rm fm}^{-2}$, $C_{22}=21.0 {\rm fm}^{-2}$, and $C_{33}=28.9 {\rm fm}^{-2}$.
The corresponding wave function is given by
\begin{eqnarray}\label{Eq442}
|\Psi_{tot}\rangle&=&0.984|F\rangle|R^{s}\rangle|[\phi_{2}\chi_{4}]\rangle+0.171|F\rangle|R^{s}\rangle|[\phi_{1}\chi_{3}]\rangle\nonumber\\
&-&0.044|F\rangle|R^{s}\rangle|[\phi_{1}\chi_{2}]\rangle.
\end{eqnarray}
Here, we notice that the mass contribution of ground state mainly comes from the $|(Q_{1}Q_{2})^{6}_{0}(\bar{Q}_{3}\bar{Q}_{4})^{\bar{6}}_{1}\rangle_{1}$ component, and the other two components are negligible.
Then we transform Eq. (\ref{Eq442}) to the meson-meson configuration via a linear transformation, and the corresponding wave function is given as:
\begin{eqnarray}\label{Eq221}
|\Psi_{tot}\rangle&=&0.494|F\rangle|R^{s}\rangle|[\psi_{1}\zeta_{2}]\rangle-0.396|F\rangle|R^{s}\rangle|[\psi_{1}\zeta_{3}]\rangle\nonumber\\
&-&0.487|F\rangle|R^{s}\rangle|[\psi_{1}\zeta_{4}]\rangle+0.111|F\rangle|R^{s}\rangle|[\psi_{2}\zeta_{2}]\rangle\nonumber\\
&-&0.246|F\rangle|R^{s}\rangle|[\psi_{2}\zeta_{3}]\rangle-0.537|F\rangle|R^{s}\rangle|[\psi_{2}\zeta_{4}]\rangle.\nonumber\\
\end{eqnarray}
Furthermore, we can sure that its rearrangement strong width ratios is:
\begin{eqnarray}
&&\Gamma_{T_{b^{2}\bar{b}\bar{c}}(16043.2,1^{+})\rightarrow B^{*}_{c}\Upsilon}:
\Gamma_{T_{b^{2}\bar{b}\bar{c}}(16043.2,1^{+})\rightarrow B_{c}\Upsilon}:\nonumber\\
&&\Gamma_{T_{b^{2}\bar{b}\bar{c}}(16043.2,1^{+})\rightarrow B^{*}_{c}\eta_{b}}=1:1.3:1.5.
\end{eqnarray}
And its radiative decay widths are:
\begin{eqnarray}
\Gamma_{T_{b^{2}\bar{b}\bar{c}}(16149.2,2^{+})\rightarrow T_{c^{2}\bar{c}\bar{b}}(16043.2,1^{+})\gamma}=435.0\, \rm{keV}, \nonumber\\
\Gamma_{T_{c^{2}\bar{c}\bar{b}}(16043.9,0^{+})\rightarrow T_{c^{2}\bar{c}\bar{b}}(16043.2,1^{+})\gamma}=10^{-6} \,\rm{keV}.\nonumber
\end{eqnarray}

Let us now focus on the internal contributions for this state and the relative lengths between the quarks.
For the kinetic energy part, the state gets 876.1 MeV, which is obviously smaller than that of the lowest meson-meson threshold $B_{c}\eta_{b}$.
The actual kinetic energy of the $b-\bar{b}$ ($b-\bar{c}$) in the $J^{P}=1^{+}$ $bb\bar{b}\bar{c}$ state is smaller than that in the $\eta_{b}$ ($B^{*}_{c}$) meson.
The reason for this can be seen in Table \ref{bbbc1}. The size of this pair is larger in the $J^{P}=1^{+}$ $bb\bar{b}\bar{c}$ state than in the meson: the distance (3,4) is 0.245 fm in this tetraquark while it is 0.148 fm in $\eta_{b}$.

Here, let us turn our discussion to the potential parts.
The potential part of this state is much smaller than that of its lowest meson-meson threshold.
Although the $V^{C}$ between quark and antiquark are attractive, the $V^{C}$ in the diquark and antiquark are repulsive.
However, relative to the $\eta_{b}$ and $B_{c}$ mesons, the $V^{C}$ in the tetraquark are less attractive.
Therefore, they still have a relatively large positive binding energy in this state.

\subsection{$cb\bar{c}\bar{b}$ state}

Finally, we investigate the $cb\bar{c}\bar{b}$ system.
Similar to the $cc\bar{c}\bar{c}$ and $bb\bar{b}\bar{b}$ systems, the $cb\bar{c}\bar{b}$ system is also a pure neutral system and has a certain C-parity.
Thus the corresponding magnetic moment is $0 \mu_{N}$ for all the ground $cb\bar{c}\bar{b}$ states.
Moreover, the Pauli principle does not impose any constraints on the wave functions of the $cb\bar{c}\bar{b}$ system.
Thus, compared to other discussed tetraquark systems, the $cb\bar{c}\bar{b}$ system has more allowed states.
There are four $J^{PC}=0^{++}$ states, four $J^{PC}=1^{+-}$ states, two $J^{PC}=1^{++}$ states, two $J^{PC}=2^{++}$ states in the $cb\bar{c}\bar{b}$ system.


Here, we now analyze the numerical results about the $cb\bar{c}\bar{b}$ system obtained from the variational method.
Here, we take the $J^{PC}=0^{++}$ $cb\bar{c}\bar{b}$ ground state as an example to discuss specifically, and others would have similar discussions.
The mass of the lowest $J^{PC}=0^{++}$ $cb\bar{c}\bar{b}$ state is 12759.3 MeV, and the corresponding binding energy $B_{T}$ is +371.8 MeV.
Thus, the state obviously has a larger mass than the lowest rearrangement meson-meson decay channel $\eta_{b}\eta_{c}$, and it should be an unstable compact tetraquark state.
Its variational parameters are given as $C_{11}=11.9 ~{\rm fm}^{-2}$, $C_{22}=11.9 ~{\rm fm}^{-2}$, and $C_{33}=22.9 ~{\rm fm}^{-2}$.
Since this state is a pure neutral state, we naturally notice that the value of $C_{11}$ is equal to $C_{22}$, which means that the distance of $(b-b)$ is equal to $(\bar{b}-\bar{b})$.
Our results also reflect these properties according to Table \ref{cbcb1}.
The corresponding wave function is given as:
\begin{eqnarray}\label{Eq449}
\Psi_{tot}&=&0.961|F\rangle|R^{s}\rangle|[\phi_{2}\chi_{5}]\rangle+0.114|F\rangle|R^{s}\rangle|[\phi_{2}\chi_{6}]\rangle\nonumber\\
&-&0.069|F\rangle|R^{s}\rangle|[\phi_{1}\chi_{5}]\rangle-0.241|F\rangle|R^{s}\rangle|[\phi_{1}\chi_{6}]\rangle.\nonumber\\
\end{eqnarray}
Based on Eq. (\ref{Eq449}), we find that its mass contribution to the ground state mainly comes from the $6\otimes \bar{6}$ component, the corresponding $3\otimes \bar{3}$ component being negligible.
Then we transform Eq. (\ref{Eq449}) to $c\bar{c}-b\bar{b}$ and $c\bar{b}-b\bar{c}$ configurations via a linear transformation, and the corresponding two wave functions are given as:
\begin{eqnarray}
|\Psi_{tot}\rangle&=&-0.830|F\rangle|R^{s}\rangle|[\psi_{1}\zeta_{5}]\rangle
+0.211|F\rangle|R^{s}\rangle|[\psi_{1}\zeta_{6}]\rangle\nonumber\\
&&-0.367|F\rangle|R^{s}\rangle|[\psi_{2}\zeta_{5}]\rangle+
0.363|F\rangle|R^{s}\rangle|[\psi_{2}\zeta_{6}]\rangle\nonumber\\
&=&-0.668|F\rangle|R^{s}\rangle|[\psi'_{1}\zeta'_{5}]\rangle
+0.333|F\rangle|R^{s}\rangle|[\psi'_{1}\zeta'_{6}]\rangle\nonumber\\
&&-0.398|F\rangle|R^{s}\rangle|[\psi'_{2}\zeta'_{5}]\rangle+
0.533|F\rangle|R^{s}\rangle|[\psi'_{2}\zeta'_{6}]\rangle.\nonumber\\
\end{eqnarray}
Further, we can certain its rearrangement strong width ratios.
For the $c\bar{b}-b\bar{c}$ decay mode
\begin{eqnarray}
&&\Gamma_{T_{cb\bar{c}\bar{b}}(12796.9,1^{+-})\rightarrow B^{*}_{c}\bar{B}^{*}_{c}}:
\Gamma_{T_{cb\bar{c}\bar{b}}(12796.9,1^{+-})\rightarrow  B^{*}_{c}\bar{B}_{c}}: \nonumber\\
&&\Gamma_{T_{cb\bar{c}\bar{b}}(12796.9,1^{+-})\rightarrow B_{c}\bar{B}^{*}_{c}}=1:3.9:3.9,
\end{eqnarray}
where both the $B^{*}_{c}\bar{B}_{c}$ and $B_{c}\bar{B}^{*}_{c}$ channels are the dominant decay modes for the $T_{cb\bar{c}\bar{b}}(12796.9,1^{+-})$ tetraquark state.
\begin{eqnarray}
&&\Gamma_{T_{cb\bar{c}\bar{b}}(12796.9,1^{+-})\rightarrow \eta_{b}J/\psi}:
\Gamma_{T_{cb\bar{c}\bar{b}}(12796.9,1^{+-})\rightarrow \Upsilon\eta_{c}}\nonumber\\
&&=1:18.4.
\end{eqnarray}
The dominant decay channel is the $\Upsilon\eta_{c}$ final states in the $c\bar{c}-b\bar{b}$ decay mode.

We also calculate the transition magnetic moments for this state:
\begin{eqnarray}
\mu_{T_{cb\bar{c}\bar{b}}(12882.4,2^{++})\rightarrow T_{cb\bar{c}\bar{b}}(12759.6,0^{++})\gamma}&=&0,\nonumber\\
\mu_{T_{cb\bar{c}\bar{b}}(12856.6,1^{++})\rightarrow T_{cb\bar{c}\bar{b}}(12759.6,0^{++})\gamma}&=&0,\nonumber\\
\mu_{T_{cb\bar{c}\bar{b}}(12796.9,1^{+-})\rightarrow T_{cb\bar{c}\bar{b}}(12759.6,0^{++})\gamma}&=&0.081,
\end{eqnarray}
which are in units of $\mu_{N}$.
Furthermore, we can obtain its radiative decay widths are:
\begin{eqnarray}\label{Eq329}
\Gamma_{T_{cb\bar{c}\bar{b}}(12882.4,2^{++})\rightarrow T_{cb\bar{c}\bar{b}}(12759.6,0^{++})\gamma}&=&0,\nonumber\\
\Gamma_{T_{cb\bar{c}\bar{b}}(12856.6,1^{++})\rightarrow T_{cb\bar{c}\bar{b}}(12759.6,0^{++})\gamma}&=&0,\nonumber\\
\Gamma_{T_{cb\bar{c}\bar{b}}(12797.3,1^{+-})\rightarrow T_{cb\bar{c}\bar{b}}(12759.6,0^{++})\gamma}&=&33.1,
\end{eqnarray}
which are in units of keV.

Let us now turn our discussion to the internal contribution for the $J^{PC}=1^{+-}$ $cb\bar{c}\bar{b}$ ground state.
For the kinetic energy part, the state obtains 858.5 MeV, which is smaller than the 1001.2 MeV of the lowest meson-meson threshold $B_{c}\eta_{b}$ according to Table \ref{cbcb1}.
As for the potential part, although the $V^{C}$ between quark and antiquark are attractive, the $V^{C}$ in the diquark and antiquark are repulsive.
However, relative to the lowest meson-meson threshold $B_{c}\eta_{b}$, the total $V^{C}$ is not attractive than the $B_{c}\eta_{b}$, which leads to this state having a relatively larger mass.

We also notice that the $V^{C}(1,3)$,  $V^{C}(2,3)$, $V^{C}(1,4)$, and $V^{C}(2,4)$ are absolutely the same, and meanwhile the distances of (1,3), (1,4), (2,3), and (2,4) are also the same.
These actually reflect $\langle\Psi_{tot}|(\textbf{R}_{1,2}\cdot\textbf{R}_{3,4})|\Psi_{tot}\rangle= \langle\Psi_{tot}|(\textbf{R}_{1,2}\cdot\textbf{R}')|\Psi_{tot}\rangle= \langle\Psi_{tot}|(\textbf{R}_{3,4}\cdot\textbf{R}')|\Psi_{tot}\rangle=0$.
Obviously, it is unreasonable that the distance of $c\bar{c}$ is exactly the same as that of the $c\bar{b}$ and $b\bar{b}$.
According to Sec IV of Ref. \cite{Noh:2021lqs}, we only consider single Gaussian form which the $l_{1}=l_{2}=l_{3}=0$ in spatial part of the total wave function is not sufficient.
These lead to the $cb\bar{c}\bar{b}$ state, which is far away from the real structures in nature.
We have reason enough to believe that the $\langle\Psi_{tot}|(\textbf{R}_{1,2}\cdot\textbf{R}_{3,4})|\Psi_{tot}\rangle$ should not be zero.
Meanwhile, considering other spatial basis would reduce the corresponding to the binding energy $B_{T}$ \cite{Noh:2021lqs}.
But these corrections would be powerless against the higher binding energy $B_{T}$ of the ground $J^{PC}=1^{+-}$ $cb\bar{c}\bar{b}$.
In conclusion, we tend to think that the $J^{PC}=1^{+-}$ $cb\bar{c}\bar{b}$ ground state should be an unstable compact state.
\\

\section{Comparison with other work}\label{sec5}

The mass spectra have been studied with different approaches such as different nonrelativistic constituent quark models, different chromomagnetic models, the relativistic quark models, the nonrelativistic chiral quark model, the diquark models, the diffusion Monte Carlo calculation, and the QCD sum rule. In addition, these fully heavy tetraquark systems have been discussed with different color structures such as the $8_{Q\bar{Q}}\otimes8_{Q\bar{Q}}$ configuration, the diquark-antiquark configuration ($3\otimes\bar{3}$ and the $6\otimes\bar{6}$) and the couplings between the above color configurations.
For comparison, we briefly list our results and other theoretical results in Table \ref{comparsion}.

Compared to other systems, there is the most extensive discussion about the $cc\bar{c}\bar{c}$ system. So we will concentrate on the $cc\bar{c}\bar{c}$ system, but other systems can be discussed in a similar way. After comparing our results with other researches, we can see that the most theoretical masses of $cc\bar{c}\bar{c}$ in ground states lie
in a wide range of $6.0-6.8$ GeV in Table \ref{comparsion}.
Our results are 6.38, 6.45, and 6.48 GeV for the  $0^{++}$, $1^{+-}$, and $2^{++}$ $cc\bar{c}\bar{c}$ ground states, respectively.
These three ground states are expected to be broad because they can all decay to charmonium pairs: $\eta_{c}\eta_{c}$, $\eta_{c}J/ \psi$, or $J/ \psi J/ \psi$ through the quark (antiquark) rearrangements.
Therefore, these types of decays are favored both dynamically and kinematically.
According to Table \ref{comparsion},
we can conclude that the obtained masses of the ground states are obviously smaller than the $X(6900)$ observed by the LHCb collaboration.
The observed $X(6900)$ is less likely to be the ground compact tetraquark state and could be a first or second radial excited $cc\bar{c}\bar{c}$ state.

Although we all use a similar Hamiltonian expression as in the nonrelativistic constituent quark model \cite{Wang:2019rdo, Liu:2019zuc, Deng:2020iqw, Zhang:2022qtp}, the spatial wave function is mostly expanded in the Gaussian basis according to Ref. \cite{Hiyama:2003cu}, while we treat the spatial function as a Gaussian function, which is convenient for use in further variational methods to handle calculations in the four-body problem.
Our results for the $cc\bar{c}\bar{c}$ system are roughly compatible with other nonrelativistic
constituent quark models, although different papers have chosen different potential forms.

It is also interesting to note that relatively larger results are also given by the QCD sum rules \cite{Wang:2021mma}, the Monte Carlo method \cite{Gordillo:2020sgc}, the diquark model \cite{Mutuk:2021hmi}, and the chiral quark model \cite{Chen:2019vrj}.
However, the results given by the QCD sum rules \cite{Wang:2021mma} are about 1 GeV below those of the constituent quark models for the $bb\bar{b}\bar{b}$ system.
In contrast, our results are obviously larger than the chromomagnetic models \cite{Zhuang:2021pci,Weng:2020jao,Wu:2016vtq,Karliner:2016zzc}, and the diquark models \cite{Faustov:2020qfm,Berezhnoy:2011xn},
where these models usually neglect the kinematic term and explicitly include confining potential contributions or adopt a diquark picture.

\begin{table*}[htp]
\caption{Comparison of the results of different methods for the $QQ\bar{Q}\bar{Q}$ tetraquark states.}\label{comparsion}
\begin{lrbox}{\tablebox}
\renewcommand\arraystretch{1.43}
\renewcommand\tabcolsep{2.9pt}
\begin{tabular}{crl|cccc|cccc|cccc}
\toprule[1.0pt]
\toprule[0.5pt]
\multicolumn{3}{l}{Systems}&\multicolumn{4}{c}{$cc\bar{c}\bar{c}$}&\multicolumn{4}{c}{$bb\bar{b}\bar{b}$}&\multicolumn{4}{c}{$cc\bar{b}\bar{b}$}\\
\multicolumn{3}{l}{$J^{P(C)}$}&\multicolumn{2}{c}{$0^{++}$}&$1^{+-}$&\multicolumn{1}{c|}{$2^{++}$}&\multicolumn{2}{c}{$0^{++}$}&$1^{+-}$&\multicolumn{1}{c|}{$2^{++}$}&\multicolumn{2}{c}{$0^{+}$}&$1^{+}$&$2^{+}$\\
\toprule[0.5pt]
\multicolumn{2}{c}{\multirow{7}*{\makecell[c]{The nonrelativistic\\ constituent\\quark models}}}&Our result&6384&6512&6452&\multicolumn{1}{c|}{6483}&19352&19240&19304&\multicolumn{1}{c|}{19328}&12920&13008&12940&12961\\
\Xcline{3-15}{0.5pt}
&&\multirow{2}*{Ref. \cite{Wang:2019rdo}}&6377&6425&6425&6432&19215&19247&19247&19249&12847&12866&12864&12868\\
&&&6371&6483&6450&6479&19243&19305&19311&19325&12886&12946&12924&12940\\
\Xcline{3-3}{0.01pt}
&&Ref. \cite{Liu:2019zuc}&6487&6518&6500&6524&19332&19338&19329&19341&12947&13039&12960&12972\\
\Xcline{3-3}{0.01pt}
&&Ref. \cite{Zhang:2022qtp}&6500&6411&6453&6475&19200&19235&19216&19225&12880&12981&12890&12902\\
\Xcline{3-3}{0.01pt}
&&Ref. \cite{Lloyd:2003yc}&6477&6695&6528&6573&-&-&-&-&-&-&-&-\\
\Xcline{3-3}{0.01pt}
&\multicolumn{1}{l}{}&\multirow{3}*{Ref. \cite{Deng:2020iqw}}&\multicolumn{2}{c}{6491}&6580&6607&\multicolumn{2}{c}{19357}&19413&19429&\multicolumn{2}{c}{12963}&13024&13041\\
\Xcline{1-2}{0.01pt}\Xcline{4-15}{0.01pt}
\multicolumn{2}{c}{\makecell[c]{ Multiquark color\\ flux-tube model}}&&\multicolumn{2}{c}{6407}&6463&6486&\multicolumn{2}{c}{19329}&19373&19387&\multicolumn{2}{c}{12906}&12946&12960\\
\Xcline{1-2}{0.01pt}\Xcline{4-15}{0.01pt}
&\multicolumn{1}{l}{}&&\multicolumn{2}{c}{6035}&6139&6194&\multicolumn{2}{c}{18834}&18890&18921&\multicolumn{2}{c}{12597}&12660&12695\\
\Xcline{3-3}{0.01pt}
\multicolumn{2}{c}{\multirow{4}*{\makecell[c]{The chromo-\\magnetic model}}}&Ref. \cite{Karliner:2016zzc}&\multicolumn{2}{c}{6192}&-&-&\multicolumn{2}{c}{18826}&-&-&\multicolumn{2}{c}{-}&-&-\\
\Xcline{3-3}{0.01pt}
&&\multirow{2}*{Ref. \cite{Wu:2016vtq}}&6899&7016&6899&6956&20155&20275&20212&20243&13496&13634&13560&13595\\
&&&6035&6253&6137&6194&18834&18954&18890&18921&12597&12734&12660&12695\\
\Xcline{3-3}{0.01pt}
&&Ref. \cite{Weng:2020jao}&6045&6271&6231&6287&18836&18981&18969&19000&12596&12712&12672&12703\\
\Xcline{3-3}{0.01pt}
&&Ref. \cite{Zhuang:2021pci}&6034&6254&6137&6194&18834&18953&18890&18921&-&-&-&-\\
\Xcline{1-15}{0.01pt}
\multicolumn{2}{c}{\makecell[c]{The Bethe-\\ Salpeter equations}}&Ref.\cite{Li:2021ygk}&\multicolumn{2}{c}{6419}&6456&6516&\multicolumn{2}{c}{19205}&19221&19253&\multicolumn{2}{c}{-}&-&-\\
\Xcline{1-15}{0.01pt}
\multicolumn{2}{c}{\multirow{2}*{\makecell[c]{The relativistic\\ quark model}}}&Ref. \cite{Faustov:2020qfm}&\multicolumn{2}{c}{6190}&6271&6367&\multicolumn{2}{c}{19314}&19320&19330&\multicolumn{2}{c}{12846}&12859&12883\\
\Xcline{3-3}{0.01pt}
&&Ref. \cite{Lu:2020cns}&6435&6542&6515&6543&19201&19255&19251&19262&-&-&-&-\\
\Xcline{1-15}{0.01pt}
\multicolumn{2}{c}{Monte Carlo method}&Ref. \cite{Gordillo:2020sgc}&\multicolumn{2}{c}{6351}&6441&6471&\multicolumn{2}{c}{19199}&19276&19289&\multicolumn{2}{c}{12865}&12908&12926\\
\Xcline{1-15}{0.01pt}
\multicolumn{2}{c}{\multirow{2}*{\makecell[c]{
The diquark model}}}&Ref. \cite{Berezhnoy:2011xn}&\multicolumn{2}{c}{5966}&6051&6223&\multicolumn{2}{c}{18754}&18808&18916&\multicolumn{2}{c}{-}&-&-\\
\Xcline{3-3}{0.01pt}
&&Ref. \cite{Mutuk:2021hmi,Mutuk:2022nkw}&\multicolumn{2}{c}{6322}&6354&6385&\multicolumn{2}{c}{19666}&19673&19680&\multicolumn{2}{c}{12401}&12409&12427\\
\Xcline{1-15}{0.01pt}
\multicolumn{2}{c}{\multirow{2}*{\makecell[c]{The QCD sum\\ rule method}}}&Ref. \cite{Wang:2021mma}&6360&6540&6470&6520&18130&18150&18140&18150&-&-&-&-\\
\Xcline{3-3}{0.01pt}
&&Ref. \cite{Wang:2017jtz,Wang:2018poa}&\multicolumn{2}{c}{5990}&6050&6090&\multicolumn{2}{c}{18840}&18840&18850&\multicolumn{2}{c}{-}&-&-\\
\Xcline{1-15}{0.01pt}
\multicolumn{2}{c}{\makecell[c]{The nonrelativistic \\ chiral quark model}}&Ref. \cite{Chen:2019vrj,Chen:2020lgj}&\multicolumn{2}{c}{6510}&6600&6708&\multicolumn{2}{c}{-}&-&-&\multicolumn{2}{c}{12684}&12737&12791\\
\Xcline{1-15}{0.01pt}
\multicolumn{2}{c}{\makecell[c]{An effective\\ potential model} }&Ref. \cite{Zhao:2020nwy}&6346&6476&6441&6475&19154&19226&19214&19232&-&-&-&-\\
\toprule[1.0pt]
\multicolumn{1}{c}{Systems}&\multicolumn{6}{c}{$cb\bar{c}\bar{b}$}&\multicolumn{4}{c}{$cc\bar{c}\bar{b}$}&\multicolumn{4}{c}{$bb\bar{b}\bar{c}$}\\
$J^{P(C)}$&\multicolumn{2}{r}{$0^{++}$\quad\quad\quad}&\multicolumn{2}{c}{$1^{+-}$}&$1^{++}$&$2^{++}$&\multicolumn{1}{c}{$0^{+}$}&\multicolumn{2}{c}{$1^{+}$}&$2^{+}$&\multicolumn{1}{c}{$0^{+}$}&\multicolumn{2}{c}{$1^{+}$}&$2^{+}$\\
\toprule[0.5pt]
\multicolumn{1}{r|}{\multirow{2}*{Our result}}&\multicolumn{2}{r}{\makecell[r]{12760\quad12851}}&12797&12856&12857&128824&9621&96246&9706&9731&16044&16043&16125&16149\\
\multicolumn{1}{r|}{}&\multicolumn{2}{r}{\makecell[r]{12989\quad13008}}&12999&13056&12960&12971&9766&9729&&&16163&16144&&\\
\toprule[0.5pt]
\multicolumn{1}{r|}{\multirow{2}*{Ref. \cite{Zhang:2022qtp}}}&\multicolumn{2}{r}{\makecell[r]{12783\quad12850}}&12802&12835&12851&12852&9665&9676&9699&9713&16061&16046&16079&16089\\
\multicolumn{1}{r|}{}&\multicolumn{2}{r}{\makecell[r]{12966\quad13035}}&12949&12964&12938&12964&9732&9718&&&16100&16089&&\\
\Xcline{1-1}{0.01pt}
\multicolumn{1}{r|}{\multirow{2}*{Ref. \cite{Liu:2019zuc}}}&\multicolumn{2}{r}{\makecell[r]{12835\quad12864}}&12852&12864&12870&12864&9740&9746&9749&9768&16158&16157&16164&16176\\
\multicolumn{1}{r|}{}&\multicolumn{2}{r}{\makecell[r]{12864\quad13050}}&13047&13052&13056&13070&9763&9757&&&16158&16167\\
\Xcline{1-1}{0.01pt}
\multicolumn{1}{r|}{\multirow{3}*{Ref. \cite{Deng:2020iqw}}}&\multicolumn{2}{r}{12894\quad\quad\quad}&\multicolumn{3}{c}{12955}&13000&\multicolumn{1}{c}{9735}&\multicolumn{2}{c}{9766}&9839&\multicolumn{1}{c}{16175}&\multicolumn{2}{c}{16179}&16274\\
\Xcline{2-15}{0.01pt}
\multicolumn{1}{r|}{}&\multicolumn{2}{r}{12829\quad\quad\quad}&\multicolumn{3}{c}{12881}&12925&\multicolumn{1}{c}{9670}&\multicolumn{2}{c}{9683}&9732&\multicolumn{1}{c}{16126}&\multicolumn{2}{c}{16130}&16182\\
\Xcline{2-15}{0.01pt}
\multicolumn{1}{r|}{}&\multicolumn{2}{r}{12354\quad\quad\quad}&\multicolumn{3}{c}{12436}&12548&\multicolumn{1}{c}{9705}&\multicolumn{2}{c}{9705}&9732&\multicolumn{1}{c}{15713}&\multicolumn{2}{c}{15729}&15806\\
\Xcline{1-1}{0.01pt}
\multicolumn{1}{r|}{\multirow{2}*{Ref. \cite{Weng:2020jao}}}&\multicolumn{2}{r}{\makecell[r]{12363\quad12509}}&12425&12477&12524&12537&9318&9335&9384&9526&15712&15719&15851&15882\\
\multicolumn{1}{r|}{}&\multicolumn{2}{r}{\makecell[r]{12682\quad12747}}&12720&12744&12703&12755&9506&9499&&&15862&15854&\\
\Xcline{1-1}{0.01pt}
\multicolumn{1}{r|}{\multirow{6}*{Ref. \cite{Wu:2016vtq}}}&\multicolumn{2}{r}{\makecell[r]{13396\quad13634}}&13478&13592&13510&13590&10144&10174&10231&10273&16832&16840&16884&16917\\
\multicolumn{1}{r|}{}&\multicolumn{2}{r}{\makecell[r]{13483\quad13553}}&13520&13555&13599&13599&10322&10282&&&16952&16915&\\
\multicolumn{1}{r|}{}&\multicolumn{2}{r}{\makecell[r]{12354\quad12592}}&12436&12550&12468&12548&9313&9343&9400&9442&15713&15729&15773&15806\\
\multicolumn{1}{r|}{}&\multicolumn{2}{r}{\makecell[r]{12441\quad12511}}&12478&12513&12557&12557&9491&9451&&&15841&15804\\
\multicolumn{1}{r|}{}&\multicolumn{2}{r}{\makecell[r]{12578\quad12620}}&12496&12583&12611&12690&&&&&\\
\multicolumn{1}{r|}{}&\multicolumn{2}{r}{\makecell[r]{12656\quad12693}}&12653&12735&12700&12700&&&&&\\
\Xcline{1-15}{0.01pt}
\multicolumn{1}{r|}{Ref. \cite{Faustov:2020qfm}}&\multicolumn{2}{r}{\makecell[r]{12813\quad12824}}&12826&12831&12831&12849&9572&9602&9619&9647&16109&16117&16117&16132\\
\Xcline{1-15}{0.01pt}
\multicolumn{1}{r|}{Ref. \cite{Gordillo:2020sgc}}&\multicolumn{2}{r}{12534\quad\quad\quad}&\multicolumn{2}{c}{12510}&12569&12582&9615&\multicolumn{2}{c}{9610}&9719&16040&\multicolumn{2}{c}{16013}&16129\\
\Xcline{1-15}{0.01pt}
\multicolumn{1}{r|}{Ref. \cite{Berezhnoy:2011xn}}&\multicolumn{2}{r}{\makecell[r]{12359\quad12471}}&12424&12488&12485&12566&-&-&-&-&-&-&-&-\\
\Xcline{1-15}{0.01pt}
\multicolumn{1}{r|}{Ref. \cite{Chen:2020lgj}}&\multicolumn{2}{r}{-\quad\quad\quad}&\multicolumn{2}{c}{-}&-&-&\multicolumn{2}{c}{9579}&9590&9613&
\multicolumn{2}{c}{16060}&16062&16068\\
\toprule[0.5pt]
\toprule[1.0pt]
\end{tabular}
\end{lrbox}\scalebox{0.905}{\usebox{\tablebox}}
\end{table*}

\section{SUMMARY}\label{sec6}
The discovery of exotic structures in the di-$J/\psi$ invariant mass spectrum from the LHCb, CMS, and ATLAS collaborations gives us strong confidence to investigate the fully heavy tetraquark system.
Thus, we use the variational method to systematically calculate the masses of all possible configurations for fully heavy tetraquarks within the framework of the constituent quark model.
Meanwhile, we also give the corresponding internal mass contributions,
the relative lengths between (anti)quarks, their lowest meson-meson thresholds, the specific wave function, the magnetic moments, the transition magnetic moments, the radiative decay widths, the rearrangement strong width ratios, and the comparisons with the two different CMI models.

To obtain the above results, we need to construct the total wave functions of the tetraquark states, including the flavor part, the color part, the spin part, and the spatial part, which is chosen to be a simple Gaussian form.
Here, we first estimate the theoretical values of traditional hadrons, which are used to compare the experimental values to prove the reliability of this model.
Before the discussing the numerical analysis, we analyze the stability condition by using only the color-spin interaction.
Then, we obtain the specific numerical values and show them in corresponding Tables and the spatial distribution of valence quarks for the $J^{PC}=0^{++}$ $bb\bar{b}\bar{b}$ ground state in Fig. \ref{1}.

For the $cc\bar{c}\bar{c}$ and $bb\bar{b}\bar{b}$ systems, there are two pure neutral systems with definite C-parity.
There are only two $J^{PC}=0^{++}$ states, one $J^{PC}=1^{+-}$ state, and one $J^{PC}=2^{++}$ state, due to the Pauli principle.
We also find that these states with different quantum numbers are all above the lowest thresholds, and have larger masses.
Since these states are pure neutral particles, the corresponding magnetic moments are all 0 for the ground $cc\bar{c}\bar{c}$ and $bb\bar{b}\bar{b}$ states.
Meanwhile, of course, the variational parameters $C_{11}$ and $C_{22}$ are the same, so the distances of the diquark and antidiquark are also the same.
Moreover, the distances between quark and antiquark are all the same according to the symmetry analysis of Eqs. (\ref{Eq17}-\ref{Eq18}).
Furthermore, three Jacobi coordinates are orthogonal to each other according to Eqs. (\ref{Eq19}-\ref{Eq21}).
Based on this, we take the $J^{PC}=0^{++}$ $bb\bar{b}\bar{b}$ ground state as an example to show the spatial distribution of four valence quarks.
As for the internal contribution, although the kinetic energy part is smaller than that of the $\eta_{b}\eta_{b}$ state, the $V^{C}$ in $\eta_{b}$ is much
more attractive relative to the $J^{PC}=0^{++}$ $bb\bar{b}\bar{b}$ ground state, which is the main reason why this state has a larger mass than the meson-meson threshold.
Similar situations also occur in other systems.

Similar to the $cc\bar{c}\bar{c}$ and $bb\bar{b}\bar{b}$ systems, the $cc\bar{b}\bar{b}$ system has the same number of the allowed ground states.
According to the specific function, their mass contribution mainly comes from the $\bar{3}\otimes3$ component within the diquark-antiquark configuration.
Furthermore, we get the relevant the values of the magnetic moments, the transition magnetic moments, and the radiative decay widths.
We also obtain the rearrangement strong width ratios within the meson-meson configuration.

As for the $cc\bar{c}\bar{b}$ and $bb\bar{b}\bar{c}$ systems, there are more allowed states due to fewer symmetry restrictions.
Considering only the hyperfine potential, we can expect to have a compact stable state for $J^{P}=1^{+}$ $bb\bar{b}\bar{c}$ configuration.
However, since the $V^{C}$ of the tetraquark are less attractive than the corresponding mesons, this state still has a mass larger than the meson-meson threshold.

In the $cb\bar{c}\bar{b}$ system, these states are also pure neutral particles, and we naturally obtain that their variational parameters $C_{11}$ and $C_{22}$ are the same.
There is no constraint from the Pauli principle, so there are four $J^{PC}=0^{++}$ states, four $J^{PC}=1^{+-}$ states, two $J^{PC}=1^{++}$ states, two $J^{PC}=2^{++}$ states.
All of the $cb\bar{c}\bar{b}$ states have larger masses relative to the lowest thresholds.
Moreover, they all have two different rearrangement strong decay modes: $c\bar{c}-b\bar{b}$ and $c\bar{b}-b\bar{c}$.

Then we compare our results with other theoretical work.
Our results are roughly compatible with other nonrelativistic constituent quark models,
although different papers have chosen different potential forms.
Meanwhile, it is also interesting to find that similar mass ranges are given by the QCD sum rules, the Monte Carlo method, and the chiral quark model.
This shows that our results are quite reasonable.

In summary, our theoretical calculations show that the masses of the  $cc\bar{c}\bar{c}$ ground states are around 6.45 GeV, which is obviously lower than 6.9 GeV.
Thus, the experimentally observed $X(6900)$ state does not seem to be a ground $cc\bar{c}\bar{c}$ tetraquark state, but could be a radially or orbitally excited state.
We also find that these lowest states all have a large positive binding energy $B_{T}$.
In other words, all these states are found to have masses above the corresponding two meson decay thresholds via the quark rearrangement.
Hence, we conclude that there is no compact bound ground fully heavy tetraquark state which is stable against the strong decay into two mesons within the constituent quark model.
Finally, we hope that more relevant experimental analyses will be able to focus on this system in the near future.
\\


\section{ACKNOWLEDGMENTS}

This work is  supported by the China National Funds for Distinguished Young Scientists under Grant No. 11825503, National Key Research and Development Program of China under Contract No. 2020YFA0406400, the 111 Project under Grant No. B20063, the National Natural Science Foundation of China under Grant No. 12247101, and the project for top-notch innovative talents of Gansu province. Z.W.L. would like to thank the support from the National Natural Science Foundation of China under Grants No. 12175091, and 11965016, and CAS Interdisciplinary Innovation Team.
\\

\section{Appendix}
In this appendix, we show the masses, binding energies, variational parameters, internal contribution, total wave functions, magnetic moments, transition magnetic moments, radiative decay widths, rearrangement strong width ratios, and the relative lengths between the quarks for the $cc\bar{b}\bar{b}$, $cc\bar{c}\bar{b}$, $bb\bar{b}\bar{c}$, and $cb\bar{c}\bar{b}$ states with different $J^{P(C)}$ quantum numbers and their lowest meson-meson thresholds.

\begin{table*}
\caption{
The masses, binding energies,  variational parameters, the internal contribution, total wave functions, magnetic moments, transition magnetic moments, radiative decay widths, rearrangement strong width ratios, and the relative lengths between quarks for the $J^{P}=0^{+}$, $1^{+}$ $cc\bar{b}\bar{b}$ states and their lowest meson-meson thresholds. The notation is the same as that of Table \ref{cccc}.
}\label{ccbb}
\begin{lrbox}{\tablebox}
\renewcommand\arraystretch{1.45}
\renewcommand\tabcolsep{2.09pt}
\begin{tabular}{c|c|ccc!{\color{black}\vrule width 1.5pt}ccc!{\color{black}\vrule width 1.5pt}c|cc|c|cc}
\midrule[1.5pt]
\toprule[0.5pt]
\multicolumn{1}{c}{$cc\bar{b}\bar{b}$}&\multicolumn{4}{r!{\color{black}\vrule width 1.5pt}}{The contribution from each term}&\multicolumn{3}{c !{\color{black}\vrule width 1.5pt}}{Relative Lengths (fm)}&\multirow{2}*{Overall}&\multicolumn{2}{c}{Present Work}&\multicolumn{2}{c}{CMI Model}\\
\Xcline{1-8}{0.5pt}\Xcline{10-13}{0.5pt}
\multicolumn{1}{r}{$J^{P}=0^{+}$}&&Value&\multicolumn{1}{r}{$B_{c}B_{c}$}&
\multicolumn{1}{r!{\color{black}\vrule width 1.5pt}}{Difference}&$(i,j)$&\multicolumn{1}{r}{Value}
&\multicolumn{1}{c!{\color{black}\vrule width 1.5pt}}{$B_{c}B_{c}$}&&Contribution&Value&Ref. \cite{Weng:2020jao}&Ref. \cite{Wu:2016vtq}\\ \Xcline{1-13}{0.5pt}
\multicolumn{2}{c}{Mass/$B_{T}$}&\multicolumn{1}{r}{12920.0}&12575.8&344.2&(1,2)&0.348
&&\multirow{8}*{\makecell[c]{$c$-quark:\\ \\$m^{eff}_{c}$}}&2$m_{c}$&3836.0&\\
\Xcline{1-5}{0.5pt}
\multirow{2}*{\makecell[c]{Variational\\ Parameters\\ (fm$^{-2}$)}}&\multirow{2}*{\makecell[c]{$C_{11}$\\$C_{22}$ \\ $C_{33}$}}&\multicolumn{1}{r}{\multirow{2}*{\makecell[c]{23.9\\10.5\\12.3}}}
&\multicolumn{1}{r}{\multirow{2}*{\makecell[c]{22.9\\22.9\\-}}}&&(1.3)&0.308
&\multicolumn{1}{c!{\color{black}\vrule width 1.5pt}}{0.235($B_{c}$)}&
&$\frac{\textbf{p}^{2}_{x_{1}}}{2m'_{1}}$&319.0&\\
&&&&&(2,3)&0.308&&&
$\frac{m_{\bar{b}}}{m_{c}+m_{\bar{b}}}\frac{\textbf{p}^{2}_{x_{3}}}{2m'_{3}}$&188.0&\\
\Xcline{1-5}{0.5pt}\multicolumn{2}{c}{Quark Mass}&\multicolumn{1}{r}{14522.0}&14522.0&0.0&(1,4)&0.308&&
&$V^{C}(12)$&\multirow{1}*{-46.9}&$\frac{1}{2}$$m_{cc}$\\
\multicolumn{2}{c}{\multirow{1}{*}{Confinement Potential}}&\multirow{1}{*}{-2420.1}&\multirow{1}{*}{-2795.5}&\multirow{1}{*}{375.4}&(2,4)&0.308&
\multicolumn{1}{c!{\color{black}\vrule width 1.5pt}}{0.235($B_{c}$)}&&\multirow{1}*{\makecell[l]{$\frac{1}{2}[V^{C}(13)+V^{C}(14)$}}&
\multirow{2}*{\makecell[l]{-95.0}}&1585.8\\
\Xcline{1-5}{0.5pt}
\multicolumn{2}{c}{\multirow{1}*{Kinetic Energy}}&\multicolumn{1}{r}{\multirow{1}*{835.9}}&
\multirow{1}*{947.3}&\multirow{1}*{-111.4}
&(3,4)&0.230&&
&$+V^{C}(23)+V^{C}(24)]$&&$\frac{m_{c}}{m_{\bar{b}}+m_{c}}m_{c\bar{b}}$&$2m_{c}$\\
\Xcline{6-8}{0.5pt}
\multicolumn{2}{c}{\multirow{1}*{CS Interaction}}&\multicolumn{1}{r}{\multirow{1}*{-7.0}}&\multirow{1}*{-98.0}&
\multirow{1}*{91.0}
&\multicolumn{2}{r}{(1,2)-(3,4):}&0.226 fm&&$-D$&-983.0&1578.7&3449.6\\
\Xcline{10-13}{0.5pt}\Xcline{1-5}{0.5pt}
\multicolumn{1}{c|}{\multirow{4}{*}{$V^{C}$}}&\multicolumn{1}{c}{\multirow{1}{*}{(1,2)}}&\multicolumn{1}{r}{-46.9}&&
&\multicolumn{2}{r}{Radius:}&0.151 fm&&\multirow{1}*{Subtotal}&\multirow{1}*{3218.1}&
\multirow{1}*{3164.5}&\multirow{1}*{3449.6}\\
\Xcline{9-13}{0.5pt}\Xcline{6-8}{1.6pt}
&\multicolumn{1}{c}{\multirow{1}{*}{(2,3)}}&\multicolumn{1}{r}{-47.5}&&\multicolumn{1}{c}{}&&&
&\multirow{8}*{\makecell[c]{$\bar{b}$-quark:\\ \\$m^{eff}_{\bar{b}}$}}&$2m_{b}$&10686.0&\\
&\multicolumn{1}{c}{\multirow{1}{*}{(1,4)}}&\multicolumn{1}{r}{-47.5}&&\multicolumn{1}{r}{}
&(1,3)&-47.5&$-414.8(B_{c})$
&&$\frac{\textbf{p}^{2}_{x_{2}}}{2m'_{2}}$&216.4&\\
\Xcline{2-5}{0.5pt}
\multicolumn{1}{c|}{\multirow{1}{*}{}}&\multicolumn{1}{c}{\multirow{1}{*}{Subtotal}}&\multicolumn{1}{r}{-454.0}&\multicolumn{1}{r}{-829.5}&\multicolumn{1}{c|}{375.3}
&\multicolumn{1}{c}{(3,4)}&-217.1&&&
$\frac{m_{c}}{m_{c}+m_{\bar{b}}}\frac{\textbf{p}^{2}_{x_{3}}}{2m'_{3}}$&67.5&\\
\Xcline{1-5}{0.5pt}
\multicolumn{2}{c}{\multirow{1}*{Total Contribution}}&\multicolumn{1}{r}{\multirow{1}*{-374.9}}&\multicolumn{1}{r}{19.8}&\multicolumn{1}{c|}{\multirow{1}*{355.1}}&(2,4)&-47.5&$-414.8(B_{c})$&
&$V^{C}(34)$&\multirow{1}*{-217.1}&$\frac{1}{2}$$m_{bb}$\\
\Xcline{1-8}{1.6pt}
\multicolumn{8}{l!{\color{black}\vrule width 1.5pt}}{Total Wave function:}&&\multirow{1}*{\makecell[l]{$\frac{1}{2}[V^{C}(13)+V^{C}(14)$}}&
\multirow{2}*{\makecell[l]{-95.0}}&4764.8\\
\multicolumn{8}{r!{\color{black}\vrule width 1.5pt}}{$\Psi_{tot}=0.259|F\rangle|R^{s}\rangle|[\phi_{2}\chi_{6}]\rangle-0.966|F\rangle|R^{s}\rangle|[\phi_{1}\chi_{5}]\rangle=-0.589|F\rangle|R^{s}\rangle|[\psi_{1}\zeta_{5}]\rangle$}&&$+V^{C}(23)+V^{C}(24)]$&&$\frac{m_{\bar{b}}}{m_{\bar{b}}+m_{c}}m_{c\bar{b}}$&$2m_{b}$\\
\multicolumn{8}{r!{\color{black}\vrule width 1.5pt}}{$+0.095|F\rangle|R^{s}\rangle|[\psi_{1}\zeta_{6}]\rangle+0.608|F\rangle|R^{s}\rangle|[\psi_{2}\zeta_{5}]\rangle+
0.524|F\rangle|R^{s}\rangle|[\psi_{2}\zeta_{6}]\rangle$}&&$-D$&-983.0&4743.6&10105.8\\
\Xcline{1-8}{0.3pt}\Xcline{10-13}{0.5pt}
\multicolumn{8}{l!{\color{black}\vrule width 1.5pt}}{The rearrangement strong width ratios :}&&
\multirow{1}*{Subtotal}&\multirow{1}*{9674.8}&9508.4&\multirow{1}*{10105.8}\\
\Xcline{9-13}{0.5pt}
\multicolumn{8}{l!{\color{black}\vrule width 1.5pt}}{$\Gamma_{T_{c^{2}\bar{b}^{2}}(12920.0,0^{++})\rightarrow B^{*}_{c}B^{*}_{c}}:\Gamma_{T_{c^{2}\bar{b}^{2}}(12920.0.0,0^{++})\rightarrow B_{c}B_{c}}=1:50$
}&\multirow{5}*{\makecell[c]{CS \\Interaction}}&\multirow{1}*{$\frac{1}{2}V^{SS}(12)$}&\multirow{1}*{9.6}&$\frac{8}{3}v_{cc}+\frac{8}{3}v_{bb}$&$\frac{8}{3}C_{cc}+\frac{8}{3}C_{bb}$\\
\Xcline{1-8}{0.3pt}
\multicolumn{8}{l!{\color{black}\vrule width 1.5pt}}{The radiative decay widths :}&&\multirow{1}*{$\frac{1}{2}V^{SS}(34)$}&5.5&9.5+5.1&14.1+7.7\\
\multicolumn{8}{l!{\color{black}\vrule width 1.5pt}}{$\Gamma_{T_{c^{2}\bar{b}^{2}}(12960.9,2^{+})\rightarrow T_{c^{2}\bar{b}^{2}}(12920.0,0^{+})\gamma}=0\, \rm{keV}$ ;}&&\multirow{2}*{\makecell[c]{$-\frac{1}{2}(V^{SS}(13)+V^{SS}(14)$\\$V^{SS}(23)+V^{SS}(24))$}}
&\multirow{2}*{-22.1}&$-\frac{32}{3}v_{c\bar{b}}$&$-\frac{32}{3}C_{c\bar{b}}$\\
\multicolumn{8}{l!{\color{black}\vrule width 1.5pt}}{$\Gamma_{T_{c^{2}\bar{b}^{2}}(12939.9,1^{+})\rightarrow T_{c^{2}\bar{b}^{2}}(12920.0,0^{+})\gamma}=3.8\, \rm{keV}$}&&&&-31.5&-35.2\\
\Xcline{10-13}{0.5pt}
\Xcline{1-8}{0.3pt}
\multicolumn{8}{l!{\color{black}\vrule width 1.5pt}}{The magnetic moments : \quad $\mu_{T_{c^{2}\bar{b}^{2}}(12920.0,0^{+})}=\langle\Psi^{0^{+}}_{tot}|\hat{\mu^{z}}|\Psi^{0^{+}}_{tot}\rangle= 0$}
&&\multirow{1}*{Subtotal}&\multirow{1}*{-7.0}&\multirow{1}*{-16.9}&-13.3\\
\Xcline{9-13}{0.5pt}\Xcline{1-8}{0.3pt}
\multicolumn{8}{l!{\color{black}\vrule width 1.5pt}}{The transition magnetic moments: }&\multicolumn{2}{l|}{\multirow{1}*{Matrix nondiagonal element}}&\multirow{1}*{-10.7}&-59.7&46.1\\
\Xcline{9-13}{0.5pt}
\multicolumn{8}{l!{\color{black}\vrule width 1.5pt}}{$\mu_{T_{c^{2}\bar{b}^{2}}(12960.9,2^{+})\rightarrow T_{c^{2}\bar{b}^{2}}(19240.0,0^{+})\gamma}=\langle\Psi^{2^{+}}_{tot}|\hat{\mu^{z}}|\Psi^{0^{+}}_{tot}\rangle= 0$}&
\multicolumn{2}{l|}{\multirow{1}*{Total contribution}}&\multirow{2}*{12920.0}&\multirow{2}*{12596.3}&\multirow{2}*{13496.0}\\
\Xcline{9-10}{1.6pt}
\multicolumn{10}{l!{\color{black}\vrule width 1.5pt}}{$\mu_{T_{c^{2}\bar{b}^{2}}(12939.9,1^{+})\rightarrow T_{c^{2}\bar{b}^{2}}(12920.0,0^{+})\gamma}=\langle\Psi^{1^{+}}_{tot}|\hat{\mu^{z}}|\Psi^{0^{+}}_{tot}\rangle=0.966\times\frac{4}{\sqrt{6}}(\mu_{c}-\mu_{\bar{b}})=0.534 \mu_{N}$}
&&&\\
\toprule[1.5pt]
\multicolumn{1}{r}{$J^{P}=1^{+}$}&&Value&\multicolumn{1}{r}{$B^{*}_{c}B_{c}$}&
\multicolumn{1}{r!{\color{black}\vrule width 1.5pt}}{Difference}&\multicolumn{3}{c !{\color{black}\vrule width 1.5pt}}{Relative Lengths (fm)}&&Contribution&Value&Ref. \cite{Weng:2020jao}&Ref. \cite{Wu:2016vtq}\\ \Xcline{1-13}{0.5pt}
\multicolumn{2}{c}{Mass/$B_{T}$}&\multicolumn{1}{r}{12939.9}&12638.4&301.5&$(i,j)$&\multicolumn{1}{r}{Value}
&\multicolumn{1}{c!{\color{black}\vrule width 1.5pt}}{$B^{*}_{c}B_{c}$}&
\multirow{7}*{\makecell[c]{$c$-quark:\\ \\$m^{eff}_{c}$}}&2$m_{c}$&3836.0&\\
\Xcline{1-8}{0.5pt}
\multirow{2}*{\makecell[c]{Variational\\ Parameters\\ (fm$^{-2}$)}}&\multirow{2}*{\makecell[c]{$C_{11}$\\$C_{22}$ \\ $C_{33}$}}&\multicolumn{1}{r}{\multirow{2}*{\makecell[c]{24.8\\10.3\\11.1}}}&
\multicolumn{1}{r}{\multirow{2}*{\makecell[c]{20.2\\22.9\\-}}}&&(1,2)&0.351&&&
&$\frac{\textbf{p}^{2}_{x_{1}}}{2m'_{1}}$&312.7&\\
&&&&&(1,3)&0.317&\multicolumn{1}{c!{\color{black}\vrule width 1.5pt}}{0.250($B^{*}_{c}$)}&&
$\frac{m_{\bar{b}}}{m_{c}+m_{\bar{b}}}\frac{\textbf{p}^{2}_{x_{3}}}{2m'_{3}}$&169.1&$\frac{1}{2}$$m_{cc}$\\
\Xcline{1-5}{0.5pt}
\multicolumn{2}{c}{Quark Mass}&\multicolumn{1}{r}{14522.0}&14522.0&0.0&(2,3)&0.317&&
&\multirow{2}*{\makecell[c]{$V^{C}(12)$\\$\frac{1}{2}[V^{C}(13)+V^{C}(14)$\\$+V^{C}(23)+V^{C}(24)]$}}
&\multirow{2}*{\makecell[c]{-43.5\\\\-83.2}}&1585.8\\
 \multicolumn{2}{c}{\multirow{1}{*}{Confinement Potential}}&\multirow{1}{*}{-2400.7}&\multirow{1}{*}{-2741.1}&\multirow{1}{*}{340.4}&(1,4)&0.317&
&&&&$\frac{m_{c}}{m_{\bar{b}}+m_{c}}m_{c\bar{b}}$&$2m_{c}$\\
\Xcline{1-5}{0.5pt}
\multicolumn{2}{c}{\multirow{1}*{Kinetic Energy}}&\multicolumn{1}{r}{\multirow{1}*{814.0}}&\multirow{1}*{891.5}&\multirow{1}*{-77.5}
&(2,4)&0.317&\multicolumn{1}{c!{\color{black}\vrule width 1.5pt}}{0.235($B_{c}$)}&&$-D$&-983.0&1587.7&3449.6\\
\Xcline{10-13}{0.5pt}
\multicolumn{2}{c}{\multirow{1}*{CS Interaction}}&\multicolumn{1}{r}{\multirow{1}*{4.6}}&\multirow{1}*{-34.0}&
\multirow{1}*{38.6}
&(3,4)&0.226&&&\multirow{1}*{Subtotal}&\multirow{1}*{3208.1}&
\multirow{1}*{3164.5}&\multirow{1}*{3449.6}\\
\Xcline{1-8}{0.5pt}\Xcline{9-13}{0.5pt}
\multicolumn{1}{c|}{\multirow{4}{*}{$V^{C}$}}&\multicolumn{1}{c}{\multirow{1}{*}{(1,2)}}&\multicolumn{1}{r}{-43.5}&&
&\multicolumn{2}{r}{(1,2)-(3,4):}&0.238 fm
&\multirow{8}*{\makecell[c]{$\bar{b}$-quark:\\ \\$m^{eff}_{\bar{b}}$}}&$2m_{b}$&10686.0&\\
&\multicolumn{1}{c}{\multirow{1}{*}{(2,3)}}&\multicolumn{1}{r}{-41.6}&&\multicolumn{1}{c!{\color{black}\vrule width 1.5pt}}{}&\multicolumn{2}{r}{Radius:}&0.157 fm&
&$\frac{\textbf{p}^{2}_{x_{2}}}{2m'_{2}}$&271.5&\\
\Xcline{6-8}{1.6pt}
&\multicolumn{1}{c}{\multirow{1}{*}{(1,4)}}&\multicolumn{1}{r}{-41.6}&&\multicolumn{1}{r}{}&(1,3)&-41.6&$-360.4(B^{*}_{c})$&&
$\frac{m_{c}}{m_{c}+m_{\bar{b}}}\frac{\textbf{p}^{2}_{x_{3}}}{2m'_{3}}$&60.7&$\frac12 m_{bb}$\\
\Xcline{2-5}{0.5pt}
\multicolumn{1}{c|}{\multirow{1}{*}{}}&\multicolumn{1}{c}{\multirow{1}{*}{Subtotal}}&\multicolumn{1}{r}{-454.0}&\multicolumn{1}{r}{-829.5}&\multicolumn{1}{c|}{375.3}
&\multicolumn{1}{c}{(3,4)}&-225.5&
&&\multirow{2}*{\makecell[c]{$V^{C}(34)$\\$\frac{1}{2}[V^{C}(13)+V^{C}(14)$\\$+V^{C}(23)+V^{C}(24)]$}}
&\multirow{2}*{\makecell[c]{-225.5\\\\-83.2}}&4764.8\\
\Xcline{1-5}{0.5pt}
\multicolumn{2}{c}{Total Contribution}&\multicolumn{1}{r}{383.9}&\multicolumn{1}{r}{82.4}&\multicolumn{1}{c|}{\multirow{1}*{301.5}}&(2,4)&-41.6&$-414.8(B_{c})$&&
&&$\frac{m_{\bar{b}}}{m_{\bar{b}}+m_{c}}m_{c\bar{b}}$&$2m_{b}$\\
\Xcline{1-8}{1.6pt}
\multicolumn{8}{l!{\color{black}\vrule width 1.5pt}}{Total Wave function: \quad $\Psi_{tot}=|F\rangle|R^{s}\rangle|[\phi_{1}\chi_{2}]\rangle=$}&&$-D$&-983.0&4743.6&10105.8\\
\Xcline{10-13}{0.5pt}
\multicolumn{8}{r!{\color{black}\vrule width 1.5pt}}{$-0.419|F\rangle|R^{s}\rangle|[\psi_{1}\zeta_{2}]\rangle-0.393|F\rangle|R^{s}\rangle|[\psi_{1}\zeta_{3}]\rangle-0.066|F\rangle|R^{s}\rangle|[\psi_{1}\zeta_{4}]\rangle$}&&\multirow{1}*{Subtotal}&\multirow{1}*{9726.5}&
\multirow{1}*{9508.4}&\multirow{1}*{10105.8}\\
\Xcline{9-13}{0.5pt}
\multicolumn{8}{r!{\color{black}\vrule width 1.5pt}}{$+0.587|F\rangle|R^{s}\rangle|[\psi_{2}\zeta_{2}]\rangle+0.557|F\rangle|R^{s}\rangle|[\psi_{2}\zeta_{3}]\rangle+0.105|F\rangle|R^{s}\rangle|[\psi_{2}\zeta_{4}]\rangle$}
&\multirow{5}*{\makecell[c]{CS \\Interaction}}&\multirow{1}*{\makecell[c]{$\frac{1}{2}V^{SS}(12)$
}}&\multirow{1}*{5.7}&$\frac83v_{cc}+\frac83v_{\bar{b}\bar{b}}$&$\frac83C_{cc}+\frac83C_{\bar{b}\bar{b}}$\\
\Xcline{1-8}{0.3pt}
\multicolumn{8}{l!{\color{black}\vrule width 1.5pt}}{The rearrangement strong decay channel: $B^{*}B_{c}$}
&&\multirow{1}*{\makecell[c]{$\frac{1}{2}V^{SS}(34)$
}}&\multirow{1}*{4.6}&9.5+5.1&14.1+7.7\\
\Xcline{1-8}{0.3pt}
\multicolumn{8}{l!{\color{black}\vrule width 1.5pt}}{The radiative decay widths: $\Gamma_{T_{c^{2}\bar{b}^{2}}(12960.9,2^{+})\rightarrow T_{c^{2}\bar{b}^{2}}(12939.9.9,1^{+})\gamma}=3.6 \, \rm{keV}$}&&\multirow{2}*{\makecell[c]{$-\frac{1}{4}(V^{SS}(13)+V^{SS}(14)$\\$+V^{SS}(23)+V^{SS}(24)$
}}&\multirow{2}*{-10.6}&$-\frac{16}{3}v_{\bar{c}\bar{b}}$&$-\frac{16}{3}C_{\bar{c}\bar{b}}$\\
\multicolumn{8}{r!{\color{black}\vrule width 1.5pt}}{$\Gamma_{T_{c^{2}\bar{b}^{2}}(12939.9,1^{+})\rightarrow T_{c^{2}\bar{b}^{2}}(12920.0,0^{+})\gamma}=3.8\, \rm{keV}$}&&\multirow{2}*{}&\multirow{2}*{}
&-15.7&-17.6\\
\Xcline{10-13}{0.5pt}\Xcline{1-8}{0.3pt}
\multicolumn{8}{l!{\color{black}\vrule width 1.5pt}}{The magnetic moments:}
&&\multirow{1}*{Subtotal}&\multirow{1}*{4.6}&\multirow{1}*{-1.2}&4.3\\
\Xcline{9-13}{0.5pt}
\multicolumn{8}{l!{\color{black}\vrule width 1.5pt}}{$\mu_{T_{c^{2}\bar{b}^{2}}(12939.9,1^{+})}=\langle\Psi^{1^{+}}_{tot}|\hat{\mu^{z}}|\Psi^{1^{+}}_{tot}\rangle=\mu_{c}+\mu_{\bar{b}}=0.490 \mu_{N}$}
&\multicolumn{2}{l|}{\multirow{1}*{Total contribution}}&\multirow{1}*{\makecell[c]{12939.9}}&\multirow{1}*{\makecell[c]{12671.6}}&\multirow{1}*{\makecell[c]{13560.0}}\\
\Xcline{9-13}{1.6pt}\Xcline{1-8}{0.3pt}
\multicolumn{13}{l}{The transition magnetic moments: $\mu_{T_{b^{2}\bar{b}^{2}}(12960.9,2^{+})\rightarrow T_{b^{2}\bar{b}^{2}}(12939.9,1^{+})\gamma}=\langle\Psi^{2^{+}}_{tot}|\hat{\mu^{z}}|\Psi^{1^{+}}_{tot}\rangle=
\mu_{c}-\mu_{\bar{b}}=0.342 \mu_{N}$}\\
\multicolumn{13}{l}{\quad\quad\quad\quad\quad\quad\quad\quad\quad\quad\quad\quad\quad\quad\quad$\mu_{T_{c^{2}\bar{b}^{2}}(12939.9,1^{+})\rightarrow T_{c^{2}\bar{b}^{2}}(12920.0,0^{+})\gamma}=\langle\Psi^{1^{+}}_{tot}|\hat{\mu^{z}}|\Psi^{0^{+}}_{tot}\rangle=0.966\times\frac{4}{\sqrt{6}}(\mu_{c}-\mu_{\bar{b}})=0.534 \mu_{N}$}\\
\toprule[0.5pt]
\toprule[1.5pt]
\end{tabular}
\end{lrbox}\scalebox{0.868}{\usebox{\tablebox}}
\end{table*}

\begin{table*}
\caption{The masses, binding energies,  variational parameters, the internal contribution, total wave functions, magnetic moments, transition magnetic moments, radiative decay widths, rearrangement strong width ratios, and the relative lengths between quarks for the $J^{P}=2^{+}$ $cc\bar{b}\bar{b}$ state and its lowest meson-meson threshold. The notation is the same as that in Table \ref{cccc}.
}\label{ccbb2}
\begin{lrbox}{\tablebox}
\renewcommand\arraystretch{1.45}
\renewcommand\tabcolsep{2.09pt}
\begin{tabular}{c|c|ccc !{\color{black}\vrule width 1.5pt} ccc !{\color{black}\vrule width 1.5pt} c|cc|c|cc}
\midrule[1.5pt]
\toprule[0.5pt]
\multicolumn{1}{c}{$cc\bar{b}\bar{b}$}&\multicolumn{4}{r !{\color{black}\vrule width 1.5pt}}{The contribution from each term}&\multicolumn{3}{c !{\color{black}\vrule width 1.5pt}}{Relative Lengths (fm)}&\multirow{2}*{Overall}&\multicolumn{2}{c}{Present Work}&\multicolumn{2}{c}{CMI Model}\\
\Xcline{1-8}{0.5pt}\Xcline{10-13}{0.5pt}
\multicolumn{1}{r}{$J^{P}=2^{+}$}&&Value&\multicolumn{1}{r}{$B^{*}_{c}B^{*}_{c}$}&
\multicolumn{1}{r !{\color{black}\vrule width 1.5pt}}{Difference}&$(i,j)$&\multicolumn{1}{r}{Value}
&\multicolumn{1}{c !{\color{black}\vrule width 1.5pt}}{$B^{*}_{c}B^{*}_{c}$}&&Contribution&Value&Ref. \cite{Weng:2020jao}&Ref. \cite{Wu:2016vtq}\\ \Xcline{1-13}{0.5pt}
\multicolumn{2}{c}{Mass/$B_{T}$}&\multicolumn{1}{r}{12960.9}&12700.9&260.0&(1,2)&0.355&&
\multirow{7}*{\makecell[c]{$c$-quark:\\ \\$m^{eff}_{c}$}}&2$m_{c}$&3836.0&\\
\Xcline{1-5}{0.5pt}
\multirow{2}*{\makecell[c]{Variational\\ Parameters\\ (fm$^{-2}$)}}&\multirow{2}*{\makecell[c]{$C_{11}$\\$C_{22}$ \\ $C_{33}$}}&\multicolumn{1}{r}{\multirow{2}*{\makecell[c]{24.5\\10.1\\10.7}}}&
\multicolumn{1}{r}{\multirow{2}*{\makecell[c]{20.2\\20.2\\-}}}&&(1.3)&0.322&
\multicolumn{1}{c!{\color{black}\vrule width 1.5pt}}{0.250($B^{*}_{c}$)}&
&$\frac{\textbf{p}^{2}_{x_{1}}}{2m'_{1}}$&306.2&\\
&&&&&(2,3)&0.322&&&
$\frac{m_{\bar{b}}}{m_{c}+m_{\bar{b}}}\frac{\textbf{p}^{2}_{x_{3}}}{2m'_{3}}$&163.1&$\frac{1}{2}$$m_{cc}$\\
\Xcline{1-5}{0.5pt}
\multicolumn{2}{c}{Quark Mass}&\multicolumn{1}{r}{14522.0}&14522.0&0.0&(1,4)&0.322&&
&\multirow{2}*{\makecell[c]{$V^{C}(12)$\\$\frac{1}{2}[V^{C}(13)+V^{C}(14)$\\$+V^{C}(23)+V^{C}(24)]$}}
&\multirow{2}*{\makecell[c]{-38.9\\\\-77.4}}&1585.8\\
\multicolumn{2}{c}{\multirow{1}{*}{Confinement Potential}}&\multirow{1}{*}{-2382.1}&\multirow{1}{*}{-2686.8}&\multirow{1}{*}{304.7}&(2,4)&0.322&
\multicolumn{1}{c!{\color{black}\vrule width 1.5pt}}{0.250($B^{*}_{c}$)}&&&&$\frac{m_{c}}{m_{\bar{b}}+m_{c}}m_{c\bar{b}}$&$2m_{c}$\\
\Xcline{1-5}{0.5pt}
\multicolumn{2}{c}{\multirow{1}*{Kinetic Energy}}&\multicolumn{1}{r}{\multirow{1}*{795.6}}&\multirow{1}*{835.7}&\multirow{1}*{-40.1}
&(3,4)&0.227&&&$-D$&-983.0&1587.7&3449.6\\
\Xcline{6-8}{0.5pt}\Xcline{10-13}{0.5pt}
\multicolumn{2}{c}{\multirow{1}*{CS Interaction}}&\multicolumn{1}{r}{\multirow{1}*{25.3}}&\multirow{1}*{30.0}&\multirow{1}*{-4.7}
&\multicolumn{2}{r}{(1,2)-(3,4):}&0.243 fm&&\multirow{1}*{Subtotal}&\multirow{1}*{3206.0}&
\multirow{1}*{3164.5}&\multirow{1}*{3449.6}\\
\Xcline{9-13}{0.5pt}\Xcline{1-5}{0.5pt}
\multicolumn{1}{c|}{\multirow{4}{*}{$V^{C}$}}&\multicolumn{1}{c}{\multirow{1}{*}{(1,2)}}&\multicolumn{1}{r}{-38.9}&&
&\multicolumn{2}{r}{Radius:}&0.160 fm&\multirow{8}*{\makecell[c]{$\bar{b}$-quark:\\ \\$m^{eff}_{\bar{b}}$}}&$2m_{b}$&10686.0&\\
\Xcline{6-8}{1.6pt}
&\multicolumn{1}{c}{\multirow{1}{*}{(2,3)}}&\multicolumn{1}{r}{-38.7}&&\multicolumn{1}{c}{}&&&&&
$\frac{\textbf{p}^{2}_{x_{2}}}{2m'_{2}}$&268.8&\\
&\multicolumn{1}{c}{\multirow{1}{*}{(1,4)}}&\multicolumn{1}{r}{-38.7}&&\multicolumn{1}{r}{}
&(1,3)&-38.7&$-360.4(B^{*}_{c})$&&
$\frac{m_{c}}{m_{c}+m_{\bar{b}}}\frac{\textbf{p}^{2}_{x_{3}}}{2m'_{3}}$&58.5&$\frac12 m_{bb}$\\
\Xcline{2-5}{0.5pt}
\multicolumn{1}{c|}{\multirow{1}{*}{}}&\multicolumn{1}{c}{\multirow{1}{*}{Subtotal}}&\multicolumn{1}{r}{-416.0}&\multicolumn{1}{r}{-720.8}&\multicolumn{1}{c|}{304.8}
&\multicolumn{1}{c}{(3,4)}&-222.4&&
&\multirow{2}*{\makecell[c]{$V^{C}(34)$\\$\frac{1}{2}[V^{C}(13)+V^{C}(14)$\\$+V^{C}(23)+V^{C}(24)]$}}
&\multirow{2}*{\makecell[c]{-222.4\\\\-77.4}}&4764.8\\
\Xcline{1-5}{0.5pt}
\multicolumn{2}{c}{\multirow{1}*{Total Contribution}}&\multicolumn{1}{r}{\multirow{1}*{404.9}}&\multicolumn{1}{r}{144.9}&\multicolumn{1}{c|}{\multirow{1}*{260.0}}&(2,4)&-38.7&$-360.4(B^{*}_{c})$&&&
&$\frac{m_{\bar{b}}}{m_{\bar{b}}+m_{c}}m_{c\bar{b}}$&$2m_{b}$\\
\Xcline{1-8}{1.6pt}
\multicolumn{8}{l!{\color{black}\vrule width 1.5pt}}{Total Wave function:}&&$-D$&-983.0&4743.6&10105.8\\
\Xcline{10-13}{0.5pt}
\multicolumn{8}{l!{\color{black}\vrule width 1.5pt}}{$\Psi_{tot}=|F\rangle|R^{s}\rangle|[\phi_{1}\chi_{2}]\rangle=0.577|F\rangle|R^{s}\rangle|[\psi_{1}\zeta_{1}]\rangle-0.816|F\rangle|R^{s}\rangle|[\psi_{2}\zeta_{1}]\rangle$}&&\multirow{1}*{Subtotal}&\multirow{1}*{9730.5}&
\multirow{1}*{9508.4}&\multirow{1}*{10105.8}\\
\Xcline{9-13}{0.5pt}\Xcline{1-8}{0.3pt}
\multicolumn{8}{l!{\color{black}\vrule width 1.5pt}}{The rearrangement strong decay channel: $B^{*}_{c}B^{*}_{c}$}&\multirow{5}*{\makecell[c]{CS \\Interaction}}&\multirow{2}*{\makecell[c]{$\frac{1}{2}V^{SS}(12)$
}}&\multirow{2}*{9.3}&$\frac83v_{cc}$&$\frac83C_{cc}$\\
\Xcline{1-8}{0.3pt}
\multicolumn{3}{l}{The radiative decay widths: }&\multicolumn{5}{r!{\color{black}\vrule width 1.5pt}}{$\Gamma_{T_{c^{2}\bar{b}^{2}}(12960.9,2^{+})\rightarrow T_{c^{2}\bar{b}^{2}}(12920.0,0^{+})\gamma}=0\, \rm{keV}\quad$}
&&&&9.5&14.1\\
\multicolumn{8}{r!{\color{black}\vrule width 1.5pt}}{$\Gamma_{T_{c^{2}\bar{b}^{2}}(12960.9,2^{+})\rightarrow T_{c^{2}\bar{b}^{2}}(12939.9.9,1^{+})\gamma}=3.6\, \rm{keV}$}&&\multirow{2}*{\makecell[c]{$\frac{1}{2}V^{SS}(34)$
}}&\multirow{2}*{4.6}&$\frac{8}{3}v_{\bar{b}\bar{b}}$&$\frac{8}{3}C_{\bar{b}\bar{b}}$\\
\Xcline{1-8}{0.3pt}
\multicolumn{8}{l!{\color{black}\vrule width 1.5pt}}{The magnetic moments:}&&\multirow{2}*{}&\multirow{2}*{}
&5.1&7.7\\
\multicolumn{8}{l!{\color{black}\vrule width 1.5pt}}{$\mu_{T_{c^{2}\bar{b}^{2}}(12969.9,2^{+})}=\langle\Psi^{2^{+}}_{tot}|\hat{\mu^{z}}|\Psi^{2^{+}}_{tot}\rangle
=2\mu_{c}+2\mu_{\bar{b}}=0.982\mu_{N}$ }
&&\multirow{2}*{\makecell[c]{$\frac{1}{4}[V^{SS}(13)+V^{SS}(14)$\\$+V^{SS}(23)+V^{SS}(24)]$
}}&\multirow{2}*{10.3}&$\frac{16}{3}v_{\bar{c}\bar{b}}$&$\frac{16}{3}C_{\bar{c}\bar{b}}$\\
\Xcline{1-8}{0.3pt}
\multicolumn{8}{l!{\color{black}\vrule width 1.5pt}}{The transition magnetic moments :}
&&\multirow{2}*{}&\multirow{2}*{}&15.7&17.6\\
\Xcline{10-13}{0.5pt}
\multicolumn{8}{l!{\color{black}\vrule width 1.5pt}}{$\mu_{T_{c^{2}\bar{b}^{2}}(12960.9,2^{+})\rightarrow T_{c^{2}\bar{b}^{2}}(19240.0,0^{+})\gamma}=\langle\Psi^{2^{+}}_{tot}|\hat{\mu^{z}}|\Psi^{0^{+}}_{tot}\rangle=0$ ;}
&&\multirow{1}*{Subtotal}&\multirow{1}*{25.3}&\multirow{1}*{30.3}&39.5\\
\Xcline{9-13}{0.5pt}
\multicolumn{8}{l!{\color{black}\vrule width 1.5pt}}{$\mu_{T_{b^{2}\bar{b}^{2}}(12960.9,2^{+})\rightarrow T_{b^{2}\bar{b}^{2}}(12939.9,1^{+})\gamma}=\langle\Psi^{2^{+}}_{tot}|\hat{\mu^{z}}|\Psi^{1^{+}}_{tot}\rangle=
\mu_{c}-\mu_{\bar{b}}=0.342 \mu_{N}$}&\multicolumn{2}{l|}{Total contribution}&\multirow{1}*{\makecell[c]{12960.9}}&\multirow{1}*{\makecell[c]{12703.1}}
&\multirow{1}*{\makecell[c]{13595.0}}\\
\toprule[0.5pt]
\toprule[1.5pt]
\end{tabular}
\end{lrbox}\scalebox{0.868}{\usebox{\tablebox}}
\end{table*}

\begin{table*}
\caption{ The masses, binding energy,  variational parameters, the internal contribution, total wave functions, magnetic moments, transition magnetic moments, radiative decay widths, rearrangement strong width ratios, and the relative lengths between quarks for the $J^{P}=0^{+}$, $1^{+}$ $cc\bar{c}\bar{b}$ states and their lowest meson-meson thresholds. The notation is the same as that in Table \ref{cccc}.
}\label{cccb1}
\begin{lrbox}{\tablebox}
\renewcommand\arraystretch{1.45}
\renewcommand\tabcolsep{1.8pt}
\begin{tabular}{c|c|ccc !{\color{black}\vrule width 1.5pt} ccc !{\color{black}\vrule width 1.5pt} c|cc|c|cc}
\midrule[1.5pt]
\toprule[0.5pt]
\multicolumn{1}{c}{$cc\bar{c}\bar{b}$}&\multicolumn{4}{r!{\color{black}\vrule width 1.5pt}}{The contribution from each term}&\multicolumn{3}{c!{\color{black}\vrule width 1.5pt}}{Relative Lengths (fm)}
&\multirow{2}*{Overall}&\multicolumn{2}{c}{Present Work}&\multicolumn{2}{c}{CMI Model}\\
\Xcline{1-8}{0.5pt}\Xcline{10-13}{0.5pt}
\multicolumn{1}{r}{$J^{P}=0^{+}$}&&Value&\multicolumn{1}{r}{$B_{c}\eta_{c}$}&
\multicolumn{1}{r!{\color{black}\vrule width 1.5pt}}{Difference}&$(i,j)$&\multicolumn{1}{r}{Value}
&\multicolumn{1}{c!{\color{black}\vrule width 1.5pt}}{$B_{c}\eta_{c}$}&&Contribution&Value&Ref. \cite{Weng:2020jao}&Ref. \cite{Wu:2016vtq}\\ \Xcline{1-13}{0.5pt}
\Xcline{1-5}{0.5pt}
\multicolumn{2}{c}{Mass/$B_{T}$}&\multicolumn{1}{r}{9620.5}&9286.4&332.1&(1,2)&0.418&&
\multirow{5}*{\makecell[c]{$c$-quark:\\ \\$m^{eff}_{c}$}}&2$m_{c}$&3836.0
&\multirow{5}*{\makecell[c]{$-\frac{1}{4}m_{cc}$\\-792.9\\$\frac{5m_{c}}{4(m_{c}+m_{\bar{c}})}m_{c\bar{c}}$\\1917.8\\
$\frac{5m_{c}}{4(m_{c}+m_{\bar{b}})}m_{c\bar{b}}$\\1973.7}}\\
\Xcline{1-5}{0.5pt}
\multirow{2}*{\makecell[c]{Variational\\ Parameters\\ (fm$^{-2}$)}}&\multirow{2}*{\makecell[c]{$C_{11}$\\$C_{22}$ \\ $C_{33}$}}&\multicolumn{1}{r}{\multirow{2}*{\makecell[c]{11.4\\7.2\\15.2}}}&
\multicolumn{1}{r}{\multirow{2}*{\makecell[c]{22.9\\15.0\\-}}}&&(1.3)&0.325&
\multicolumn{1}{c!{\color{black}\vrule width 1.5pt}}{0.290($\eta_{c}$)}&
&$\frac{\textbf{p}^{2}_{x_{1}}}{2m'_{1}}+\frac{m_{c}+m_{\bar{b}}}{3m_{c}+m_{\bar{b}}}\frac{\textbf{p}^{2}_{x_{3}}}{2m'_{3}}$
&451.6&\\
&&&&&(2,3)&0.325&&&
\multirow{3}*{\makecell[c]{$V^{C}(12)$\\$\frac{1}{2}[V^{C}(13)+V^{C}(23)]$\\$\frac{1}{2}[V^{C}(14)+V^{C}(24)]$\\$-D$}}
&\multirow{3}*{\makecell[c]{-12.6\\-91.8\\-74.8\\-983.0}}&&\\
\Xcline{1-5}{0.5pt}
\multicolumn{2}{c}{Quark Mass}&\multicolumn{1}{r}{11097.0}&11097.0&0.0&(1,4)&0.336&&
&\multirow{2}*{}
&\multirow{2}*{}&&$2m_{c}$\\
\multicolumn{2}{c}{\multirow{1}{*}{Confinement Potential}}&\multirow{1}{*}{-2280.0}&\multirow{1}{*}{-2618.0}&\multirow{1}{*}{338.0}&(2,4)&0.336&
\multicolumn{1}{c!{\color{black}\vrule width 1.5pt}}{0.235($B_{c}$)}&&&&&3449.6\\
\Xcline{1-5}{0.5pt}\Xcline{10-13}{0.5pt}
\multicolumn{2}{c}{\multirow{1}*{Kinetic Energy}}&\multicolumn{1}{r}{\multirow{1}*{810.3}}&\multirow{1}*{931.2}&\multirow{1}*{-120.9}
&(3,4)&0.333&&&\multirow{1}*{Subtotal}&\multirow{1}*{3125.4}&
\multirow{1}*{3098.6}&\multirow{1}*{3449.6}\\
\Xcline{6-8}{0.5pt}\Xcline{9-13}{0.5pt}
\multicolumn{2}{c}{\multirow{1}*{CS Interaction}}&\multicolumn{1}{r}{\multirow{1}*{18.2}}&\multirow{1}*{-123.8}&\multirow{1}*{142.0}
&\multicolumn{2}{r}{(1,2)-(3,4):}&0.204 fm&&\multirow{4}*{\makecell[c]{$m_{\bar{c}}$\\
$\frac{m_{\bar{b}}}{m_{\bar{c}}+m_{\bar{b}}}\frac{\textbf{p}^{2}_{x_{2}}}{2m'_{2}}+\frac{m_{c}}{3m_{c}+m_{\bar{b}}}\frac{\textbf{p}^{2}_{x_{3}}}{2m'_{3}}$
\\$\frac{1}{2}V^{C}(34)$
\\$\frac{1}{2}[V^{C}(13)+V^{C}(23)]$
\\$-\frac{1}{2}D$}}
&\multirow{4}*{\makecell[c]{1918.0\\235.0\\16.0\\-91.9\\ -491.5}}&
\multirow{1}*{$\frac{-m_{\bar{c}}}{4(m_{\bar{b}}+m_{\bar{c}})}m_{\bar{c}\bar{b}}$}\\
\Xcline{6-8}{1.6pt}\Xcline{1-5}{0.5pt}
\multicolumn{1}{c|}{\multirow{3}{*}{$V^{C}$}}&\multicolumn{1}{c}{\multirow{1}{*}{(1,2)}}&\multicolumn{1}{r}{-12.6}&
&\multicolumn{1}{r}{}&(1,3)&\multicolumn{1}{r}{-91.9}&-237.2($\eta_{c}$)
&\multirow{4}*{\makecell[c]{$\bar{c}$-quark:\\ \\$m^{eff}_{\bar{c}}$}}
&&&\multirow{1}*{-400.5}\\
\multicolumn{1}{c|}{\multirow{1}{*}{}}&\multicolumn{1}{c}{(1,4)}&\multicolumn{1}{r}{-74.8}&&\multicolumn{1}{r}{}&(2,3)&\multicolumn{1}{r}{-91.9}&
&&&&$\frac{5m_{c}}{4(m_{c}+m_{\bar{b}})}m_{c\bar{b}}$&$m_{\bar{c}}$\\
\Xcline{2-5}{0.5pt}
\multicolumn{1}{c|}{\multirow{1}{*}{}}&\multicolumn{1}{c}{\multirow{1}{*}{Subtotal}}&\multicolumn{1}{r}{-314.0}&\multicolumn{1}{r}{-652.0}&\multicolumn{1}{c|}{338.0}
&(2,4)&-74.8&$-414.8(B_{c})$
&&&&1917.8&1724.8\\
\Xcline{10-13}{0.5pt}\Xcline{1-5}{0.5pt}
\multicolumn{2}{c}{\multirow{1}*{Total Contribution}}&\multicolumn{1}{r}{514.4}&\multicolumn{1}{r}{155.4}&\multicolumn{1}{c|}{359.0}
&(3,4)&32.0&
&&\multirow{1}*{Subtotal}&\multirow{1}*{1585.6}&
\multirow{1}*{1517.3}&\multirow{1}*{1724.8}\\
\Xcline{9-13}{0.5pt}\Xcline{1-8}{1.6pt}
\multicolumn{8}{l!{\color{black}\vrule width 1.5pt}}{Total Wave function:}&\multirow{4}*{\makecell[c]{$\bar{b}$-quark:\\ \\$m^{eff}_{\bar{b}}$}}&\multirow{4}*{\makecell[c]{$m_{\bar{b}}$\\
$\frac{m_{\bar{c}}}{m_{\bar{c}}+m_{\bar{b}}}\frac{\textbf{p}^{2}_{x_{2}}}{2m'_{2}}+\frac{m_{c}}{3m_{c}+m_{\bar{b}}}\frac{\textbf{p}^{2}_{x_{3}}}{2m'_{3}}$
\\$\frac{1}{2}V^{C}(34)$
\\$\frac{1}{2}[V^{C}(14)+V^{C}(24)]$
\\$-\frac{1}{2}D$}}
&\multirow{4}*{\makecell[c]{5343.0\\123.6\\16.0\\-74.8\\ -491.5}}&\multirow{1}*{$\frac{-m_{\bar{b}}}{4(m_{\bar{b}}+m_{\bar{c}})}m_{\bar{c}\bar{b}}$}
\\
\multicolumn{8}{r!{\color{black}\vrule width 1.5pt}}{$\Psi_{tot}=0.401|F\rangle|R^{s}\rangle|[\phi_{1}\chi_{5}]\rangle-0.916|F\rangle|R^{s}\rangle|[\phi_{2}\chi_{6}]\rangle=-0.574|F\rangle|R^{s}\rangle|[\psi_{1}\zeta_{5}]\rangle$}&&&&-1203.5&\\
\multicolumn{8}{r!{\color{black}\vrule width 1.5pt}}{$-0.532|F\rangle|R^{s}\rangle|[\psi_{1}\zeta_{6}]\rangle+0.019|F\rangle|R^{s}\rangle|[\psi_{2}\zeta_{5}]\rangle-
0.622|F\rangle|R^{s}\rangle|[\psi_{2}\zeta_{6}]\rangle$}&
&&&$\frac{5m_{\bar{b}}}{4(m_{c}+m_{\bar{b}})}m_{c\bar{b}}$&$m_{\bar{b}}$\\
\Xcline{1-8}{0.3pt}
\multicolumn{8}{l!{\color{black}\vrule width 1.5pt}}{The rearrangement strong decay channel:}
&&&&5929.5&5052.9\\
\Xcline{10-13}{0.5pt}
\multicolumn{8}{l!{\color{black}\vrule width 1.5pt}}{$\Gamma_{T_{c^{2}\bar{b}^{2}}(9620.5,0^{+})\rightarrow B^{*}_{c}J/\psi}:\Gamma_{T_{c^{2}\bar{b}^{2}}(9620.5,0^{+})\rightarrow B_{c}\eta_{c}}=1:1.2$}&&
\multirow{1}*{Subtotal}&\multirow{1}*{4916.3}&
\multirow{1}*{4726.0}&\multirow{1}*{5052.9}\\
\Xcline{9-13}{0.5pt}\Xcline{1-8}{0.3pt}
\multicolumn{8}{l!{\color{black}\vrule width 1.5pt}}{The radiative decay widths:\quad  $\Gamma_{T_{c^{2}\bar{c}\bar{b}}(9730.5,2^{+})\rightarrow T_{c^{2}\bar{c}\bar{b}}(9620.5,0^{+})\gamma}=0 \rm{keV}$}&
\multirow{3}*{\makecell[c]{CS \\Interaction}}&\multirow{3}*{\makecell[c]{$\frac{3}{4}V^{SS}(12)$\\ \\$\frac{3}{4}V^{SS}(34)$}}
&\multirow{3}*{\makecell[c]{10.8\\ \\7.3}}
&\multirow{3}*{\makecell[c]{$4v_{cc}$\\14.3\\$4v_{\bar{c}\bar{b}}$\\7.9}}
&\multirow{3}*{\makecell[c]{$4C_{cc}$\\21.2\\$4C_{\bar{c}\bar{b}}$\\13.2}}
\\
\multicolumn{8}{r!{\color{black}\vrule width 1.5pt}}{$\Gamma_{T_{c^{2}\bar{c}\bar{b}}(9624.6,1^{+})\rightarrow T_{c^{2}\bar{c}\bar{b}}(9620.5,0^{+})\gamma}=0.007 \rm{keV}$\quad }&&&&\\
\Xcline{1-8}{0.3pt}
\multicolumn{8}{l!{\color{black}\vrule width 1.5pt}}{The magnetic moments: $\mu_{T_{c^{2}\bar{c}\bar{b}}(9620.5,0^{+})}=
\langle\Psi^{0^{+}}_{tot}|\hat{\mu^{z}}|\Psi^{0^{+}}_{tot}\rangle=0$}
&&&&&\\
\Xcline{10-13}{0.5pt}\Xcline{1-8}{0.3pt}
\multicolumn{8}{l!{\color{black}\vrule width 1.5pt}}{The transition magnetic moments:}
&&\multirow{1}*{Subtotal}&\multirow{1}*{18.2}&\multirow{1}*{22.1}&34.4\\
\Xcline{9-13}{0.5pt}
\multicolumn{8}{l!{\color{black}\vrule width 1.5pt}}{$\mu_{T_{c^{2}\bar{c}\bar{b}}(9730.5,2^{+})\rightarrow T_{c^{2}\bar{c}\bar{b}}(9620.5,0^{+})\gamma}=\langle\Psi^{2^{+}}_{tot}|\hat{\mu^{z}}|\Psi^{0^{+}}_{tot}\rangle=0$ }&\multicolumn{2}{l|}{\multirow{1}*{Matrix nondiagonal element}}&\multirow{1}*{-25.0}&-46.2&-117.7\\
\Xcline{9-13}{0.5pt}
\multicolumn{8}{l!{\color{black}\vrule width 1.5pt}}{$\mu_{T_{c^{2}\bar{c}\bar{b}}(9624.6,1^{+})\rightarrow T_{c^{2}\bar{c}\bar{b}}(9620.5,0^{+})\gamma}=\langle\Psi^{1^{+}}_{tot}|\hat{\mu^{z}}|\Psi^{0^{+}}_{tot}\rangle=-0.096  \mu_{N}$}&\multicolumn{2}{l|}{\multirow{1}*{Total contribution}}&\multirow{1}*{9620.5}&\multirow{1}*{9317.5}&\multirow{1}*{10144.0}\\
\toprule[1.5pt]
\multicolumn{1}{r}{$J^{P}=1^{+}$}&&Value&\multicolumn{1}{r}{$B^{*}_{c}\eta_{c}$}&
\multicolumn{1}{r!{\color{black}\vrule width 1.5pt}}{Difference}&\multicolumn{3}{c!{\color{black}\vrule width 1.5pt}}{Relative Lengths (fm)}&&Contribution&Value&Ref. \cite{Weng:2020jao}&Ref. \cite{Wu:2016vtq}\\ \Xcline{1-13}{0.5pt}
\Xcline{1-5}{0.5pt}
\multicolumn{2}{c}{Mass/$B_{T}$}&\multicolumn{1}{r}{9624.6}&9349.0&275.6&$(i,j)$&\multicolumn{1}{r}{Value}&\multicolumn{1}{c!{\color{black}\vrule width 1.5pt}}{$B^{*}_{c}\eta_{c}$}&
\multirow{5}*{\makecell[c]{$c$-quark:\\ \\$m^{eff}_{c}$}}&2$m_{c}$&3836.0
&\multirow{5}*{\makecell[c]{$-\frac{1}{4}m_{cc}$\\-792.9\\$\frac{5m_{c}}{4(m_{c}+m_{\bar{c}})}m_{c\bar{c}}$\\1917.8\\
$\frac{5m_{c}}{4(m_{c}+m_{\bar{b}})}m_{c\bar{b}}$\\1973.7}}\\
\Xcline{1-8}{0.5pt}
\multirow{2}*{\makecell[c]{Variational\\ Parameters\\ (fm$^{-2}$)}}&\multirow{2}*{\makecell[c]{$C_{11}$\\$C_{22}$ \\ $C_{33}$}}&\multicolumn{1}{r}{\multirow{2}*{\makecell[c]{11.1\\6.9\\15.3}}}&
\multicolumn{1}{r}{\multirow{2}*{\makecell[c]{20.2\\15.0\\-}}}&&(1.2)&0.429&&
&$\frac{\textbf{p}^{2}_{x_{1}}}{2m'_{1}}+\frac{m_{c}+m_{\bar{b}}}{3m_{c}+m_{\bar{b}}}\frac{\textbf{p}^{2}_{x_{3}}}{2m'_{3}}$
&442.8&\\
&&&&&(1,3)&0.328&\multicolumn{1}{c!{\color{black}\vrule width 1.5pt}}{0.290($\eta_{c}$)}&&
\multirow{3}*{\makecell[c]{$V^{C}(12)$\\$\frac{1}{2}[V^{C}(13)+V^{C}(23)]$\\$\frac{1}{2}[V^{C}(14)+V^{C}(24)]$\\$-D$}}
&\multirow{3}*{\makecell[c]{-17.4\\-87.0\\-68.9\\-983.0}}&&\\
\Xcline{1-5}{0.5pt}
\multicolumn{2}{c}{Quark Mass}&\multicolumn{1}{r}{11097.0}&11097.0&0.0&(2,3)&0.328&&
&\multirow{2}*{}
&\multirow{2}*{}&&$2m_{c}$\\
\multicolumn{2}{c}{\multirow{1}{*}{Confinement Potential}}&\multirow{1}{*}{-2266.2}&\multirow{1}{*}{-2563.6}&\multirow{1}{*}{297.4}&(1,4)&0.340&&&&&&3449.6\\
\Xcline{1-5}{0.5pt}\Xcline{10-13}{0.5pt}
\multicolumn{2}{c}{\multirow{1}*{Kinetic Energy}}&\multicolumn{1}{r}{\multirow{1}*{795.2}}&\multirow{1}*{875.4}&\multirow{1}*{-80.2}
&(2,4)&0.340&\multicolumn{1}{c!{\color{black}\vrule width 1.5pt}}{0.250($B^{*}_{c}$)}&&\multirow{1}*{Subtotal}&\multirow{1}*{3122.5}&
\multirow{1}*{3098.6}&\multirow{1}*{3449.6}\\
\Xcline{9-13}{0.5pt}
\multicolumn{2}{c}{\multirow{1}*{CS Interaction}}&\multicolumn{1}{r}{\multirow{1}*{8.0}}&\multirow{1}*{-59.9}&
\multirow{1}*{67.9}
&(3,4)&0.338&&&\multirow{4}*{\makecell[c]{$m_{\bar{c}}$\\
$\frac{m_{\bar{b}}}{m_{\bar{c}}+m_{\bar{b}}}\frac{\textbf{p}^{2}_{x_{2}}}{2m'_{2}}+\frac{m_{c}}{3m_{c}+m_{\bar{b}}}\frac{\textbf{p}^{2}_{x_{3}}}{2m'_{3}}$
\\$\frac{1}{2}V^{C}(34)$
\\$\frac{1}{2}[V^{C}(13)+V^{C}(23)]$
\\$-\frac{1}{2}D$}}
&\multirow{4}*{\makecell[c]{1918.0\\230.3\\14.5\\-87.0\\ -491.5}}&
\multirow{1}*{$\frac{-m_{\bar{c}}}{4(m_{\bar{b}}+m_{\bar{c}})}m_{\bar{c}\bar{b}}$}\\
\Xcline{1-8}{0.5pt}
\multicolumn{1}{c|}{\multirow{4}{*}{$V^{C}$}}&\multicolumn{1}{c}{\multirow{1}{*}{(1,2)}}&\multicolumn{1}{r}{-17.4}&&
&\multicolumn{2}{r}{(1,2)-(3,4):}&0.204 fm
&\multirow{4}*{\makecell[c]{$\bar{c}$-quark:\\ \\$m^{eff}_{\bar{c}}$}}
&&&\multirow{1}*{-400.5}\\
\Xcline{6-8}{1.6pt}
\multicolumn{1}{c|}{}&\multicolumn{1}{c}{(2,3)}&\multicolumn{1}{r}{-87.0}&&\multicolumn{1}{r}{}&&&&
&&&$\frac{5m_{c}}{4(m_{c}+m_{\bar{b}})}m_{c\bar{b}}$&$m_{\bar{c}}$\\
\multicolumn{1}{c|}{\multirow{1}{*}{}}&\multicolumn{1}{c}{(1,4)}&\multicolumn{1}{r}{-68.9}
&&\multicolumn{1}{r}{}&(2,4)&-68.9&$-360.4(B^{*}_{c})$&
&&&1917.8&1724.8\\
\Xcline{10-13}{0.5pt}\Xcline{2-5}{0.5pt}
\multicolumn{1}{c|}{\multirow{1}{*}{}}&\multicolumn{1}{c}{\multirow{1}{*}{Subtotal}}&\multicolumn{1}{r}{-300.2}&\multicolumn{1}{r}{-597.6}&\multicolumn{1}{c|}{297.4}
&(3,4)&29.0&
&&\multirow{1}*{Subtotal}&\multirow{1}*{1584.3}&
\multirow{1}*{1517.3}&\multirow{1}*{1724.8}\\
\Xcline{9-13}{0.5pt}\Xcline{1-5}{0.5pt}
\multicolumn{2}{c}{\multirow{1}*{Total Contribution}}&\multicolumn{1}{r}{503.0}&\multicolumn{1}{r}{218.0}&\multicolumn{1}{c|}{285.0}
&(1,3)&-87.0&$-237.2(\eta_{c})$&
\multirow{4}*{\makecell[c]{$\bar{b}$-quark:\\ \\$m^{eff}_{\bar{b}}$}}&\multirow{4}*{\makecell[c]{$m_{\bar{b}}$\\
$\frac{m_{\bar{c}}}{m_{\bar{c}}+m_{\bar{b}}}\frac{\textbf{p}^{2}_{x_{2}}}{2m'_{2}}+\frac{m_{c}}{3m_{c}+m_{\bar{b}}}\frac{\textbf{p}^{2}_{x_{3}}}{2m'_{3}}$
\\$\frac{1}{2}V^{C}(34)$
\\$\frac{1}{2}[V^{C}(14)+V^{C}(24)]$
\\$-\frac{1}{2}D$}}
&\multirow{4}*{\makecell[c]{5343.0\\122.1\\14.5\\-68.9\\ -491.5}}&\multirow{1}*{$\frac{-m_{\bar{b}}}{4(m_{\bar{b}}+m_{\bar{c}})}m_{\bar{c}\bar{b}}$}
\\
\Xcline{1-8}{1.6pt}
\multicolumn{8}{l!{\color{black}\vrule width 1.5pt}}{Total Wave function:}&&&&&-1203.5&\\
\multicolumn{8}{r!{\color{black}\vrule width 1.5pt}}{$\Psi_{tot}=0.220|F\rangle|R^{s}\rangle|[\phi_{1}\chi_{3}]\rangle+0.968|F\rangle|R^{s}\rangle|[\phi_{2}\chi_{4}]\rangle+0.118|F\rangle|R^{s}\rangle|[\phi_{1}\chi_{2}]\rangle$}&&&&$\frac{5m_{\bar{b}}}{4(m_{c}+m_{\bar{b}})}m_{c\bar{b}}$&$m_{\bar{b}}$\\
\multicolumn{8}{r!{\color{black}\vrule width 1.5pt}}{$=0.494|F\rangle|R^{s}\rangle|[\psi_{1}\zeta_{2}]\rangle
-0.396|F\rangle|R^{s}\rangle|[\psi_{1}\zeta_{3}]\rangle-0.487|F\rangle|R^{s}\rangle|[\psi_{1}\zeta_{4}]\rangle$}&&&&5929.5&5052.9\\
\Xcline{10-13}{0.5pt}
\multicolumn{8}{r!{\color{black}\vrule width 1.5pt}}{$+0.111|F\rangle|R^{s}\rangle|[\psi_{2}\zeta_{2}]\rangle-0.246|F\rangle|R^{s}\rangle|[\psi_{2}\zeta_{3}]\rangle
-0.537|F\rangle|R^{s}\rangle|[\psi_{2}\zeta_{4}]\rangle$}
&&
\multirow{1}*{Subtotal}&\multirow{1}*{4919.2}&
\multirow{1}*{4726.0}&\multirow{1}*{5052.9}\\
\Xcline{9-13}{0.5pt}
\Xcline{1-8}{0.3pt}
\multicolumn{8}{l!{\color{black}\vrule width 1.5pt}}{The rearrangement strong decay channel: \quad $\Gamma_{T_{c^{2}\bar{c}\bar{b}}(9624.6,1^{+})\rightarrow B^{*}_{c}J/\psi}:$}
&\multirow{3}*{\makecell[c]{CS \\Interaction}}&\multirow{3}*{\makecell[c]{$-\frac{1}{4}V^{SS}(12)$\\ \\$\frac{3}{4}V^{SS}(34)$}}
&\multirow{3}*{\makecell[c]{-2.3\\ \\10.4}}
&\multirow{3}*{\makecell[c]{$-\frac{4}{3}v_{\bar{c}\bar{b}}$\\-2.6\\$4v_{cc}$\\14.2}}
&\multirow{3}*{\makecell[c]{$-\frac{4}{3}C_{\bar{c}\bar{b}}$\\-4.4\\$4C_{cc}$\\21.2}}
\\
\multicolumn{8}{l!{\color{black}\vrule width 1.5pt}}{$\Gamma_{T_{c^{2}\bar{c}\bar{b}}(9624.6,1^{+})\rightarrow B_{c}J/\psi}:
\Gamma_{T_{c^{2}\bar{c}\bar{b}}(9624.6,1^{+})\rightarrow B^{*}_{c}\eta_{c}}=1:0.8:1.2$}
&&&&\\
\Xcline{1-8}{0.3pt}
\multicolumn{8}{l!{\color{black}\vrule width 1.5pt}}{The radiative decay widths: \quad $\Gamma_{T_{c^{2}\bar{c}\bar{b}}(9730.5,2^{+})\rightarrow T_{c^{2}\bar{c}\bar{b}}(9624.6,1^{+})\gamma}=145.0 \rm{keV}$}
&&&&&\\
\Xcline{10-13}{0.5pt}
\multicolumn{8}{r!{\color{black}\vrule width 1.5pt}}{$\Gamma_{T_{c^{2}\bar{c}\bar{b}}(9624.6,1^{+})\rightarrow T_{c^{2}\bar{c}\bar{b}}(9620.5,0^{+})\gamma}=0.007 \rm{keV}$}
&&\multirow{1}*{Subtotal}&\multirow{1}*{8.0}&\multirow{1}*{11.6}&16.8\\
\Xcline{9-13}{0.5pt}\Xcline{1-8}{0.3pt}
\multicolumn{8}{l!{\color{black}\vrule width 1.5pt}}{The magnetic moments: \quad $\mu_{T_{c^{2}\bar{c}\bar{b}}(9624.6,1^{+})}=\langle\Psi^{1^{+}}_{tot}|\hat{\mu^{z}}|\Psi^{1^{+}}_{tot}\rangle=-0.233\mu_{N}$}
&\multicolumn{2}{l|}{\multirow{1}*{Matrix nondiagonal element}}&\multirow{1}*{-9.3}&-18.4&-100.0\\
\Xcline{9-13}{0.5pt}\Xcline{1-8}{0.3pt}
\multicolumn{8}{l!{\color{black}\vrule width 1.5pt}}{The transition magnetic moments:}
&\multicolumn{2}{l|}{\multirow{1}*{Total contribution}}&\multirow{1}*{9624.6}&\multirow{1}*{9335.1}&\multirow{1}*{10144}\\
\Xcline{9-13}{1.6pt}
\multicolumn{13}{l}{$\mu_{T_{c^{2}\bar{c}\bar{b}}(9624.6,1^{+})\rightarrow T_{c^{2}\bar{c}\bar{b}}(9620.5,0^{+})\gamma}=\langle\Psi^{1^{+}}_{tot}|\hat{\mu^{z}}|\Psi^{0^{+}}_{tot}\rangle=-0.096
 \mu_{N}$
\quad
$\mu_{T_{c^{2}\bar{c}\bar{b}}(9730.5,2^{+})\rightarrow T_{c^{2}\bar{c}\bar{b}}(9624.6,1^{+})\gamma}=\langle\Psi^{2^{+}}_{tot}|\hat{\mu^{z}}|\Psi^{1^{+}}_{tot}\rangle=-0.294 \mu_{N}$}\\
\toprule[0.5pt]
\toprule[1.5pt]
\end{tabular}
\end{lrbox}\scalebox{0.868}{\usebox{\tablebox}}
\end{table*}

\begin{table*}
\caption{ The masses, binding energy,  variational parameters, the internal contribution, total wave functions, magnetic moments, transition magnetic moments, radiative decay widths, rearrangement strong width ratios, and the relative lengths between quarks for the $J^{P}=0^{+}$, $1^{+}$ $bb\bar{b}\bar{c}$ states and their lowest meson-meson thresholds. The notation is the same as that in Table \ref{cccc}.
}\label{bbbc1}
\begin{lrbox}{\tablebox}
\renewcommand\arraystretch{1.45}
\renewcommand\tabcolsep{1.8pt}
\begin{tabular}{cc|ccc !{\color{black}\vrule width 1.5pt} ccc !{\color{black}\vrule width 1.5pt}c|cc|c|cc}
\toprule[1.5pt]
\toprule[0.5pt]
\multicolumn{1}{c}{$bb\bar{b}\bar{c}$}&\multicolumn{4}{r!{\color{black}\vrule width 1.5pt}}{The contribution from each term}&\multicolumn{3}{c!{\color{black}\vrule width 1.5pt}}{Relative Lengths (fm)}&\multirow{2}*{Overall}&\multicolumn{2}{c}{Present Work}&\multicolumn{2}{c}{CMI Model}\\
\Xcline{1-8}{0.5pt}\Xcline{10-13}{0.5pt}
\multicolumn{1}{r}{$J^{P}=0^{+}$}&&Value&\multicolumn{1}{r}{$B_{c}\eta_{b}$}&
\multicolumn{1}{r!{\color{black}\vrule width 1.5pt}}{Difference}&$(i,j)$&\multicolumn{1}{r}{Vaule}
&\multicolumn{1}{c!{\color{black}\vrule width 1.5pt}}{$B_{c}\eta_{b}$}&&Contribution&Value&Ref. \cite{Weng:2020jao}&Ref. \cite{Wu:2016vtq}\\ \Xcline{1-13}{0.5pt}
\Xcline{1-5}{0.5pt}
\multicolumn{2}{c}{Mass/$B_{T}$}&\multicolumn{1}{r}{16043.9}&15676.9&367.0&(1,2)&0.242&&
\multirow{5}*{\makecell[c]{$b$-quark:\\ \\$m^{eff}_{b}$}}&2$m_{b}$&10686.0
&\multirow{5}*{\makecell[c]{$-\frac{1}{4}m_{bb}$\\-2382.4\\$\frac{5m_{b}}{4(m_{b}+m_{\bar{b}})}m_{b\bar{b}}$\\5903.1\\
$\frac{5m_{b}}{4(m_{b}+m_{\bar{c}})}m_{b\bar{c}}$\\5929.5}}\\
\Xcline{1-5}{0.5pt}
\multirow{2}*{\makecell[c]{Variational\\ Parameters\\ (fm$^{-2}$)}}&\multirow{2}*{\makecell[c]{$C_{11}$\\$C_{22}$ \\ $C_{33}$}}&\multicolumn{1}{r}{\multirow{2}*{\makecell[c]{12.5\\21.7\\28.7}}}&
\multicolumn{1}{r}{\multirow{2}*{\makecell[c]{22.9\\58.8\\-}}}&&(1.3)&0.239&
\multicolumn{1}{c!{\color{black}\vrule width 1.5pt}}{0.148($\eta_{b}$)}&
&$\frac{\textbf{p}^{2}_{x_{1}}}{2m'_{1}}+\frac{m_{c}+m_{\bar{b}}}{3m_{c}+m_{\bar{b}}}\frac{\textbf{p}^{2}_{x_{3}}}{2m'_{3}}$
&393.8&\\
&&&&&(2,3)&0.239&&&
\multirow{3}*{\makecell[c]{$V^{C}(12)$\\$\frac{1}{2}[V^{C}(13)+V^{C}(23)]$\\$\frac{1}{2}[V^{C}(14)+V^{C}(24)]$\\$-D$}}
&\multirow{3}*{\makecell[c]{97.7\\-251.3\\-228.3\\-983.0}}&&\\
\Xcline{1-5}{0.5pt}
\multicolumn{2}{c}{Quark Mass}&\multicolumn{1}{r}{17947.0}&17947.0&0.0&(1,4)&0.249&&
&\multirow{2}*{}
&\multirow{2}*{}&&$2m_{b}$\\
\multicolumn{2}{c}{\multirow{1}{*}{Confinement Potential}}&\multirow{1}{*}{-2786.7}&\multirow{1}{*}{-3259.9}&\multirow{1}{*}{473.2}&(2,4)&0.249&
\multicolumn{1}{c!{\color{black}\vrule width 1.5pt}}{0.235($B_{c}$)}&&&&&10105.8\\
\Xcline{1-5}{0.5pt}\Xcline{10-13}{0.5pt}
\multicolumn{2}{c}{\multirow{1}*{Kinetic Energy}}&\multicolumn{1}{r}{\multirow{1}*{883.3}}&\multirow{1}*{1101.6}&\multirow{1}*{-218.3}
&(3,4)&0.318&&&\multirow{1}*{Subtotal}&\multirow{1}*{9714.9}&
\multirow{1}*{9450.2}&\multirow{1}*{10105.8}\\
\Xcline{6-8}{0.5pt}\Xcline{9-13}{0.5pt}
\multicolumn{2}{c}{\multirow{1}*{CS Interaction}}&\multicolumn{1}{r}{\multirow{1}*{15.5}}&\multirow{1}*{-111.8}&
\multirow{1}*{127.3}
&\multicolumn{2}{r}{(1,2)-(3,4):}&0.148 fm&&\multirow{4}*{\makecell[c]{$m_{\bar{b}}$\\
$\frac{m_{\bar{c}}}{m_{\bar{b}}+m_{\bar{c}}}\frac{\textbf{p}^{2}_{x_{2}}}{2m'_{2}}+\frac{m_{b}}{3m_{b}+m_{\bar{c}}}\frac{\textbf{p}^{2}_{x_{3}}}{2m'_{3}}$
\\$\frac{1}{2}V^{C}(34)$
\\$\frac{1}{2}[V^{C}(13)+V^{C}(23)]$
\\$-\frac{1}{2}D$}}
&\multirow{4}*{\makecell[c]{5343.0\\183.8\\20.4\\-251.3\\ -491.5}}&
\multirow{1}*{$\frac{-m_{\bar{b}}}{4(m_{\bar{c}}+m_{\bar{b}})}m_{\bar{c}\bar{b}}$}\\
\Xcline{6-8}{1.6pt}\Xcline{1-5}{0.5pt}
\multicolumn{1}{c|}{\multirow{3}{*}{$V^{C}$}}&\multicolumn{1}{c}{\multirow{1}{*}{(1,2)}}&\multicolumn{1}{r}{97.7}&
&\multicolumn{1}{r}{}&(1,3)&\multicolumn{1}{r}{-251.3}&-879.1($\eta_{b}$)
&\multirow{4}*{\makecell[c]{$\bar{b}$-quark:\\ \\$m^{eff}_{\bar{b}}$}}
&&&\multirow{1}*{-1203.5}\\
\multicolumn{1}{c|}{\multirow{1}{*}{}}&\multicolumn{1}{c}{(1,4)}&\multicolumn{1}{r}{-225.8}&&\multicolumn{1}{r}{}&(2,3)&\multicolumn{1}{r}{-251.3}&
&&&&$\frac{5m_{b}}{4(m_{b}+m_{\bar{c}})}m_{b\bar{b}}$&$m_{\bar{b}}$\\
\Xcline{2-5}{0.5pt}
\multicolumn{1}{c|}{\multirow{1}{*}{}}&\multicolumn{1}{c}{\multirow{1}{*}{Subtotal}}&\multicolumn{1}{r}{-820.7}&\multicolumn{1}{r}{-1293.9}&\multicolumn{1}{c|}{473.2}
&(2,4)&-228.3&$-414.8(B_{c})$
&&&&5903.1&5052.9\\
\Xcline{10-13}{0.5pt}\Xcline{1-5}{0.5pt}
\multicolumn{2}{c}{\multirow{1}*{Total Contribution}}&\multicolumn{1}{r}{78.1}&\multicolumn{1}{r}{-304.1}&\multicolumn{1}{c|}{382.2}
&(3,4)&40.8&
&&\multirow{1}*{Subtotal}&\multirow{1}*{4804.4}&
\multirow{1}*{4699.6}&\multirow{1}*{5052.9}\\
\Xcline{9-13}{0.5pt}\Xcline{1-8}{1.6pt}
\multicolumn{8}{l!{\color{black}\vrule width 1.5pt}}{Total Wave function:}
&\multirow{4}*{\makecell[c]{$\bar{c}$-quark:\\ \\$m^{eff}_{\bar{c}}$}}&\multirow{4}*{\makecell[c]{$m_{\bar{c}}$\\
$\frac{m_{\bar{b}}}{m_{\bar{b}}+m_{\bar{c}}}\frac{\textbf{p}^{2}_{x_{2}}}{2m'_{2}}+\frac{m_{b}}{3m_{b}+m_{\bar{c}}}\frac{\textbf{p}^{2}_{x_{3}}}{2m'_{3}}$
\\$\frac{1}{2}V^{C}(34)$
\\$\frac{1}{2}[V^{C}(14)+V^{C}(24)]$
\\$-\frac{1}{2}D$}}
&\multirow{4}*{\makecell[c]{1918.0\\305.7\\20.4\\-228.3\\ -491.5}}&\multirow{1}*{$\frac{-m_{\bar{c}}}{4(m_{\bar{c}}+m_{\bar{b}})}m_{\bar{b}\bar{c}}$}
\\
\multicolumn{8}{l!{\color{black}\vrule width 1.5pt}}{$\Psi_{tot}=0.308|F\rangle|R^{s}\rangle|[\phi_{1}\chi_{5}]\rangle-0.951|F\rangle|R^{s}\rangle|[\phi_{2}\chi_{6}]\rangle=-0.543|F\rangle|R^{s}\rangle|[\psi_{1}\zeta_{5}]\rangle$}&&&&&-400.5&\\
\multicolumn{8}{r!{\color{black}\vrule width 1.5pt}}{$-0.584|F\rangle|R^{s}\rangle|[\psi_{1}\zeta_{6}]\rangle-0.056|F\rangle|R^{s}\rangle|[\psi_{2}\zeta_{5}]\rangle-
0.602|F\rangle|R^{s}\rangle|[\psi_{2}\zeta_{6}]\rangle$.}&&&&$\frac{5m_{\bar{c}}}{4(m_{b}+m_{\bar{c}})}m_{b\bar{c}}$&$m_{\bar{c}}$\\
\Xcline{1-8}{0.3pt}
\multicolumn{8}{l!{\color{black}\vrule width 1.5pt}}{The rearrangement strong decay channel:}
&&&&1973.8.8&1724.8\\
\Xcline{10-13}{0.5pt}
\multicolumn{8}{l!{\color{black}\vrule width 1.5pt}}{$\Gamma_{T_{b^{2}\bar{b}\bar{c}}(16043.9,0^{+})\rightarrow B^{*}_{c}\Upsilon}:\Gamma_{T_{b^{2}\bar{b}\bar{c}}(16043.9,0^{+})\rightarrow B_{c}\eta_{b}}=1:1.4$.}&&
\multirow{1}*{Subtotal}&\multirow{1}*{1524.3}&
\multirow{1}*{1573.3}&\multirow{1}*{1724.8}\\
\Xcline{9-13}{0.5pt}\Xcline{1-8}{0.3pt}
\multicolumn{8}{l!{\color{black}\vrule width 1.5pt}}{The radiative decay widths: \quad $\Gamma_{T_{b^{2}\bar{b}\bar{c}}(16149.2,2^{+})\rightarrow T_{b^{2}\bar{b}\bar{c}}(16043.9,0^{+})\gamma}=0\, \rm{keV}$}&
\multirow{3}*{\makecell[c]{CS \\Interaction}}&\multirow{3}*{\makecell[c]{$\frac{3}{4}V^{SS}(12)$\\ \\$\frac{3}{4}V^{SS}(34)$}}
&\multirow{3}*{\makecell[c]{7.9\\ \\7.7}}
&\multirow{3}*{\makecell[c]{$4v_{bb}$\\7.7\\$4v_{\bar{c}\bar{b}}$\\7.9}}
&\multirow{3}*{\makecell[c]{$4C_{bb}$\\11.6\\$4C_{\bar{c}\bar{b}}$\\13.2}}
\\
\multicolumn{8}{r!{\color{black}\vrule width 1.5pt}}{$\Gamma_{T_{c^{2}\bar{c}\bar{b}}(16043.9,0^{+})\rightarrow T_{c^{2}\bar{c}\bar{b}}(16043.2,1^{+})\gamma}=10^{-6} \, \rm{keV}$\quad}&&&&\\
\Xcline{1-8}{0.3pt}
\multicolumn{8}{l!{\color{black}\vrule width 1.5pt}}{The magnetic moments: \quad $\mu_{T_{b^{2}\bar{b}\bar{c}}(16043.9,0^{+})}=
\langle\Psi^{0^{+}}_{tot}|\hat{\mu^{z}}|\Psi^{0^{+}}_{tot}\rangle= 0$}
&&&&&\\
\Xcline{10-13}{0.5pt}\Xcline{1-8}{0.3pt}
\multicolumn{8}{l!{\color{black}\vrule width 1.5pt}}{The transition magnetic moments:}
&&\multirow{1}*{Subtotal}&\multirow{1}*{15.5}&\multirow{1}*{15.5}&24.8\\
\Xcline{9-13}{0.5pt}
\multicolumn{8}{l!{\color{black}\vrule width 1.5pt}}{$\mu_{T_{b^{2}\bar{b}\bar{c}}(16149.2,2^{+})\rightarrow T_{b^{2}\bar{b}\bar{c}}(16043.9,0^{+})\gamma}=\langle\Psi^{2^{+}}_{tot}|\hat{\mu^{z}}|\Psi^{0^{+}}_{tot}\rangle= 0$ }&\multicolumn{2}{l|}{\multirow{1}*{Matrix nondiagonal element}}&\multirow{1}*{-15.2}&-26.7&43.7\\
\Xcline{9-13}{0.5pt}
\multicolumn{8}{l!{\color{black}\vrule width 1.5pt}}{$\mu_{T_{b^{2}\bar{b}\bar{c}}(16043.9,0^{+})\rightarrow T_{b^{2}\bar{b}\bar{c}}(16043.2,1^{+})\gamma}=\langle\Psi^{0^{+}}_{tot}|\hat{\mu^{z}}|\Psi^{1^{+}}_{tot}\rangle=0.096
 \mu_{N}$}&\multicolumn{2}{l|}{\multirow{1}*{Total contribution}}&\multirow{1}*{16043.9}&\multirow{1}*{15711.9}&\multirow{1}*{16932.0}\\
\toprule[1.5pt]
\multicolumn{1}{r}{$J^{P}=1^{+}$}&&Value&\multicolumn{1}{r}{$B^{*}_{c}\eta_{b}$}&
\multicolumn{1}{r!{\color{black}\vrule width 1.5pt}}{Difference}&\multicolumn{3}{c!{\color{black}\vrule width 1.5pt}}{Relative Lengths (fm)}&&Contribution&Value&Ref. \cite{Weng:2020jao}&Ref. \cite{Wu:2016vtq}\\ \Xcline{1-13}{0.5pt}
\Xcline{1-5}{0.5pt}
\multicolumn{2}{c}{Mass/$B_{T}$}&\multicolumn{1}{r}{16043.2}&15739.5&303.7&$(i,j)$&\multicolumn{1}{r}{Value}
&\multicolumn{1}{c!{\color{black}\vrule width 1.5pt}}{$B^{*}_{c}\eta_{b}$}&
\multirow{5}*{\makecell[c]{$b$-quark:\\ \\$m^{eff}_{b}$}}&2$m_{b}$&10686.0
&\multirow{5}*{\makecell[c]{$-\frac{1}{4}m_{bb}$\\-2382.4\\$\frac{5m_{b}}{4(m_{b}+m_{\bar{b}})}m_{b\bar{b}}$\\5903.1\\
$\frac{5m_{b}}{4(m_{b}+m_{\bar{c}})}m_{b\bar{c}}$\\5929.5}}\\
\Xcline{1-5}{0.5pt}\Xcline{6-8}{0.5pt}
\multirow{2}*{\makecell[c]{Variational\\ Parameters\\ (fm$^{-2}$)}}&\multirow{2}*{\makecell[c]{$C_{11}$\\$C_{22}$ \\ $C_{33}$}}&\multicolumn{1}{r}{\multirow{2}*{\makecell[c]{12.4\\21.0\\28.9}}}&
\multicolumn{1}{r}{\multirow{2}*{\makecell[c]{20.2\\57.4\\-}}}&&(1.2)&0.245&&
&$\frac{\textbf{p}^{2}_{x_{1}}}{2m'_{1}}+\frac{m_{\bar{b}}+m_{\bar{c}}}{3m_{b}+m_{\bar{c}}}\frac{\textbf{p}^{2}_{x_{3}}}{2m'_{3}}$
&387.9&\\
&&&&&(1,3)&0.240&\multicolumn{1}{c!{\color{black}\vrule width 1.5pt}}{0.148($\eta_{b}$)}&&
\multirow{3}*{\makecell[c]{$V^{C}(12)$\\$\frac{1}{2}[V^{C}(13)+V^{C}(23)]$\\$\frac{1}{2}[V^{C}(14)+V^{C}(24)]$\\$-D$}}
&\multirow{3}*{\makecell[c]{94.5\\-248.2\\-225.8\\-983.0}}&&\\
\Xcline{1-5}{0.5pt}
\multicolumn{2}{c}{Quark Mass}&\multicolumn{1}{r}{17947.0}&17947.0&0.0&(2,3)&0.240&&
&\multirow{2}*{}
&\multirow{2}*{}&&$2m_{b}$\\
\multicolumn{2}{c}{\multirow{1}{*}{Confinement Potential}}&\multirow{1}{*}{-2779.8}&\multirow{1}{*}{-3205.5}&\multirow{1}{*}{425.7}&(1,4)&0.250&&&&&&10105.8\\
\Xcline{1-5}{0.5pt}\Xcline{10-13}{0.5pt}
\multicolumn{2}{c}{\multirow{1}*{Kinetic Energy}}&\multicolumn{1}{r}{\multirow{1}*{876.1}}&\multirow{1}*{961.5}&\multirow{1}*{-123.3}
&(2,4)&0.250&\multicolumn{1}{c!{\color{black}\vrule width 1.5pt}}{0.250($B^{*}_{c}$)}&&\multirow{1}*{Subtotal}&\multirow{1}*{9711.4}&
\multirow{1}*{9450.2}&\multirow{1}*{10105.8}\\
\Xcline{9-13}{0.5pt}
\multicolumn{2}{c}{\multirow{1}*{CS Interaction}}&\multicolumn{1}{r}{\multirow{1}*{4.8}}&\multirow{1}*{-47.8}&
\multirow{1}*{52.6}
&(3,4)&0.320&
&&\multirow{4}*{\makecell[c]{$m_{\bar{b}}$\\
$\frac{m_{\bar{c}}}{m_{\bar{b}}+m_{\bar{c}}}\frac{\textbf{p}^{2}_{x_{2}}}{2m'_{2}}+\frac{m_{b}}{3m_{b}+m_{\bar{c}}}\frac{\textbf{p}^{2}_{x_{3}}}{2m'_{3}}$
\\$\frac{1}{2}V^{C}(34)$
\\$\frac{1}{2}[V^{C}(13)+V^{C}(23)]$
\\$-\frac{1}{2}D$}}
&\multirow{4}*{\makecell[c]{5342.0\\183.8\\19.9\\-248.2\\ -491.5}}&
\multirow{1}*{$\frac{-m_{\bar{b}}}{4(m_{\bar{c}}+m_{\bar{b}})}m_{\bar{b}\bar{c}}$}\\
\Xcline{1-8}{0.5pt}
\multicolumn{1}{c|}{\multirow{4}{*}{$V^{C}$}}&\multicolumn{1}{c}{\multirow{1}{*}{(1,2)}}&\multicolumn{1}{r}{94.5}&&
&\multicolumn{2}{r}{(1,2)-(3,4):}&0.148 fm
&\multirow{4}*{\makecell[c]{$\bar{b}$-quark:\\ \\$m^{eff}_{\bar{b}}$}}
&&&\multirow{1}*{-1203.5}\\
\Xcline{6-8}{1.6pt}
\multicolumn{1}{c|}{}&\multicolumn{1}{c}{(2,3)}&\multicolumn{1}{r}{-248.2}&&\multicolumn{1}{r}{}&&&
&&&&$\frac{5m_{b}}{4(m_{b}+m_{\bar{c}})}m_{b\bar{c}}$&$m_{\bar{b}}$\\
\multicolumn{1}{c|}{\multirow{1}{*}{}}&\multicolumn{1}{c}{(1,4)}&\multicolumn{1}{r}{-225.8}
&&\multicolumn{1}{r}{}&(2,4)&-225.8&$-360.4(B^{*}_{c})$&
&&&5903.1&5052.9\\
\Xcline{10-13}{0.5pt}\Xcline{2-5}{0.5pt}
\multicolumn{1}{c|}{\multirow{1}{*}{}}&\multicolumn{1}{c}{\multirow{1}{*}{Subtotal}}&\multicolumn{1}{r}{-813.7}&\multicolumn{1}{r}{-1239.5}&\multicolumn{1}{c|}{425.7}
&(3,4)&39.7&
&&\multirow{1}*{Subtotal}&\multirow{1}*{4807.0}&
\multirow{1}*{4699.6}&\multirow{1}*{5052.9}\\
\Xcline{9-13}{0.5pt}\Xcline{1-5}{0.5pt}
\multicolumn{2}{c}{\multirow{1}*{Total Contribution}}&\multicolumn{1}{r}{\multirow{1}*{67.1}}&\multicolumn{1}{r}{-241.5}&\multicolumn{1}{c|}{\multirow{1}*{308.6}}
&(1,3)&-248.2&$-879.1(\eta_{b})$
&\multirow{4}*{\makecell[c]{$\bar{c}$-quark:\\ \\$m^{eff}_{\bar{c}}$}}&\multirow{4}*{\makecell[c]{$m_{\bar{c}}$\\
$\frac{m_{\bar{b}}}{m_{\bar{b}}+m_{\bar{c}}}\frac{\textbf{p}^{2}_{x_{2}}}{2m'_{2}}+\frac{m_{b}}{3m_{b}+m_{\bar{c}}}\frac{\textbf{p}^{2}_{x_{3}}}{2m'_{3}}$
\\$\frac{1}{2}V^{C}(34)$
\\$\frac{1}{2}[V^{C}(14)+V^{C}(24)]$
\\$-\frac{1}{2}D$}}
&\multirow{4}*{\makecell[c]{1918.0\\304.4\\19.9\\-225.8\\ -491.5}}&\multirow{1}*{$\frac{-m_{\bar{c}}}{4(m_{\bar{c}}+m_{\bar{b}})}m_{\bar{b}\bar{c}}$}
\\
\Xcline{1-8}{1.6pt}
\multicolumn{8}{l!{\color{black}\vrule width 1.5pt}}{Total Wave function:}&&&&-400.5&\\
\multicolumn{8}{r!{\color{black}\vrule width 1.5pt}}{$\Psi_{tot}=0.171|F\rangle|R^{s}\rangle|[\phi_{1}\chi_{3}]\rangle+0.984|F\rangle|R^{s}\rangle|[\phi_{2}\chi_{4}]\rangle+0.044|F\rangle|R^{s}\rangle|[\phi_{1}\chi_{2}]\rangle$}&&&&$\frac{5m_{\bar{c}}}{4(m_{b}+m_{\bar{c}})}m_{b\bar{c}}$&$m_{\bar{c}}$\\
\multicolumn{8}{r!{\color{black}\vrule width 1.5pt}}{$=0.431|F\rangle|R^{s}\rangle|[\psi_{1}\zeta_{2}]\rangle
-0.467|F\rangle|R^{s}\rangle|[\psi_{1}\zeta_{3}]\rangle-0.500|F\rangle|R^{s}\rangle|[\psi_{1}\zeta_{4}]\rangle$}
&&&&1973.8&1724.8\\
\Xcline{10-13}{0.5pt}
\multicolumn{8}{r!{\color{black}\vrule width 1.5pt}}{$+0.240|F\rangle|R^{s}\rangle|[\psi_{2}\zeta_{2}]\rangle-0.188|F\rangle|R^{s}\rangle|[\psi_{2}\zeta_{3}]\rangle
-0.502|F\rangle|R^{s}\rangle|[\psi_{2}\zeta_{4}]\rangle$}
&&
\multirow{1}*{Subtotal}&\multirow{1}*{1525.0}&
\multirow{1}*{1573.3}&\multirow{1}*{1724.8}\\
\Xcline{9-13}{0.5pt}\Xcline{1-8}{0.3pt}
\multicolumn{8}{l!{\color{black}\vrule width 1.5pt}}{The rearrangement strong decay channel:\quad $\Gamma_{T_{b^{2}\bar{b}\bar{c}}(16043.2,1^{+})\rightarrow B^{*}_{c}\Upsilon}:$ }
&\multirow{3}*{\makecell[c]{CS \\Interaction}}&\multirow{3}*{\makecell[c]{$-\frac{1}{4}V^{SS}(12)$\\ \\$\frac{3}{4}V^{SS}(34)$}}
&\multirow{3}*{\makecell[c]{-2.6\\ \\7.5}}
&\multirow{3}*{\makecell[c]{$-\frac{4}{3}v_{\bar{b}\bar{c}}$\\-2.6\\$4v_{bb}$\\7.7}}
&\multirow{3}*{\makecell[c]{$-\frac{4}{3}C_{\bar{b}\bar{c}}$\\-4.4\\$4C_{bb}$\\11.6}}
\\
\multicolumn{8}{l!{\color{black}\vrule width 1.5pt}}{$\Gamma_{T_{b^{2}\bar{b}\bar{c}}(16043.2,1^{+})\rightarrow B_{c}\Upsilon}:
\Gamma_{T_{b^{2}\bar{b}\bar{c}}(16043.2,1^{+})\rightarrow B^{*}_{c}\eta_{b}}=1:1.3:1.5$}
&&&&\\
\Xcline{1-8}{0.3pt}
\multicolumn{8}{r!{\color{black}\vrule width 1.5pt}}{The radiative decay widths: $\Gamma_{T_{b^{2}\bar{b}\bar{c}}(16149.2,2^{+})\rightarrow T_{c^{2}\bar{c}\bar{b}}(16043.2,1^{+})\gamma}=435.0 \rm{keV}$}
&&&&&\\
\Xcline{10-13}{0.5pt}
\multicolumn{8}{r!{\color{black}\vrule width 1.5pt}}{$\Gamma_{T_{c^{2}\bar{c}\bar{b}}(16043.9,0^{+})\rightarrow T_{c^{2}\bar{c}\bar{b}}(16043.2,1^{+})\gamma}=10^{-6} \rm{keV}$}
&&\multirow{1}*{Subtotal}&\multirow{1}*{4.8}&\multirow{1}*{5.0}&7.2\\
\Xcline{9-13}{0.5pt}\Xcline{1-8}{0.3pt}
\multicolumn{8}{l!{\color{black}\vrule width 1.5pt}}{The magnetic moments: \quad $\mu_{T_{b^{2}\bar{b}\bar{c}}(19043.2,1^{+})}=\langle\Psi^{1^{+}}_{tot}|\hat{\mu^{z}}|\Psi^{1^{+}}_{tot}\rangle=-0.346\mu_{N}$}
&\multicolumn{2}{l|}{\multirow{1}*{Matrix nondiagonal element}}&\multirow{1}*{-5.0}&-8.6&24.3\\
\Xcline{9-13}{0.5pt}
\Xcline{1-8}{0.3pt}
\multicolumn{8}{l!{\color{black}\vrule width 1.5pt}}{The transition magnetic moments:}
&\multicolumn{2}{l|}{\multirow{1}*{Total contribution}}&\multirow{1}*{16048.1}&\multirow{1}*{15719.1}&\multirow{1}*{16915.0}\\
\Xcline{9-13}{1.6pt}
\multicolumn{13}{l}{$\mu_{T_{b^{2}\bar{b}\bar{c}}(16043.9,0^{+})\rightarrow T_{b^{2}\bar{b}\bar{c}}(16043.2,1^{+})\gamma}=\langle\Psi^{0^{+}}_{tot}|\hat{\mu^{z}}|\Psi^{1^{+}}_{tot}\rangle=0.201\mu_{N}$ \quad
$\mu_{T_{b^{2}\bar{b}\bar{c}}(16149.2,2^{+})\rightarrow T_{b^{2}\bar{b}\bar{c}}(16043.2,1^{+})\gamma}=\langle\Psi^{2^{+}}_{tot}|\hat{\mu^{z}}|\Psi^{1^{+}}_{tot}\rangle=0.329\mu_{N}$}\\
\toprule[0.5pt]
\toprule[1.5pt]
\end{tabular}
\end{lrbox}\scalebox{0.868}{\usebox{\tablebox}}
\end{table*}

\begin{table*}
\caption{ The masses, binding energy,  variational parameters, the internal contribution, total wave functions, magnetic moments, transition magnetic moments, radiative decay widths, rearrangement strong width ratios, and the relative lengths between quarks for the $J^{P}=2^{+}$ $cc\bar{c}\bar{b}$ and $bb\bar{b}\bar{c}$ states and their lowest meson-meson thresholds. The notation is the same as that in Table \ref{cccc}.
}\label{cccb2}
\begin{lrbox}{\tablebox}
\renewcommand\arraystretch{1.45}
\renewcommand\tabcolsep{1.5pt}
\begin{tabular}{c|c|ccc !{\color{black}\vrule width 1.5pt} ccc !{\color{black}\vrule width 1.5pt} c|cc|c|cc}
\midrule[1.5pt]
\toprule[0.5pt]
\multicolumn{1}{c}{$cc\bar{c}\bar{b}$}&\multicolumn{4}{r!{\color{black}\vrule width 1.5pt}}{The contribution from each term}&\multicolumn{3}{c !{\color{black}\vrule width 1.5pt}}{Relative Lengths (fm)}&\multirow{2}*{Overall}&\multicolumn{2}{c}{Present Work}&\multicolumn{2}{c}{CMI Model}\\
\Xcline{1-8}{0.5pt}\Xcline{10-13}{0.5pt}
\multicolumn{1}{r}{$J^{P}=2^{+}$}&&Value&\multicolumn{1}{r}{$B^{*}_{c}J/\psi$}&
\multicolumn{1}{r!{\color{black}\vrule width 1.5pt}}{Difference}&$(i,j)$&\multicolumn{1}{r}{Vaule}
&\multicolumn{1}{c!{\color{black}\vrule width 1.5pt}}{$B_{c}J/\psi$}&&Contribution&Value&Ref. \cite{Weng:2020jao}&Ref. \cite{Wu:2016vtq}\\ \Xcline{1-13}{0.5pt}
\Xcline{1-5}{0.5pt}
\multicolumn{2}{c}{Mass/$B_{T}$}&\multicolumn{1}{r}{9730.5}&9442.7&287.8&(1,2)&0.378&&
\multirow{5}*{$c$-quark}&2$m_{c}$&3836.0
&\multirow{5}*{\makecell[c]{$\frac{1}{2}m_{cc}$\\1585.8\\$\frac{m_{c}}{4(m_{c}+m_{\bar{c}})}m_{c\bar{c}}$\\767.1\\
$\frac{m_{c}}{4(m_{c}+m_{\bar{b}})}m_{c\bar{b}}$\\789.3}}\\
\Xcline{1-5}{0.5pt}
\multirow{2}*{\makecell[c]{Variational\\ Parameters\\ (fm$^{-2}$)}}&\multirow{2}*{\makecell[c]{$C_{11}$\\$C_{22}$ \\ $C_{33}$}}&\multicolumn{1}{r}{\multirow{2}*{\makecell[c]{13.7\\8.9\\9.1}}}&
\multicolumn{1}{r}{\multirow{2}*{\makecell[c]{20.2\\12.5\\-}}}&&(1.3)&0.350&
\multicolumn{1}{c!{\color{black}\vrule width 1.5pt}}{0.318($J/\psi$)}&
&$\frac{\textbf{p}^{2}_{x_{1}}}{2m'_{1}}+\frac{m_{c}+m_{\bar{b}}}{3m_{c}+m_{\bar{b}}}\frac{\textbf{p}^{2}_{x_{3}}}{2m'_{3}}$
&408.0&\\
&&&&&(2,3)&0.350&&&
\multirow{3}*{\makecell[c]{$V^{C}(12)$\\$\frac{1}{2}[V^{C}(13)+V^{C}(23)]$\\$\frac{1}{2}[V^{C}(14)+V^{C}(24)]$\\$-D$}}
&\multirow{3}*{\makecell[c]{-14.3\\-21.9\\-17.3\\-983.0}}&&\\
\Xcline{1-5}{0.5pt}
\multicolumn{2}{c}{Quark Mass}&\multicolumn{1}{r}{11097.0}&11097.0&0.0&(1,4)&0.359&&
&\multirow{2}*{}
&\multirow{2}*{}&&$2m_{c}$\\
\multicolumn{2}{c}{\multirow{1}{*}{Confinement Potential}}&\multirow{1}{*}{-2158.4}&\multirow{1}{*}{-2490.6}&\multirow{1}{*}{332.2}&(2,4)&0.359&
\multicolumn{1}{c!{\color{black}\vrule width 1.5pt}}{0.250($B^{*}_{c}$)}&&&&&3449.6\\
\Xcline{1-5}{0.5pt}\Xcline{10-13}{0.5pt}
\multicolumn{2}{c}{\multirow{1}*{Kinetic Energy}}&\multicolumn{1}{r}{\multirow{1}*{763.9}}&\multirow{1}*{799.3}&\multirow{1}*{-35.4}
&(3,4)&0.304&&&\multirow{1}*{Subtotal}&\multirow{1}*{3207.5}&
\multirow{1}*{3142.2}&\multirow{1}*{3449.6}\\
\Xcline{6-8}{0.5pt}\Xcline{9-13}{0.5pt}
\multicolumn{2}{c}{\multirow{1}*{CS Interaction}}&\multicolumn{1}{r}{\multirow{1}*{28.0}}&\multirow{1}*{36.9}&
\multirow{1}*{-8.9}&
\multicolumn{2}{r}{(1,2)-(3,4):}&0.264 fm&&\multirow{4}*{\makecell[c]{$m_{\bar{c}}$\\
$\frac{m_{\bar{b}}}{m_{\bar{c}}+m_{\bar{b}}}\frac{\textbf{p}^{2}_{x_{2}}}{2m'_{2}}+\frac{m_{c}}{3m_{c}+m_{\bar{b}}}\frac{\textbf{p}^{2}_{x_{3}}}{2m'_{3}}$
\\$\frac{1}{2}V^{C}(34)$
\\$\frac{1}{2}[V^{C}(13)+V^{C}(23)]$
\\$-\frac{1}{2}D$}}
&\multirow{4}*{\makecell[c]{1918.0\\244.7\\-49.8\\-21.9\\ -491.5}}&
\multirow{1}*{$\frac{m_{\bar{c}}}{4(m_{\bar{c}}+m_{\bar{c}})}m_{c\bar{c}}$}\\
\Xcline{6-8}{1.6pt}\Xcline{1-5}{0.5pt}
\multicolumn{1}{c|}{\multirow{3}{*}{$V^{C}$}}&\multicolumn{1}{c}{\multirow{1}{*}{(1,2)}}&\multicolumn{1}{r}{-14.3}&
&\multicolumn{1}{r}{}&(1,3)&\multicolumn{1}{r}{-21.9}&-164.2($J/\psi$)
&\multirow{4}*{$\bar{c}$-quark}
&&&\multirow{1}*{767.1}\\
\multicolumn{1}{c|}{\multirow{1}{*}{}}&\multicolumn{1}{c}{(1,4)}&\multicolumn{1}{r}{-17.3}&&\multicolumn{1}{r}{}&(2,3)&\multicolumn{1}{r}{-21.9}&&
&&&$\frac{m_{\bar{c}}}{4(m_{\bar{c}}+m_{\bar{b}})}m_{\bar{c}\bar{b}}$&$m_{\bar{c}}$\\
\Xcline{2-5}{0.5pt}
\multicolumn{1}{c|}{\multirow{1}{*}{}}&\multicolumn{1}{c}{\multirow{1}{*}{Subtotal}}&\multicolumn{1}{r}{-192.4}&\multicolumn{1}{r}{-524.6}&\multicolumn{1}{c|}{332.2}
&(2,4)&-17.3&$-360.4(B^{*}_{c})$
&&&&801.1&1724.8\\
\Xcline{10-13}{0.5pt}\Xcline{1-5}{0.5pt}
\multicolumn{2}{c}{\multirow{1}*{Total Contribution}}&\multicolumn{1}{r}{599.5}&\multicolumn{1}{r}{311.7}&\multicolumn{1}{c|}{287.8}
&(3,4)&-99.6&
&&\multirow{1}*{Subtotal}&\multirow{1}*{1599.5}&
\multirow{1}*{1568.2}&\multirow{1}*{1724.8}\\
\Xcline{9-13}{0.5pt}\Xcline{1-8}{1.6pt}
\multicolumn{8}{l!{\color{black}\vrule width 1.5pt}}{Total Wave function:}&\multirow{4}*{$\bar{b}$-quark}&\multirow{4}*{\makecell[c]{$m_{\bar{b}}$\\
$\frac{m_{\bar{c}}}{m_{\bar{c}}+m_{\bar{b}}}\frac{\textbf{p}^{2}_{x_{2}}}{2m'_{2}}+\frac{m_{c}}{3m_{c}+m_{\bar{b}}}\frac{\textbf{p}^{2}_{x_{3}}}{2m'_{3}}$
\\$\frac{1}{2}V^{C}(34)$
\\$\frac{1}{2}[V^{C}(14)+V^{C}(24)]$
\\$-\frac{1}{2}D$}}
&\multirow{4}*{\makecell[c]{5343.0\\111.2\\-49.8\\-17.4\\ -491.5}}&\multirow{1}*{$\frac{m_{\bar{b}}}{4(m_{\bar{b}}+m_{\bar{c}})}m_{\bar{c}\bar{b}}$}
\\
\multicolumn{8}{l!{\color{black}\vrule width 1.5pt}}{$\Psi_{tot}=|F\rangle|R^{s}\rangle|[\phi_{1}\chi_{1}]\rangle=0.577|F\rangle|R^{s}\rangle|[\psi_{1}\zeta_{1}]\rangle-0.816|F\rangle|R^{s}\rangle|[\psi_{2}\zeta_{1}]\rangle$}&&&&&2407.0&\\
\Xcline{1-8}{0.3pt}
\multicolumn{8}{l!{\color{black}\vrule width 1.5pt}}{The rearrangement strong decay channel:  $B^{*}_{c}J/\psi$.}&&&&$\frac{m_{\bar{b}}}{4(m_{c}+m_{\bar{b}})}m_{c\bar{b}}$&$m_{\bar{b}}$\\
\Xcline{1-8}{0.3pt}
\multicolumn{8}{l!{\color{black}\vrule width 1.5pt}}{The radiative decay widths:}
&&&&2371.8&5052.9\\
\Xcline{10-13}{0.5pt}
\multicolumn{8}{l!{\color{black}\vrule width 1.5pt}}{$\Gamma_{T_{c^{2}\bar{c}\bar{b}}(9730.5,2^{+})\rightarrow T_{c^{2}\bar{c}\bar{b}}(9620.5,0^{+})\gamma}=0\, \rm{keV}$}&&
\multirow{1}*{Subtotal}&\multirow{1}*{4895.5}&
\multirow{1}*{4778.8}&\multirow{1}*{5052.9}\\
\Xcline{9-13}{0.5pt}
\multicolumn{8}{l!{\color{black}\vrule width 1.5pt}}{$\Gamma_{T_{c^{2}\bar{c}\bar{b}}(9730.5,2^{+})\rightarrow T_{c^{2}\bar{c}\bar{b}}(9624.6,1^{+})\gamma}=145.0\, \rm{keV}$}&
\multirow{4}*{\makecell[c]{CS \\Interaction}}&\multirow{4}*{\makecell[c]{$\frac{1}{2}V^{SS}(12)$\\ $\frac{1}{2}V^{SS}(34)$\\$\frac{1}{4}(V^{SS}(13)+V^{SS}(23))$\\$\frac{1}{4}(V^{SS}(14)+V^{SS}(24))$}}
&\multirow{4}*{\makecell[c]{5.6\\8.5\\9.5\\4.3}}
&\multirow{4}*{\makecell[c]{$\frac{8}{3}v_{cc}+\frac{8}{3}v_{\bar{c}\bar{b}}$\\9.5+5.2\\$\frac{8}{3}v_{c\bar{c}}+\frac{8}{3}v_{c\bar{b}}$\\14.2+7.9}}
&\multirow{4}*{\makecell[c]{$\frac{8}{3}C_{cc}+\frac{8}{3}C_{\bar{c}\bar{b}}$\\14.1+8.8\\$\frac{8}{3}C_{c\bar{c}}+\frac{8}{3}C_{c\bar{b}}$\\14.1+8.8}}
\\
\Xcline{1-8}{0.3pt}
\multicolumn{8}{l!{\color{black}\vrule width 1.5pt}}{The magnetic moments:}&&&&\\
\multicolumn{8}{l!{\color{black}\vrule width 1.5pt}}{$\mu_{T_{c^{2}\bar{c}\bar{b}}(9730.5,2^{+})}=
\langle\Psi^{2^{+}}_{tot}|\hat{\mu^{z}}|\Psi^{2^{+}}_{tot}\rangle=2\mu_{c}+\mu_{\bar{c}}+\mu_{\bar{b}}=0.464 \mu_{N}$}
&&&&&\\
\Xcline{1-8}{0.3pt}
\multicolumn{8}{l!{\color{black}\vrule width 1.5pt}}{The transition magnetic moments:}
&&&&\\
\Xcline{10-13}{0.5pt}
\multicolumn{8}{l!{\color{black}\vrule width 1.5pt}}{$\mu_{T_{c^{2}\bar{c}\bar{b}}(9730.5,2^{+})\rightarrow T_{c^{2}\bar{c}\bar{b}}(9620.5,0^{+})\gamma}=\langle\Psi^{2^{+}}_{tot}|\hat{\mu^{z}}|\Psi^{0^{+}}_{tot}\rangle=0$ }
&&\multirow{1}*{Subtotal}&\multirow{1}*{28.0}&\multirow{1}*{36.8}&45.9\\
\Xcline{9-13}{0.5pt}
\multicolumn{8}{l!{\color{black}\vrule width 1.5pt}}{$\mu_{T_{c^{2}\bar{c}\bar{b}}(9730.5,2^{+})\rightarrow T_{c^{2}\bar{c}\bar{b}}(9624.6,1^{+})\gamma}=\langle\Psi^{2^{+}}_{tot}|\hat{\mu^{z}}|\Psi^{1^{+}}_{tot}\rangle=-0.294  \mu_{N}$}&\multicolumn{2}{l|}{\multirow{1}*{Total contribution}}&\multirow{1}*{9730.5}&\multirow{1}*{9526.0}&\multirow{1}*{10273.2}\\
\toprule[1.5pt]
\multicolumn{1}{r}{$bb\bar{b}\bar{c}$ $J^{P}=2^{+}$}&&Value&\multicolumn{1}{r}{$B^{*}_{c}\Upsilon$}&
\multicolumn{1}{r !{\color{black}\vrule width 1.5pt}}{Difference}&\multicolumn{3}{c !{\color{black}\vrule width 1.5pt}}{Relative Lengths (fm)}&&Contribution&Value&Ref. \cite{Weng:2020jao}&Ref. \cite{Wu:2016vtq}\\ \Xcline{1-13}{0.5pt}
\Xcline{1-5}{0.5pt}
\multicolumn{2}{c}{Mass/$B_{T}$}&\multicolumn{1}{r}{16149.2}&15819.4&329.8&$(i,j)$&Value&\multicolumn{1}{c !{\color{black}\vrule width 1.5pt}}{$B^{*}_{c}\Upsilon$}&
\multirow{5}*{$b$-quark}&2$m_{c}$&10686.0
&\multirow{5}*{\makecell[c]{$\frac{1}{2}m_{bb}$\\4764.8\\$\frac{m_{b}}{4(m_{b}+m_{\bar{b}})}m_{b\bar{b}}$\\2361.2\\
$\frac{m_{b}}{4(m_{b}+m_{\bar{c}})}m_{b\bar{c}}$\\2371.8}}\\
\Xcline{1-8}{0.5pt}
\multirow{2}*{\makecell[c]{Variational\\ Parameters\\ (fm$^{-2}$)}}&\multirow{2}*{\makecell[c]{$C_{11}$\\$C_{22}$ \\ $C_{33}$}}&\multicolumn{1}{r}{\multirow{2}*{\makecell[c]{14.4\\28.6\\16.9}}}&
\multicolumn{1}{r}{\multirow{2}*{\makecell[c]{20.2\\49.7\\-}}}&&(1.2)&0.210&
&
&$\frac{\textbf{p}^{2}_{x_{1}}}{2m'_{1}}+\frac{m_{b}+m_{\bar{c}}}{3m_{b}+m_{\bar{c}}}\frac{\textbf{p}^{2}_{x_{3}}}{2m'_{3}}$
&404.4&\\
&&&&&(1,3)&0.256&\multicolumn{1}{c!{\color{black}\vrule width 1.5pt}}{0.160($\Upsilon$)}&&
\multirow{3}*{\makecell[c]{$V^{C}(12)$\\$\frac{1}{2}[V^{C}(13)+V^{C}(23)]$\\$\frac{1}{2}[V^{C}(14)+V^{C}(24)]$\\$-D$}}
&\multirow{3}*{\makecell[c]{-257.5\\-85.5\\-77.6\\-983.0}}&&\\
\Xcline{1-5}{0.5pt}
\multicolumn{2}{c}{Quark Mass}&\multicolumn{1}{r}{17947.0}&17497.0&0.0&(2,3)&0.256&&
&\multirow{2}*{}
&\multirow{2}*{}&&$2m_{b}$\\
\multicolumn{2}{c}{\multirow{1}{*}{Confinement Potential}}&\multirow{1}{*}{-2659.7}&\multirow{1}{*}{-3123.1}&\multirow{1}{*}{463.4}
&(1,4)&0.266&&&&&&10105.8\\
\Xcline{10-13}{0.5pt}\Xcline{1-5}{0.5pt}
\multicolumn{2}{c}{\multirow{1}*{Kinetic Energy}}&\multicolumn{1}{r}{\multirow{1}*{838.2}}&\multirow{1}*{961.5}&\multirow{1}*{-123.3}
&(2,4)&0.266&\multicolumn{1}{c !{\color{black}\vrule width 1.5pt}}{0.250($B^{*}_{c}$)}&&\multirow{1}*{Subtotal}&\multirow{1}*{9527.1}&
\multirow{1}*{9497.8}&\multirow{1}*{10105.8}\\
\Xcline{9-13}{0.5pt}
\multicolumn{2}{c}{\multirow{1}*{CS Interaction}}&\multicolumn{1}{r}{\multirow{1}*{23.8}}&\multirow{1}*{34.0}&
\multirow{1}*{-10.2}
&(3,4)&0.296&&&\multirow{4}*{\makecell[c]{$m_{\bar{b}}$\\
$\frac{m_{\bar{c}}}{m_{\bar{c}}+m_{\bar{b}}}\frac{\textbf{p}^{2}_{x_{2}}}{2m'_{2}}+\frac{m_{b}}{3m_{b}+m_{\bar{c}}}\frac{\textbf{p}^{2}_{x_{3}}}{2m'_{3}}$
\\$\frac{1}{2}V^{C}(34)$
\\$\frac{1}{2}[V^{C}(13)+V^{C}(23)]$
\\$-\frac{1}{2}D$}}
&\multirow{4}*{\makecell[c]{5343.0\\146.6\\-55.1\\-85.5\\ -491.5}}&
\multirow{1}*{$\frac{m_{\bar{b}}}{4(m_{\bar{b}}+m_{\bar{b}})}m_{b\bar{b}}$}\\
\Xcline{1-8}{0.5pt}
\multicolumn{1}{c|}{\multirow{4}{*}{$V^{C}$}}&\multicolumn{1}{c}{\multirow{1}{*}{(1,2)}}&\multicolumn{1}{r}{-257.5}&&
&\multicolumn{2}{r}{(1,2)-(3,4):}&0.194 fm&\multirow{4}*{$\bar{b}$-quark}
&&&\multirow{1}*{2382.4}\\
\Xcline{6-8}{1.6pt}
&\multicolumn{1}{c}{(2,3)}&\multicolumn{1}{r}{-85.5}&&\multicolumn{1}{r}{}&&&&
&&&$\frac{m_{\bar{b}}}{4(m_{\bar{b}}+m_{\bar{c}})}m_{\bar{b}\bar{c}}$&$m_{\bar{b}}$\\
\multicolumn{1}{c|}{\multirow{1}{*}{}}&\multicolumn{1}{c}{(1,4)}&\multicolumn{1}{r}{-77.6}
&&\multicolumn{1}{r}{}&(2,4)&-77.6&$-360.4(B^{*}_{c})$&
&&&2360.6&5052.9\\
\Xcline{10-13}{0.5pt}\Xcline{2-5}{0.5pt}
\multicolumn{1}{c|}{\multirow{1}{*}{}}&\multicolumn{1}{c}{\multirow{1}{*}{Subtotal}}&\multicolumn{1}{r}{-693.7}&\multicolumn{1}{r}{-1157.1}&\multicolumn{1}{c|}{463.4}
&(3,4)&-110.1&
&&\multirow{1}*{Subtotal}&\multirow{1}*{4857.5}&
\multirow{1}*{4743.0}&\multirow{1}*{5052.9}\\
\Xcline{9-13}{0.5pt}\Xcline{1-5}{0.5pt}
\multicolumn{2}{c}{\multirow{1}*{Total Contribution}}&\multicolumn{1}{r}{\multirow{1}*{168.2}}&\multicolumn{1}{r}{-161.6}&\multicolumn{1}{c|}{\multirow{1}*{329.8}}
&(1,3)&-85.5&$-796.7(\Upsilon)$&
\multirow{4}*{$\bar{c}$-quark}&\multirow{4}*{\makecell[c]{$m_{\bar{c}}$\\
$\frac{m_{\bar{b}}}{m_{\bar{c}}+m_{\bar{b}}}\frac{\textbf{p}^{2}_{x_{2}}}{2m'_{2}}+\frac{m_{b}}{3m_{b}+m_{\bar{c}}}\frac{\textbf{p}^{2}_{x_{3}}}{2m'_{3}}$
\\$\frac{1}{2}V^{C}(34)$
\\$\frac{1}{2}[V^{C}(14)+V^{C}(24)]$
\\$-\frac{1}{2}D$}}
&\multirow{4}*{\makecell[c]{1918.0\\287.2\\-55.1\\-77.6\\ -491.5}}&\multirow{1}*{$\frac{m_{\bar{c}}}{4(m_{\bar{b}}+m_{\bar{c}})}m_{\bar{c}\bar{b}}$}
\\
\Xcline{1-8}{1.6pt}
\multicolumn{8}{l!{\color{black}\vrule width 1.5pt}}{Total Wave function:}&&&&&801.3&\\
\multicolumn{8}{l!{\color{black}\vrule width 1.5pt}}{$\Psi_{tot}=|F\rangle|R^{s}\rangle|[\phi_{1}\chi_{1}]\rangle=0.577|F\rangle|R^{s}\rangle|[\psi_{1}\zeta_{1}]\rangle-0.816|F\rangle|R^{s}\rangle|[\psi_{2}\zeta_{1}]\rangle$}&
&&&$\frac{m_{\bar{c}}}{4(m_{b}+m_{\bar{c}})}m_{b\bar{c}}$&$m_{\bar{c}}$\\
\Xcline{1-8}{0.3pt}
\multicolumn{8}{l!{\color{black}\vrule width 1.5pt}}{The rearrangement strong decay channel: $B^{*}_{c}\Upsilon$.}
&&&&789.3&1724.8\\
\Xcline{10-13}{0.5pt}\Xcline{1-8}{0.3pt}
\multicolumn{8}{l!{\color{black}\vrule width 1.5pt}}{The radiative decay widths: \quad  $\Gamma_{T_{b^{2}\bar{b}\bar{c}}(16149.2,2^{+})\rightarrow T_{b^{2}\bar{b}\bar{c}}(16043.9,0^{+})\gamma}=0\, \rm{keV}$}&&
\multirow{1}*{Subtotal}&\multirow{1}*{1581.2}&
\multirow{1}*{1590.6}&\multirow{1}*{1724.8}\\
\Xcline{9-13}{0.5pt}
\multicolumn{8}{r!{\color{black}\vrule width 1.5pt}}{$\Gamma_{T_{b^{2}\bar{b}\bar{c}}(16149.2,2^{+})\rightarrow T_{c^{2}\bar{c}\bar{b}}(16043.2,1^{+})\gamma}=435.0\, \rm{keV}$}&
\multirow{4}*{\makecell[c]{CS \\Interaction}}&\multirow{4}*{\makecell[c]{$\frac{1}{2}V^{SS}(12)$\\ $\frac{1}{2}V^{SS}(34)$\\$\frac{1}{4}(V^{SS}(13)+V^{SS}(23))$\\$\frac{1}{4}(V^{SS}(14)+V^{SS}(24))$}}
&\multirow{4}*{\makecell[c]{5.9\\6.4\\4.6\\6.9}}
&\multirow{4}*{\makecell[c]{$\frac{8}{3}v_{bb}+\frac{8}{3}v_{\bar{c}\bar{b}}$\\5.1+5.2\\$\frac{8}{3}v_{b\bar{b}}+\frac{8}{3}v_{b\bar{c}}$\\7.7+7.9}}
&\multirow{4}*{\makecell[c]{$\frac{8}{3}C_{bb}+\frac{8}{3}C_{\bar{c}\bar{b}}$\\7.7+8.8\\$\frac{8}{3}C_{b\bar{b}}+\frac{8}{3}C_{b\bar{c}}$\\7.7+8.8}}
\\\Xcline{1-8}{0.3pt}
\multicolumn{8}{l!{\color{black}\vrule width 1.5pt}}{The magnetic moments:}&&&&\\
\multicolumn{8}{l!{\color{black}\vrule width 1.5pt}}{$\mu_{T_{c^{2}\bar{c}\bar{b}}(16149.2,2^{+})}=
\langle\Psi^{2^{+}}_{tot}|\hat{\mu^{z}}|\Psi^{2^{+}}_{tot}\rangle=2\mu_{b}+\mu_{\bar{b}}+\mu_{\bar{c}}=-0.472\mu_{N}$}
&&&&&\\\Xcline{1-8}{0.3pt}
\multicolumn{8}{l!{\color{black}\vrule width 1.5pt}}{The transition magnetic moments:}
&&&&\\
\Xcline{10-13}{0.5pt}
\multicolumn{8}{l!{\color{black}\vrule width 1.5pt}}{$\mu_{T_{b^{2}\bar{b}\bar{c}}(16149.2,2^{+})\rightarrow T_{b^{2}\bar{b}\bar{c}}(16043.2,1^{+})\gamma}=\langle\Psi^{2^{+}}_{tot}|\hat{\mu^{z}}|\Psi^{1^{+}}_{tot}\rangle=0.329 \mu_{N}$}
&&\multirow{1}*{Subtotal}&\multirow{1}*{23.8}&\multirow{1}*{25.9}&33.1\\
\Xcline{9-13}{0.5pt}
\multicolumn{8}{l!{\color{black}\vrule width 1.5pt}}{$\mu_{T_{b^{2}\bar{b}\bar{c}}(16149.2,2^{+})\rightarrow T_{b^{2}\bar{b}\bar{c}}(16043.9,0^{+})\gamma}=\langle\Psi^{2^{+}}_{tot}|\hat{\mu^{z}}|\Psi^{0^{+}}_{tot}\rangle=0$}&\multicolumn{2}{l|}{\multirow{1}*{Total contribution}}&\multirow{1}*{16149.2}&\multirow{1}*{15882.3}&\multirow{1}*{16917.0}\\
\toprule[0.5pt]
\toprule[1.5pt]
\end{tabular}
\end{lrbox}\scalebox{0.864}{\usebox{\tablebox}}
\end{table*}

\begin{table*}
\caption{ The masses, binding energies,  variational parameters, the internal contribution, total wave functions, magnetic moments, transition magnetic moments, radiative decay widths, rearrangement strong width ratios, and the relative lengths between quarks for the the $J^{PC}=0^{++}$, $2^{++}$ $cb\bar{c}\bar{b}$  states and their lowest meson-meson thresholds. The notation is the same as that in Table \ref{cccc}.
}\label{cbcb1}
\begin{lrbox}{\tablebox}
\renewcommand\arraystretch{1.45}
\renewcommand\tabcolsep{1.4pt}
\begin{tabular}{c|c|ccc !{\color{black}\vrule width 1.5pt} ccc !{\color{black}\vrule width 1.5pt} c|cc|c|cc}
\midrule[1.5pt]
\toprule[0.5pt]
\multicolumn{1}{c}{$cb\bar{c}\bar{b}$}&\multicolumn{4}{r!{\color{black}\vrule width 1.5pt}}{The contribution from each term}&\multicolumn{3}{c!{\color{black}\vrule width 1.5pt}}{Relative Lengths (fm)}&\multirow{2}*{Overall}&\multicolumn{2}{c}{Present Work}&\multicolumn{2}{c}{CMI Model}\\
\Xcline{1-8}{0.5pt}\Xcline{10-13}{0.5pt}
\multicolumn{1}{r}{$J^{PC}=0^{++}$}&&Value&\multicolumn{1}{r}{$\eta_{b}\eta_{c}$}&
\multicolumn{1}{r!{\color{black}\vrule width 1.5pt}}{Difference}&$(i,j)$&\multicolumn{1}{r}{Value}
&\multicolumn{1}{c!{\color{black}\vrule width 1.5pt}}{$\eta_{b}\eta_{c}$}&&Contribution&Value&Ref. \cite{Weng:2020jao}&Ref. \cite{Wu:2016vtq}\\ \Xcline{1-13}{0.5pt}
\Xcline{1-5}{0.5pt}
\multicolumn{2}{c}{Mass/$B_{T}$}&\multicolumn{1}{r}{12759.6}&12387.5&372.1&(1,2)&0.315&&
\multirow{5}*{\makecell[c]{$c$-quark:\\ \\$m^{eff}_{c}$}}&\multirow{1}*{\makecell[c]{$m_{c}$\\
$\frac{m_{b}}{m_{c}+m_{b}}\frac{\textbf{p}^{2}_{x_{1}}}{2m'_{1}}$\\
$\frac{m_{b}}{2m_{c}+2m_{b}}\frac{\textbf{p}^{2}_{x_{3}}}{2m'_{3}}$
\\$\frac{1}{2}V^{C}(12)$\\$\frac{1}{2}[V^{C}(13)+V^{C}(14)]$\\$-1/2D$
}}&
\multirow{1}*{\makecell[c]{1918.0
}}&
\multirow{4}*{\makecell[c]{$\frac{-m_{c}}{4(m_{c}+m_{b})}m_{cb}$\\-400.5\\
$\frac{5m_{c}}{8(m_{c}+m_{\bar{b}})}m_{c\bar{b}}$\\986.7
\\$\frac{5m_{c}}{8(m_{c}+m_{\bar{c}})}m_{c\bar{c}}$}}\\
\Xcline{1-5}{0.5pt}
\multirow{2}*{\makecell[c]{Variational\\ Parameters\\ (fm$^{-2}$)}}&\multirow{2}*{\makecell[c]{$C_{11}$\\$C_{22}$ \\ $C_{33}$}}&\multicolumn{1}{r}{\multirow{2}*{\makecell[c]{12.7\\12.7\\23.6}}}&
\multicolumn{1}{r}{\multirow{2}*{\makecell[c]{15.0\\57.4\\-}}}&&(1.3)&0.277&
\multicolumn{1}{c!{\color{black}\vrule width 1.5pt}}{0.290($\eta_{c}$)}&
&&193.7&\\
&&&&&(2,3)&0.277&&&
\multirow{3}*{}&\multirow{1}*{\makecell[c]{139.6\\21.3\\-172.6}}&\\
\Xcline{1-5}{0.5pt}
\multicolumn{2}{c}{Quark Mass}&\multicolumn{1}{r}{14522.0}&14522.0&0.0&(1,4)&0.277&&
&\multirow{2}*{}
&\multirow{3}*{}&&$m_{c}$\\
\multicolumn{2}{c}{\multirow{1}{*}{Confinement Potential}}&\multirow{1}{*}{-2571.3}&\multirow{1}{*}{-3082.3}&\multirow{1}{*}{511.0}&(2,4)&0.277&
\multicolumn{1}{c!{\color{black}\vrule width 1.5pt}}{0.148($\eta_{b}$)}&&&-491.5&958.9&1724.8\\
\Xcline{1-5}{0.5pt}\Xcline{10-13}{0.5pt}
\multicolumn{2}{c}{\multirow{1}*{Kinetic Energy}}&\multicolumn{1}{r}{\multirow{1}*{905.8}}&\multirow{1}*{1085.5}&\multirow{1}*{-179.7}
&(3,4)&0.315&&&\multirow{1}*{Subtotal}&\multirow{1}*{1608.5}&
\multirow{1}*{1545.1}&\multirow{1}*{1724.8}\\
\Xcline{6-8}{0.3pt}\Xcline{9-13}{0.5pt}
\multicolumn{2}{c}{\multirow{1}*{CS Interaction}}&\multicolumn{1}{r}{\multirow{1}*{-81.1}}&\multirow{1}*{-137.6}&
\multirow{1}*{56.6}
&\multicolumn{2}{r}{(1,2)-(3,4):}&0.164 fm&&\multirow{5}*{\makecell[c]{$m_{\bar{b}}$\\
$\frac{m_{c}}{m_{c}+m_{b}}\frac{\textbf{p}^{2}_{x_{1}}}{2m'_{1}}$\\
$\frac{m_{c}}{2m_{c}+2m_{b}}\frac{\textbf{p}^{2}_{x_{3}}}{2m'_{3}}$
\\$\frac{1}{2}V^{C}(12)$\\$\frac{1}{2}[V^{C}(23)+V^{C}(24)]$\\$-1/2D$}}
&\multirow{1}*{\makecell[c]{5343.0}}&
\multirow{4}*{\makecell[c]{$\frac{-m_{b}}{4(m_{c}+m_{b})}m_{cb}$\\-1203.8\\
$\frac{5m_{b}}{8(m_{c}+m_{\bar{b}})}m_{c\bar{b}}$\\2965.5
\\$\frac{5m_{b}}{8(m_{b}+m_{\bar{b}})}m_{b\bar{b}}$}}\\
\Xcline{6-8}{1.6pt}\Xcline{1-5}{0.5pt}
\multicolumn{1}{c|}{\multirow{3}{*}{$V^{C}$}}&\multicolumn{1}{c}{\multirow{1}{*}{(1,2)}}&\multicolumn{1}{r}{42.6}&
&\multicolumn{1}{r}{}&(1,3)&\multicolumn{1}{r}{-172.6}&-237.2($\eta_{c}$)
&\multirow{4}*{\makecell[c]{$b$-quark:\\ \\$m^{eff}_{b}$}}
&&\multirow{1}*{\makecell[c]{69.5}}&\multirow{1}*{}\\
\multicolumn{1}{c|}{\multirow{1}{*}{}}&\multicolumn{1}{c}{(1,4)}&\multicolumn{1}{r}{-172.6}&&\multicolumn{1}{r}{}&(2,3)&\multicolumn{1}{r}{-172.6}&&
&&\multirow{1}*{\makecell[c]{50.1\\21.3\\-172.6}}&&\\
\Xcline{2-5}{0.5pt}
\multicolumn{1}{c|}{\multirow{1}{*}{}}&\multicolumn{1}{c}{\multirow{1}{*}{Subtotal}}&\multicolumn{1}{r}{-605.3}&\multicolumn{1}{r}{-1116.3}&\multicolumn{1}{c|}{511.0}
&(2,4)&-172.6&$-879.1(\eta_{b})$
&&&&&$m_{b}$\\
\Xcline{1-5}{0.5pt}
\multicolumn{2}{c}{\multirow{1}*{Total Contribution}}&219.3&\multicolumn{1}{r}{-168.5}&\multicolumn{1}{c|}{387.8}
&(3,4)&42.6&
&&&-491.5&2951.6&5052.9\\
\Xcline{10-13}{0.5pt}\Xcline{1-8}{1.6pt}
\multicolumn{8}{l !{\color{black}\vrule width 1.5pt}}{Total Wave function:}
&&\multirow{1}*{Subtotal}&\multirow{1}*{4819.8}&
\multirow{1}*{4713.3}&\multirow{1}*{5052.9}\\
\Xcline{9-13}{0.5pt}
\multicolumn{8}{r!{\color{black}\vrule width 1.5pt}}{$\Psi_{tot}=0.961|F\rangle|R^{s}\rangle|[\phi_{2}\chi_{5}]\rangle+0.114|F\rangle|R^{s}\rangle|[\phi_{2}\chi_{6}]\rangle-0.069|F\rangle|R^{s}\rangle|[\phi_{1}\chi_{5}]\rangle$}&
\multirow{5}*{\makecell[c]{CS \\Interaction}}&\multirow{1}*{\makecell[c]{$-\frac{1}{4}(V^{SS}(12)+V^{SS}(34))$
}}
&\multirow{1}*{\makecell[c]{-5.3
}}
&\multirow{1}*{\makecell[c]{$-\frac{8}{3}v_{cb}-\frac{20}{3}v_{\bar{c}\bar{c}}$}}
&\multirow{1}*{\makecell[c]{$-\frac{8}{3}C_{cb}-\frac{20}{3}C_{\bar{c}\bar{c}}$}}
\\
\multicolumn{8}{r!{\color{black}\vrule width 1.5pt}}{$-0.241|F\rangle|R^{s}\rangle|[\phi_{1}\chi_{6}]\rangle=0.211|F\rangle|R^{s}\rangle|[\psi_{1}\zeta_{6}]\rangle-0.830|F\rangle|R^{s}\rangle|[\psi_{1}\zeta_{5}]\rangle$}&
&\multirow{1}*{\makecell[c]{$-\frac{5}{4}V^{SS}(13)$}}&\multirow{1}*{\makecell[c]{-33.3}}
&\multirow{1}*{\makecell[c]{-5.1-35.5}}&\multirow{1}*{\makecell[c]{-8.8-35.3}}\\
\multicolumn{8}{r!{\color{black}\vrule width 1.5pt}}{$-0.367|F\rangle|R^{s}\rangle|[\psi_{2}\zeta_{5}]\rangle+
0.363|F\rangle|R^{s}\rangle|[\psi_{2}\zeta_{6}]\rangle=0.333|F\rangle|R^{s}\rangle|[\psi'_{1}\zeta'_{6}]\rangle$}&&
\multirow{1}*{\makecell[c]{$-\frac{5}{4}V^{SS}(24)$}}&\multirow{1}*{\makecell[c]{-10.1}}&
\multirow{1}*{\makecell[c]{$-\frac{20}{3}v_{b\bar{b}}-\frac{40}{3}v_{b\bar{c}}$}}&
\multirow{1}*{\makecell[c]{$-\frac{20}{3}C_{b\bar{b}}-\frac{40}{3}C_{b\bar{c}}$}}\\
\multicolumn{8}{r!{\color{black}\vrule width 1.5pt}}{$-0.668|F\rangle|R^{s}\rangle|[\psi'_{1}\zeta'_{5}]\rangle-0.398|F\rangle|R^{s}\rangle|[\psi'_{2}\zeta'_{5}]\rangle+
0.533|F\rangle|R^{s}\rangle|[\psi'_{2}\zeta'_{6}]\rangle$}
&&\multirow{1}*{\makecell[c]{$-\frac{5}{4}(V^{SS}(14)+V^{SS}(23))$}}
&\multirow{1}*{\makecell[c]{-32.5}}&\multirow{1}*{\makecell[c]{-19.2-39.3}}&\multirow{1}*{\makecell[c]{-19.3-44.3}}\\
\Xcline{10-13}{0.5pt}\Xcline{1-8}{0.3pt}
\multicolumn{8}{l!{\color{black}\vrule width 1.5pt}}{The rearrangement strong decay channel:}&&\multirow{1}*{Subtotal}&\multirow{1}*{-81.1}&
\multirow{1}*{-99.2}&\multirow{1}*{-107.5}\\
\Xcline{9-13}{0.5pt}
\multicolumn{8}{l!{\color{black}\vrule width 1.5pt}}{$\Gamma_{T_{cb\bar{c}\bar{b}}(12759.6,0^{++})\rightarrow \eta_{c}\eta_{b}}:
\Gamma_{T_{cb\bar{c}\bar{b}}(12759.6,0^{++})\rightarrow J/\psi\Upsilon}=21:1$}
&\multicolumn{2}{l|}{\multirow{1}*{Matrix nondiagonal element}}&\multirow{1}*{-15.8}&-80.7&-51.9\\
\Xcline{9-13}{0.5pt}
\multicolumn{8}{l!{\color{black}\vrule width 1.5pt}}{$\Gamma_{T_{cb\bar{c}\bar{b}}(12759.6,0^{++})\rightarrow B_{c}B_{c}}:
\Gamma_{T_{cb\bar{c}\bar{b}}(12759.6,0^{++})\rightarrow B^{*}_{c}B^{*}_{c}}=7.6:1$}
&\multicolumn{2}{l|}{\multirow{1}*{Total contribution}}&\multirow{1}*{12759.6}&\multirow{1}*{12336.1}&\multirow{1}*{13396.0}\\
\Xcline{9-13}{1.6pt}\Xcline{1-8}{0.3pt}
\multicolumn{5}{l|}{The radiative decay widths:}&\multicolumn{8}{l}{The magnetic moments: $T_{cb\bar{c}\bar{b}}(12759.6,0^{++})=
<\Psi^{0^{++}}_{tot}|\hat{\mu^{z}}|\Psi^{0^{++}}_{tot}>=0$}\\
\Xcline{6-13}{0.3pt}
\multicolumn{5}{l|}{$\Gamma_{T_{cb\bar{c}\bar{b}}(12797.3,1^{+-})\rightarrow T_{cb\bar{c}\bar{b}}(12759.6,0^{++})\gamma}=33.1$\,keV}
&\multicolumn{8}{l}{The transition magnetic moments: \quad
$\mu_{T_{cb\bar{c}\bar{b}}(12882.4,2^{++})\rightarrow T_{cb\bar{c}\bar{b}}(12759.6,0^{++})\gamma}=<\Psi^{2^{++}}_{tot}|\hat{\mu^{z}}|\Psi^{0^{++}}_{tot}>=0$ }\\
\multicolumn{5}{l|}{$\Gamma_{T_{cb\bar{c}\bar{b}}(12856.6,1^{++})\rightarrow T_{cb\bar{c}\bar{b}}(12759.6,0^{++})\gamma}=0$}
&\multicolumn{8}{l}{$\mu_{T_{cb\bar{c}\bar{b}}(12856.6,1^{++})\rightarrow T_{cb\bar{c}\bar{b}}(12759.6,0^{++})\gamma}=<\Psi^{1^{++}}_{tot}|\hat{\mu^{z}}|\Psi^{0^{++}}_{tot}>=0$}\\
\multicolumn{5}{l|}{$\Gamma_{T_{cb\bar{c}\bar{b}}(12882.4,2^{++})\rightarrow T_{cb\bar{c}\bar{b}}(12759.6,0^{++})\gamma}=0$}
&\multicolumn{8}{l}{$\mu_{T_{cb\bar{c}\bar{b}}(12796.9,1^{+-})\rightarrow T_{cb\bar{c}\bar{b}}(12759.6,0^{++})\gamma}=<\Psi^{1^{+-}}_{tot}|\hat{\mu^{z}}|\Psi^{0^{++}}_{tot}>=0.226\times2\sqrt{\frac{1}{6}}(\mu_{c}-\mu_{b})=0.081\mu_{N}$}\\
\toprule[1.5pt]
\multicolumn{1}{r}{$J^{PC}=2^{++}$}&&Value&\multicolumn{1}{r}{$\Upsilon J/\psi$}&
\multicolumn{1}{r!{\color{black}\vrule width 1.5pt}}{Difference}&\multicolumn{3}{c!{\color{black}\vrule width 1.5pt}}{Relative Lengths (fm)}&
&Contribution&Value&Ref. \cite{Weng:2020jao}&Ref. \cite{Wu:2016vtq}\\ \Xcline{1-13}{0.5pt}
\multicolumn{2}{c}{Mass/$B_{T}$}&\multicolumn{1}{r}{12882.4}&12561.1&321.3&$(i,j)$&\multicolumn{1}{r}{Value}
&\multicolumn{1}{c!{\color{black}\vrule width 1.5pt}}{$\Upsilon J/\psi$}&
\multirow{5}*{\makecell[c]{$c$-quark:\\ \\$m^{eff}_{c}$}}&\multirow{1}*{\makecell[c]{$m_{c}$\\
$\frac{m_{b}}{m_{c}+m_{b}}\frac{\textbf{p}^{2}_{x_{1}}}{2m'_{1}}$\\
$\frac{m_{b}}{2m_{c}+2m_{b}}\frac{\textbf{p}^{2}_{x_{3}}}{2m'_{3}}$
\\$\frac{1}{2}V^{C}(12)$\\$\frac{1}{2}[V^{C}(13)+V^{C}(14)]$\\$-1/2D$
}}&
\multirow{1}*{\makecell[c]{1918.0
}}&
\multirow{4}*{\makecell[c]{$\frac{-m_{c}}{4(m_{c}+m_{b})}m_{cb}$\\-400.5\\
$\frac{5m_{c}}{8(m_{c}+m_{\bar{b}})}m_{c\bar{b}}$\\986.7
\\$\frac{5m_{c}}{8(m_{c}+m_{\bar{c}})}m_{c\bar{c}}$}}\\
\Xcline{1-8}{0.5pt}
\multirow{2}*{\makecell[c]{Variational\\ Parameters\\ (fm$^{-2}$)}}&\multirow{2}*{\makecell[c]{$C_{11}$\\$C_{22}$ \\ $C_{33}$}}&\multicolumn{1}{r}{\multirow{2}*{\makecell[c]{11.0\\11.0\\21.0}}}&
\multicolumn{1}{r}{\multirow{2}*{\makecell[c]{12.5\\49.7\\-}}}&&(1.3)&0.340&
&
&&167.1&\\
&&&&&(1,3)&0.296&\multicolumn{1}{c!{\color{black}\vrule width 1.5pt}}{0.318($J/\psi$)}&&
\multirow{3}*{}&\multirow{1}*{\makecell[c]{124.2\\14.0\\-137.7}}&\\
\Xcline{1-5}{0.5pt}
\multicolumn{2}{c}{Quark Mass}&\multicolumn{1}{r}{14522.0}&14522.0&0.0&(2,3)&0.296&&
&\multirow{2}*{}
&\multirow{3}*{}&&$m_{c}$\\
\multicolumn{2}{c}{\multirow{1}{*}{Confinement Potential}}&\multirow{1}{*}{-2460.7}&\multirow{1}{*}{-2926.9}&\multirow{1}{*}{466.2}&(1,4)&0.296&&&&-491.5&958.9&1724.8\\
\Xcline{1-5}{0.5pt}\Xcline{10-13}{0.5pt}
\multicolumn{2}{c}{\multirow{1}*{Kinetic Energy}}&\multicolumn{1}{r}{\multirow{1}*{791.7}}&\multirow{1}*{925.1}&\multirow{1}*{-133.4}
&(2,4)&0.296&\multicolumn{1}{c!{\color{black}\vrule width 1.5pt}}{0.160($\Upsilon$)}&&\multirow{1}*{Subtotal}&\multirow{1}*{1594.1}&
\multirow{1}*{1545.1}&\multirow{1}*{1724.8}\\
\Xcline{9-13}{0.5pt}
\multicolumn{2}{c}{\multirow{1}*{CS Interaction}}&\multicolumn{1}{r}{\multirow{1}*{29.5}}&\multirow{1}*{41.0}&
\multirow{1}*{-11.5}
&(3,4)&0.340&&&\multirow{5}*{\makecell[c]{$m_{\bar{b}}$\\
$\frac{m_{c}}{m_{c}+m_{b}}\frac{\textbf{p}^{2}_{x_{1}}}{2m'_{1}}$\\
$\frac{m_{c}}{2m_{c}+2m_{b}}\frac{\textbf{p}^{2}_{x_{3}}}{2m'_{3}}$
\\$\frac{1}{2}V^{C}(12)$\\$\frac{1}{2}[V^{C}(23)+V^{C}(24)]$\\$-1/2D$}}
&\multirow{1}*{\makecell[c]{5343.0}}&
\multirow{4}*{\makecell[c]{$\frac{-m_{b}}{4(m_{c}+m_{b})}m_{cb}$\\-1203.8\\
$\frac{5m_{b}}{8(m_{c}+m_{\bar{b}})}m_{c\bar{b}}$\\2965.5
\\$\frac{5m_{b}}{8(m_{b}+m_{\bar{b}})}m_{b\bar{b}}$}}\\
\Xcline{6-8}{0.5pt}\Xcline{1-5}{0.5pt}
\multicolumn{1}{c|}{\multirow{4}{*}{$V^{C}$}}&\multicolumn{1}{c}{\multirow{1}{*}{(1,2)}}&\multicolumn{1}{r}{28.0}&&
&\multicolumn{2}{r}{(1,2)-(3,4):}&0.174 fm&\multirow{4}*{\makecell[c]{$b$-quark:\\ \\$m^{eff}_{b}$}}
&&\multirow{1}*{\makecell[c]{60.0}}&\multirow{1}*{}\\
\Xcline{6-8}{1.6pt}
&\multicolumn{1}{c}{(2,3)}&\multicolumn{1}{r}{-137.7}&&\multicolumn{1}{r}{}&&&&&&\multirow{1}*{\makecell[c]{44.6\\14.0\\-137.7}}&&\\
\multicolumn{1}{c|}{\multirow{1}{*}{}}&\multicolumn{1}{c}{(1,4)}&\multicolumn{1}{r}{-137.7}
&&\multicolumn{1}{r}{}&(2,4)&-137.7&$-796.7(\Upsilon)$&&&&
&$m_{b}$\\
\Xcline{2-5}{0.5pt}
\multicolumn{1}{c|}{\multirow{1}{*}{}}&\multicolumn{1}{c}{\multirow{1}{*}{Subtotal}}&\multicolumn{1}{r}{-494.7}&\multicolumn{1}{r}{-960.9}&\multicolumn{1}{c|}{466.2}
&(3,4)&28.0&
&&&-491.5&2951.6&5052.9\\
\Xcline{10-13}{0.5pt}\Xcline{1-5}{0.5pt}
\multicolumn{2}{c}{\multirow{1}*{Total Contribution}}&\multicolumn{1}{r}{\multirow{1}*{326.5}}&\multicolumn{1}{r}{5.1}&\multicolumn{1}{c|}{\multirow{1}*{321.4}}
&(1,3)&-137.7&$-164.2(J/\psi)$&
&\multirow{1}*{Subtotal}&\multirow{1}*{4832.4}&
\multirow{1}*{4713.3}&\multirow{1}*{5052.9}\\
\Xcline{9-13}{0.5pt}\Xcline{1-8}{1.6pt}
\multicolumn{7}{l}{Total Wave function:}&&
\multirow{5}*{\makecell[c]{CS \\Interaction}}&\multirow{1}*{\makecell[c]{$-\frac{1}{4}(V^{SS}(12)+V^{SS}(34))$
}}
&\multirow{1}*{\makecell[c]{-4.7
}}
&\multirow{1}*{\makecell[c]{$-\frac{8}{3}v_{cb}+\frac{10}{3}v_{\bar{c}\bar{c}}$}}
&\multirow{1}*{\makecell[c]{$-\frac{8}{3}C_{cb}+\frac{10}{3}C_{\bar{c}\bar{c}}$}}
\\
\multicolumn{8}{r !{\color{black}\vrule width 1.5pt}}{$\Psi_{tot}=0.999|F\rangle|R^{s}\rangle|[\phi_{2}\chi_{1}]\rangle-0.027|F\rangle|R^{s}\rangle|[\phi_{1}\chi_{1}]\rangle=-0.801|F\rangle|R^{s}\rangle|[\psi_{1}\zeta_{1}]\rangle$}
&&\multirow{1}*{\makecell[c]{$\frac{5}{8}V^{SS}(13)$}}&\multirow{1}*{\makecell[c]{15.2}}
&\multirow{1}*{\makecell[c]{-5.1+17.7}}&\multirow{1}*{\makecell[c]{-8.8+17.7}}\\
\multicolumn{8}{r !{\color{black}\vrule width 1.5pt}}{
$-0.599|F\rangle|R^{s}\rangle|[\psi_{2}\zeta_{1}]\rangle=-0.832|F\rangle|R^{s}\rangle|[\psi'_{1}\zeta'_{1}]\rangle+0.555|F\rangle|R^{s}\rangle|[\psi'_{2}\zeta'_{1}]\rangle$}&&
\multirow{1}*{\makecell[c]{$\frac{5}{8}V^{SS}(24)$}}&\multirow{1}*{\makecell[c]{4.5}}&
\multirow{1}*{\makecell[c]{$\frac{10}{3}v_{b\bar{b}}+\frac{20}{3}v_{b\bar{c}}$}}&
\multirow{1}*{\makecell[c]{$\frac{10}{3}C_{b\bar{b}}+\frac{20}{3}C_{b\bar{c}}$}}\\
\Xcline{1-8}{0.3pt}
\multicolumn{8}{l!{\color{black}\vrule width 1.5pt}}{
The rearrangement strong decay channel: $\Upsilon J/\psi$  and  $B^{*}_{c}B^{*}_{c}$
}
&&\multirow{1}*{\makecell[c]{$\frac{5}{8}(V^{SS}(14)+V^{SS}(23))$}}
&\multirow{1}*{\makecell[c]{14.7}}&\multirow{1}*{\makecell[c]{9.6+19.7}}&\multirow{1}*{\makecell[c]{9.6+22.0}}\\
\Xcline{10-13}{0.5pt}
\Xcline{1-8}{0.3pt}
\multicolumn{8}{l !{\color{black}\vrule width 1.5pt}}{The magnetic moments: \quad  $T_{cb\bar{c}\bar{b}}(12882.4,2^{++})=
\langle\Psi^{2^{++}}_{tot}|\hat{\mu^{z}}|\Psi^{2^{++}}_{tot}\rangle=0$}
&&\multirow{1}*{Subtotal}&\multirow{1}*{29.5}&
\multirow{1}*{41.7}&\multirow{1}*{40.5}\\
\Xcline{9-13}{0.5pt}
\Xcline{1-8}{0.3pt}
\multicolumn{8}{l!{\color{black}\vrule width 1.5pt}}{The radiative decay widths:
}
&\multicolumn{2}{l|}{\multirow{1}*{Matrix nondiagonal element}}&\multirow{1}*{0.1}&-29.1&3.2\\
\Xcline{9-13}{0.5pt}
\multicolumn{8}{l!{\color{black}\vrule width 1.5pt}}{$\Gamma_{T_{cb\bar{c}\bar{b}}(12882.4,2^{++})\rightarrow T_{cb\bar{c}\bar{b}}(12797.3,1^{+-})\gamma}=4.9$ keV }
&\multicolumn{2}{l|}{\multirow{1}*{Total contribution}}&
\multirow{1}*{12882.4}&\multirow{1}*{12529.4}&\multirow{1}*{13599.0}\\
\Xcline{9-13}{1.6pt}
\multicolumn{5}{l}{
$\Gamma_{T_{cb\bar{c}\bar{b}}(12882.4,2^{++})\rightarrow T_{cb\bar{c}\bar{b}}(12759.6,0^{++})\gamma}=0$ }
&
\multicolumn{5}{l}{$\Gamma_{T_{cb\bar{c}\bar{b}}(12882.4,2^{++})\rightarrow T_{cb\bar{c}\bar{b}}(12856.6,1^{++})\gamma}=0$}
\\
\toprule[0.3pt]
\multicolumn{4}{l}{
The transition magnetic moments: }&
\multicolumn{9}{r}{
$\mu_{T_{cb\bar{c}\bar{b}}(12882.4,2^{++})\rightarrow
T_{cb\bar{c}\bar{b}}(12759.6,1^{+-})\gamma}=
\langle\Psi^{2^{++}}_{tot}|\hat{\mu^{z}}|\Psi^{1^{+-}}_{tot}\rangle=0.774\times(\mu_{c}+\mu_{b})+0.226\times(\mu_{c}-\mu_{b})=0.345\mu_{N}$
}\\
\multicolumn{8}{l}{$\mu_{T_{cb\bar{c}\bar{b}}(12882.4,2^{++})\rightarrow
T_{cb\bar{c}\bar{b}}(12856.6,1^{++})\gamma}=\langle\Psi^{2^{++}}_{tot}|\hat{\mu^{z}}|\Psi^{1^{++}}_{tot}\rangle=0$ }&
\multicolumn{5}{l}{
$\mu_{T_{cb\bar{c}\bar{b}}(12882.4,2^{++})\rightarrow
T_{cb\bar{c}\bar{b}}(12759.3,0^{++})\gamma}=\langle\Psi^{2^{++}}_{tot}|\hat{\mu^{z}}|\Psi^{0^{++}}_{tot}\rangle=0$}\\
\toprule[0.5pt]
\toprule[1.5pt]
\end{tabular}
\end{lrbox}\scalebox{0.868}{\usebox{\tablebox}}
\end{table*}

\begin{table*}
\caption{
 The masses, binding energies,  variational parameters, the internal contribution, total wave functions, magnetic moments, transition magnetic moments, radiative decay widths, rearrangement strong width ratios, and the relative lengths between quarks for the the $J^{PC}=1^{+-}$, $1^{++}$ $cb\bar{c}\bar{b}$  states and their lowest meson-meson thresholds. The notation it the same as that in Table \ref{cccc}.
}\label{cbcb2}
\begin{lrbox}{\tablebox}
\renewcommand\arraystretch{1.45}
\renewcommand\tabcolsep{1.4pt}
\begin{tabular}{c|c|ccc!{\color{black}\vrule width 1.5pt} ccc!{\color{black}\vrule width 1.5pt}c|cc|c|cc}
\midrule[1.5pt]
\toprule[0.5pt]
\multicolumn{1}{c}{$cb\bar{c}\bar{b}$}&\multicolumn{4}{r!{\color{black}\vrule width 1.5pt}}{The contribution from each term}&\multicolumn{3}{c!{\color{black}\vrule width 1.5pt}}{Relative Length (fm)}&\multirow{2}*{Overall}&\multicolumn{2}{c}{Present Work}&\multicolumn{2}{c}{CMI Model}\\
\Xcline{1-8}{0.5pt}\Xcline{10-13}{0.5pt}
\multicolumn{1}{r}{$J^{PC}=1^{+-}$}&&Value&\multicolumn{1}{r}{$\Upsilon\eta_{c}$}&
Difference&$(i,j)$&\multicolumn{1}{r}{Value}
&$\Upsilon\eta_{c}$&&Contribution&Value&Ref. \cite{Weng:2020jao}&Ref. \cite{Wu:2016vtq}\\ \Xcline{1-13}{0.5pt}
\Xcline{1-5}{0.5pt}
\multicolumn{2}{c}{Mass/$B_{T}$}&\multicolumn{1}{r}{12796.9}&12467.4&329.5&(1,2)&0.331&&
\multirow{5}*{\makecell[c]{$c$-quark:\\ \\$m^{eff}_{c}$}}&\multirow{1}*{\makecell[c]{$m_{c}$\\
$\frac{m_{b}}{m_{c}+m_{b}}\frac{\textbf{p}^{2}_{x_{1}}}{2m'_{1}}$\\
$\frac{m_{b}}{2m_{c}+2m_{b}}\frac{\textbf{p}^{2}_{x_{3}}}{2m'_{3}}$
\\$\frac{1}{2}V^{C}(12)$\\$\frac{1}{2}[V^{C}(13)+V^{C}(14)]$\\$-1/2D$
}}&
\multirow{1}*{\makecell[c]{1918.0
}}&
\multirow{4}*{\makecell[c]{$\frac{-m_{c}}{4(m_{c}+m_{b})}m_{cb}$\\-400.5\\
$\frac{5m_{c}}{8(m_{c}+m_{\bar{b}})}m_{c\bar{b}}$\\986.7
\\$\frac{5m_{c}}{8(m_{c}+m_{\bar{c}})}m_{c\bar{c}}$}}\\
\Xcline{1-5}{0.5pt}
\multirow{2}*{\makecell[c]{Variational\\ Parameters\\ (fm$^{-2}$)}}&\multirow{2}*{\makecell[c]{$C_{11}$\\$C_{22}$ \\ $C_{33}$}}&\multicolumn{1}{r}{\multirow{2}*{\makecell[c]{11.9\\11.9\\22.9}}}&
\multicolumn{1}{r}{\multirow{2}*{\makecell[c]{15.0\\49.7\\-}}}&&(1.3)&0.289&
0.290($\eta_{c}$)&
&&180.6&\\
&&&&&(2,3)&0.289&&&
\multirow{3}*{}&\multirow{1}*{\makecell[c]{135.3\\17.8\\-158.1}}&\\
\Xcline{1-5}{0.5pt}
\multicolumn{2}{c}{Quark Mass}&\multicolumn{1}{r}{14522.0}&14522.0&0.0&(1,4)&0.289&&
&\multirow{2}*{}
&\multirow{3}*{}&&$m_{c}$\\
\multicolumn{2}{c}{\multirow{1}{*}{Confinement Potential}}&\multirow{1}{*}{-2527.1}&\multirow{1}{*}{-3000.0}&\multirow{1}{*}{472.9}&(2,4)&0.289&
0.160($\Upsilon$)&&&-491.5&958.9&1724.8\\
\Xcline{1-5}{0.5pt}\Xcline{10-13}{0.5pt}
\multicolumn{2}{c}{\multirow{1}*{Kinetic Energy}}&\multicolumn{1}{r}{\multirow{1}*{858.5}}&\multicolumn{1}{r}{1001.2}&
\multirow{1}*{-142.7}&(3,4)&0.331&&&\multirow{1}*{Subtotal}&\multirow{1}*{1602.1}&
\multirow{1}*{1545.1}&\multirow{1}*{1724.8}\\
\Xcline{6-8}{0.3pt}\Xcline{9-13}{0.5pt}
\multicolumn{2}{c}{\multirow{1}*{CS Interaction}}&\multicolumn{1}{r}{\multirow{1}*{-41.4}}&\multicolumn{1}{r}{-55.8}&
\multirow{1}*{14.4}
&\multicolumn{2}{r}{(1,2)-(3,4):}&0.168 fm&&\multirow{5}*{\makecell[c]{$m_{\bar{b}}$\\
$\frac{m_{c}}{m_{c}+m_{b}}\frac{\textbf{p}^{2}_{x_{1}}}{2m'_{1}}$\\
$\frac{m_{c}}{2m_{c}+2m_{b}}\frac{\textbf{p}^{2}_{x_{3}}}{2m'_{3}}$
\\$\frac{1}{2}V^{C}(12)$\\$\frac{1}{2}[V^{C}(23)+V^{C}(24)]$\\$-1/2D$}}
&\multirow{1}*{\makecell[c]{5343.0}}&
\multirow{4}*{\makecell[c]{$\frac{-m_{b}}{4(m_{c}+m_{b})}m_{cb}$\\-1203.8\\
$\frac{5m_{b}}{8(m_{c}+m_{\bar{b}})}m_{c\bar{b}}$\\2965.5
\\$\frac{5m_{b}}{8(m_{b}+m_{\bar{b}})}m_{b\bar{b}}$}}\\
\Xcline{1-5}{0.5pt}\Xcline{6-8}{1.6pt}
\multicolumn{1}{c|}{\multirow{3}{*}{$V^{C}$}}&\multicolumn{1}{c}{\multirow{1}{*}{(1,2)}}&\multicolumn{1}{r}{35.6}&
&\multicolumn{1}{r}{}&\multicolumn{1}{c}{\multirow{1}{*}{(1,3)}}&\multicolumn{1}{r}{-158.1}&-237.2($\eta_{c}$)&\multirow{4}*{\makecell[c]{$b$-quark:\\ \\$m^{eff}_{b}$}}
&&\multirow{1}*{\makecell[c]{64.8}}&\multirow{1}*{}\\
\multicolumn{1}{c|}{\multirow{1}{*}{}}&\multicolumn{1}{c}{(1,4)}&\multicolumn{1}{r}{-158.1}&&\multicolumn{1}{r}{}&(2,3)&\multicolumn{1}{r}{-158.1}&&&&\multirow{1}*{\makecell[c]{48.6\\17.8\\-158.1}}&&\\
\Xcline{2-5}{0.5pt}
\multicolumn{1}{c|}{\multirow{1}{*}{}}&\multicolumn{1}{c}{\multirow{1}{*}{Subtotal}}&\multicolumn{1}{r}{-561.1}&\multicolumn{1}{r}{-1034.0}&\multicolumn{1}{c|}{472.9}
&(2,4)&-158.1&$-796.7(\Upsilon)$&&&&&$m_{b}$\\
\Xcline{1-5}{0.5pt}
\multicolumn{2}{c}{\multirow{1}*{Total Contribution}}&\multicolumn{1}{r}{\multirow{1}*{256.0}}&\multicolumn{1}{r}{-88.6}&\multicolumn{1}{c|}{\multirow{1}*{344.6}}
&(3,4)&35.6&&&&-491.5&2951.6&5052.9\\
\Xcline{1-8}{1.6pt}\Xcline{10-13}{0.5pt}
\multicolumn{5}{l}{Total Wave function:}&&&
&&\multirow{1}*{Subtotal}&\multirow{1}*{4824.7}&
\multirow{1}*{4713.3}&\multirow{1}*{5052.9}\\
\Xcline{9-13}{0.5pt}
\multicolumn{8}{r!{\color{black}\vrule width 1.5pt}}{$\Psi_{tot}=0.877|F\rangle|R^{s}\rangle|[\phi_{1}\chi_{2}]\rangle-0.064|F\rangle|R^{s}\rangle|[\phi_{2}\chi_{2}]\rangle+0.320|F\rangle|R^{s}\rangle|[\phi_{1}\chi_{3}]\rangle$}&
&\multirow{1}*{\makecell[c]{$-\frac{1}{4}[V^{SS}(12)+V^{SS}(34)]$
}}
&\multirow{1}*{\makecell[c]{-5.0
}}
&\multirow{1}*{\makecell[c]{$-\frac{8}{3}v_{cb}-\frac{10}{3}v_{\bar{c}\bar{c}}$}}
&\multirow{1}*{\makecell[c]{$-\frac{8}{3}C_{cb}-\frac{10}{3}C_{\bar{c}\bar{c}}$}}
\\
\multicolumn{8}{r!{\color{black}\vrule width 1.5pt}}{$+0.320|F\rangle|R^{s}\rangle|[\phi_{1}\chi_{4}]\rangle+0.105|F\rangle|R^{s}\rangle|[\phi_{2}\chi_{3}]\rangle+0.105|F\rangle|R^{s}\rangle|[\phi_{2}\chi_{4}]\rangle$}
&\multirow{3}*{\makecell[c]{CS \\Interaction}}&\multirow{1}*{\makecell[c]{$-\frac{5}{8}V^{SS}(13)$}}&\multirow{1}*{\makecell[c]{-16.0}}
&\multirow{1}*{\makecell[c]{-5.2-17.7}}&\multirow{1}*{\makecell[c]{-8.8-17.7}}\\
\multicolumn{8}{r!{\color{black}\vrule width 1.5pt}}{$=0.211|F\rangle|R^{s}\rangle|[\psi_{1}\zeta_{2}]\rangle
+0.854|F\rangle|R^{s}\rangle|[\psi_{1}\zeta_{3}]\rangle+0.223|F\rangle|R^{s}\rangle|[\psi_{2}\zeta_{2}]\rangle$}
&&
\multirow{1}*{\makecell[c]{$-\frac{5}{8}V^{SS}(24)$}}&\multirow{1}*{\makecell[c]{-4.8}}&
\multirow{1}*{\makecell[c]{$-\frac{10}{3}v_{b\bar{b}}-\frac{20}{3}v_{b\bar{c}}$}}&
\multirow{1}*{\makecell[c]{$-\frac{10}{3}C_{b\bar{b}}-\frac{20}{3}C_{b\bar{c}}$}}\\
\multicolumn{8}{r!{\color{black}\vrule width 1.5pt}}{$+0.420|F\rangle|R^{s}\rangle|[\psi_{2}\zeta_{3}]\rangle$}
&&\multirow{1}*{\makecell[c]{$-\frac{5}{8}[V^{SS}(14)+V^{SS}(23)]$}}
&\multirow{1}*{\makecell[c]{-15.6}}&\multirow{1}*{\makecell[c]{-9.6-19.7}}&\multirow{1}*{\makecell[c]{-9.7-22.0}}\\
\Xcline{10-13}{0.5pt}
\multicolumn{8}{r!{\color{black}\vrule width 1.5pt}}{$=-0.481|F\rangle|R^{s}\rangle|[\psi'_{1}\zeta'_{2}]\rangle
-0.481|F\rangle|R^{s}\rangle|[\psi'_{1}\zeta'_{3}]\rangle+0.283|F\rangle|R^{s}\rangle|[\psi'_{1}\zeta'_{4}]\rangle$}
&&\multirow{1}*{Subtotal}&\multirow{1}*{-41.4}&
\multirow{1}*{-52.2}&\multirow{1}*{-58.1}\\
\Xcline{9-13}{0.5pt}
\multicolumn{8}{r!{\color{black}\vrule width 1.5pt}}{$-0.395|F\rangle|R^{s}\rangle|[\psi'_{2}\zeta'_{2}]\rangle-0.395|F\rangle|R^{s}\rangle|[\psi'_{2}\zeta'_{3}]\rangle+0.382|F\rangle|R^{s}\rangle|[\psi'_{2}\zeta'_{4}]\rangle$}
&\multicolumn{2}{l|}{\multirow{1}*{Matrix nondiagonal element}}&\multirow{1}*{-15.3}&-54.7&-19.6\\
\Xcline{1-8}{0.3pt}\Xcline{9-13}{0.5pt}
\multicolumn{8}{l!{\color{black}\vrule width 1.5pt}}{The rearrangement strong decay channel:}
&\multicolumn{2}{l|}{\multirow{1}*{Total contribution}}&\multirow{1}*{12796.9}&\multirow{1}*{12409.9}&\multirow{1}*{13478.0}\\
\Xcline{9-13}{1.6pt}
\multicolumn{13}{l}{$\Gamma_{T_{cb\bar{c}\bar{b}}(12796.9,1^{+-})\rightarrow \eta_{b}J/\psi}:\Gamma_{T_{cb\bar{c}\bar{b}}(12796.9,1^{+-})\rightarrow \Upsilon\eta_{c}}=1:18.4$
\quad$\Gamma_{T_{cb\bar{c}\bar{b}}(12796.9,1^{+-})\rightarrow B^{*}_{c}\bar{B}^{*}_{c}}:
\Gamma_{T_{cb\bar{c}\bar{b}}(12796.9,1^{+-})\rightarrow  B^{*}_{c}\bar{B}_{c}}: \Gamma_{T_{cb\bar{c}\bar{b}}(12796.9,1^{+-})\rightarrow B_{c}\bar{B}^{*}_{c}}=1:3.9:3.9$}
&\\
\Xcline{1-13}{0.3pt}
\multicolumn{5}{l|}{The radiative decay widths:}
&\multicolumn{8}{l}{The magnetic moments: \quad $T_{cb\bar{c}\bar{b}}(12796.9,1^{+-})=\langle\Psi^{1^{+-}}_{tot}|\hat{\mu^{z}}|\Psi^{1^{+-}}_{tot}\rangle=0$}\\
\Xcline{6-13}{0.3pt}
\multicolumn{5}{l|}{$\Gamma_{T_{cb\bar{c}\bar{b}}(12797.3,1^{+-})\rightarrow T_{cb\bar{c}\bar{b}}(12759.6,0^{++})\gamma}=33.1$ keV }
&\multicolumn{8}{l}{The transition magnetic moments: \quad $\mu_{T_{cb\bar{c}\bar{b}}(12882.4,2^{++})\rightarrow
T_{cb\bar{c}\bar{b}}(12759.6,1^{+-})\gamma}=0.345\mu_{N}$
}\\
\multicolumn{5}{l|}{$\Gamma_{T_{cb\bar{c}\bar{b}}(12856.6,1^{++})\rightarrow T_{cb\bar{c}\bar{b}}(12796.9,1^{+-})\gamma}=0.1$ keV }
&\multicolumn{8}{l}{$\mu_{T_{cb\bar{c}\bar{b}}(12856.6,1^{++})\rightarrow T_{cb\bar{c}\bar{b}}(12796.9,1^{+-})\gamma}=\langle\Psi^{1^{++}}_{tot}|\hat{\mu^{z}}|\Psi^{1^{+-}}_{tot}\rangle=0.113(\mu_{c}+\mu_{b})=0.036\mu_{N}$}\\
\multicolumn{5}{l|}{$\Gamma_{T_{cb\bar{c}\bar{b}}(12882.4,2^{++})\rightarrow T_{cb\bar{c}\bar{b}}(12796.9,1^{+-})\gamma}=4.9$ keV }
&\multicolumn{8}{l}{$\mu_{T_{cb\bar{c}\bar{b}}(12796.9,1^{+-})\rightarrow T_{cb\bar{c}\bar{b}}(12759.6,0^{++})\gamma}
=\langle\Psi^{1^{+-}}_{tot}|\hat{\mu^{z}}|\Psi^{0^{++}}_{tot}\rangle=0.226\times2\sqrt{\frac{1}{6}}(\mu_{c}-\mu_{b})=0.081\mu_{N}$}\\
\toprule[1.5pt]
\multicolumn{1}{r}{$J^{PC}=1^{++}$}&&Value&\multicolumn{1}{r}{$\Upsilon J/\psi$}&
\multicolumn{1}{r!{\color{black}\vrule width 1.5pt}}{Difference}&\multicolumn{3}{c!{\color{black}\vrule width 1.5pt}}{Relative Lengths (fm)}
&&Contribution&Value&Ref. \cite{Weng:2020jao}&Ref. \cite{Wu:2016vtq}\\ \Xcline{1-13}{0.5pt}
\multicolumn{2}{c}{Mass/$B_{T}$}&\multicolumn{1}{r}{12856.6}&12561.1&295.5&
$(i,j)$&\multicolumn{1}{r}{Value}&$\Upsilon J/\psi$&
\multirow{5}*{\makecell[c]{$c$-quark:\\ \\$m^{eff}_{c}$}}&\multirow{1}*{\makecell[c]{$m_{c}$\\
$\frac{m_{b}}{m_{c}+m_{b}}\frac{\textbf{p}^{2}_{x_{1}}}{2m'_{1}}$\\
$\frac{m_{b}}{2m_{c}+2m_{b}}\frac{\textbf{p}^{2}_{x_{3}}}{2m'_{3}}$
\\$\frac{1}{2}V^{C}(12)$\\$\frac{1}{2}[V^{C}(13)+V^{C}(14)]$\\$-1/2D$
}}&
\multirow{1}*{\makecell[c]{1918.0
}}&
\multirow{4}*{\makecell[c]{$\frac{-m_{c}}{4(m_{c}+m_{b})}m_{cb}$\\-400.5\\
$\frac{5m_{c}}{8(m_{c}+m_{\bar{b}})}m_{c\bar{b}}$\\986.7
\\$\frac{5m_{c}}{8(m_{c}+m_{\bar{c}})}m_{c\bar{c}}$}}\\
\Xcline{1-8}{0.5pt}
\multirow{2}*{\makecell[c]{Variational\\ Parameters\\ (fm$^{-2}$)}}&\multirow{2}*{\makecell[c]{$C_{11}$\\$C_{22}$ \\ $C_{33}$}}&\multicolumn{1}{r}{\multirow{2}*{\makecell[c]{11.4\\11.4\\21.5}}}&
\multicolumn{1}{r}{\multirow{2}*{\makecell[c]{12.5\\49.7\\-}}}&&(1.2)&0.333&&
&&174.0&\\
&&&&&(1,3)&0.291&$0.318(J/\psi)$&&
\multirow{3}*{}&\multirow{1}*{\makecell[c]{127.2\\16.0\\-146.5}}&\\
\Xcline{1-5}{0.5pt}
\multicolumn{2}{c}{Quark Mass}&\multicolumn{1}{r}{14522.0}&14522.0&0.0&(2,3)&0.291&&
&\multirow{2}*{}
&\multirow{3}*{}&&$m_{c}$\\
\multicolumn{2}{c}{\multirow{1}{*}{Confinement Potential}}&\multirow{1}{*}{-2488.0}&\multirow{1}{*}{-2926.9}&\multirow{1}{*}{438.9}&(1,4)&0.291&&&&-491.5&958.9&1724.8\\
\Xcline{1-5}{0.5pt}\Xcline{10-13}{0.5pt}
\multicolumn{2}{c}{\multirow{1}*{Kinetic Energy}}&\multicolumn{1}{r}{\multirow{1}*{818.6}}&\multirow{1}*{925.1}&\multirow{1}*{-106.5}
&(2,4)&0.291&$0.160(\Upsilon)$&&\multirow{1}*{Subtotal}&\multirow{1}*{1597.2}&
\multirow{1}*{1545.1}&\multirow{1}*{1724.8}\\
\Xcline{9-13}{0.5pt}
\multicolumn{2}{c}{\multirow{1}*{CS Interaction}}&\multicolumn{1}{r}{\multirow{1}*{10.0}}&\multirow{1}*{41.0}&\multirow{1}*{-31.0}
&(3,4)&0.333&&&\multirow{5}*{\makecell[c]{$m_{\bar{b}}$\\
$\frac{m_{c}}{m_{c}+m_{b}}\frac{\textbf{p}^{2}_{x_{1}}}{2m'_{1}}$\\
$\frac{m_{c}}{2m_{c}+2m_{b}}\frac{\textbf{p}^{2}_{x_{3}}}{2m'_{3}}$
\\$\frac{1}{2}V^{C}(12)$\\$\frac{1}{2}[V^{C}(23)+V^{C}(24)]$\\$-1/2D$}}
&\multirow{1}*{\makecell[c]{5343.0}}&
\multirow{4}*{\makecell[c]{$\frac{-m_{b}}{4(m_{c}+m_{b})}m_{cb}$\\-1203.8\\
$\frac{5m_{b}}{8(m_{c}+m_{\bar{b}})}m_{c\bar{b}}$\\2965.5
\\$\frac{5m_{b}}{8(m_{b}+m_{\bar{b}})}m_{b\bar{b}}$}}\\
\Xcline{1-5}{0.5pt}\Xcline{6-8}{0.3pt}
\multicolumn{1}{c|}{\multirow{4}{*}{$V^{C}$}}&\multicolumn{1}{c}{\multirow{1}{*}{(1,2)}}&\multicolumn{1}{r}{32.0}&&
&\multicolumn{2}{r}{(1,2)-(3,4):}&0.172 fm&\multirow{4}*{\makecell[c]{$b$-quark:\\ \\$m^{eff}_{b}$}}
&&\multirow{1}*{\makecell[c]{62.5}}&\multirow{1}*{}\\
\Xcline{6-8}{1.6pt}
\multicolumn{1}{c|}{\multirow{1}{*}{}}&\multicolumn{1}{c}{(2,3)}&\multicolumn{1}{r}{-146.5}
&&\multicolumn{1}{r}{}&&&&&&\multirow{1}*{\makecell[c]{45.7\\16.0\\-146.5}}&&\\
\multicolumn{1}{c|}{\multirow{1}{*}{}}&\multicolumn{1}{c}{(1,4)}&\multicolumn{1}{r}{-146.5}
&&\multicolumn{1}{r}{}&(2,4)&-146.5&$-796.7(\Upsilon)$&&&&&$m_{b}$\\
\Xcline{2-5}{0.5pt}
\multicolumn{1}{c|}{\multirow{1}{*}{}}&\multicolumn{1}{c}{\multirow{1}{*}{Subtotal}}&\multicolumn{1}{r}{-522.0}&\multicolumn{1}{r}{-960.9}&\multicolumn{1}{c|}{438.9}
&(3,4)&32.0&
&&&-491.5&2951.6&5052.9\\
\Xcline{1-5}{0.5pt}\Xcline{10-13}{0.5pt}
\multicolumn{2}{c}{\multirow{1}*{Total Contribution}}&\multicolumn{1}{r}{\multirow{1}*{306.6}}&\multicolumn{1}{r}{5.1}&\multicolumn{1}{c|}{\multirow{1}*{301.5}}
&(1,3)&-146.5&$-164.2(J/\psi)$&&\multirow{1}*{Subtotal}&\multirow{1}*{4829.2}&
\multirow{1}*{4713.3}&\multirow{1}*{5052.9}\\
\Xcline{1-8}{1.6pt}\Xcline{9-13}{0.5pt}
\multicolumn{5}{l}{Total Wave function:}&&
&&
\multirow{5}*{\makecell[c]{CS \\Interaction}}&\multirow{1}*{\makecell[c]{$\frac{1}{4}[V^{SS}(12)+V^{SS}(34)]$
}}
&\multirow{1}*{\makecell[c]{4.9
}}
&\multirow{1}*{\makecell[c]{$\frac{8}{3}v_{cb}+\frac{10}{3}v_{\bar{c}\bar{c}}$}}
&\multirow{1}*{\makecell[c]{$\frac{8}{3}C_{cb}+\frac{10}{3}C_{\bar{c}\bar{c}}$}}
\\
\multicolumn{8}{r!{\color{black}\vrule width 1.5pt}}{$\Psi_{tot}=0.693|F\rangle|R^{s}\rangle|[\phi_{2}\chi_{3}]\rangle-0.693|F\rangle|R^{s}\rangle|[\phi_{2}\chi_{4}]\rangle+0.139|F\rangle|R^{s}\rangle|[\phi_{1}\chi_{3}]\rangle$}
&&\multirow{1}*{\makecell[c]{$\frac{5}{8}V^{SS}(13)$}}&\multirow{1}*{\makecell[c]{15.5}}
&\multirow{1}*{\makecell[c]{5.2+17.7}}&\multirow{1}*{\makecell[c]{8.8+17.7}}\\
\multicolumn{8}{r!{\color{black}\vrule width 1.5pt}}{$-0.139|F\rangle|R^{s}\rangle|[\phi_{1}\chi_{4}]\rangle=0.686|F\rangle|R^{s}\rangle|[\psi_{1}\zeta_{4}]\rangle
+0.727|F\rangle|R^{s}\rangle|[\psi_{2}\zeta_{4}]\rangle$}&&
\multirow{1}*{\makecell[c]{$\frac{5}{8}V^{SS}(24)$}}&\multirow{1}*{\makecell[c]{4.6}}&
\multirow{1}*{\makecell[c]{$\frac{10}{3}v_{b\bar{b}}-\frac{20}{3}v_{b\bar{c}}$}}&
\multirow{1}*{\makecell[c]{$\frac{10}{3}C_{b\bar{b}}-\frac{20}{3}C_{b\bar{c}}$}}\\
\multicolumn{8}{r!{\color{black}\vrule width 1.5pt}}{$=-0.480|F\rangle|R^{s}\rangle|[\psi'_{1}\zeta'_{2}]\rangle
-0.480|F\rangle|R^{s}\rangle|[\psi'_{1}\zeta'_{3}]\rangle-0.395|F\rangle|R^{s}\rangle|[\psi'_{2}\zeta'_{2}]\rangle$}
&&\multirow{1}*{\makecell[c]{$-\frac{5}{8}[V^{SS}(14)+V^{SS}(23)]$}}
&\multirow{1}*{\makecell[c]{-15.0}}&\multirow{1}*{\makecell[c]{9.6-19.7}}&\multirow{1}*{\makecell[c]{9.7-22.0}}\\
\Xcline{10-13}{0.5pt}
\multicolumn{8}{r!{\color{black}\vrule width 1.5pt}}{$-0.395|F\rangle|R^{s}\rangle|[\psi'_{2}\zeta'_{3}]\rangle$}
&&\multirow{1}*{Subtotal}&\multirow{1}*{10.0}&
\multirow{1}*{12.9}&\multirow{1}*{14.1}\\
\Xcline{9-13}{0.5pt}\Xcline{1-8}{0.3pt}
\multicolumn{8}{l!{\color{black}\vrule width 1.5pt}}{The magnetic moments: $T_{cb\bar{c}\bar{b}}(12856.6,1^{++})=\langle\Psi^{1^{++}}_{tot}|\hat{\mu^{z}}|\Psi^{1^{++}}_{tot}\rangle=0$}
&\multicolumn{2}{l|}{\multirow{1}*{Matrix nondiagonal element}}&\multirow{1}*{-6.3}&-6.1&-59.5\\
\Xcline{1-8}{0.3pt}\Xcline{9-13}{0.5pt}
\multicolumn{5}{l|}{The radiative decay widths:}&&&
&\multicolumn{2}{l|}{\multirow{1}*{Total contribution}}&\multirow{1}*{12856.6}&\multirow{1}*{12523.6}&\multirow{1}*{13510.0}\\
\Xcline{9-13}{1.6pt}
\multicolumn{5}{l|}{$\Gamma_{T_{cb\bar{c}\bar{b}}(12856.6,1^{++})\rightarrow T_{cb\bar{c}\bar{b}}(12759.6,0^{++})\gamma}=0$ keV}
&\multicolumn{8}{l}{The transition magnetic moments:\quad $\mu_{T_{cb\bar{c}\bar{b}}(12882.4,2^{++})\rightarrow
T_{cb\bar{c}\bar{b}}(12856.6,1^{++})\gamma}=\langle\Psi^{2^{++}}_{tot}|\hat{\mu^{z}}|\Psi^{1^{++}}_{tot}\rangle=0$
}\\
\multicolumn{5}{l|}{$\Gamma_{T_{cb\bar{c}\bar{b}}(12856.6,1^{++})\rightarrow T_{cb\bar{c}\bar{b}}(12796.9,1^{+-})\gamma}=0.1$\, keV}
&\multicolumn{8}{l}{$\mu_{T_{cb\bar{c}\bar{b}}(12856.6,1^{++})\rightarrow T_{cb\bar{c}\bar{b}}(12759.6,0^{++})\gamma}=\langle\Psi^{1^{++}}_{tot}|\hat{\mu^{z}}|\Psi^{0^{++}}_{tot}\rangle=0$ }\\
\multicolumn{5}{l|}{$\Gamma_{T_{cb\bar{c}\bar{b}}(12882.4,2^{++})\rightarrow T_{cb\bar{c}\bar{b}}(12856.6,1^{++})\gamma}=0$\, keV}
&\multicolumn{8}{l}{$\mu_{T_{cb\bar{c}\bar{b}}(12856.6,1^{++})\rightarrow T_{cb\bar{c}\bar{b}}(12796.9,1^{+-})\gamma}=\langle\Psi^{1^{++}}_{tot}|\hat{\mu^{z}}|\Psi^{1^{+-}}_{tot}\rangle=0.113(\mu_{c}+\mu_{b})=0.036\mu_{N}$}\\
\Xcline{1-13}{0.3pt}
\multicolumn{13}{l}{The radiative decay widths: \quad $J/\psi\Upsilon$  \quad $\Gamma_{T_{cb\bar{c}\bar{b}}(12856.6,1^{++})\rightarrow  B^{*}_{c}\bar{B}_{c}}: \Gamma_{T_{cb\bar{c}\bar{b}}(12856.6,1^{++})\rightarrow B_{c}\bar{B}^{*}_{c}}=1:1$}
\\
\toprule[0.5pt]
\toprule[1.5pt]
\end{tabular}
\end{lrbox}\scalebox{0.868}{\usebox{\tablebox}}
\end{table*}


\begin{thebibliography}{300}
\bibitem{Gell-Mann:1964ewy}
M.~Gell-Mann,
A Schematic Model of Baryons and Mesons,
Phys. Lett. \textbf{8} (1964), 214-215.

\bibitem{Zweig:1964ruk}
G.~Zweig,
An SU(3) model for strong interaction symmetry and its breaking. Version 1,
CERN-TH-401.

\bibitem{Zweig:1964jf}
G.~Zweig,
An SU(3) model for strong interaction symmetry and its breaking. Version 2,
CERN-TH-412.

\bibitem{Choi:2003ue}
S.~K.~Choi {\it et al.} [Belle Collaboration],
Observation of a narrow charmonium - like state in exclusive $B^{+-} \rightarrow K^{+-} \pi^+ \pi^- J / \psi$ decays,
Phys.\ Rev.\ Lett.\  {\bf 91}, 262001 (2003).
\bibitem{Acosta:2003zx}
D.~Acosta {\it et al.} [CDF Collaboration],
Observation of the narrow state $X(3872) \to J/\psi \pi^+ \pi^-$ in $\bar{p}p$ collisions at $\sqrt{s} = 1.96$ TeV,
Phys.\ Rev.\ Lett.\  {\bf 93}, 072001 (2004).
\bibitem{Abazov:2004kp}
V.~M.~Abazov {\it et al.} [D0 Collaboration],
Observation and properties of the $X(3872)$ decaying to $J/\psi \pi^+ \pi^-$ in $p\bar{p}$ collisions at $\sqrt{s} = 1.96$ TeV,
Phys.\ Rev.\ Lett.\  {\bf 93}, 162002 (2004).


\bibitem{Swanson:2006st}
  E.~S.~Swanson,
  The New heavy mesons: A Status report,
  Phys.\ Rept.\  {\bf 429}, 243 (2006).

\bibitem{Zhu:2007wz}
  S.~L.~Zhu,
  New hadron states,
  Int.\ J.\ Mod.\ Phys.\ E {\bf 17} (2008) 283.
\bibitem{Voloshin:2007dx}
  M.~B.~Voloshin,
  Charmonium,
  Prog.\ Part.\ Nucl.\ Phys.\  {\bf 61} (2008) 455.
\bibitem{Drenska:2010kg}
  N.~Drenska, R.~Faccini, F.~Piccinini, A.~Polosa, F.~Renga and C.~Sabelli,
  New Hadronic Spectroscopy,
  Riv.\ Nuovo Cim.\  {\bf 33} (2010) 633.

\bibitem{Esposito:2014rxa}
  A.~Esposito, A.~L.~Guerrieri, F.~Piccinini, A.~Pilloni and A.~D.~Polosa,
  Four-Quark Hadrons: an Updated Review,
  Int.\ J.\ Mod.\ Phys.\ A {\bf 30} (2015) 1530002.

\bibitem{Chen:2016qju}
  H.~X.~Chen, W.~Chen, X.~Liu and S.~L.~Zhu,
  The hidden-charm pentaquark and tetraquark states,
  Phys.\ Rept.\  {\bf 639} (2016) 1.
\bibitem{Chen:2016heh}
  R.~Chen, X.~Liu and S.~L.~Zhu,
  Hidden-charm molecular pentaquarks and their charm-strange partners,
  Nucl.\ Phys.\ A {\bf 954} (2016) 406.
\bibitem{Hosaka:2016pey}
  A.~Hosaka, T.~Iijima, K.~Miyabayashi, Y.~Sakai and S.~Yasui,
  Exotic hadrons with heavy flavors: X, Y, Z, and related states,
  PTEP {\bf 2016} (2016) no.6,  062C01.

\bibitem{Richard:2016eis}
  J.~M.~Richard,
  Exotic hadrons: review and perspectives,
  Few Body Syst.\  {\bf 57} (2016) no.12,  1185.

\bibitem{Lebed:2016hpi}
R.~F.~Lebed, R.~E.~Mitchell and E.~S.~Swanson,
Heavy-Quark QCD Exotica,
Prog. Part. Nucl. Phys. \textbf{93}, 143-194 (2017).

\bibitem{Esposito:2016noz}
A.~Esposito, A.~Pilloni and A.~D.~Polosa,
Multiquark Resonances,
Phys. Rept. \textbf{668}, 1-97 (2017).

\bibitem{Ali:2017jda}
A.~Ali, J.~S.~Lange and S.~Stone,
Exotics: Heavy Pentaquarks and Tetraquarks,
Prog. Part. Nucl. Phys. \textbf{97}, 123-198 (2017).

\bibitem{Brambilla:2019esw}
N.~Brambilla, S.~Eidelman, C.~Hanhart, A.~Nefediev, C.~P.~Shen, C.~E.~Thomas, A.~Vairo and C.~Z.~Yuan,
The $XYZ$ states: experimental and theoretical status and perspectives,
Phys. Rept. \textbf{873}, 1-154 (2020).

\bibitem{Liu:2019zoy}
Y.~R.~Liu, H.~X.~Chen, W.~Chen, X.~Liu and S.~L.~Zhu,
Pentaquark and Tetraquark states,
Prog. Part. Nucl. Phys. \textbf{107} (2019), 237-320.

\bibitem{Chen:2022asf}
H.~X.~Chen, W.~Chen, X.~Liu, Y.~R.~Liu and S.~L.~Zhu,
An updated review of the new hadron states,
Rept. Prog. Phys. \textbf{86} (2023) no.2, 026201





\bibitem{Aaij:2015tga}
  R.~Aaij {\it et al.} [LHCb Collaboration],
  Observation of $J/\psi p$ Resonances Consistent with Pentaquark States in $\Lambda_b^0 \to J/\psi K^- p$ Decays,
  Phys.\ Rev.\ Lett.\  {\bf 115}, 072001 (2015).
\bibitem{Aaij:2016phn}
  R.~Aaij {\it et al.} [LHCb Collaboration],
  Model-independent evidence for $J/\psi p$ contributions to $\Lambda_b^0\to J/\psi p K^-$ decays,
  Phys.\ Rev.\ Lett.\  {\bf 117}, no. 8, 082002 (2016).
\bibitem{Aaij:2019vzc}
  R.~Aaij {\it et al.} [LHCb Collaboration],
  Observation of a narrow pentaquark state, $P_c(4312)^+$, and of two-peak structure of the $P_c(4450)^+$,
  Phys.\ Rev.\ Lett.\  {\bf 122}, no. 22, 222001 (2019).

\bibitem{Aubert:2003fg}
  B.~Aubert {\it et al.} [BaBar Collaboration],
  Observation of a narrow meson decaying to $D_s^+ \pi^0$ at a mass of 2.32 GeV/c$^2$,
  Phys.\ Rev.\ Lett.\  {\bf 90}, 242001 (2003).

\bibitem{Godfrey:1985xj}
  S.~Godfrey and N.~Isgur,
  Mesons in a Relativized Quark Model with Chromodynamics,
  Phys.\ Rev.\ D {\bf 32}, 189 (1985).

\bibitem{Barnes:2003dj}
T.~Barnes, F.~E.~Close and H.~J.~Lipkin,
Implications of a DK molecule at 2.32-GeV,
Phys. Rev. D \textbf{68} (2003), 054006.

\bibitem{Cheng:2003kg}
H.~Y.~Cheng and W.~S.~Hou,
B decays as spectroscope for charmed four quark states,
Phys. Lett. B \textbf{566} (2003), 193-200.

\bibitem{Szczepaniak:2003vy}
A.~P.~Szczepaniak,
Description of the D*(s)(2320) resonance as the D pi atom,
Phys. Lett. B \textbf{567} (2003), 23-26.

\bibitem{Besson:2003cp}
  D.~Besson {\it et al.} [CLEO Collaboration],
  Observation of a narrow resonance of mass 2.46 GeV/c$^2$ decaying to $D^{*+}_s \pi^0$ and confirmation of the $D^*_{sJ}(2317)$ state,
  Phys.\ Rev.\ D {\bf 68}, 032002 (2003).

\bibitem{Dmitrasinovic:2004cu}
V.~Dmitrasinovic,
$D^{*+}_{s}$ (2317) and $D^{*+}_{s}$ (2460): Tetraquarks bound by the t Hooft instanton-induced interaction?,
Phys. Rev. D \textbf{70} (2004), 096011.
\bibitem{Chen:2004dy}
Y.~Q.~Chen and X.~Q.~Li,
A Comprehensive four-quark interpretation of D(s)(2317), D(s)(2457) and D(s)(2632),
Phys. Rev. Lett. \textbf{93} (2004), 232001.
\bibitem{Liu:2004kd}
Y.~R.~Liu, S.~L.~Zhu, Y.~B.~Dai and C.~Liu,
D+(sJ)(2632): An Excellent candidate of tetraquarks,
Phys. Rev. D \textbf{70} (2004), 094009.
\bibitem{Faessler:2007us}
A.~Faessler, T.~Gutsche, V.~E.~Lyubovitskij and Y.~L.~Ma,
D* K molecular structure of the D(s1)(2460) meson,
Phys. Rev. D \textbf{76} (2007), 114008.
\bibitem{Ebert:2010af}
D.~Ebert, R.~N.~Faustov and V.~O.~Galkin,
Masses of tetraquarks with open charm and bottom,
Phys. Lett. B \textbf{696} (2011), 241-245.
\bibitem{Xiao:2016hoa}
c.~J.~Xiao, D.~Y.~Chen and Y.~L.~Ma,
Radiative and pionic transitions from the $D_{s1}(2460)$ to the $D_{s0}^\ast(2317)$,
Phys. Rev. D \textbf{93} (2016) no.9, 094011.
\bibitem{Kong:2021ohg}
S.~Y.~Kong, J.~T.~Zhu, D.~Song and J.~He,
Heavy-strange meson molecules and possible candidates Ds0*(2317), Ds1(2460), and X0(2900),
Phys. Rev. D \textbf{104} (2021) no.9, 094012.

\bibitem{Fu:2021wde}
H.~L.~Fu, H.~W.~Grie\ss{}hammer, F.~K.~Guo, C.~Hanhart and U.~G.~Mei\ss{}ner,
Update on strong and radiative decays of the $D_{s0}^*(2317)$ and $D_{s1}(2460)$ and their bottom cousins,
Eur. Phys. J. A \textbf{58} (2022) no.4, 70.



\bibitem{Aaij:2020hon}
R.~Aaij \textit{et al.} [LHCb],
A model-independent study of resonant structure in $B^+\to D^+D^-K^+$ decays,
Phys. Rev. Lett. \textbf{125}, 242001 (2020).

\bibitem{LHCb:2020pxc}
R.~Aaij \textit{et al.} [LHCb],
Amplitude analysis of the $B^+\to D^+D^-K^+$ decay,
Phys. Rev. D \textbf{102} (2020), 112003.


\bibitem{Liu:2020orv}
X.~H.~Liu, M.~J.~Yan, H.~W.~Ke, G.~Li and J.~J.~Xie,
Triangle singularity as the origin of $X_0(2900)$ and $X_1(2900)$ observed in $B^+\to D^+ D^- K^+$,
Eur. Phys. J. C \textbf{80} (2020) no.12, 1178.
\bibitem{Karliner:2020vsi}
M.~Karliner and J.~L.~Rosner,
First exotic hadron with open heavy flavor: $cs\bar u\bar d$ tetraquark,
Phys. Rev. D \textbf{102} (2020) no.9, 094016.
\bibitem{Wang:2020xyc}
Z.~G.~Wang,
Analysis of the $X_0(2900)$ as the scalar tetraquark state via the QCD sum rules,
Int. J. Mod. Phys. A \textbf{35} (2020) no.30, 2050187.
\bibitem{Huang:2020ptc}
Y.~Huang, J.~X.~Lu, J.~J.~Xie and L.~S.~Geng,
Strong decays of ${\bar{D}}^{*}K^{*}$ molecules and the newly observed $X_{0,1}$ states,
Eur. Phys. J. C \textbf{80} (2020) no.10, 973.
\bibitem{Burns:2020epm}
T.~J.~Burns and E.~S.~Swanson,
Kinematical cusp and resonance interpretations of the $X(2900)$,
Phys. Lett. B \textbf{813} (2021), 136057.
\bibitem{Hu:2020mxp}
M.~W.~Hu, X.~Y.~Lao, P.~Ling and Q.~Wang,
$X_0$(2900) and its heavy quark spin partners in molecular picture,
Chin. Phys. C \textbf{45} (2021) no.2, 021003.
\bibitem{Wang:2020prk}
G.~J.~Wang, L.~Meng, L.~Y.~Xiao, M.~Oka and S.~L.~Zhu,
Mass spectrum and strong decays of tetraquark ${\bar{c}}{\bar{s}} qq$ states,
Eur. Phys. J. C \textbf{81} (2021) no.2, 188.
\bibitem{Xue:2020vtq}
Y.~Xue, X.~Jin, H.~Huang and J.~Ping,
Tetraquarks with open charm flavor,
Phys. Rev. D \textbf{103} (2021) no.5, 054010.
\bibitem{Agaev:2021knl}
S.~S.~Agaev, K.~Azizi and H.~Sundu,
Vector resonance $X_1$(2900) and its structure,
Nucl. Phys. A \textbf{1011} (2021), 122202.
\bibitem{Ozdem:2022ydv}
U.~\"Ozdem and K.~Azizi,
Magnetic moment of the $X_1(2900)$ state in the diquark-antidiquark picture,
[arXiv:2202.11466 [hep-ph]].


\bibitem{DelFabbro:2004ta}
A.~Del Fabbro, D.~Janc, M.~Rosina and D.~Treleani,
Production and detection of doubly charmed tetraquarks,
Phys. Rev. D \textbf{71} (2005), 014008.


\bibitem{Ebert:2007rn}
D.~Ebert, R.~N.~Faustov, V.~O.~Galkin and W.~Lucha,
Masses of tetraquarks with two heavy quarks in the relativistic quark model,
Phys. Rev. D \textbf{76} (2007), 114015.

\bibitem{Hyodo:2012pm}
T.~Hyodo, Y.~R.~Liu, M.~Oka, K.~Sudoh and S.~Yasui,
Production of doubly charmed tetraquarks with exotic color configurations in electron-positron collisions,
Phys. Lett. B \textbf{721} (2013), 56-60.

\bibitem{Ikeda:2013vwa}
Y.~Ikeda, B.~Charron, S.~Aoki, T.~Doi, T.~Hatsuda, T.~Inoue, N.~Ishii, K.~Murano, H.~Nemura and K.~Sasaki,
Charmed tetraquarks $T_{cc}$ and $T_{cs}$ from dynamical lattice QCD simulations,
Phys. Lett. B \textbf{729} (2014), 85-90.

\bibitem{Esposito:2013fma}
A.~Esposito, M.~Papinutto, A.~Pilloni, A.~D.~Polosa and N.~Tantalo,
Doubly charmed tetraquarks in $B_c$ and $\Xi_{bc}$ decays,
Phys. Rev. D \textbf{88} (2013) no.5, 054029.



\bibitem{Aaij:2017ueg}
R.~Aaij {\it et al.} [LHCb Collaboration],
Observation of the doubly charmed baryon $\Xi_{cc}^{++}$,
Phys.\ Rev.\ Lett.\  {\bf 119}, no. 11, 112001 (2017).

\bibitem{Luo:2017eub}
  S.~Q.~Luo, K.~Chen, X.~Liu, Y.~R.~Liu and S.~L.~Zhu,
  Exotic tetraquark states with the $qq\bar{Q}\bar{Q}$ configuration,
  Eur.\ Phys.\ J.\ C {\bf 77}, no. 10, 709 (2017).


\bibitem{Karliner:2017qjm}
  M.~Karliner and J.~L.~Rosner,
  Discovery of doubly-charmed $\Xi_{cc}$ baryon implies a stable ($b b \bar{u} \bar{d}$) tetraquark,
  Phys.\ Rev.\ Lett.\  {\bf 119}, no. 20, 202001 (2017).

\bibitem{Eichten:2017ffp}
  E.~J.~Eichten and C.~Quigg,
  Heavy-quark symmetry implies stable heavy tetraquark mesons $Q_iQ_j \bar q_k \bar q_l$,
  Phys.\ Rev.\ Lett.\  {\bf 119}, no. 20, 202002 (2017).

\bibitem{Ali:2018xfq}
A.~Ali, Q.~Qin and W.~Wang,
Discovery potential of stable and near-threshold doubly heavy tetraquarks at the LHC,
Phys. Lett. B \textbf{785}, 605-609 (2018).

\bibitem{Park:2018wjk}
W.~Park, S.~Noh and S.~H.~Lee,
Masses of the doubly heavy tetraquarks in a constituent quark model,
Acta Phys. Polon. B \textbf{50}, 1151-1157 (2019).

\bibitem{Guo:2021yws}
T.~Guo, J.~Li, J.~Zhao and L.~He,
Mass spectra of doubly heavy tetraquarks in an improved chromomagnetic interaction model,
Phys. Rev. D \textbf{105} (2022) no.1, 014021.
\bibitem{Lu:2020rog}
Q.~F.~L\"u, D.~Y.~Chen and Y.~B.~Dong,
Masses of doubly heavy tetraquarks $T_{QQ^\prime}$ in a relativized quark model,
Phys. Rev. D \textbf{102} (2020) no.3, 034012.
\bibitem{Park:2013fda}
W.~Park and S.~H.~Lee,
Color spin wave functions of heavy tetraquark states,
Nucl. Phys. A \textbf{925} (2014), 161-184.
\bibitem{Noh:2021lqs}
S.~Noh, W.~Park and S.~H.~Lee,
The Doubly-heavy Tetraquarks ($qq'\bar{Q}\bar{Q'}$) in a Constituent Quark Model with a Complete Set of Harmonic Oscillator Bases,
Phys. Rev. D \textbf{103} (2021), 114009.

\bibitem{Deng:2018kly}
C.~Deng, H.~Chen and J.~Ping,
Systematical investigation on the stability of doubly heavy tetraquark states,
Eur. Phys. J. A \textbf{56} (2020) no.1, 9.

\bibitem{LHCb:2021vvq}
R.~Aaij \textit{et al.} [LHCb],
Observation of an exotic narrow doubly charmed tetraquark,
[arXiv:2109.01038].

\bibitem{Lu:2021kut}
Q.~F.~L\"u, D.~Y.~Chen, Y.~B.~Dong and E.~Santopinto,
Triply-heavy tetraquarks in an extended relativized quark model,
Phys. Rev. D \textbf{104} (2021) no.5, 054026.
\bibitem{Chen:2016ont}
  K.~Chen, X.~Liu, J.~Wu, Y.~R.~Liu and S.~L.~Zhu,
  Triply heavy tetraquark states with the $QQ\bar{Q}\bar{q}$ configuration,
  Eur.\ Phys.\ J.\ A {\bf 53}, no. 1, 5 (2017).
\bibitem{Xing:2019wil}
Y.~Xing,
Weak decays of triply heavy tetraquarks ${b{\bar{c}}}{b{\bar{q}}}$,
Eur. Phys. J. C \textbf{80} (2020) no.1, 57.
\bibitem{Jiang:2017tdc}
J.~F.~Jiang, W.~Chen and S.~L.~Zhu,
Triply heavy $QQ\bar Q\bar q$ tetraquark states,
Phys. Rev. D \textbf{96} (2017) no.9, 094022.
\bibitem{Weng:2021ngd}
X.~Z.~Weng, W.~Z.~Deng and S.~L.~Zhu,
Triply heavy tetraquark states,
Phys. Rev. D \textbf{105} (2022) no.3, 034026.
\bibitem{Chao:1980dv}
K.~T.~Chao,
The (cc) - ($\bar{c}\bar{c}$) (Diquark - Anti-Diquark) States in $e^+ e^-$ Annihilation,
Z. Phys. C \textbf{7} (1981), 317.
\bibitem{Ader:1981db}
J.~P.~Ader, J.~M.~Richard and P.~Taxil,
DO NARROW HEAVY MULTI - QUARK STATES EXIST?,
Phys. Rev. D \textbf{25} (1982), 2370.

\bibitem{Badalian:1985es}
A.~M.~Badalian, B.~L.~Ioffe and A.~V.~Smilga,
FOUR QUARK STATES IN THE HEAVY QUARK SYSTEM,
Nucl. Phys. B \textbf{281} (1987), 85


\bibitem{Heller:1985cb}
L.~Heller and J.~A.~Tjon,
On Bound States of Heavy $Q^2 \bar{Q}^2$ Systems,
Phys. Rev. D \textbf{32}, 755 (1985).

\bibitem{Zouzou:1986qh}
S.~Zouzou, B.~Silvestre-Brac, C.~Gignoux and J.~M.~Richard,
FOUR QUARK BOUND STATES,
Z. Phys. C \textbf{30} (1986), 457


\bibitem{Lloyd:2003yc}
R.~J.~Lloyd and J.~P.~Vary,
All charm tetraquarks,
Phys. Rev. D \textbf{70}, 014009 (2004).
\bibitem{Karliner:2016zzc}
M.~Karliner, S.~Nussinov and J.~L.~Rosner,
$Q Q \bar Q \bar Q$ states: masses, production, and decays,
Phys. Rev. D \textbf{95}, no.3, 034011 (2017).
\bibitem{Anwar:2017toa}
M.~N.~Anwar, J.~Ferretti, F.~K.~Guo, E.~Santopinto and B.~S.~Zou,
Spectroscopy and decays of the fully-heavy tetraquarks,
Eur. Phys. J. C \textbf{78}, no.8, 647 (2018).

\bibitem{Bai:2016int}
Y.~Bai, S.~Lu and J.~Osborne,
Beauty-full Tetraquarks,
Phys. Lett. B \textbf{798}, 134930 (2019).

\bibitem{Debastiani:2017xcr}
V.~R.~Debastiani and F.~S.~Navarra,
Spectroscopy of the All-Charm Tetraquark,
PoS \textbf{Hadron2017}, 238 (2018).

\bibitem{Chen:2016jxd}
W.~Chen, H.~X.~Chen, X.~Liu, T.~G.~Steele and S.~L.~Zhu,
Hunting for exotic doubly hidden-charm/bottom tetraquark states,
Phys. Lett. B \textbf{773} (2017), 247-251.




\bibitem{CMS:2016liw}
V.~Khachatryan \textit{et al.} [CMS],
Observation of $\Upsilon$(1S) pair production in proton-proton collisions at $ \sqrt{s}=8 $ TeV,
JHEP \textbf{05} (2017), 013.

\bibitem{CMS:2020qwa}
A.~M.~Sirunyan \textit{et al.} [CMS],
Measurement of the $\Upsilon$(1S) pair production cross section and search for resonances decaying to $\Upsilon$(1S)$\mu^+\mu^-$ in proton-proton collisions at $\sqrt{s} =$ 13 TeV,
Phys. Lett. B \textbf{808} (2020), 135578.


\bibitem{LHCb:2018uwm}
R.~Aaij \textit{et al.} [LHCb],
Search for beautiful tetraquarks in the $\Upsilon(1S)\mu^+\mu^-$ invariant-mass spectrum,
JHEP \textbf{10} (2018), 086.


\bibitem{LHCb:2020bwg}
R.~Aaij \textit{et al.} [LHCb],
Observation of structure in the $J /\psi$ -pair mass spectrum,
Sci. Bull. \textbf{65} (2020) no.23, 1983-1993.
\bibitem{CMS:2023owd}
A.~Hayrapetyan \textit{et al.} [CMS],
Observation of new structure in the J/$\psi$J/$\psi$ mass spectrum in proton-proton collisions at $\sqrt{s}$ = 13 TeV,
[arXiv:2306.07164 [hep-ex]].

\bibitem{Zhang:2022toq}
J.~Zhang \textit{et al.} [CMS],
Recent CMS results on exotic resonances,
PoS \textbf{ICHEP2022}, 775
[arXiv:2212.00504 [hep-ex]].

\bibitem{Xu:2022rnl}
Y.~Xu [ATLAS],
ATLAS Results on Exotic Hadronic Resonances,
Acta Phys. Polon. Supp. \textbf{16} (2023) no.3, 21
[arXiv:2209.12173 [hep-ex]].



\bibitem{Zhao:2020nwy}
J.~Zhao, S.~Shi and P.~Zhuang,
Fully-heavy tetraquarks in a strongly interacting medium,
Phys. Rev. D \textbf{102} (2020) no.11, 114001.

\bibitem{Chen:2020xwe}
H.~X.~Chen, W.~Chen, X.~Liu and S.~L.~Zhu,
Strong decays of fully-charm tetraquarks into di-charmonia,
Sci. Bull. \textbf{65} (2020), 1994-2000.

\bibitem{liu:2020eha}
M.~S.~liu, F.~X.~Liu, X.~H.~Zhong and Q.~Zhao,
Full-heavy tetraquark states and their evidences in the LHCb di-$J/\psi$ spectrum,
[arXiv:2006.11952 [hep-ph]].


\bibitem{Lu:2020cns}
Q.~F.~L\"u, D.~Y.~Chen and Y.~B.~Dong,
Masses of fully heavy tetraquarks $QQ {\bar{Q}} {\bar{Q}}$ in an extended relativized quark model,
Eur. Phys. J. C \textbf{80} (2020) no.9, 871.



\bibitem{Zhao:2020zjh}
Z.~Zhao, K.~Xu, A.~Kaewsnod, X.~Liu, A.~Limphirat and Y.~Yan,
Study of charmoniumlike and fully-charm tetraquark spectroscopy,
Phys. Rev. D \textbf{103} (2021) no.11, 116027.

\bibitem{Bedolla:2019zwg}
M.~A.~Bedolla, J.~Ferretti, C.~D.~Roberts and E.~Santopinto,
Spectrum of fully-heavy tetraquarks from a diquark+antidiquark perspective,
Eur. Phys. J. C \textbf{80} (2020) no.11, 1004

\bibitem{Liu:2021rtn}
F.~X.~Liu, M.~S.~Liu, X.~H.~Zhong and Q.~Zhao,
Higher mass spectra of the fully-charmed and fully-bottom tetraquarks,
Phys. Rev. D \textbf{104} (2021) no.11, 116029.

\bibitem{Ke:2021iyh}
H.~W.~Ke, X.~Han, X.~H.~Liu and Y.~L.~Shi,
Tetraquark state $X(6900)$ and the interaction between diquark and antidiquark,
Eur. Phys. J. C \textbf{81} (2021) no.5, 427.

\bibitem{Karliner:2020dta}
M.~Karliner and J.~L.~Rosner,
Interpretation of structure in the di- $J/\psi$ spectrum,
Phys. Rev. D \textbf{102} (2020) no.11, 114039.

\bibitem{Faustov:2020qfm}
R.~N.~Faustov, V.~O.~Galkin and E.~M.~Savchenko,
Masses of the $QQ\bar Q\bar Q$ tetraquarks in the relativistic diquark--antidiquark picture,
Phys. Rev. D \textbf{102} (2020), 114030.

\bibitem{Wan:2020fsk}
B.~D.~Wan and C.~F.~Qiao,
Gluonic tetracharm configuration of $X (6900)$,
Phys. Lett. B \textbf{817} (2021), 136339.
\bibitem{Gong:2020bmg}
C.~Gong, M.~C.~Du, Q.~Zhao, X.~H.~Zhong and B.~Zhou,
Nature of X(6900) and its production mechanism at LHCb,
Phys. Lett. B \textbf{824} (2022), 136794.
\bibitem{Zhu:2020snb}
J.~W.~Zhu, X.~D.~Guo, R.~Y.~Zhang, W.~G.~Ma and X.~Q.~Li,
A possible interpretation for $X(6900)$ observed in four-muon final state by LHCb -- A light Higgs-like boson?,
[arXiv:2011.07799 [hep-ph]].










\bibitem{Wang:2020wrp}
J.~Z.~Wang, D.~Y.~Chen, X.~Liu and T.~Matsuki,
Producing fully charm structures in the $J/\psi$ -pair invariant mass spectrum,
Phys. Rev. D \textbf{103} (2021) no.7, 071503.

\bibitem{Wang:2022jmb}
J.~Z.~Wang and X.~Liu,
Improved understanding of the peaking phenomenon existing in the new di-J/\ensuremath{\psi} invariant mass spectrum from the CMS Collaboration,
Phys. Rev. D \textbf{106} (2022) no.5, 054015.

\bibitem{Santowsky:2021bhy}
N.~Santowsky and C.~S.~Fischer,
Four-quark states with charm quarks in a two-body Bethe\textendash{}Salpeter approach,
Eur. Phys. J. C \textbf{82} (2022) no.4, 313.

\bibitem{Zhuang:2021pci}
Z.~Zhuang, Y.~Zhang, Y.~Ma and Q.~Wang,
Lineshape of the compact fully heavy tetraquark,
Phys. Rev. D \textbf{105} (2022) no.5, 054026.


\bibitem{Lotstedt:2023crt}
E.~L\"otstedt, L.~Wang, R.~Yoshida, Y.~Zhang and K.~Yamanouchi,
Error-mitigated quantum computing of Heisenberg spin chain dynamics,
Phys. Scripta \textbf{98} (2023) no.3, 035111.


\bibitem{Debastiani:2017msn}
V.~R.~Debastiani and F.~S.~Navarra,
A non-relativistic model for the $[cc][\bar{c}\bar{c}]$ tetraquark,
Chin. Phys. C \textbf{43} (2019) no.1, 013105.
\bibitem{Weng:2020jao}
X.~Z.~Weng, X.~L.~Chen, W.~Z.~Deng and S.~L.~Zhu,
Systematics of fully heavy tetraquarks,
Phys. Rev. D \textbf{103} (2021) no.3, 034001.
\bibitem{Wang:2019rdo}
G.~J.~Wang, L.~Meng and S.~L.~Zhu,
Spectrum of the fully-heavy tetraquark state $QQ\bar Q' \bar Q'$,
Phys. Rev. D \textbf{100} (2019) no.9, 096013.
\bibitem{Hughes:2017xie}
C.~Hughes, E.~Eichten and C.~T.~H.~Davies,
Searching for beauty-fully bound tetraquarks using lattice nonrelativistic QCD,
Phys. Rev. D \textbf{97} (2018) no.5, 054505.
\bibitem{Richard:2018yrm}
J.~M.~Richard, A.~Valcarce and J.~Vijande,
Few-body quark dynamics for doubly heavy baryons and tetraquarks,
Phys. Rev. C \textbf{97} (2018) no.3, 035211.
\bibitem{Jin:2020jfc}
X.~Jin, Y.~Xue, H.~Huang and J.~Ping,
Full-heavy tetraquarks in constituent quark models,
Eur. Phys. J. C \textbf{80}, no.11, 1083 (2020).


\bibitem{An:2022fvs}
H.~T.~An, S.~Q.~Luo, Z.~W.~Liu and X.~Liu,
Fully heavy pentaquark states in constituent quark model,
Phys. Rev. D \textbf{105} (2022) no.7, 7.


\bibitem{Park:2015nha}
W.~Park, A.~Park and S.~H.~Lee,
Dibaryons in a constituent quark model,
Phys. Rev. D \textbf{92} (2015) no.1, 014037.

\bibitem{Park:2016cmg}
W.~Park, A.~Park and S.~H.~Lee,
Dibaryons with two strange quarks and total spin zero in a constituent quark model,
Phys. Rev. D \textbf{93} (2016) no.7, 074007.

\bibitem{Park:2016mez}
A.~Park, W.~Park and S.~H.~Lee,
Dibaryons with two strange quarks and one heavy flavor in a constituent quark model,
Phys. Rev. D \textbf{94} (2016) no.5, 054027.

\bibitem{Park:2017jbn}
W.~Park, A.~Park, S.~Cho and S.~H.~Lee,
$P_c(4380)$ in a constituent quark model,
Phys. Rev. D \textbf{95} (2017) no.5, 054027.

\bibitem{Park:2018oib}
W.~Park, S.~Cho and S.~H.~Lee,
Where is the stable pentaquark?,
Phys. Rev. D \textbf{99} (2019) no.9, 094023.



\bibitem{Zhou:2022gra}
H.~Y.~Zhou, F.~L.~Wang, Z.~W.~Liu and X.~Liu,
Probing the electromagnetic properties of the \ensuremath{\Sigma}c(*)D(*)-type doubly charmed molecular pentaquarks,
Phys. Rev. D \textbf{106} (2022) no.3, 034034.

\bibitem{Kumar:2005ei}
S.~Kumar, R.~Dhir and R.~C.~Verma,
Magnetic moments of charm baryons using effective mass and screened charge of quarks,
J. Phys. G \textbf{31} (2005) no.2, 141-147

\bibitem{Wu:2016vtq}
J.~Wu, Y.~R.~Liu, K.~Chen, X.~Liu and S.~L.~Zhu,
Heavy-flavored tetraquark states with the $QQ\bar{Q}\bar{Q}$ configuration,
Phys. Rev. D \textbf{97} (2018) no.9, 094015.




\bibitem{Zhang:2021yul}
W.~X.~Zhang, H.~Xu and D.~Jia,
Masses and magnetic moments of hadrons with one and two open heavy quarks: Heavy baryons and tetraquarks,
Phys. Rev. D \textbf{104} (2021) no.11, 114011.

\bibitem{Wang:2016dzu}
G.~J.~Wang, R.~Chen, L.~Ma, X.~Liu and S.~L.~Zhu,
Magnetic moments of the hidden-charm pentaquark states,
Phys. Rev. D \textbf{94} (2016) no.9, 094018.

\bibitem{Wang:2022tib}
F.~L.~Wang, H.~Y.~Zhou, Z.~W.~Liu and X.~Liu,
What can we learn from the electromagnetic properties of hidden-charm molecular pentaquarks with single strangeness?,
Phys. Rev. D \textbf{106} (2022) no.5, 054020.

\bibitem{Wang:2022ugk}
F.~L.~Wang, H.~Y.~Zhou, Z.~W.~Liu and X.~Liu,
Exploring the electromagnetic properties of the $\Xi_c^{(\prime,\,*)} \bar D_s^*$ and $\Omega_c^{(*)} \bar D_s^*$ molecular states,
[arXiv:2210.02809 [hep-ph]].




\bibitem{Li:1994cy}
Z.~P.~Li,
The Threshold pion photoproduction of nucleons in the chiral quark model,
Phys. Rev. D \textbf{50} (1994), 5639-5646.

\bibitem{Deng:2016ktl}
W.~J.~Deng, H.~Liu, L.~C.~Gui and X.~H.~Zhong,
Spectrum and electromagnetic transitions of bottomonium,
Phys. Rev. D \textbf{95} (2017) no.7, 074002.

\bibitem{Deng:2016stx}
W.~J.~Deng, H.~Liu, L.~C.~Gui and X.~H.~Zhong,
Charmonium spectrum and their electromagnetic transitions with higher multipole contributions,
Phys. Rev. D \textbf{95} (2017) no.3, 034026.


\bibitem{Xiao:2017udy}
L.~Y.~Xiao, K.~L.~Wang, Q.~f.~Lu, X.~H.~Zhong and S.~L.~Zhu,
Strong and radiative decays of the doubly charmed baryons,
Phys. Rev. D \textbf{96}, no.9, 094005 (2017).

\bibitem{Lu:2017meb}
Q.~F.~L\"u, K.~L.~Wang, L.~Y.~Xiao and X.~H.~Zhong,
Mass spectra and radiative transitions of doubly heavy baryons in a relativized quark model,
Phys. Rev. D \textbf{96}, no.11, 114006 (2017).

\bibitem{Wang:2017kfr}
K.~L.~Wang, Y.~X.~Yao, X.~H.~Zhong and Q.~Zhao,
Strong and radiative decays of the low-lying $S$- and $P$-wave singly heavy baryons,
Phys. Rev. D \textbf{96}, no.11, 116016 (2017).

\bibitem{Yao:2018jmc}
Y.~X.~Yao, K.~L.~Wang and X.~H.~Zhong,
Strong and radiative decays of the low-lying $D$-wave singly heavy baryons,
Phys. Rev. D \textbf{98}, no.7, 076015 (2018).


\bibitem{Li:1997gd}
Z.~p.~Li, H.~x.~Ye and M.~h.~Lu,
An Unified approach to pseudoscalar meson photoproductions off nucleons in the quark model,
Phys. Rev. C \textbf{56} (1997), 1099-1113.

\bibitem{Brodsky:1968ea}
S.~J.~Brodsky and J.~R.~Primack,
The Electromagnetic Interactions of Composite Systems,
Annals Phys. \textbf{52} (1969), 315-365.

\bibitem{Zhao:2002id}
Q.~Zhao, J.~S.~Al-Khalili, Z.~P.~Li and R.~L.~Workman,
Pion photoproduction on the nucleon in the quark model,
Phys. Rev. C \textbf{65} (2002), 065204.






\bibitem{An:2020vku}
H.~T.~An, K.~Chen and X.~Liu,
Manifestly exotic pentaquarks with a single heavy quark,
Phys. Rev. D \textbf{105} (2022) no.3, 034018.

\bibitem{Guo:2021mja}
T.~Guo, J.~Li, J.~Zhao and L.~He,
Mass spectra and decays of open-heavy tetraquark states,
Phys. Rev. D \textbf{105} (2022) no.5, 054018.


\bibitem{Weng:2021hje}
X.~Z.~Weng, W.~Z.~Deng and S.~L.~Zhu,
Doubly heavy tetraquarks in an extended chromomagnetic model,
Chin. Phys. C \textbf{46} (2022) no.1, 013102.

\bibitem{Deng:2020iqw}
C.~Deng, H.~Chen and J.~Ping,
Towards the understanding of fully-heavy tetraquark states from various models,
Phys. Rev. D \textbf{103} (2021) no.1, 014001.

\bibitem{Liu:2019zuc}
M.~S.~Liu, Q.~F.~L\"u, X.~H.~Zhong and Q.~Zhao,
All-heavy tetraquarks,
Phys. Rev. D \textbf{100} (2019) no.1, 016006.

\bibitem{Zhang:2022qtp}
J.~Zhang, J.~B.~Wang, G.~Li, C.~S.~An, C.~R.~Deng and J.~J.~Xie,
Spectrum of the S-wave fully-heavy tetraquark states,
[arXiv:2209.13856 [hep-ph]].


\bibitem{Hiyama:2003cu}
E.~Hiyama, Y.~Kino and M.~Kamimura,
Gaussian expansion method for few-body systems,
Prog. Part. Nucl. Phys. \textbf{51} (2003), 223-307.


\bibitem{Wang:2021mma}
Q.~N.~Wang, Z.~Y.~Yang and W.~Chen,
Exotic fully-heavy $Q\bar QQ\bar Q$ tetraquark states in $\mathbf{8}_{[Q\bar{Q}]}\otimes \mathbf{8}_{[Q\bar{Q}]}$ color configuration,
Phys. Rev. D \textbf{104} (2021) no.11, 114037.



\bibitem{Gordillo:2020sgc}
M.~C.~Gordillo, F.~De Soto and J.~Segovia,
Diffusion Monte Carlo calculations of fully-heavy multiquark bound states,
Phys. Rev. D \textbf{102} (2020) no.11, 114007.



\bibitem{Mutuk:2021hmi}
H.~Mutuk,
Nonrelativistic treatment of fully-heavy tetraquarks as diquark-antidiquark states,
Eur. Phys. J. C \textbf{81} (2021) no.4, 367.

\bibitem{Chen:2019vrj}
X.~Chen,
Fully-heavy tetraquarks: $bb\bar{c}\bar{c}$ and $bc\bar{b}\bar{c}$,
Phys. Rev. D \textbf{100} (2019) no.9, 094009.



\bibitem{Berezhnoy:2011xn}
A.~V.~Berezhnoy, A.~V.~Luchinsky and A.~A.~Novoselov,
Tetraquarks Composed of 4 Heavy Quarks,
Phys. Rev. D \textbf{86} (2012), 034004.

\bibitem{Li:2021ygk}
Q.~Li, C.~H.~Chang, G.~L.~Wang and T.~Wang,
Mass spectra and wave functions of $T_{QQ\bar{Q}\bar{Q}}$ tetraquarks,
Phys. Rev. D \textbf{104} (2021) no.1, 014018.

\bibitem{Mutuk:2022nkw}
H.~Mutuk,
Spectrum of $cc\bar{b}\bar{b}$, $bc\bar{c}\bar{c}$, and $bc\bar{b}\bar{b}$ tetraquark states in the dynamical diquark model,
Phys. Lett. B \textbf{834} (2022), 137404.


\bibitem{Wang:2017jtz}
Z.~G.~Wang,
Analysis of the $QQ\bar{Q}\bar{Q}$ tetraquark states with QCD sum rules,
Eur. Phys. J. C \textbf{77} (2017) no.7, 432.

\bibitem{Wang:2018poa}
Z.~G.~Wang and Z.~Y.~Di,
Analysis of the vector and axialvector $QQ\bar{Q}\bar{Q}$ tetraquark states with QCD sum rules,
Acta Phys. Polon. B \textbf{50} (2019), 1335.


\bibitem{Chen:2020lgj}
X.~Chen,
Fully-charm tetraquarks: $cc\bar{c}\bar{c}$,
[arXiv:2001.06755 [hep-ph]].























\end{thebibliography}
\end{document}